\pdfoutput=1

\documentclass[11pt,twoside,a4paper,cmspaper,final,collab]{cms-tdr}
\def\svnVersion{21ecadc-D}\def\svnDate{2020/03/30}\def\cmsCernNoTag{CERN-EP-2019-028}\def\cmsCernDate{\today}\def\cmsMessage{"Published in the European Physical Journal C as \href{http://dx.doi.org/10.1140/epjc/s10052-020-7917-7}{\doi{10.1140/epjc/s10052-020-7917-7}.}"}

\begin{document}\cmsNoteHeader{TOP-18-004}

\hyphenation{had-ron-i-za-tion}
\hyphenation{cal-or-i-me-ter}
\hyphenation{de-vices}

\RCS$Revision$
\RCS$HeadURL$
\RCS$Id$

\newcommand{\Wjets}{\PW+jets\xspace}
\newcommand{\Zjets}{\cPZ+jets\xspace}
\newcommand{\ee}{\EE}
\newcommand{\pp}{\ensuremath{\Pp\Pp}\xspace}
\newcommand{\mumu}{\ensuremath{\PGmp\PGmm}\xspace}
\newcommand{\emu}{\ensuremath{\Pepm\PGmmp}\xspace}
\newcommand{\MadSpin} {{\textsc{MadSpin}}}
\newcommand{\MGaMCatNLO}{\textsc{mg}5\_\text{a}\textsc{mc@nlo}\xspace}
\newcommand{\mg}{\MGaMCatNLO\xspace}
\newcommand{\xsectheo}{\ensuremath{832\;\substack{+20\\-29}{\text{(scale)}} \pm 35({\mathrm{PDF}}}+\ensuremath{\alpS)\unit{pb}}\xspace}
\newcommand{\mb}{\unit{mb}}
\newcommand{\tW}{\ensuremath{\cPqt\PW}\xspace}

\newcommand{\mtt} {\ensuremath{M(\ttbar)}\xspace}
\newcommand{\ptt} {\ensuremath{\pt(\PQt)}\xspace}
\newcommand{\yt} {\ensuremath{y(\PQt)}\xspace}
\newcommand{\ytt} {\ensuremath{y(\ttbar)}\xspace}
\newcommand{\detatt} {\ensuremath{\Delta\eta(\PQt,\PAQt)}\xspace}
\newcommand{\dphitt} {\ensuremath{\Delta \phi(\PQt,\PAQt)}\xspace}
\newcommand{\pttt} {\ensuremath{\pt(\ttbar)}\xspace}
\newcommand{\nj} {\ensuremath{N_{\text{jet}}}\xspace}
\newcommand{\ytptt} {\ensuremath{[\yt, \ptt]}\xspace}
\newcommand{\mttyt} {\ensuremath{[\mtt, \yt]}\xspace}
\newcommand{\mttytt} {\ensuremath{[\mtt, \ytt]}\xspace}
\newcommand{\mttdetatt} {\ensuremath{[\mtt, \detatt]}\xspace}
\newcommand{\mttdphitt} {\ensuremath{[\mtt, \dphitt]}\xspace}
\newcommand{\mttpttt} {\ensuremath{[\mtt, \pttt]}\xspace}
\newcommand{\mttptt} {\ensuremath{[\mtt, \ptt]}\xspace}
\newcommand{\njmttytttwo} {\ensuremath{[N^{0,1+}_{\text{jet}}, \mtt, \ytt]}\xspace}
\newcommand{\njmttyttthree} {\ensuremath{[N^{0,1,2+}_{\text{jet}}, \mtt, \ytt]}\xspace}
\newcommand{\ptttmttytt} {\ensuremath{[\pttt, \mtt, \ytt]}\xspace}
\newcommand{\PowPyt} {{\POWHEG\ + \PYTHIA}\xspace}
\newcommand{\PowHer} {{\POWHEG\ + \HERWIGpp}\xspace}
\newcommand{\aMCPyt} {{\MGaMCatNLO\ + \PYTHIA}\xspace}
\newcommand{\PowPytSh} {`POW+PYT'\xspace}
\newcommand{\PowHerSh} {`POW+HER'\xspace}
\newcommand{\aMCPytSh} {`MG5+PYT'\xspace}
\newcommand{\as} {\alpS}
\newcommand{\asmz} {\ensuremath{\as(m_{\cPZ})}\xspace}
\newcommand{\mt} {\ensuremath{m_{\PQt}^{\text{pole}}}\xspace}
\newcommand{\mtmc} {\ensuremath{m_{\PQt}^{\text{MC}}}\xspace}
\newcommand{\qmin} {\ensuremath{Q^2_\text{min}}\xspace}
\newcommand{\cj} {CJ15\xspace}
\newcommand{\ct} {CT14\xspace}
\newcommand{\herapdf} {HERAPDF2.0\xspace}
\newcommand{\jr} {JR14\xspace}
\newcommand{\mmht} {MMHT2014\xspace}
\newcommand{\chisq}{\ensuremath{\chi^2}\xspace}
\newcommand{\ndf}{dof\xspace}
\newcommand{\chisqndf}{\ensuremath{\chi^2}/\ndf\xspace}
\newcommand{\xfitter} {\textsc{xFitter}\xspace}
\newcommand{\applgrid} {\textsc{ApplGrid}\xspace}
\newcommand{\qcdnum} {\textsc{qcdnum}\xspace}
\newcommand{\amcfast} {{\textsc{aMCfast}}\xspace}
\newcommand{\lhapdf} {{\textsc{lhapdf}}\xspace}
\newlength\cmsTabSkip\setlength\cmsTabSkip{3pt}

\newcommand{\newref}[1] {{\color{blue} #1}}

\newcommand{\cmsTable}[1]{\resizebox{\textwidth}{!}{#1}}
\newcommand{\cmsTallTable}[1]{\resizebox*{!}{\textheight}{#1}}
\ifthenelse{\boolean{cms@external}}{\newcommand{\cmsLongTable}[1]{\hbox{\hss\resizebox{660pt}{!}{#1}}}}{\newcommand{\cmsLongTable}[1]{\resizebox{660pt}{!}{#1}}}
\ifthenelse{\boolean{cms@external}}{\newcommand{\cmsSwitchTable}[1]{#1}}{\newcommand{\cmsSwitchTable}[1]{\resizebox{\textwidth}{!}{#1}}}

\cmsNoteHeader{TOP-18-004} \title{Measurement of \ttbar normalised multi-differential cross sections in $\pp$ collisions at $\sqrt{s}=13\TeV$, and simultaneous determination of the strong coupling strength, top quark pole mass, and parton distribution functions}
\titlerunning{Top quark pair production at 13\TeV}

\date{\today}

\abstract{Normalised multi-differential cross sections for top quark pair (\ttbar) production are measured in
proton-proton collisions at a centre-of-mass energy of 13\TeV using events containing two oppositely charged leptons. The analysed data were recorded with the CMS detector in 2016 and correspond to an integrated luminosity of 35.9\fbinv. The double-differential \ttbar cross section is measured as a function of the kinematic properties of the top quark and of the \ttbar system at parton level in the full phase space. A triple-differential measurement is performed as a function of the invariant mass and rapidity of the \ttbar system and the multiplicity of additional jets at particle level. The data are compared to predictions of Monte Carlo event generators that complement next-to-leading-order (NLO) quantum chromodynamics (QCD) calculations with parton showers. Together with a fixed-order NLO QCD calculation, the triple-differential measurement is used to extract values of the strong coupling strength \alpS and the top quark pole mass (\mt) using several sets of parton distribution functions (PDFs). The measurement of \mt exploits the sensitivity of the \ttbar invariant mass distribution to \mt near the production threshold. Furthermore, a simultaneous fit of the PDFs, \alpS, and \mt is performed at NLO, demonstrating that the new data have significant impact on the gluon PDF, and at the same time allow an accurate determination of \alpS and \mt.
The values $\asmz = 0.1135{}^{+0.0021}_{-0.0017}$ and $\mt = 170.5 \pm 0.8 \GeV$ are extracted, which account for experimental and theoretical uncertainties, the latter being estimated from NLO scale variations. Possible effects from Coulomb and soft-gluon resummation near the \ttbar production threshold are neglected
in these parameter extractions. A rough estimate of these 
effects indicates an expected correction of \mt of the order of $+1 \GeV$,
which can be regarded as additional theoretical uncertainty in the current \mt extraction.
}

\hypersetup{pdfauthor={CMS Collaboration},pdftitle={Measurements of normalised multi-differential cross sections for top quark pair production in pp collisions at sqrt(s)=13 TeV, and simultaneous determination of the strong coupling strength, top quark pole mass, and parton distribution functions},pdfsubject={CMS},pdfkeywords={CMS, physics, top, dilepton, unfolding, QCD, PDF}}

\maketitle \section{Introduction}

Measurements of top quark pair (\ttbar) production are important for
checking the validity of the standard model (SM) and searching for new phenomena.
In particular, the large data set
delivered by the CERN LHC allows precise measurements of the \ttbar production cross section as a function of \ttbar kinematic observables.
These can be used to check the most recent predictions of perturbative quantum chromodynamics (QCD)
and to constrain input parameters, some of which are fundamental to the SM.
At the LHC, top quarks are predominantly produced via gluon-gluon fusion.
Using measurements of the production cross section in a global fit of
parton distribution functions (PDFs)
can help determine the gluon distribution at large values of $x$~\cite{Czakon:2013tha,Guzzi:2014wia,Czakon:2016olj}, where $x$ is the fraction of the proton momentum carried by a parton. Furthermore, measurements of the cross section
as a function of the \ttbar invariant mass, from the threshold to the TeV region,
provide high sensitivity for constraining the top quark pole mass, \mt, which is defined as the pole of the top quark propagator (see e.g. Refs.~\cite{Novikov:1977dq,Tarrach:1980up,Smith:1996xz}).
At LHC energies, a large fraction of \ttbar events is produced with additional hard
jets in the final state. Events containing such additional jets constitute important backgrounds
for interesting but rare SM processes such as the associated production of a Higgs boson and \ttbar,
as well as for searches for new physics associated with \ttbar production, and must therefore be well understood.
Within the SM, processes with extra jets can also be used to constrain the strong coupling strength, \as, at the scale of the top quark mass. Furthermore, the production of \ttbar in association with extra jets provides additional sensitivity to \mt since gluon radiation depends on \mt through threshold and cone effects~\cite{Alioli:2013mxa}.

{\tolerance=1200
Differential cross sections for \ttbar production
have been measured previously in proton-antiproton collisions
at the Tevatron at a centre-of-mass energy of 1.96\TeV \cite{Aaltonen:2009iz,Abazov:2014vga}
and in proton-proton (\pp) collisions at the LHC
at $\sqrt{s} = 7$\TeV~\cite{bib:ATLAS,bib:TOP-11-013_paper,bib:ATLASnew,Aad:2015eia,Aaboud:2016iot},
8\TeV~\cite{Aaboud:2016iot,Khachatryan:2015oqa,Aad:2015mbv,Aad:2015hna,Khachatryan:2015fwh,Khachatryan:2149620,Sirunyan:2017azo,Aad:2019mkw},
and 13\TeV~\cite{Khachatryan:2016mnb,Sirunyan:2018wem,Aaboud:2016xii,Aaboud:2016syx,TOP-16-007,Sirunyan:2018ucr}.
A milestone was reached in three CMS analyses~\cite{Sirunyan:2017azo,Khachatryan:2016mnb,Sirunyan:2018wem}, where
the \ttbar production dynamics was probed with double-differential cross sections.
The first analysis~\cite{Sirunyan:2017azo} used data recorded at $\sqrt{s}=8$\TeV by the CMS experiment in 2012.
Only $\ttbar$ decays where, after the decay of each top quark
into a bottom quark and a {\PW} boson, both of the {\PW} bosons decay leptonically were considered.
Specifically, the $\Pepm\PGmmp$ decay mode ($\Pe\PGm$) was selected,
requiring two oppositely charged leptons and at least two jets.
Our present paper provides a new measurement, following the procedures of Ref.~\cite{Sirunyan:2017azo}.
It is based on data taken by the CMS experiment in 2016 at $\sqrt{s}=13$\TeV, corresponding to
an integrated luminosity of $35.9 \pm 0.9\fbinv$.
In addition to $\Pe\PGm$, the decay modes
$\Pep\Pem$  ($\Pe\Pe$) and $\PGmp\PGmm$ ($\PGm\PGm$) are also selected,
roughly doubling thereby the total number of expected \ttbar signal events.
Our latest measurement complements the analyses~\cite{Khachatryan:2016mnb,Sirunyan:2018wem},
based on data taken at $\sqrt{s}=13$\TeV, but using \ttbar\ decays
in the $\ell\text{+jets}\,$ ($\ell=\Pe,\PGm$) final state.
\par}

As in the previous work \cite{Sirunyan:2017azo},
measurements are performed of the normalised double-differential $\ttbar$ production cross section
as a function of observables describing the kinematic properties of the top quark, top antiquark,
and the \ttbar system:
the transverse momentum of the top quark, \ptt, rapidity of the top quark, \yt;
the transverse momentum, \pttt,
the rapidity, \ytt,
and invariant mass, \mtt, of the \ttbar system;
the pseudorapidity difference between the top quark and antiquark, \detatt, and
the angle between the top quark and antiquark in the transverse plane, \dphitt.
{When referring to the kinematic variables \ptt and \yt, we use only the parameters of the top quark
and not of the top antiquark, to avoid double counting of events.}
In all, the double-differential \ttbar cross section is measured as a function of six different pairs of kinematic variables.
As demonstrated in Ref.~\cite{Sirunyan:2017azo},
the different combinations of kinematic variables are sensitive to different aspects
of the QCD calculations.

For the first time at the LHC, the triple-differential cross section is measured as a function of \mtt, \ytt, and \nj, where \nj is the number of extra jets not arising from the decay of the \ttbar system.
For this purpose, the kinematic reconstruction algorithm is optimised to determine the invariant mass of the \ttbar system in an unbiased way.
As will be shown below, the triple-differential measurements provide tight constraints
on the parametrised gluon PDF, as well as on \as, and \mt.
Previous studies of additional jet activity in \ttbar events at the LHC can be found
in Refs.~\cite{Khachatryan:2015mva,Aaboud:2016omn,Sirunyan:2018wem}. The \as and \mt parameters
were also extracted from measurements of the total inclusive \ttbar production cross sections in Refs.~\cite{Abazov:2011pta,Chatrchyan:2013haa,Aad:2014kva,Abazov:2016ekt,Khachatryan:2016mqs,Sirunyan:2018goh}.

The measurements are defined at parton level and must therefore be corrected for effects
of hadronisation, and detector resolution and inefficiency.
A regularised unfolding process is performed simultaneously in bins of the two or three variables
in which the cross sections are measured.
The normalised differential \ttbar cross section is determined by dividing the
distributions by the measured total inclusive \ttbar production cross section, where
the latter is evaluated by integrating over all bins in the respective observables.

The parton-level results are compared with theoretical predictions obtained with the
generators \POWHEG~(version~2)~\cite{bib:powheg,bib:powheg3} and \MGaMCatNLO~\cite{Alwall:2014hca},
interfaced to \PYTHIA~\cite{Sjostrand:2006za,Sjostrand:2007gs} for parton showering, hadronisation, and multiple-parton interactions (MPIs).
They are also compared to theoretical predictions obtained at
next-to-leading-order (NLO) QCD using several sets of PDFs,t
after applying corrections for non-perturbative (NP) effects.

The structure of the paper is as follows:
Section~\ref{sec:cms} contains a brief description of the CMS detector.
Details of the event simulation are given in Section~\ref{sec:simulation}.
The event selection, kinematic reconstruction, and comparison between data and simulation are described in Section~\ref{sec:sel}.
The unfolding procedure is detailed in Section \ref{sec:unfold},
the method to determine the differential cross sections is presented in Section~\ref{sec:xsec},
and the assessment of the systematic uncertainties is discussed in Section~\ref{sec:systematics}.
We show the results of the measurement and their comparison to theoretical predictions in Section~\ref{sec:results}.
Section~\ref{sec:asmt} presents the extraction of \as and \mt from the measured \ttbar cross section, using several sets of PDFs,
and Section~\ref{sec:qcd} presents the simultaneous fit of the PDFs, \as, and \mt to the data.
Finally, Section~\ref{sec:concl} provides a summary.

 \section{The CMS detector}
\label{sec:cms}
The central feature of the CMS apparatus is a superconducting solenoid of 6\unit{m} internal diameter, providing a magnetic field of 3.8\unit{T}. Within the solenoid volume are a silicon pixel and strip tracker, a lead tungstate crystal electromagnetic calorimeter (ECAL), and a brass and scintillator hadron calorimeter, each composed of a barrel and two endcap sections. Forward calorimeters extend the $\eta$ coverage provided by the barrel and endcap detectors. Muons are measured in gas-ionisation detectors embedded in the steel flux-return yoke outside the solenoid. Events of interest are selected using a two-tiered trigger system~\cite{Khachatryan:2016bia}. The first level, composed of custom hardware processors, uses information from the calorimeters and muon detectors to select events at a rate of around 100\unit{kHz} within a time interval of less than 4\mus. The second level, known as the high-level trigger (HLT), consists of a farm of processors running a version of the full event reconstruction software optimised for fast processing, and reduces the event rate to around 1\unit{kHz} before data storage. A more detailed description of the CMS detector, together with a definition of the coordinate system used and the relevant kinematic variables, can be found in Ref.~\cite{bib:Chatrchyan:2008zzk}.

 \section{Event simulation}
\label{sec:simulation}
Simulations of physics processes are performed with Monte Carlo (MC) event generators and serve three
purposes:
firstly, to obtain representative
SM predictions of \ttbar\ production cross sections to be compared to the results of this analysis.
Secondly, when interfacing generated \ttbar\ signal events with a detector simulation, to determine
corrections for the effects of {hadronisation,} reconstruction and selection efficiencies, and resolutions
that are to be applied to the data.
Thirdly, when interfacing generated background processes to the detector simulation, to obtain predictions
for the backgrounds.
All MC programs used in this analysis perform the event generation in several steps:
matrix-element (ME) level, parton showering matched to ME, hadronisation, and underlying event, including
multiparton interaction (MPI).
The \ttbar\ signal processes are simulated with ME calculations at NLO in QCD.
For all simulations the proton structure is described by the NNPDF 3.0 NLO PDF set with $\asmz = 0.118$~\cite{bib:NNPDF}
where $m_{\cPZ} = 91$\GeV is the \cPZ boson mass~\cite{Patrignani:2016xqp},
and the value of the top quark mass parameter is fixed to $\mtmc=172.5\GeV$.
For the default signal simulation,
the \POWHEG\ (version~2)~\cite{Frixione:2007nw,bib:powheg,bib:powheg2} generator is taken.
The $h_\text{damp}$ parameter of \POWHEG, which regulates the damping of real emissions in the NLO calculation
when matching to the parton shower, is set to $h_\text{damp} = 1.581 \mtmc$~\cite{bib:CMS:2016kle}. The \PYTHIA program (version~8.2)~\cite{Sjostrand:2007gs} with the CUETP8M2T4 tune~\cite{bib:CMS:2016kle,bib:CUETP8tune,Skands:2014pea}
is used to model parton showering, hadronisation and MPIs.
An alternative sample is generated using the \MGaMCatNLO (version 2.2.2)~\cite{Alwall:2014hca} generator,
including up to two extra partons at the ME level at NLO.
In this setup, referred to as \MGaMCatNLO\ + \PYTHIA, \MadSpin~\cite{bib:madspin}
is used to model the decays of the top quarks while preserving their spin correlation.
The events are matched to \PYTHIA
using the FxFx prescription~\cite{Frederix:2012ps}.
A second alternative sample is generated with \POWHEG\
and interfaced with \HERWIGpp~(version~2.7.1)~\cite{bib:herwigpp} using the EE5C tune~\cite{bib:EE5Ctune}.

{\tolerance=1200
The main background contributions originate from single top quarks produced in association
with a {\PW} boson (\tW), \cPZ/$\cPgg^{*}$ bosons produced with additional jets (\Zjets), {\PW} boson production with additional jets (\Wjets)
and diboson (\PW\PW, \PW\cPZ, and \cPZ\cPZ) events.
Other backgrounds are negligible.
For all background samples, the NNPDF3.0~\cite{bib:NNPDF} PDF set is used and parton showering, hadronisation, and MPIs
are simulated with \PYTHIA.
Single top quark production is simulated with \POWHEG\ (version~1)~\cite{bib:powheg1,bib:powheg3}
using the CUETP8M2T4 tune in \PYTHIA with the $h_\text{damp}$ parameter set to 172.5\GeV in \POWHEG.
The \Zjets process is simulated at NLO using \MGaMCatNLO\ with up to two additional partons at ME level and matched to \PYTHIA using the FxFx prescription.
The \Wjets process is simulated at leading order (LO) using \MGaMCatNLO\
with up to four additional partons at ME level and matched to \PYTHIA using the MLM prescription~\cite{Alwall:2007fs}.
Diboson events are simulated with \PYTHIA.
Predictions are normalised based on their theoretical cross sections and the integrated luminosity of the data sample.
{The cross sections are calculated to approximate next-to-NLO (NNLO) for single top quark in the {\PQt}{\PW} channel~\cite{bib:twchan}, NNLO for \Zjets
and \Wjets~\cite{Li:2012wna},
and NLO for diboson production~\cite{bib:mcfm:diboson}. }
The \ttbar\ simulation is normalised to a cross section of \xsectheo\ calculated with the \textsc{Top++} (version~2.0) program~\cite{Czakon:2011xx}
at NNLO including resummation of next-to-next-to-leading-logarithm (NNLL) soft-gluon terms assuming $\mt = 172.5\GeV$ {and the proton structure described by the CT14 NNLO PDF set~\cite{Dulat:2015mca}.}
\par}

To model the effect of additional \pp interactions within the same bunch crossing (pileup), simulated minimum bias interactions are added to the simulated data.
Events in the simulation are then weighted to reproduce the pileup distribution in the data, which is estimated from the measured bunch-to-bunch instantaneous
luminosity assuming a total inelastic \pp cross section of 69.2\mb~\cite{Aaboud:2016mmw}.

In all cases, the interactions of particles with the CMS detector are simulated using \GEANTfour (version~9.4)~\cite{bib:geant}.

 \section{Event selection and \texorpdfstring{\ttbar}{ttbar} kinematic reconstruction}
\label{sec:sel}
The event selection procedure follows closely the one reported in Ref.~\cite{Sirunyan:2018ucr}.
Events are selected that correspond to the decay topology
where both top quarks decay into a {\PW} boson and a {\cPqb} quark, and each of the {\PW} bosons decays directly into an electron or a muon and a neutrino.
This defines the signal, while
all other \ttbar\ events, including those with at least one electron or muon originating from the decay of a $\PGt$ lepton
are regarded as background.
The signal comprises three distinct final state channels: the same-flavour channels corresponding to two electrons (\ee) or two muons (\mumu)
and the different-flavour channel corresponding to one electron and one muon (\emu).
Final results are derived by combining the three channels.

At HLT level, events are selected either by single-lepton or dilepton triggers.
The former require the presence of at least one electron or muon
and the latter the presence of either two electrons, two muons, or an electron and a muon.
For the single-electron and -muon triggers, \pt thresholds of 27 and 24\GeV are applied, respectively.
The same-flavour dilepton triggers require either an electron pair with $\pt > 23\GeV$ for the leading electron and
$\pt > 12 \GeV$ for the subleading electron
or a muon pair with $\pt > 17 \GeV$ for the leading muon and $\pt > 8 \GeV$ for the subleading muon.
Here leading and subleading refers to the electron or muon
with the highest and second-highest \pt, respectively, in the event.
The different-flavour dilepton triggers require either an electron with  $\pt > 23 \GeV$
and a muon with $\pt > 8 \GeV$, or a muon with $\pt > 23 \GeV$ and an electron with $\pt > 8 \GeV$.

Events are reconstructed using a particle-flow (PF) algorithm~\cite{bib:Sirunyan:2017ulk},
{which aims to identify and reconstruct each individual particle in an event with an optimised combination of information from the various elements of the CMS detector. }
Charged hadrons from pileup
are subtracted on an event-by-event basis.
Subsequently, the remaining neutral-hadron component from pileup is accounted for through jet energy corrections~\cite{bib:PUSubtraction}.

Electron candidates are reconstructed from a combination of the track momentum at the main interaction vertex,
the corresponding energy deposition in the ECAL, and the energy sum of all bremsstrahlung photons associated with the track~\cite{bib:ele2013}.
The electron candidates are required to have $\pt > 25 \GeV$ for the leading candidate and $\pt > 20 \GeV$ for the subleading candidate and $\abs{\eta} < 2.4$. Electron candidates with ECAL clusters in the region between the barrel and endcap ($1.44 <\abs{\eta_{\mathrm{cluster}}}< 1.57$) are excluded, because the reconstruction of an electron object in this region is not optimal.
A relative isolation criterion $I_{\text{rel}} < 0.06$ is applied, where $I_{\text{rel}}$ is defined as the \pt sum of all neutral hadron, charged hadron, and photon candidates
within a distance of 0.3 from the electron in $\eta$--$\phi$~space, divided by the \pt of the electron candidate.
In addition, electron identification requirements are applied to reject misidentified electron candidates and candidates originating from photon conversions.
Muon candidates are reconstructed using the track information from the tracker and the muon system.
They are required to have $\pt > 25\GeV$ for the leading candidate and $\pt > 20 \GeV$ for the subleading candidate and $\abs{\eta}<2.4$.
An isolation requirement of $I_{\text{rel}} < 0.15$ is applied to muon candidates, including particles
within a distance of 0.4 from the muon in $\eta$--$\phi$~space.
In addition, muon identification requirements are applied to reject misidentified muon candidates and candidates originating from in-flight decay processes.
For both electron and muon candidates, a correction is applied to $I_{\text{rel}}$ to suppress residual pileup effects.

Jets are reconstructed by clustering the PF candidates using the anti-\kt clustering algorithm~\cite{bib:antikt,bib:Cacciari:2011ma}
with a distance parameter $R = 0.4$. The jet energies are corrected following the procedures described in Ref.~\cite{Khachatryan:2016kdb} and applied  to the data taken by the CMS experiment in 2016. After correcting for all residual energy depositions from charged and neutral
particles from pileup, \pt- and $\eta$-dependent jet energy corrections are applied to correct for the detector response.
A jet is selected if it has $\pt > 30\GeV$ and $\abs{\eta} < 2.4$.
Jets are rejected if the distance in $\eta$--$\phi$~space between the jet and the closest lepton is less than 0.4.
Jets originating from the hadronisation of {\cPqb} quarks ({\cPqb} jets) are identified with an algorithm~\cite{Sirunyan:2017ezt}
that uses secondary vertices together with track-based lifetime information
to construct a {\cPqb} tagging discriminant.
The chosen working point has a {\cPqb} jet tagging efficiency of ${\approx}80$--90\%
and a mistagging efficiency of  ${\approx}10\%$ for jets originating from gluons, as well as \PQu, \PQd, or \PQs quarks,
and ${\approx}30$--40\% for jets originating from {\PQc} quarks.

The missing transverse momentum vector \ptvecmiss is defined as the projection
on the plane perpendicular to the beams of the negative vector sum
of the momenta of all PF candidates in an event.
Its magnitude is referred to as \ptmiss.
Jet energy corrections are propagated to improve the
determination of \ptvecmiss.

Events are selected offline if they contain exactly two isolated electrons or muons of opposite electric charge.
Furthermore, they need to contain at least two jets and at least one of these jets must be {\cPqb} tagged.
Events with an invariant mass of the lepton pair, $M(\ell\ell)$, smaller than 20\GeV are removed
in order to suppress contributions from heavy-flavour resonance decays and low-mass Drell--Yan processes.
Backgrounds from \Zjets processes in the \ee and \mumu channels are further suppressed by requiring $\abs{m_{\cPZ}-M(\ell\ell)} > 15\GeV$, and $\ptmiss > 40\GeV$.
The remaining background contribution from \tW, \Zjets, \Wjets, diboson and \ttbar\ events from decay channels other
than that of the signal are estimated from the simulation.

In this analysis, the \ttbar production cross section is also measured as a function of the extra jet multiplicity, \nj.
Extra jets (also referred to as additional jets) are jets arising primarily from hard QCD radiation and not from the top quark decays.
At generator level, the extra jets are defined in dilepton \ttbar events as jets with $\pt > 30\GeV$, $\abs{\eta} < 2.4$, built of particles except neutrinos
using the anti-\kt clustering algorithm~\cite{bib:antikt,bib:Cacciari:2011ma}
with a distance parameter $R = 0.4$, and isolated from the charged leptons (i.e.\ $\Pe$ or $\PGm$)
and {\cPqb} quarks originating from the top quark decays by a minimal distance of 0.4 in $\eta$--$\phi$~space.
The charged leptons and {\cPqb} quarks are taken directly after {\PW} and top quark decays, respectively.
At reconstruction level the extra jets are defined in dilepton \ttbar candidate events as jets with $\pt > 30\GeV$, $\abs{\eta} < 2.4$,
and isolated from the leptons and {\cPqb} jets originating from the top quark decays by the same minimal distance in $\eta$--$\phi$~space.

\begin{figure*}
    \centering
    \includegraphics[width=0.49\textwidth]{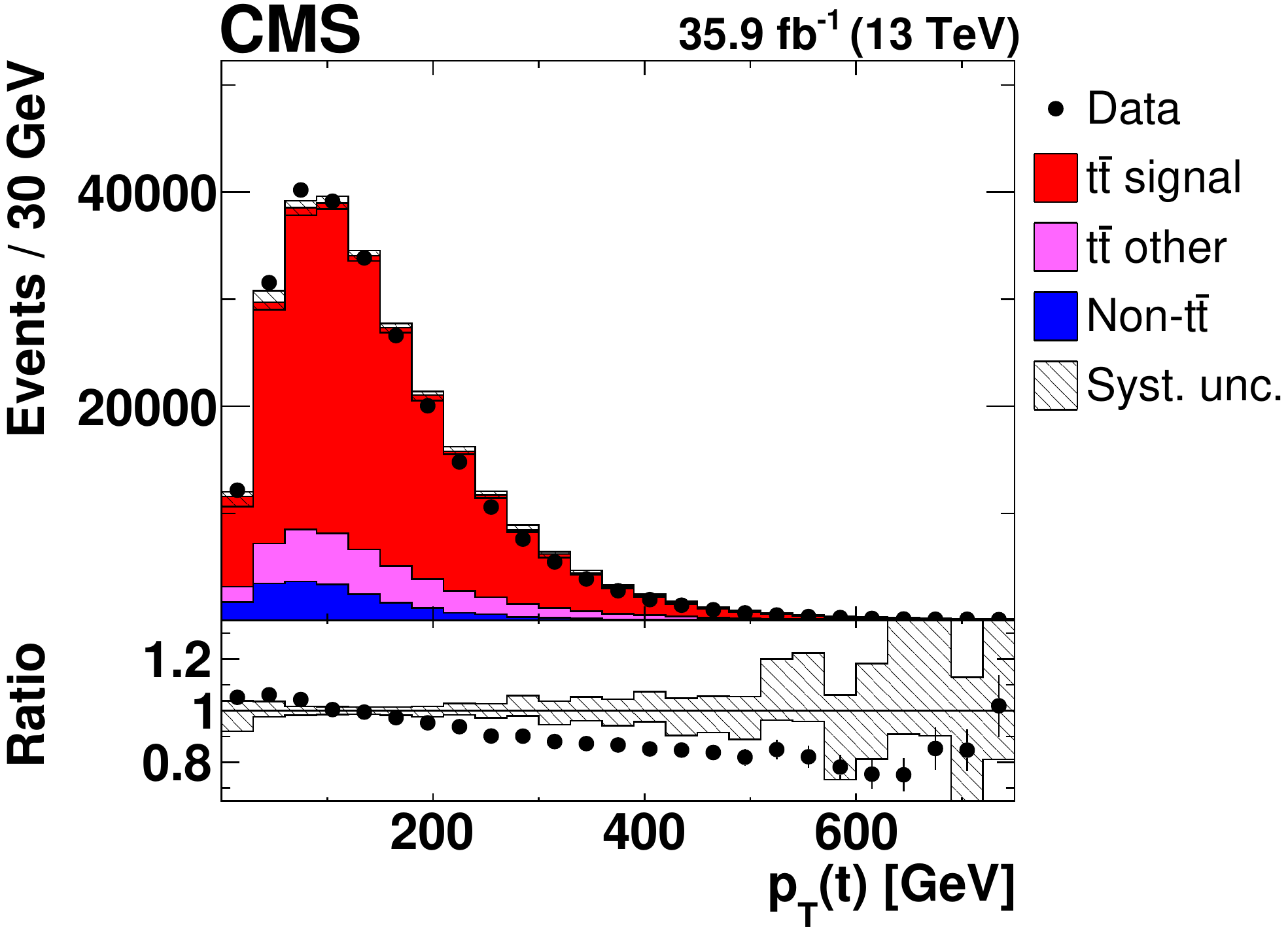}
    \includegraphics[width=0.49\textwidth]{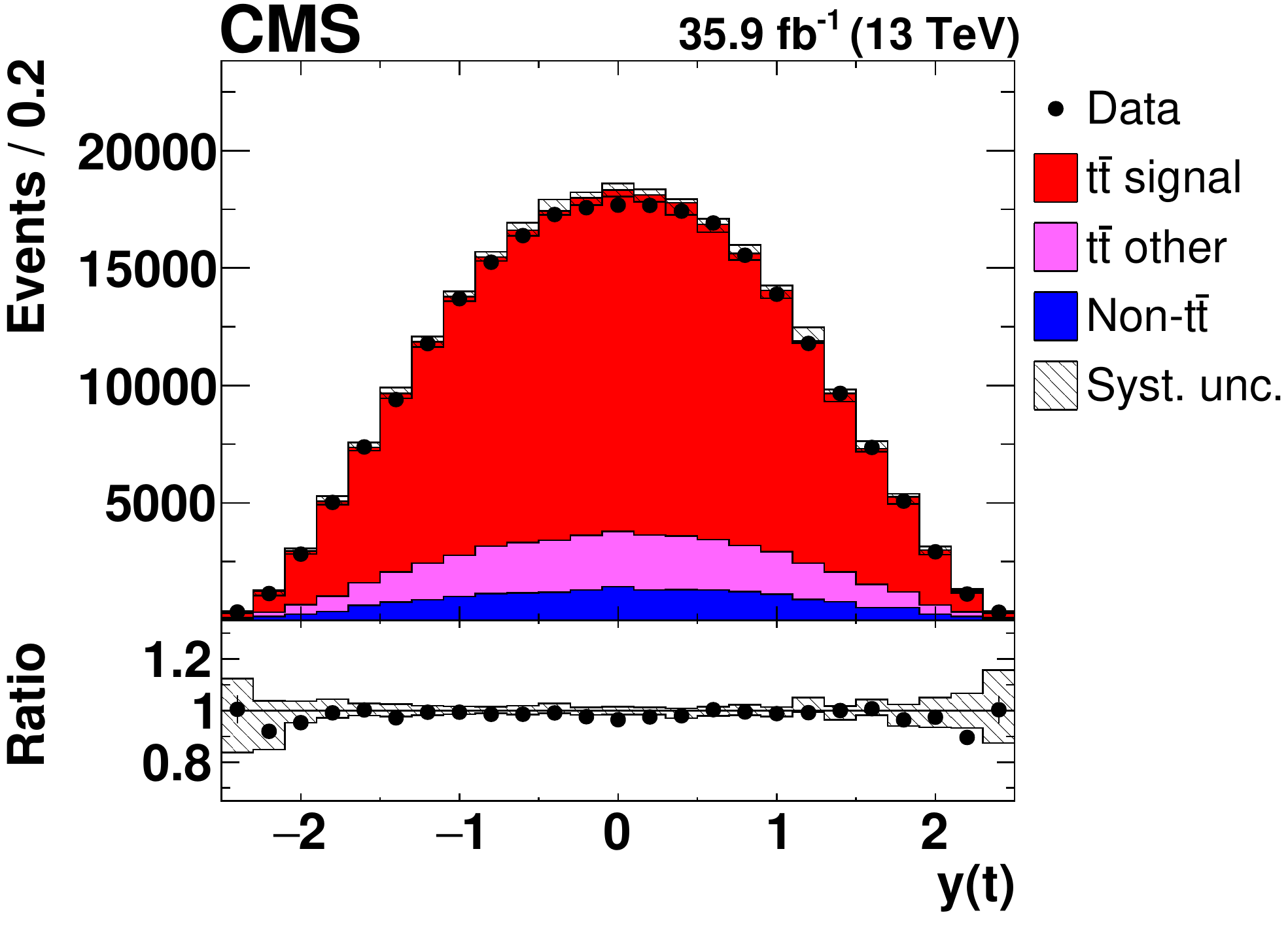}
    \includegraphics[width=0.49\textwidth]{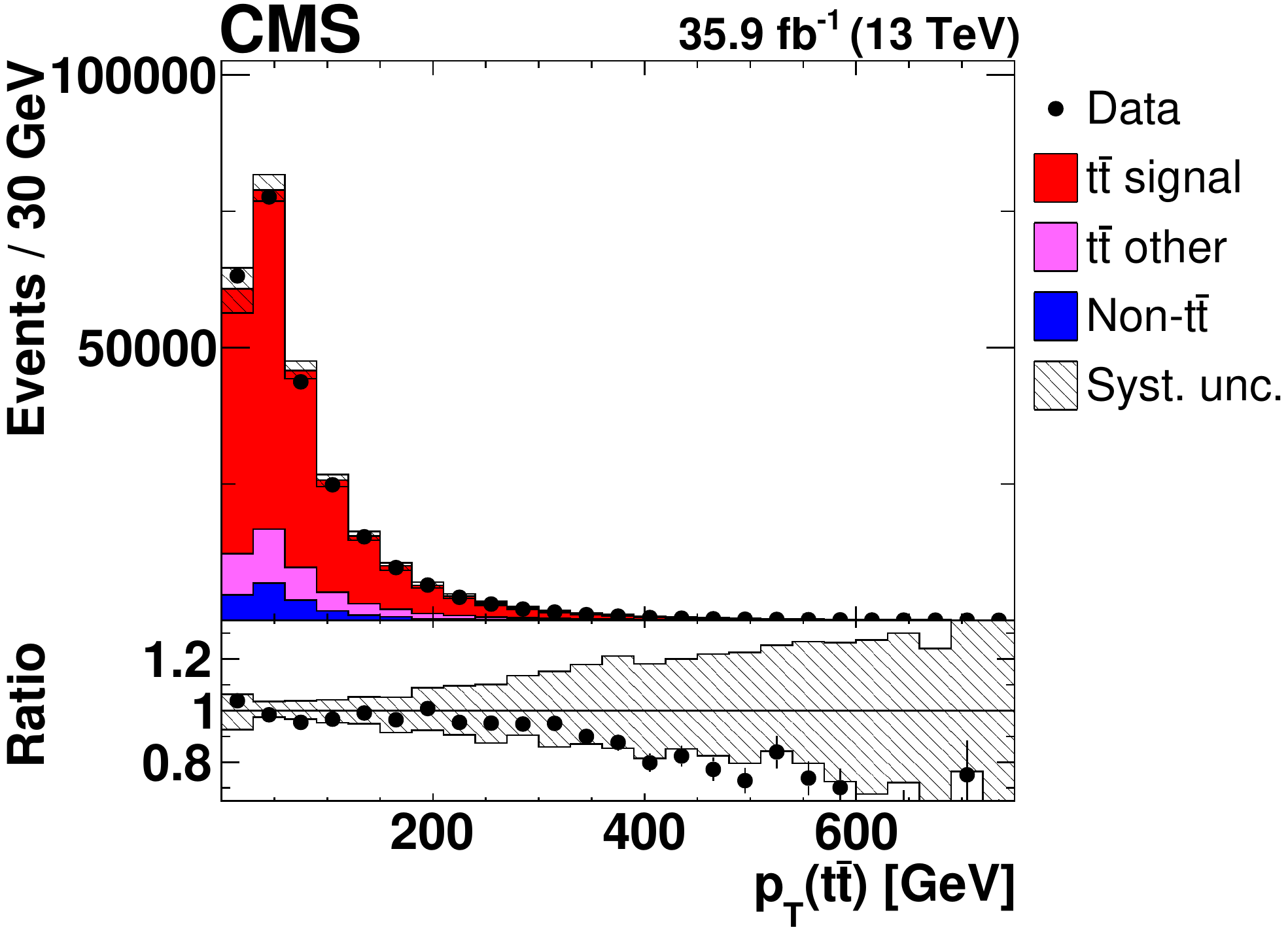}
    \includegraphics[width=0.49\textwidth]{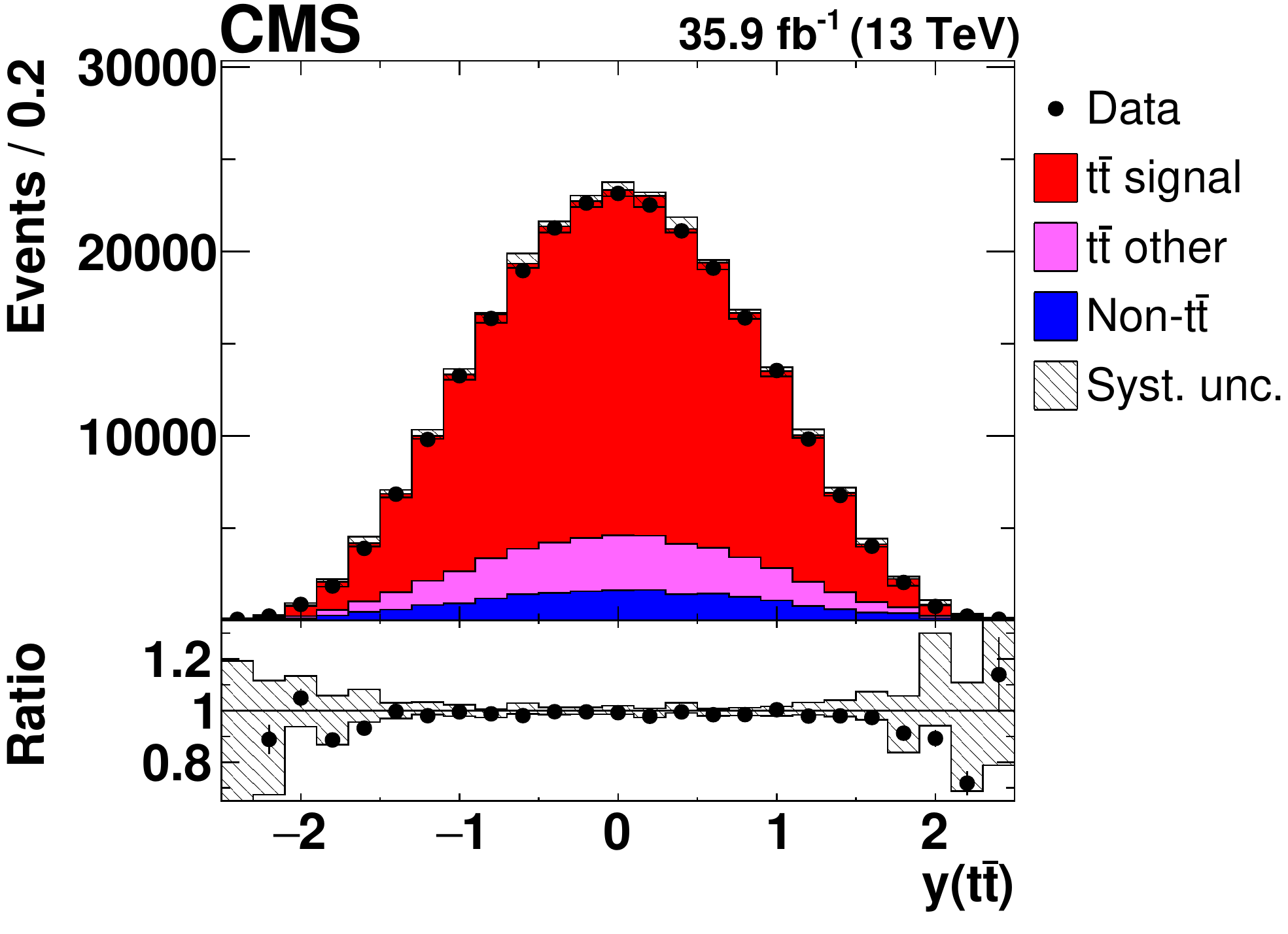}
    \includegraphics[width=0.49\textwidth]{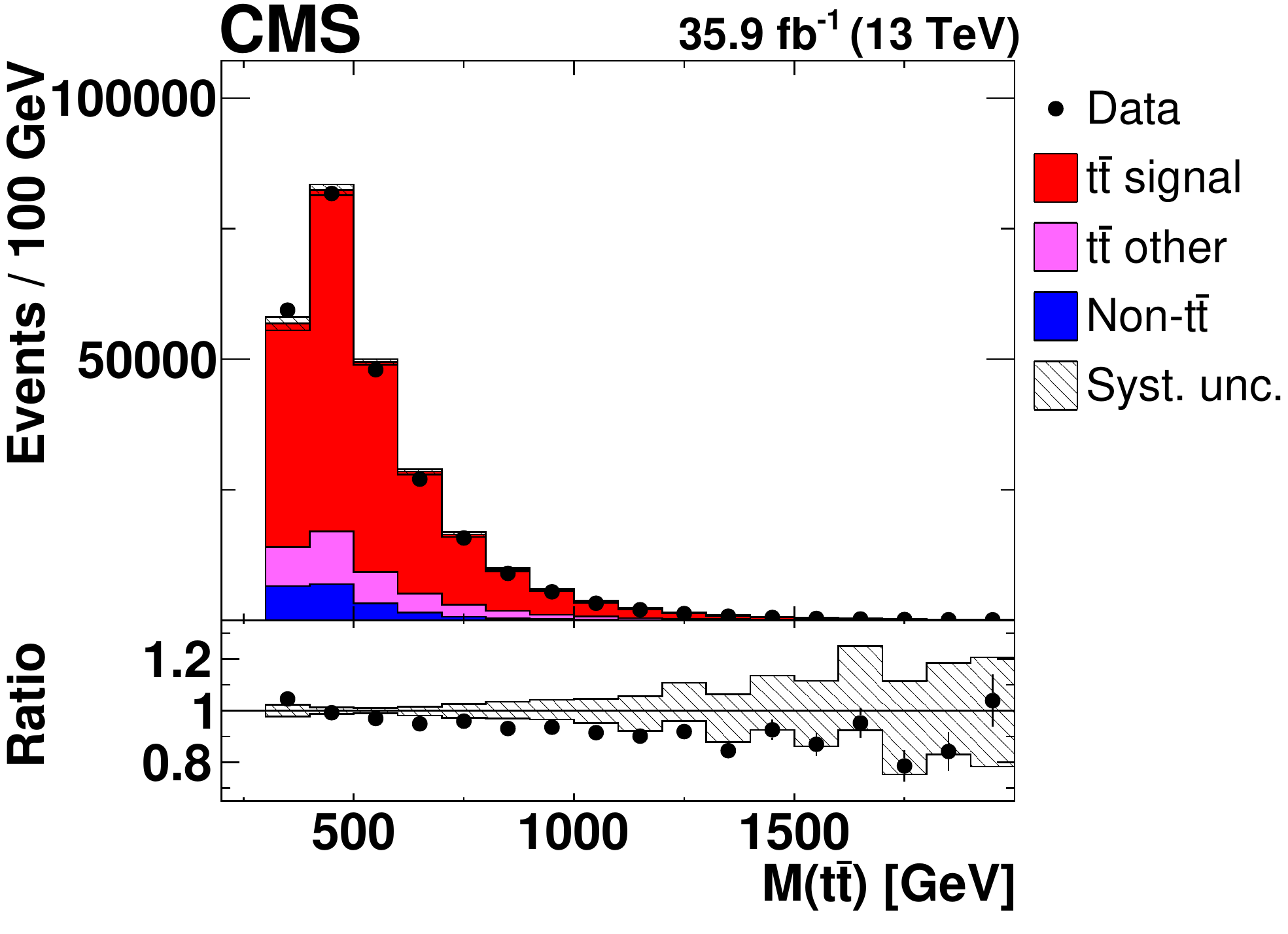}
    \includegraphics[width=0.49\textwidth]{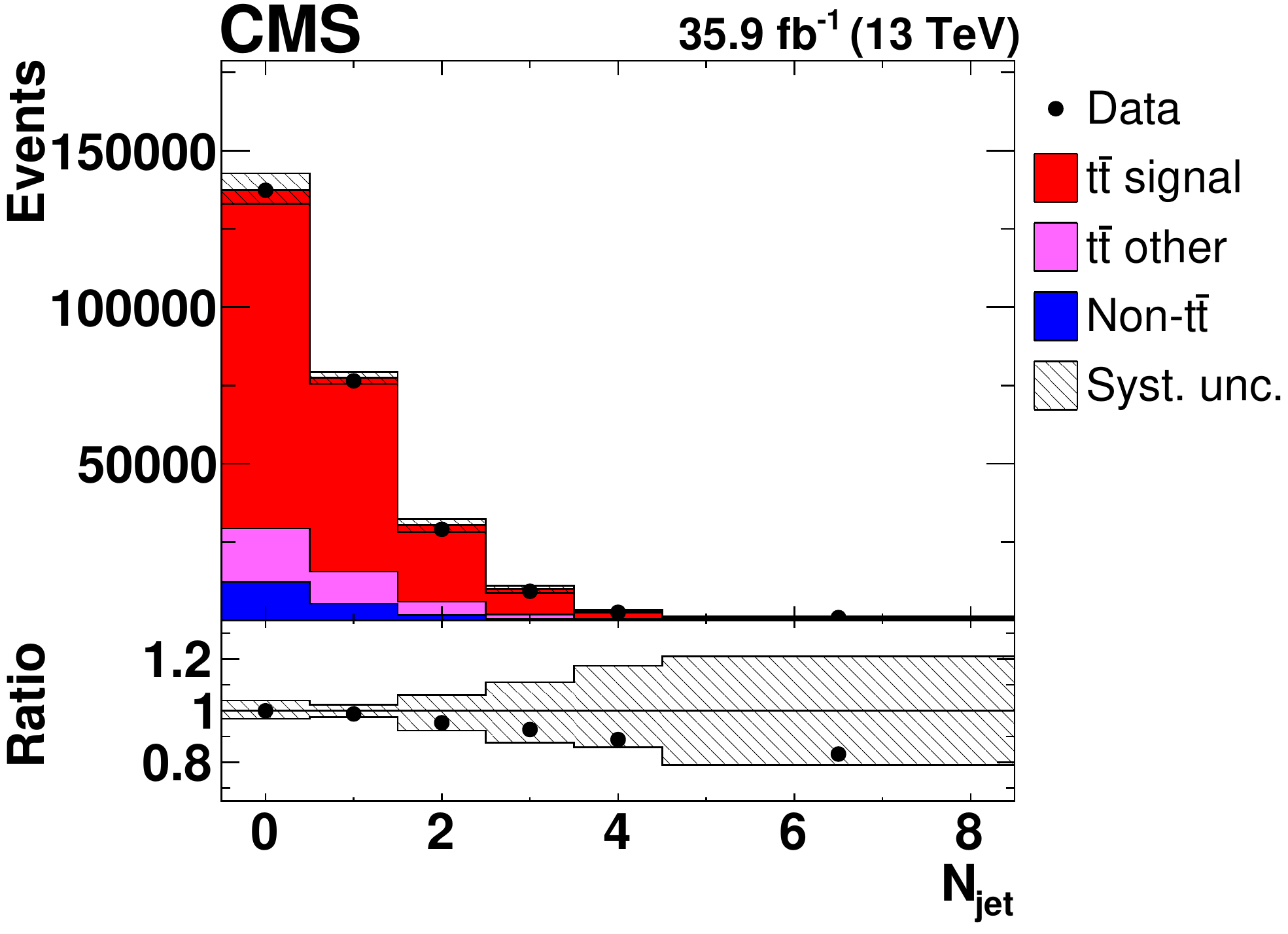}
    \caption{Distributions of $\pt(\PQt)$ (upper left), $y(\PQt)$ (upper right), $\pt(\ttbar)$ (middle left), $y(\ttbar)$ (middle right), \mtt (lower left),
    and \nj (lower right) in selected events after the kinematic reconstruction, at detector level.
    The experimental data with the vertical bars corresponding to their statistical uncertainties
    are plotted together with distributions of simulated signal and different background processes.
    The hatched regions correspond to the estimated shape uncertainties in the signal and backgrounds (as detailed in Section~\ref{sec:systematics}).
    The lower panel in each plot shows the ratio of the observed data event yields to those expected in the simulation.}
    \label{fig:cp}
\end{figure*}

The \ttbar kinematic properties are determined from the four-momenta of the decay products using a
kinematic reconstruction method~\cite{Khachatryan:2015oqa}.
The three-momenta of the neutrino ($\PGn$) and of the antineutrino (${\PAGn}$) are not directly measured, but they can be reconstructed by imposing
the following six kinematic constraints: the conservation in the event of the total transverse momentum vector,
and the masses of the {\PW} bosons, top quark, and top antiquark.
The reconstructed top quark and antiquark masses
are required to be 172.5\GeV.
The \ptvecmiss in the event is assumed to originate solely from the two neutrinos in the top quark and antiquark decay chains.
To resolve the ambiguity due to multiple algebraic solutions of the equations for the neutrino momenta,
the solution with the smallest invariant mass of the \ttbar system is taken.
The reconstruction is performed 100 times, each time randomly smearing the measured energies and directions
of the reconstructed leptons and jets within their resolution.
This smearing procedure recovers certain events that initially yield no solution because of measurement uncertainties.
The three-momenta of the two neutrinos are determined as a weighted average over all smeared solutions.
For each solution, the weight is calculated based on the expected true spectrum of the invariant mass of a lepton and a {\cPqb} jet
stemming from the decay of a top quark and taking
the product of the two weights for the top quark and antiquark decay chains.
All possible lepton-jet combinations in the event that satisfy the requirement on the invariant mass of the lepton and jet $M_{\ell {\PQb}} < 180$\GeV are considered.
Combinations are ranked based on the presence of \cPqb-tagged jets in the assignments, \ie a combination with
both leptons assigned to \cPqb-tagged jets is preferred over those with one or no \cPqb-tagged jet.
Among assignments with equal number of \cPqb-tagged jets, the one with the highest sum of weights is chosen.
Events with no solution after smearing are discarded.
The efficiency of the kinematic reconstruction, defined as the number of events where a solution
is found divided by the total number of selected \ttbar\ events, is studied in data and simulation and consistent results are observed.
The efficiency is about 90\% for signal events.
After applying the full event selection and the kinematic reconstruction of the \ttbar\ system,
150\,410 events are observed in the \emu channel,
34\,890 events in the \ee channel,
and 70\,346 events in the \mumu channel.
Combining all decay channels, the estimated signal fraction in data is 80.6\%.
Figure~\ref{fig:cp} shows the distributions of the reconstructed top quark and \ttbar kinematic variables and
of the multiplicity of additional jets in the events.
In general, the data are reasonably well described by the simulation,
however some trends are visible, in particular for $\pt(\PQt)$, where the simulation
predicts a somewhat harder spectrum than that observed in data, {as reported in previous differential \ttbar cross section measurements~\cite{Khachatryan:2015oqa,Khachatryan:2015fwh,Sirunyan:2017azo,Khachatryan:2016mnb,Sirunyan:2018wem,TOP-16-007,Sirunyan:2018ucr}.}

\begin{figure*}
    \centering
    \includegraphics[width=0.49\textwidth]{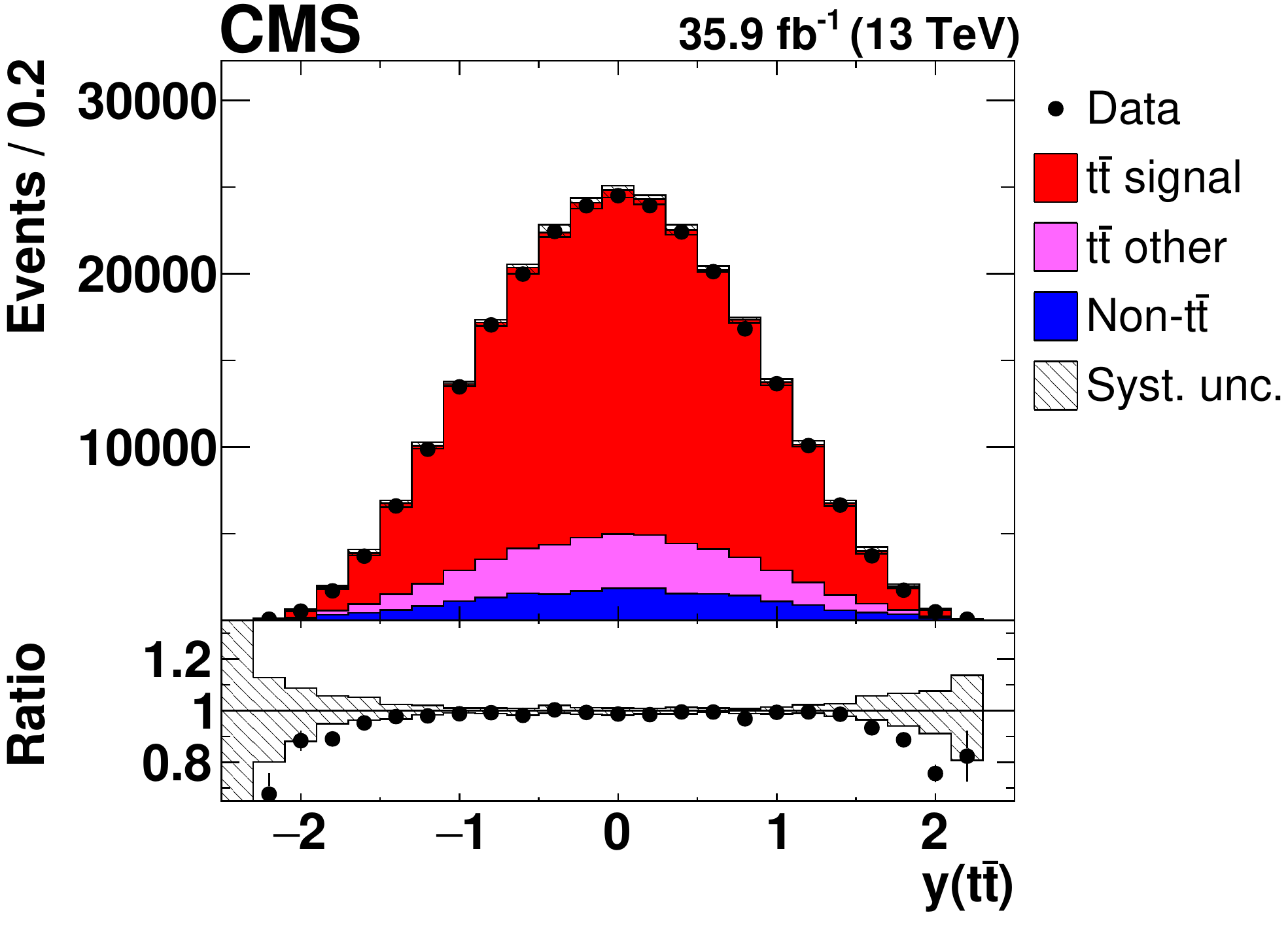}
    \includegraphics[width=0.49\textwidth]{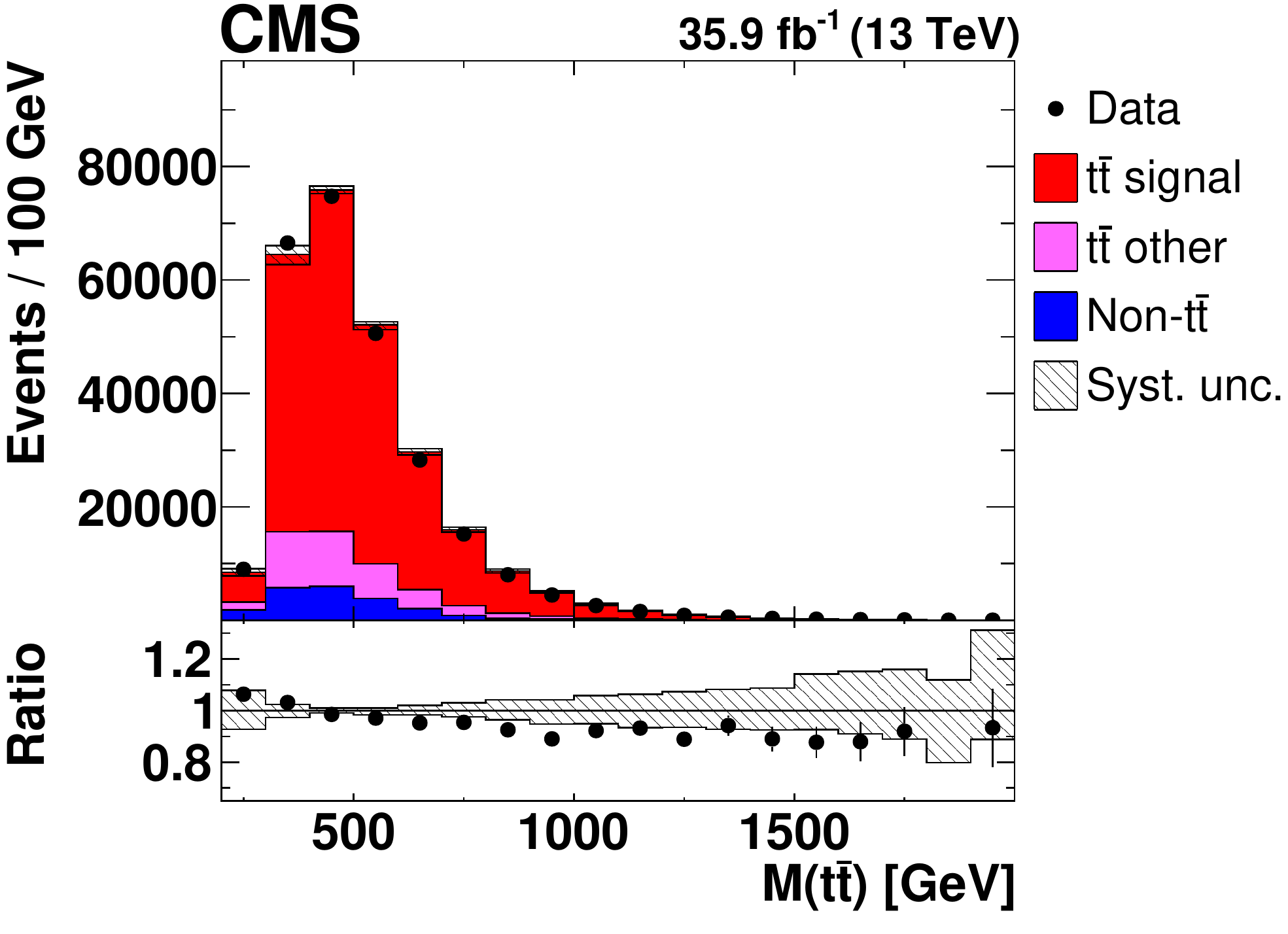}
    \caption{Distributions of $y(\ttbar)$ (left) and \mtt (right)
    in selected events after the loose kinematic reconstruction.
    Details can be found in the caption of Fig.~\ref{fig:cp}.
    }
    \label{fig:cp:loosekr}
\end{figure*}

The \mtt value obtained using the full kinematic reconstruction
described above is highly sensitive to the value of the top quark mass used as a kinematic constraint.
Since one of the objectives of this analysis is to extract the top quark mass from the differential \ttbar measurements,
exploiting the \mtt distribution in particular,
an alternative algorithm is employed, which reconstructs the \ttbar kinematic variables without using the top quark mass constraint.
This algorithm is referred to as the ``loose kinematic reconstruction''.
In this algorithm, the $\PGn\PAGn$ system is reconstructed rather than  the $\PGn$ and $\PAGn$ separately.
Consequently, it can only be used to reconstruct the total $\ttbar$ system but not the top quark and antiquark
separately.
As in the full kinematic reconstruction, all possible lepton-jet combinations in the event that satisfy the requirement
on the invariant mass of the lepton and jet $M_{\ell {\PQb}} < 180$\GeV are considered.
Combinations are ranked based on the presence of \cPqb-tagged jets in the assignments, but
among combinations with equal number of \cPqb-tagged jets, the ones with the highest-\pt jets are chosen.
The kinematic variables of the $\PGn\PAGn$ system are derived as follows:
its \ptvec is set equal to \ptvecmiss, while its
unknown longitudinal momentum and energy are set equal to
the longitudinal momentum and energy of the lepton pair.
Additional constraints are applied on the invariant mass of the neutrino pair,  $M(\PGn\PAGn) \ge 0$,
and on the invariant mass of the {\PW} bosons, $M({\PW}^{+}{\PW}^{-}) \ge 2M_{\PW}$, which have only minor effects on the performance of the reconstruction.
The method yields similar \ttbar kinematic resolutions and reconstruction efficiency as for the full kinematic reconstruction.
In this analysis, the loose kinematic reconstruction is exclusively used to measure triple-differential cross sections as a function
of \mtt, \ytt, and extra jet multiplicity, which are exploited to determine QCD parameters{, as well as the distributions used to cross-check the results.}
Figure~\ref{fig:cp:loosekr} shows the distributions of the reconstructed \ttbar invariant mass and rapidity using the loose kinematic reconstruction.
These distributions are similar to the ones obtained using the full kinematic reconstruction (as shown in Fig.~\ref{fig:cp}).
Towards forward rapidities $|y(\ttbar)|\ge 1.5$
a trend is visible in which the MC simulations predict
more events than observed in the data. However, the differences between simulations and data are still compatible
within the estimated shape uncertainties in the signal and backgrounds.

 \section{Signal extraction and unfolding}
\label{sec:unfold}
The number of signal events in data
is extracted by subtracting the expected number of background events from
the observed number of events for each bin of the observables.
All expected
background numbers are obtained directly from the MC simulations (see Section~\ref{sec:simulation})
except for $\ttbar$ final states other than the signal. The latter are dominated by events
in which one or both of the intermediate {\PW} bosons decay into $\PGt$ leptons with subsequent decay into an electron or muon.
These events arise from the same \ttbar production process as the signal and thus the normalisation of this background
is fixed to that of the signal. For each bin the number of events obtained after the subtraction of other background sources
is multiplied by the ratio of the number of selected \ttbar signal events to the total number of selected \ttbar events
(\ie the signal and all other \ttbar events) in simulation.

The numbers of signal events obtained after background subtraction are
corrected for detector effects, using the \textsc{TUnfold} package~\cite{Schmitt:2012kp}.
{The event yields in the \ee, \mumu and \emu channels are added together, and the unfolding is performed.
It is verified that the measurements in the separate channels yield consistent results.}
The response matrix plays a key role in this unfolding procedure.
An element of this matrix specifies the probability for an event originating
from one bin of the true distribution to be observed in a specific bin of the reconstructed
observables.
The response matrix includes the effects of acceptance,
detector efficiencies, and resolutions.
The response matrix is defined such that the true level corresponds
to the full phase space (with no kinematic restrictions) for \ttbar production at parton level.
At the detector level, the number of bins used
is typically a few times larger than the number of bins used at generator level.
The response matrix is taken from the signal simulation.
The generalised inverse of the response matrix is used
to obtain the distribution of unfolded event numbers from the measured distribution by applying a \chisq minimisation technique.
An additional \chisq term is included representing Tikhonov regularisation~\cite{Tikhonov:1963}.
The regularisation reduces the effect of the statistical fluctuations present in the measured distribution
on the high-frequency content of the unfolded spectrum.
The regularisation strength is chosen such that the global correlation coefficient is minimal~\cite{Schmitt:2016orm}.
For the measurements presented here, this choice results in a small contribution from the regularisation term to the total $\chi^2$, on the order of a few percent.
The unfolding of multidimensional distributions is performed by internally mapping the multi-dimensional arrays to one-dimensional arrays~\cite{Schmitt:2012kp}.

\section{Cross section determination}
\label{sec:xsec}

The normalised cross sections for \ttbar production are measured in the full \ttbar kinematic phase space at parton level.
The number of unfolded signal events $\hat{M}^\text{unf}_{i}$ in bins $i$ of kinematic variables
is used to define the normalised cross sections as a function of several (two or three) variables
\begin{equation}
\frac{\sigma_i}{\sigma} = \frac{1}{\sigma} \, \frac{\hat{M}^\text{unf}_{i}} {\mathcal{B} \, \mathcal{L}},
\label{eq:xsecdef}
\end{equation}
where the total cross section $\sigma$ is evaluated by summing $\sigma_{i}$ over all bins,
$\mathcal{B}$ is the branching ratio of \ttbar into \ee, \mumu, and \emu final states and $\mathcal{L}$ is the integrated luminosity of the data sample.
For presentation purposes, the measured cross sections are
divided by the bin width of the first variable. They present single-differential
cross sections as a function of the first variable in different ranges of the second or
second and third variables and are referred to as double- or triple-differential
cross sections, respectively.
The bin widths are chosen based on the resolutions of the kinematic variables,
such that the purity and the stability of each bin is generally above 20\%.
For a given bin, the purity is defined as the fraction of events in the \ttbar signal simulation that are generated and reconstructed in the same bin
with respect to the total number of events reconstructed in that bin.
To evaluate the stability, the number of events in the \ttbar signal simulation that are generated and reconstructed in a given bin are divided
by the total number of reconstructed events generated in the bin.

The cross section determination based on the signal extraction and unfolding described in Section~\ref{sec:unfold} has been validated with closure tests. 
Large numbers of pseudo-data sets were generated from the \ttbar signal MC simulations and analysed as if they were real data.  
The normalised differential cross sections are found to be unbiased 
and the confidence intervals based on the nominal measurements and the estimated $\pm 1 \sigma$ uncertainties provide correct coverage probability. 
Any residual non-closure between generated and measured cross sections is found to be small compared to the statistical uncertainties of the measurements 
and is therefore neglected. A further closure test has been performed by unfolding pseudo-data sets generated 
using reweighted signal MCs for the detector corrections. 
The reweighting is performed as a function of the differential cross section kinematic observables and is used to introduce controlled shape variations, 
e.g. making the \ptt spectrum harder or softer.  
This test is sensitive to the stability of the unfolding with respect to the underlying physics model in the simulation.
The effect on the unfolded cross sections is negligible
for reweightings that lead to shape changes that are comparable to the observed differences between data and nominal MC distributions.

\section{Systematic uncertainties}
\label{sec:systematics}
The systematic uncertainties in the measured differential cross sections are categorised into two classes:
experimental uncertainties arising from imperfect modelling of the detector response, and theoretical uncertainties
arising from the modelling of the signal and background processes.
Each source of systematic uncertainty
is assessed by changing in the simulation the corresponding efficiency, resolution, or scale by its uncertainty, using
a prescription similar to the one followed in Ref.~\cite{Sirunyan:2018ucr}.
For each change made, the cross section determination is repeated,
and the difference with respect to the nominal result in each bin is taken as the systematic uncertainty.
\subsection{Experimental uncertainties}
To account for the pileup uncertainty, the value of the total $\Pp\Pp$ inelastic cross section,
which is used to estimate the mean number of additional \pp interactions, is varied by
$\pm4.6\%$, corresponding to the uncertainty in the measurement of this cross section~\cite{Aaboud:2016mmw}.

The efficiencies of the dilepton triggers are measured with independent
triggers based on a \ptmiss requirement.
Scale factors, defined as the ratio of the trigger efficiencies in data and simulation,
are calculated in bins of lepton $\eta$ and \pt. They are applied to the simulation
and varied within their uncertainties.
The uncertainties from the modelling of lepton identification and isolation efficiencies are determined using the ``tag-and-probe'' method
with \Zjets event samples~\cite{Chatrchyan:2011cm,bib:TOP-15-003_paper}.
The differences of these efficiencies between data and simulation in bins of $\eta$ and \pt are generally less than 10\% for electrons,
and negligible for muons. The uncertainty is estimated by varying the corresponding scale factors in the simulation within their uncertainties.
An implicit assumption made in the analysis is that the scale factors derived from the \Zjets sample
are applicable for the \ttbar samples, where the efficiency for lepton isolation is reduced
due to the typically larger number of jets present in the events.
An additional uncertainty of 1\% is added to take into account a possible violation of this assumption.
This uncertainty is verified with studies with \ttbar enriched samples using a similar event selection as for the present analysis.
In these studies the lepton isolation criteria are relaxed for one lepton and the efficiency for passing the criteria
is measured both in data and simulation.

The uncertainty arising from the jet energy scale (JES) is determined by varying the twenty-six sources of uncertainty in the JES in bins of \pt and $\eta$
and taking the quadrature sum of the effects~\cite{Khachatryan:2016kdb}.
These uncertainties also include several sources related to pileup, that contribute
a smaller part of all JES related uncertainties.
The JES variations are also propagated to the uncertainties in \ptvecmiss.
The uncertainty from the jet energy resolution (JER) is determined by the variation
of the simulated JER by $\pm$~1 standard deviation in different $\eta$ regions~\cite{Khachatryan:2016kdb}.
An additional uncertainty in the calculation of \ptvecmiss is estimated by varying the energies of reconstructed particles not clustered into jets.

The uncertainty due to imperfect modelling of the {\cPqb} tagging efficiency
is determined by varying the measured scale factor for {\cPqb} tagging efficiencies within its uncertainties~\cite{Sirunyan:2017ezt}.

The uncertainty in the integrated luminosity of the 2016 data sample recorded by CMS is 2.5\%~\cite{bib:CMS-PAS-LUM-17-001}
and is applied simultaneously to the normalisation of all simulated distributions.

\subsection{Theoretical uncertainties}
The uncertainties of the modelling of the \ttbar signal events are evaluated with appropriate variations of the nominal
simulation based on \PowPyt and the CUETP8M2T4 tune
(see Section~\ref{sec:simulation} for details).
The studies presented in~\cite{bib:CMS:2016kle} show that the nominal simulation provides a reasonable
prediction of differential \ttbar production cross sections at $\sqrt{s}=8$\TeV and
$\sqrt{s}=13$\TeV, also for events with additional jets, and thus can be used as a solid basis
for evaluating theoretical uncertainties in the present analysis.

The uncertainty arising from missing higher-order terms in the simulation of the signal process at ME level is assessed
by varying the renormalisation{, $\mu_\mathrm{r}$,} and factorisation{, $\mu_\mathrm{f}$,} scales in the \POWHEG\ simulation up and down by factors of two with respect to the nominal values.
In the \POWHEG\ sample, the nominal scales are defined as ${\mu_\mathrm{r} = \mu_\mathrm{f} =} \sqrt{\smash[b]{m^2_{\text{t}} + p^2_{\mathrm{T,\PQt}}}}$, where $p_{\mathrm{T,\PQt}}$
denotes the \pt of the top quark in the \ttbar\ rest frame.
In total, three variations are applied: one with the factorisation scale fixed, one with the renormalisation scale fixed,
and one with both scales varied up and down coherently together.
The maximum of the resulting measurement variations is taken as the final uncertainty.
In the parton-shower simulation, the corresponding uncertainty
is estimated by varying the scale up and down by factors of 2 for initial-state radiation and $\sqrt{2}$ for final-state radiation, as suggested in Ref.~\cite{Skands:2014pea}.

The uncertainty from the choice of PDF is assessed by reweighting the signal simulation
according to the prescription provided for the CT14 NLO set~\cite{Dulat:2015mca}.
An additional uncertainty is independently derived by varying the $\alpS$ value within its uncertainty in the PDF set.
The dependence of the measurement on the assumed top quark mass parameter $\mtmc$ value is estimated by varying $\mtmc$ in the simulation by $\pm$1\GeV around the central value of 172.5\GeV.

The uncertainty originating from the scheme used to match the ME-level calculation to the parton-shower simulation
is derived by varying the $h_\text{damp}$ parameter in \POWHEG\ in the range $0.996\mtmc < h_\text{damp} < 2.239\mtmc$, according to the tuning results from Ref.~\cite{bib:CMS:2016kle}.

The uncertainty related to modelling of the underlying event is estimated by varying the parameters used
to derive the CUETP8M2T4 tune in the default setup.
The default setup in \PYTHIA includes a model of colour reconnection based on MPIs with early resonance decays switched off.
The analysis is repeated with three other models of colour reconnection within \PYTHIA: the MPI-based scheme with early resonance decays
switched on, a gluon-move scheme~\cite{Argyropoulos:2014zoa}, and a QCD-inspired scheme~\cite{Christiansen:2015yqa}.
The total uncertainty from colour reconnection modelling is estimated by taking the maximum deviation from the nominal result.

The uncertainty from the knowledge of the {\cPqb} quark fragmentation function is assessed by varying the Bowler--Lund function
within its uncertainties \cite{Bowler:1981sb}.
In addition, the analysis is repeated using the Peterson model for {\cPqb} quark fragmentation~\cite{Peterson:1982ak}{,
and the final uncertainty is determined, separately for each measurement bin, as an envelope of the variations of the normalised
cross section resulting from all variations of the {\cPqb} quark fragmentation function.}
{An uncertainty from the semileptonic branching ratios of {\cPqb} hadrons is estimated by varying them according to the
world average uncertainties~\cite{Patrignani:2016xqp}.}
As \ttbar\ events producing electrons or muons originating from the decay of $\PGt$ leptons are considered to be background,
the measured differential cross sections are sensitive to the branching ratios
of $\PGt$ leptons decaying into electrons or muons assumed in the simulation.
Hence, an uncertainty is determined by varying the branching ratios by 1.5\%~\cite{Patrignani:2016xqp} in the simulation.

The normalisations of all non-\ttbar backgrounds are varied up and down by $\pm30\%$ taken from measurements as explained in Ref.~\cite{bib:TOP-15-003_paper}.

The total systematic uncertainty in each measurement bin is estimated by adding all the contributions described above in quadrature,
separately for positive and negative cross section variations.
If a systematic uncertainty results in two cross section variations of the same sign, the largest one is taken, while the opposite variation is set to zero.
 \section{Results of the measurement}
\label{sec:results}

The normalised differential cross sections of \ttbar production are measured in the full phase space at parton level
for top quarks (after radiation and before the top quark and antiquark decays)
and at particle level for additional jets in the events, for the following variables:
\begin{enumerate}
    \item double-differential cross sections as a function of pair of variables:
    \begin{itemize}
        \item $\abs{\yt}$ and \ptt,
        \item \mtt and $\abs{\yt}$,
        \item \mtt and $\abs{\ytt}$,
        \item \mtt and \detatt,
        \item \mtt and \dphitt,
        \item \mtt and \pttt and
        \item \mtt and \ptt.
    \end{itemize}
    These cross sections are denoted in the following as \ytptt, etc.
    \item triple-differential cross sections as a function of \nj, \mtt, and \ytt. These cross sections are measured separately using two ($\nj = 0$ and $\nj \ge 1$) and three ($\nj = 0$, $\nj = 1$, and $\nj \ge 2$) bins of \nj, for the particle-level jets.
     {These cross sections are denoted as \njmttytttwo and \njmttyttthree, respectively.}
\end{enumerate}
The pairs of variables for the double-differential cross sections are chosen in order to obtain representative combinations
that are sensitive to different aspects of the \ttbar production dynamics, mostly following
the previous measurement~\cite{Sirunyan:2017azo}. The variables for the triple-differential cross sections are chosen in order to enhance sensitivity to the PDFs, \as, and \mt.
In particular, the combination of \ytt and \mtt variables provides sensitivity for the PDFs, as demonstrated
in~\cite{Sirunyan:2017azo}, the \nj distribution for \as and \mtt for \mt.

{\tolerance=1200
The numerical values of the measured cross sections and their uncertainties are provided in Appendix~\ref{sec:apptab}.
In general, the total uncertainties for all measured cross sections
are about 5--10\%, but exceed 20\% in some regions of phase space, such as the last \nj range of the \njmttyttthree distribution.
The total uncertainties are dominated by the systematic uncertainties
receiving similar contributions from the experimental and theoretical systematic sources.
The largest experimental systematic uncertainty is associated with the JES.
Both the JES and signal modelling systematic uncertainties are also affected
by the statistical uncertainties in the simulated samples that are used for the evaluation of these uncertainties.
The cross sections measured in the \ee, \mumu and \emu channels separately are compatible with each other.
\par}

\begin{figure*}
    \centering
    \includegraphics[width=1.00\textwidth]{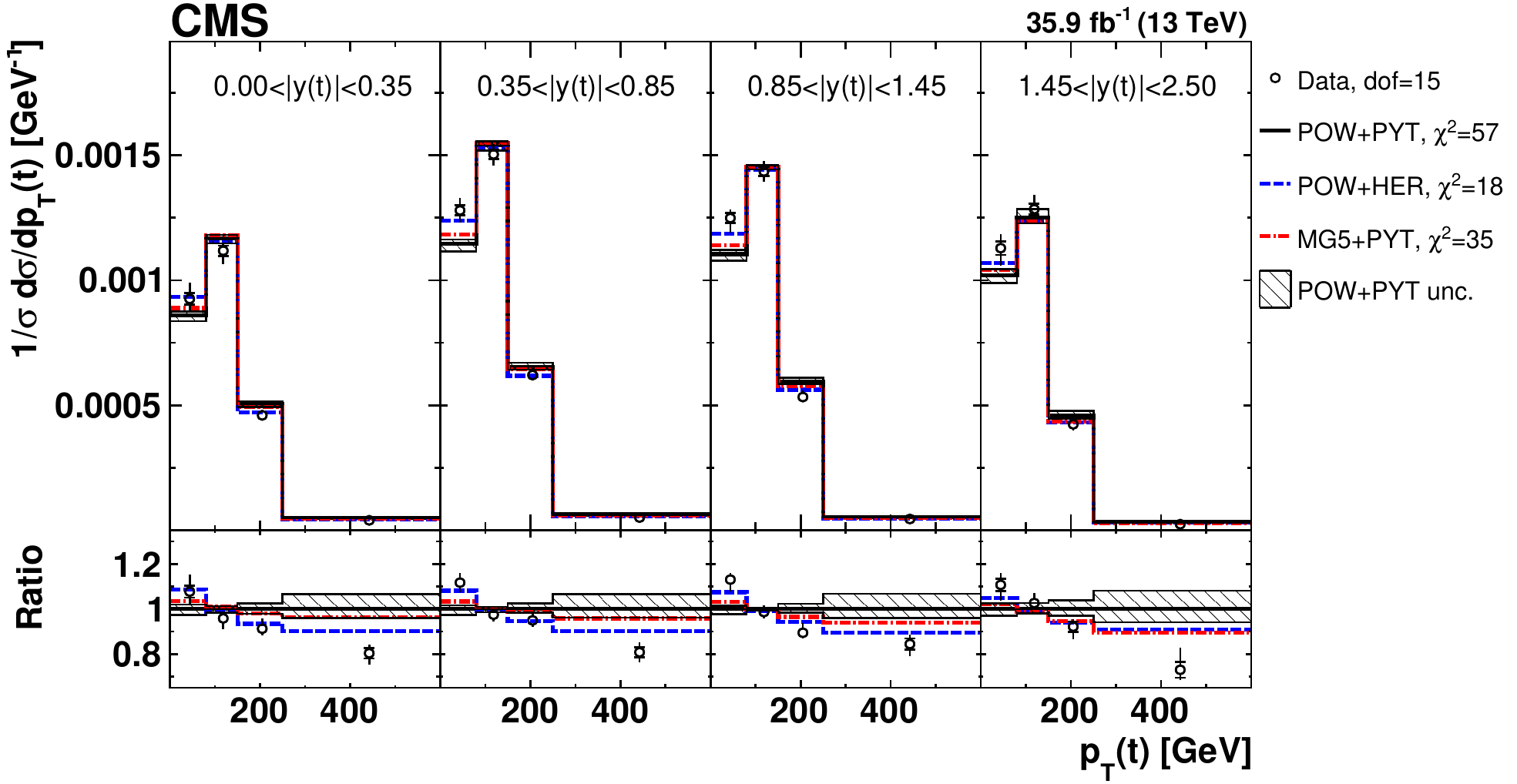}
    \caption{Comparison of the measured \ytptt cross sections {to the theoretical predictions calculated using \PowPyt (\PowPytSh),
\PowHer (\PowHerSh), and \aMCPyt (\aMCPytSh) event generators.}
        The inner vertical bars on the data points represent the statistical uncertainties and the full bars include also the systematic uncertainties added in quadrature.
        For each MC model, values of \chisq which take into account the bin-to-bin
        correlations and \ndf for the comparison with the data are reported. The hatched regions correspond to the theoretical uncertainties in \PowPyt (see Section~\ref{sec:systematics}).
        In the lower panel, the ratios of the data and other simulations to the \PowPytSh predictions are shown.}
    \label{fig:xsec-mc-ytptt}
\end{figure*}

\begin{figure*}
    \centering
    \includegraphics[width=1.00\textwidth]{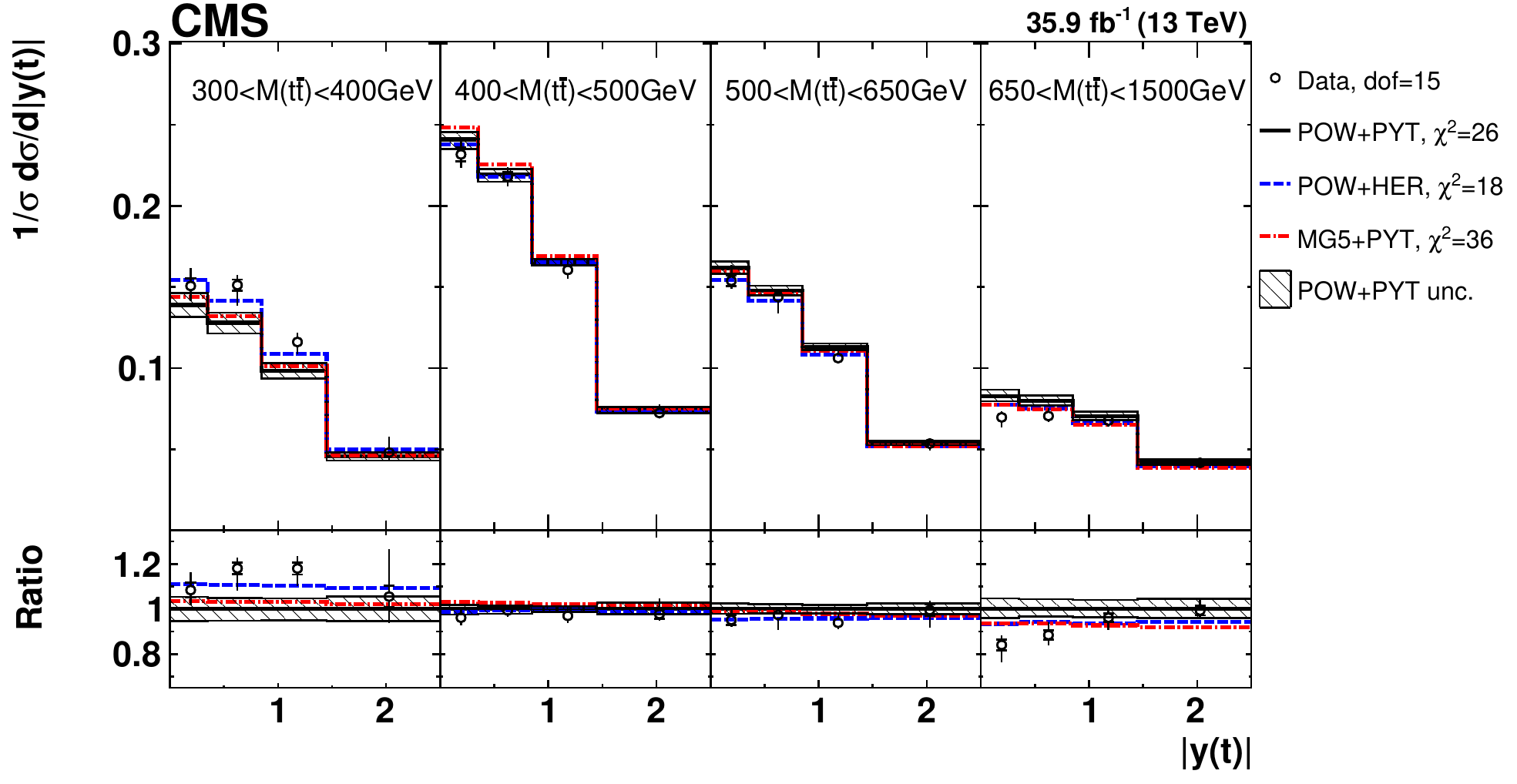}
    \caption{Comparison of the measured \mttyt cross sections {to the theoretical predictions calculated using MC event generators} (further details
can be found in the Fig.~\ref{fig:xsec-mc-ytptt} caption).}
    \label{fig:xsec-mc-mttyt}
\end{figure*}

\begin{figure*}
    \centering
    \includegraphics[width=1.00\textwidth]{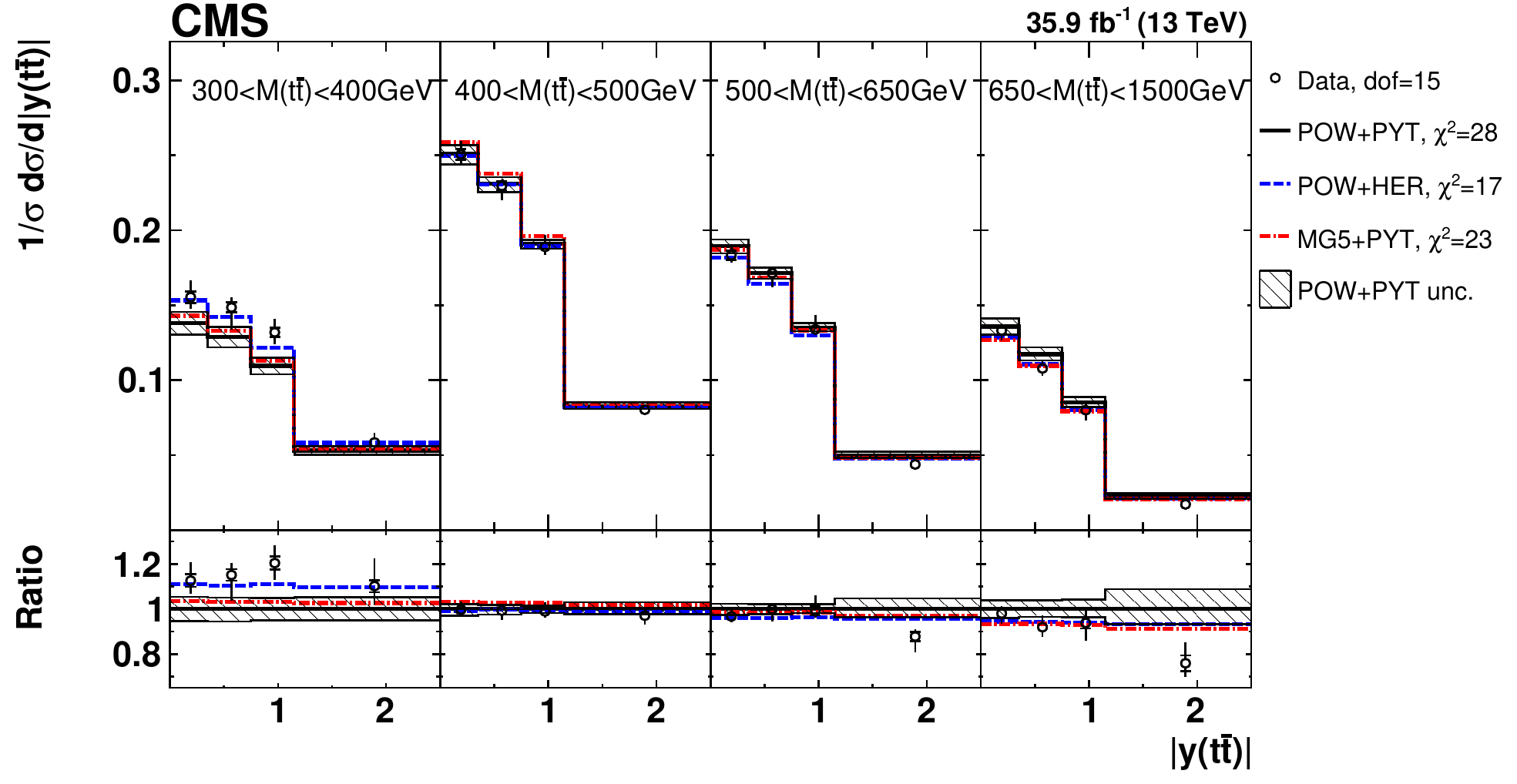}
    \caption{Comparison of the measured \mttytt cross sections {to the theoretical predictions calculated using MC event generators} (further details can be found in the Fig.~\ref{fig:xsec-mc-ytptt} caption).}
    \label{fig:xsec-mc-mttytt}
\end{figure*}

\begin{figure*}
    \centering
    \includegraphics[width=1.00\textwidth]{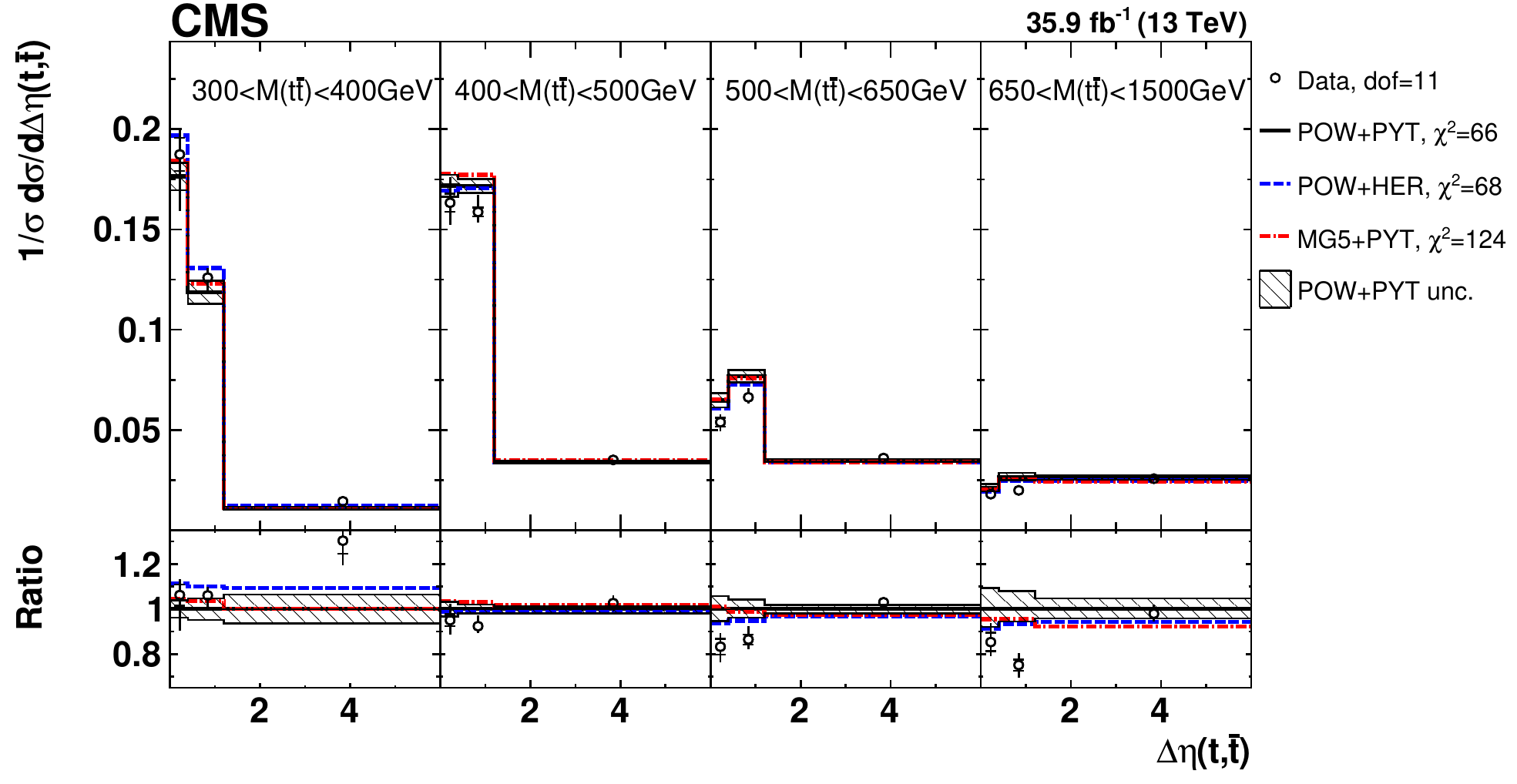}
    \caption{Comparison of the measured \mttdetatt cross sections {to the theoretical predictions calculated using MC event generators} (further details can be found in the Fig.~\ref{fig:xsec-mc-ytptt} caption).}
    \label{fig:xsec-mc-mttdetatt}
\end{figure*}

\begin{figure*}
    \centering
    \includegraphics[width=1.00\textwidth]{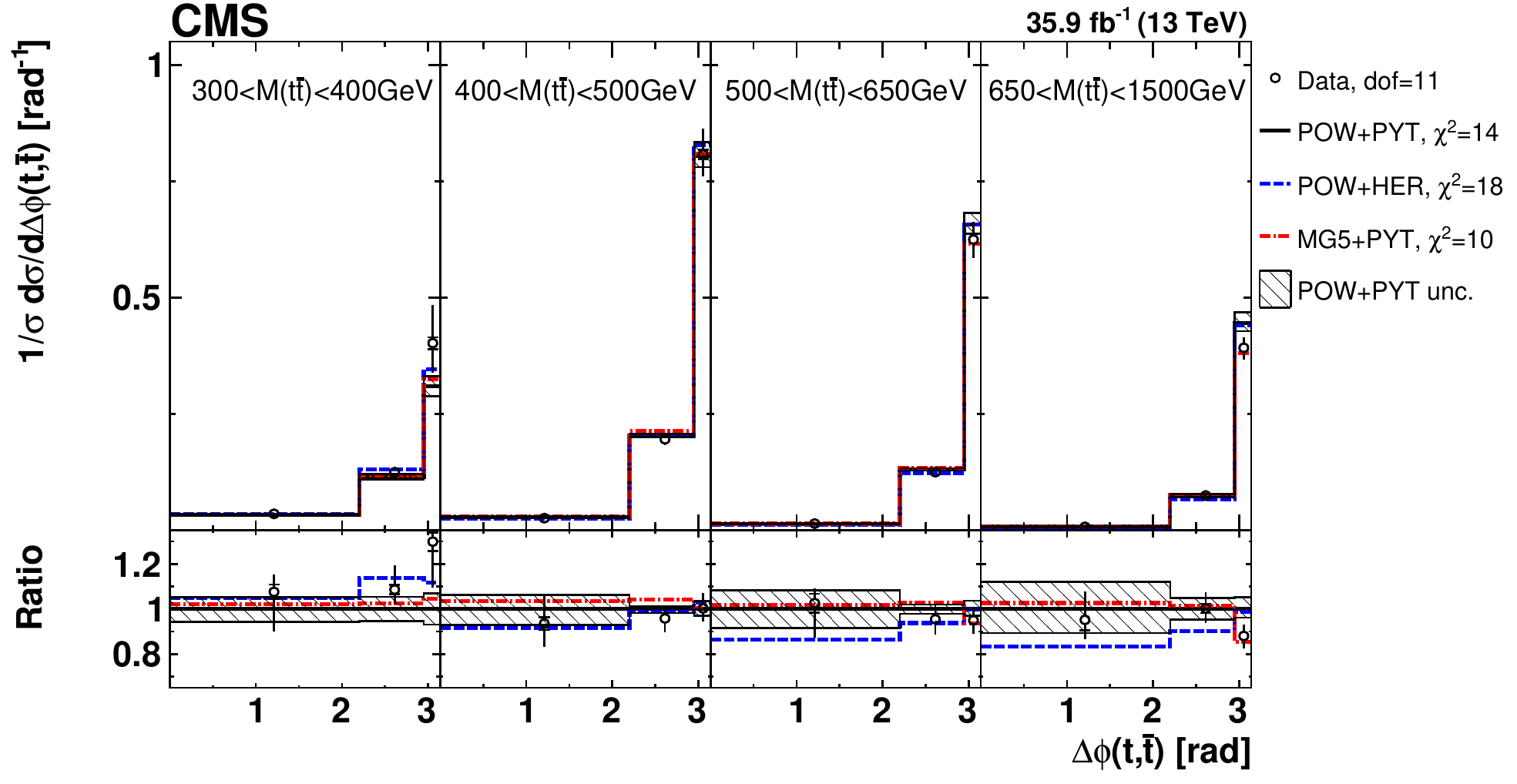}
    \caption{Comparison of the measured \mttdphitt cross sections {to the theoretical predictions calculated using MC event generators} (further details can be found in the Fig.~\ref{fig:xsec-mc-ytptt} caption).}
    \label{fig:xsec-mc-mttdphitt}
\end{figure*}

\begin{figure*}
    \centering
    \includegraphics[width=1.00\textwidth]{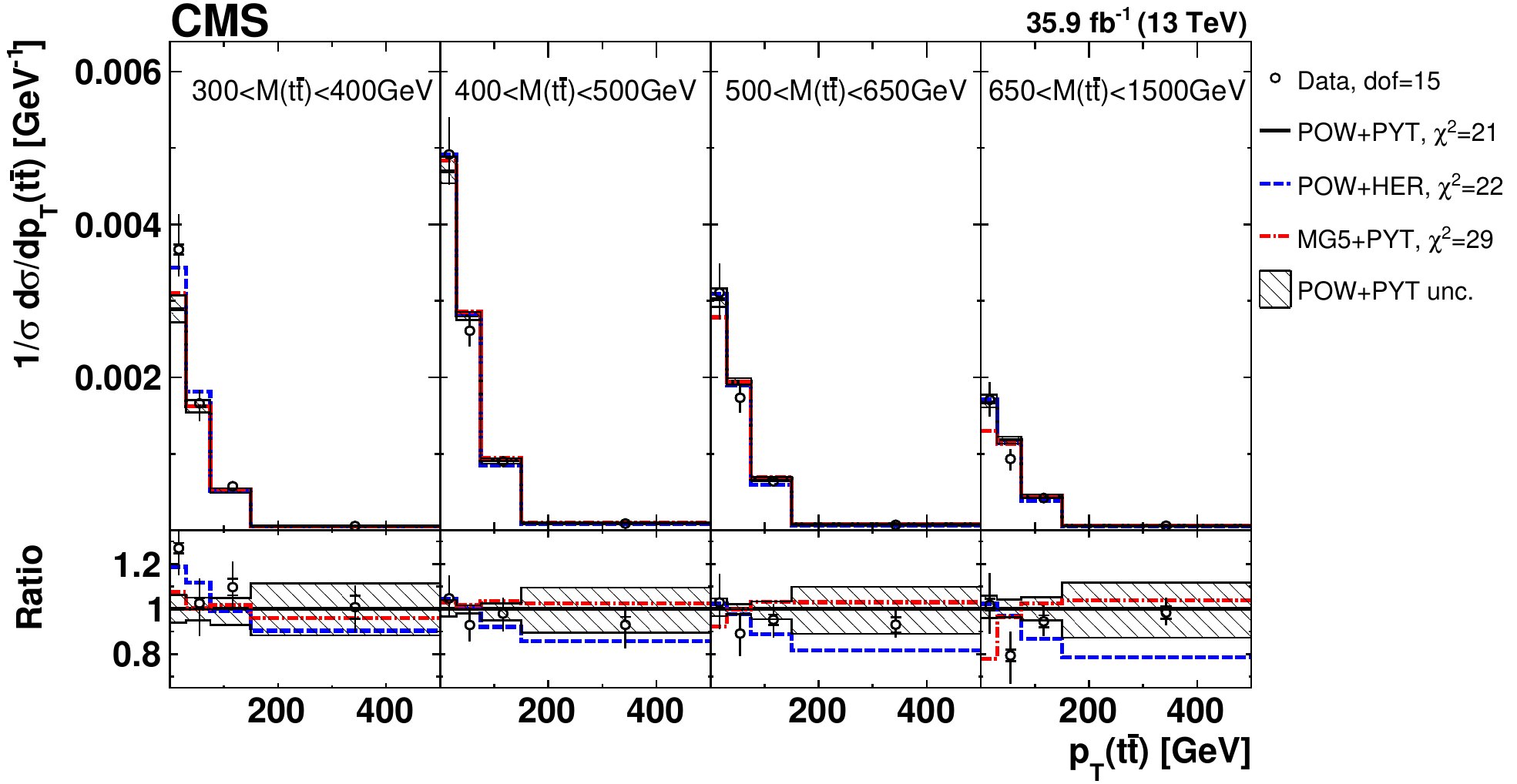}
    \caption{Comparison of the measured \mttpttt cross sections {to the theoretical predictions calculated using MC event generators} (further details can be found in the Fig.~\ref{fig:xsec-mc-ytptt} caption).}
    \label{fig:xsec-mc-mttpttt}
\end{figure*}

\begin{figure*}
    \centering
    \includegraphics[width=1.00\textwidth]{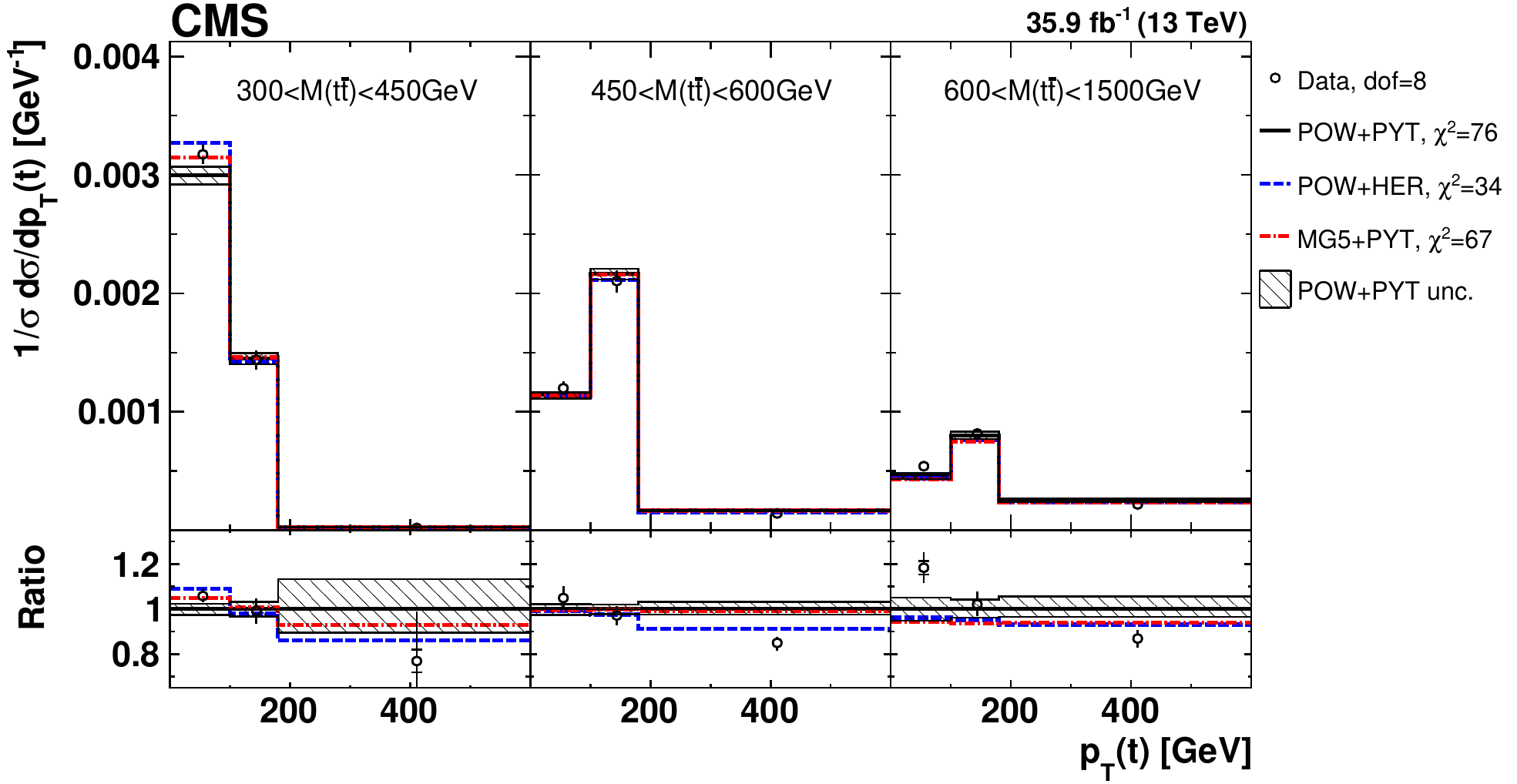}
    \caption{Comparison of the measured \mttptt cross sections {to the theoretical predictions calculated using MC event generators} (further details can be found in the Fig.~\ref{fig:xsec-mc-ytptt} caption).}
    \label{fig:xsec-mc-mttptt}
\end{figure*}

\begin{figure*}
    \centering
    \includegraphics[width=1.00\textwidth]{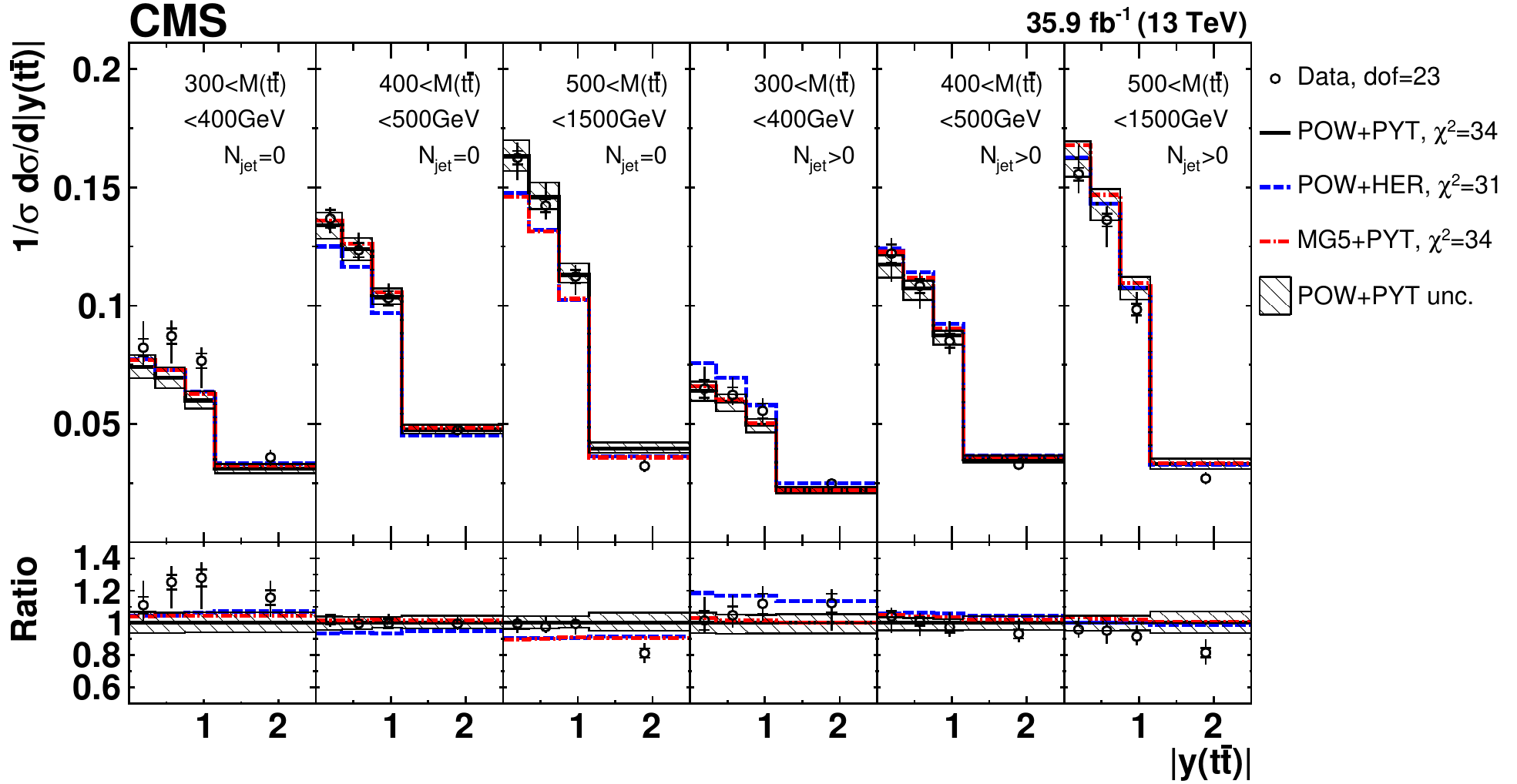}
    \caption{Comparison of the measured \njmttytttwo cross sections {to the theoretical predictions calculated using MC event generators} (further details can be found in the Fig.~\ref{fig:xsec-mc-ytptt} caption).}
    \label{fig:xsec-mc-nj2mttytt}
\end{figure*}

\begin{figure*}
    \centering
    \includegraphics[width=1.00\textwidth]{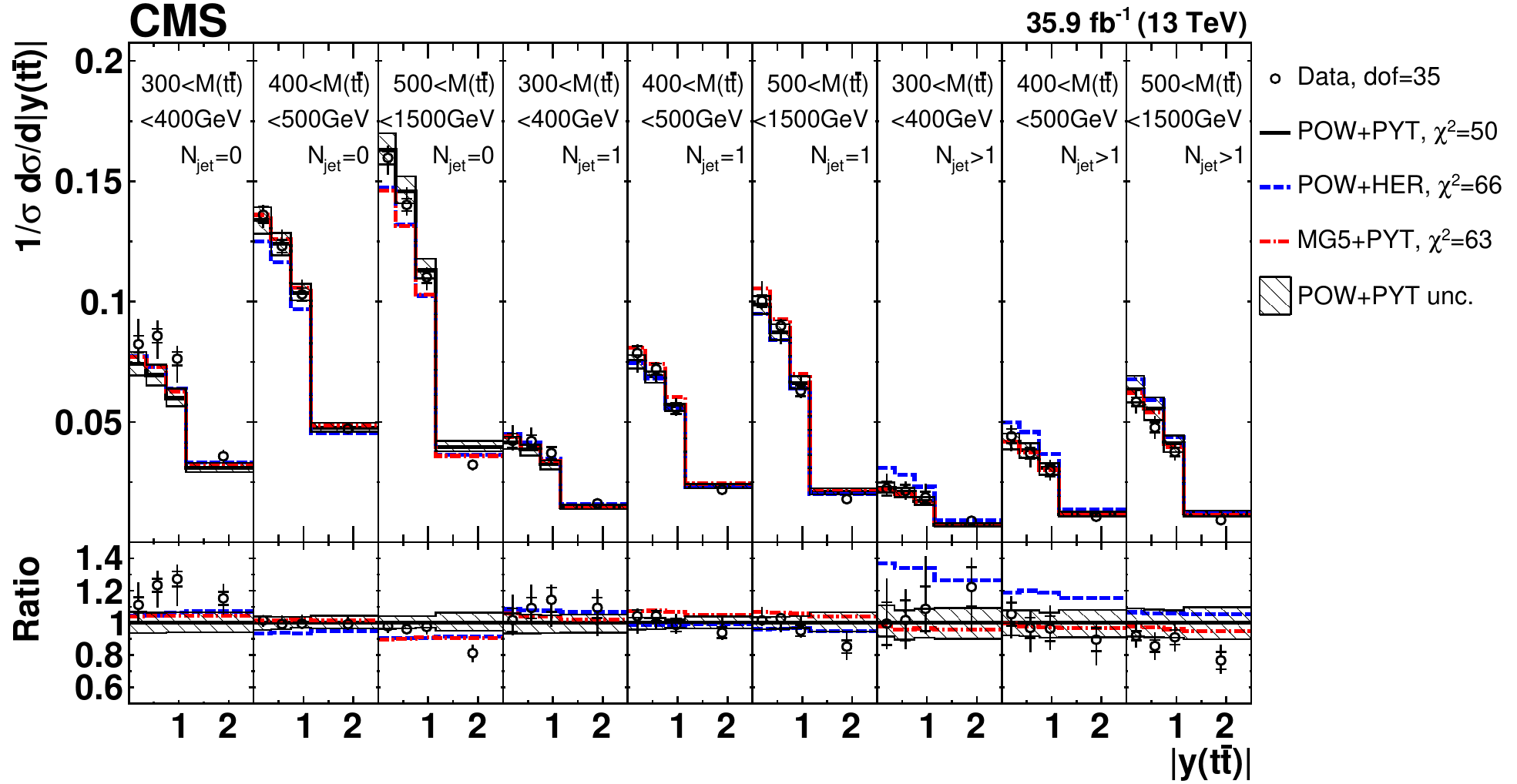}
    \caption{Comparison of the measured \njmttyttthree cross sections {to the theoretical predictions calculated using MC event generators} (further details can be found in the Fig.~\ref{fig:xsec-mc-ytptt} caption).}
    \label{fig:xsec-mc-nj3mttytt}
\end{figure*}

{\tolerance=9600
In Figs.~\ref{fig:xsec-mc-ytptt}--\ref{fig:xsec-mc-nj3mttytt}, the measured cross sections are compared {to three theoretical predictions based on MC simulations:}
\PowPyt (\PowPytSh), \PowHer (\PowHerSh), and \aMCPyt (\aMCPytSh).
{The \PowPytSh and \PowHerSh theoretical predictions differ by the parton-shower method, hadronisation and event tune (\pt-ordered parton showering, string hadronisation model and CUETP8M2T4 tune in \PowPytSh, or angular ordered parton showering, cluster hadronisation model and EE5C tune in \PowHerSh), while the \PowPytSh and \aMCPytSh predictions adopt different matrix elements (inclusive \ttbar production at NLO in \PowPytSh, or \ttbar with up to two extra partons at NLO in \aMCPytSh) and different methods for matching with parton shower (correcting the first parton shower emission to the NLO result in \PowPytSh, or subtracting from the exact NLO result its parton shower approximation in \aMCPytSh).}
For each comparison, a \chisq and the number of degrees of freedom (\ndf) are reported. The \chisq value
is calculated taking into account the statistical and systematic data uncertainties, while ignoring uncertainties of the predictions:
\par}
\begin{equation}
\label{eq:chi2nm1}
\chisq = \mathbf{R}^{T}_{N-1} \mathbf{Cov}^{-1}_{N-1} \mathbf{R}_{N-1},
\end{equation}
where $\mathbf{R}_{N-1}$ is the column vector of the residuals calculated as the difference of the measured cross sections and the corresponding predictions
obtained by discarding one of the $N$ bins, and $\mathbf{Cov}_{N-1}$ is the $(N-1)\times(N-1)$ submatrix obtained from the full covariance matrix by
discarding the corresponding row and column.
The matrix $\mathbf{Cov}_{N-1}$ obtained in this way is invertible, while the original covariance matrix $\mathbf{Cov}$ is singular because for normalised cross sections one degree of freedom is lost, as can be deduced from Eq.~(\ref{eq:xsecdef}).
The covariance matrix $\mathbf{Cov}$ is calculated as:
\begin{equation}
\label{eq:covmat}
\mathbf{Cov} = \mathbf{Cov}^\text{unf} + \mathbf{Cov}^\text{syst},
\end{equation}
where $\mathbf{Cov}^\text{unf}$ and $\mathbf{Cov}^\text{syst}$ are the covariance matrices corresponding to the statistical uncertainties from the unfolding,
and the systematic uncertainties, respectively.
The systematic covariance matrix $\mathbf{Cov}^\text{syst}$ is calculated as:
\begin{equation}
    \mathbf{Cov}^\text{syst}_{ij} = \sum_{k,l} \frac{1}{N_k} C_{j,k,l}C_{i,k,l},\\ \quad 1 \le i \le N, \quad 1 \le j \le N,
    \label{eq:chi2}
    \end{equation}
where $C_{i,k,l}$ stands for the systematic uncertainty from variation $l$ of source $k$ in the $i$th bin, and $N_k$ is the number of variations for source $k$.
The sums run over all sources of the systematic uncertainties and all corresponding variations.
Most of the systematic uncertainty sources in this analysis consist of positive and negative variations and thus have $N_k = 2$, whilst several model uncertainties (the model of colour reconnection and the {\cPqb} quark fragmentation function) consist of more than two variations, a property which is accounted for in Eq.~(\ref{eq:chi2}).
All systematic uncertainties are treated as additive,
i.e.\ the relative uncertainties are used to scale the corresponding measured value in the construction of $\mathbf{Cov}^\text{syst}$.
This treatment is consistent with the cross section normalisation and makes the \chisq in Eq.~(\ref{eq:chi2nm1}) independent of which of the $N$ bins is excluded.
A multiplicative treatment of uncertainties has been tested as well, and consistent results were obtained.
The cross section measurements for different multi-differential distributions are statistically and systematically correlated.
No attempt is made to quantify the correlations between bins from different multi-differential
distributions. Thus, quantitative comparisons between theoretical predictions and the data can only be made for each single
set of multi-differential cross sections.

In Fig.~\ref{fig:xsec-mc-ytptt}, the \ptt distribution is compared in different ranges of $\abs{\yt}$ to {predictions from \PowPytSh, \PowHerSh, and \aMCPytSh.}
The data distribution is softer than that of the predictions over the entire $\yt$ range.
Only \PowHerSh describes the data well, while the other two simulations predict a harder \ptt distribution than measured in the data over the entire $\yt$ range.
The disagreement is strongest for \PowPytSh.

Figures~\ref{fig:xsec-mc-mttyt} and \ref{fig:xsec-mc-mttytt} illustrate the distributions of $\abs{\yt}$ and $\abs{\ytt}$ in different \mtt ranges compared to the same set of MC models.
The shapes of the \yt and \ytt distributions are reasonably well described by all models, except for the largest \mtt range, where all theoretical predictions are more central than the data for \yt and less central for \ytt. The \mtt distribution is softer in the data than {in the theoretical predictions.}
The latter trend is the strongest for \PowPytSh, being consistent with the disagreement for the \ptt distribution (as shown in Fig.~\ref{fig:xsec-mc-ytptt}).
The best agreement for both \mttyt and \mttytt cross sections is provided by \PowHerSh.

In Fig.~\ref{fig:xsec-mc-mttdetatt}, the $\detatt$ distribution is compared in the same \mtt ranges {to the theoretical predictions.}
For all generators, there is a discrepancy between the data and simulation for the medium and high \mtt bins, where the predicted $\detatt$ values are too low.
The disagreement is the strongest for \aMCPytSh.

Figures~\ref{fig:xsec-mc-mttdphitt} and \ref{fig:xsec-mc-mttpttt} illustrate the comparison of the distributions of \dphitt and \pttt in the same \mtt ranges {to the theoretical predictions.}
Both these distributions are sensitive to gluon radiation. All MC models describe the data well within uncertainties, except for \aMCPytSh,  which predicts a \pttt distribution in the last \mtt bin of the \mttpttt cross sections
that is too hard.

In Fig.~\ref{fig:xsec-mc-mttptt}, the \ptt distribution is compared in different \mtt ranges {to the theoretical predictions.}
None of the MC generators is able to describe the data, generally predicting a too hard \ptt distribution. The discrepancy
is larger at high \mtt values where the softer \ptt spectrum in the data must be kinematically correlated
with the larger $\detatt$ values (as shown in Fig.~\ref{fig:xsec-mc-mttdetatt}), compared to the predictions.
The disagreement is the strongest for \PowPytSh. While the \PowHerSh simulation
is able to reasonably describe the \ptt distribution in the entire range of \ytt (as shown in Fig.~\ref{fig:xsec-mc-ytptt}),
it does not provide a good description in all ranges of \mtt, in particular predicting a too hard \ptt distribution at high \mtt.

Figures~\ref{fig:xsec-mc-nj2mttytt} and \ref{fig:xsec-mc-nj3mttytt} illustrate the triple-differential cross sections as a function of $\abs{\ytt}$ in different \mtt and \nj ranges, measured using two or three bins of \nj.
For the \njmttytttwo measurement, all MC models describe the data well.
For the \njmttyttthree measurement, only \PowPytSh is in satisfactory agreement with the data.
In particular, \PowHerSh predicts too high a cross section for $\nj > 1$, while \aMCPytSh provides the worst description of the \mtt distribution for $\nj = 1$.

\begin{figure*}
    \centering
    \includegraphics[width=0.80\textwidth]{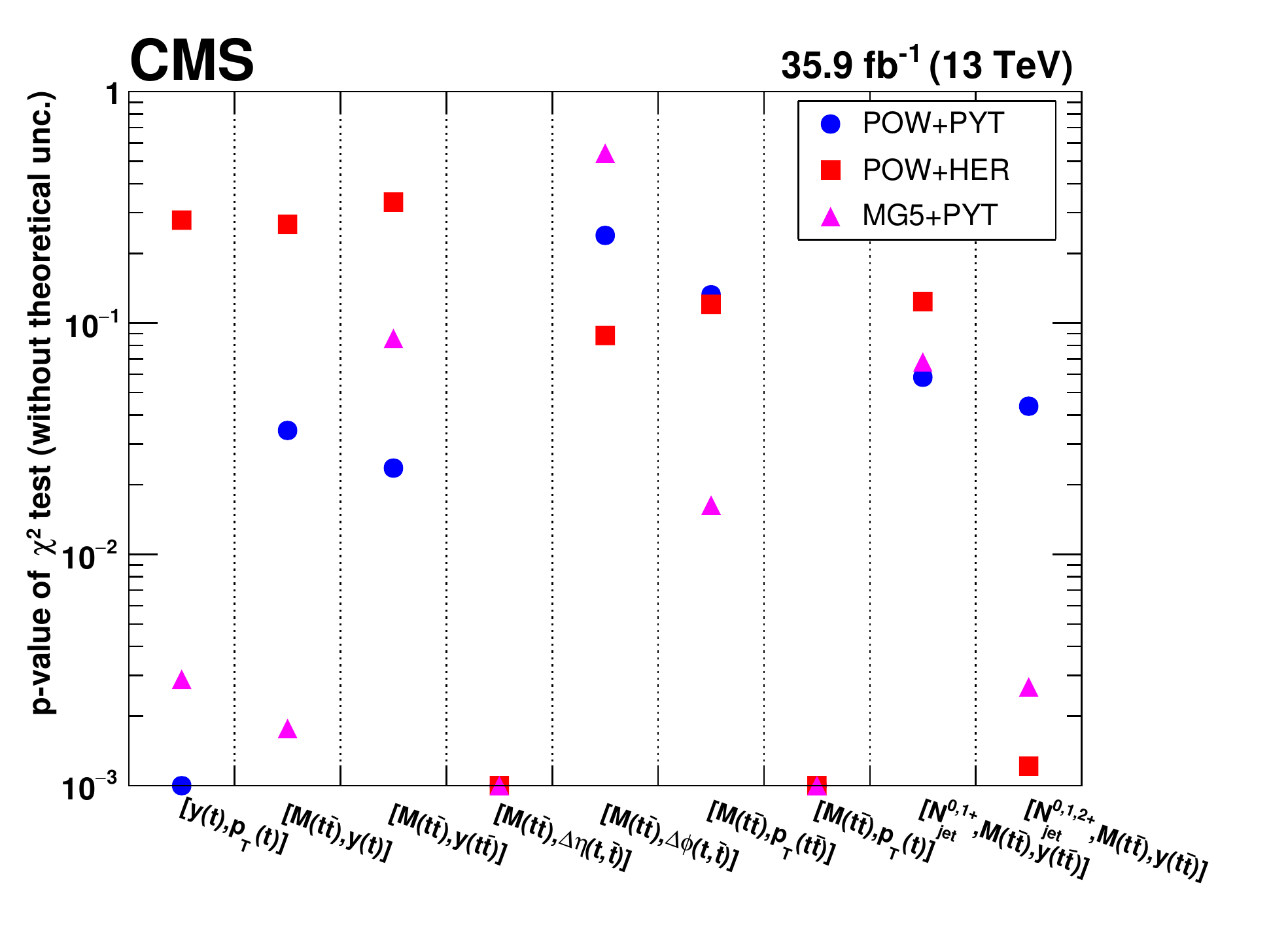}
    \caption{Assessment of compatibility of various MC predictions with the data. The plot show the $p$-values of \chisq-tests between data and predictions. Only the data uncertainties are taken into account in the \chisq-tests while uncertainties on the theoretical calculations are ignored. Points with $p \le 0.001$ are shown at $p = 0.001$.}
    \label{fig:ptval-mc}
\end{figure*}

All obtained \chisq values, ignoring theoretical uncertainties, are listed in Table~\ref{tab:chi2mc}.
The corresponding $p$-values are visualised in Fig.~\ref{fig:ptval-mc}.
From these values one can conclude that
none of the central predictions of the considered MC generators is able to provide predictions that
correctly describe all distributions.
In particular, for \mttdetatt and \mttptt the \chisq values are relatively large for all MC generators.
In total, the best agreement with the data is provided by \PowPytSh and \PowHerSh,
with \PowPytSh better describing the measurements probing \nj and radiation, and \PowHerSh better describing the ones involving probes of the \pt distribution.

\begin{table*}
    \topcaption{The \chisq values (taking into account data uncertainties and ignoring theoretical uncertainties) and dof of the measured cross sections with respect {to the predictions of various MC generators.}}
    \label{tab:chi2mc}
    \renewcommand{\arraystretch}{1.4}
    \centering
    \begin{tabular}{lcccc}
        \multirow{1}{*}{Cross section} & \multirow{2}{*}{\ndf} & \multicolumn{3}{c}{\chisq} \\
        \cline{3-5}
        {variables}&&\PowPytSh    & \PowHerSh   & \aMCPytSh\\
        \hline
        \ytptt      & 15 & 57 & 18 & 35 \\
        \mttyt      & 15 & 26 & 18 & 36 \\
        \mttytt      & 15 & 28 & 17 & 23 \\
        \mttdetatt      & 11 & 66 & 68 & 124 \\
        \mttdphitt      & 15 & 14 & 18 & 10 \\
        \mttpttt      & 15 & 21 & 22 & 29 \\
        \mttptt      & 15 & 77 & 34 & 68 \\
        \njmttytttwo      & 23 & 34 & 31 & 34 \\
        \njmttyttthree      & 35 & 50 & 66 & 63 \\
    \end{tabular}
\end{table*}

\section{\texorpdfstring{Extraction of \as and \mt from \njmttytttwo cross sections using external PDFs}{Extraction of as and mt from njmttytt cross sections using external PDFs}}
\label{sec:asmt}

To extract \as and \mt, the measured triple-differential cross sections are compared to fixed-order NLO predictions that do not have variable parameters,
except for the factorisation and renormalisation scales. These predictions provide a simpler assessment of theoretical uncertainties than
predictions from MC event generators. The latter complement fixed-order computations with parton showers, thus accounting for important QCD corrections beyond fixed order, but complicating thereby the interpretation of the extracted parameters, because the modelling of the showers can involve different PDFs and \as values.
Furthermore, for PDF fits using these data (to be discussed in Section~\ref{sec:qcd}), fast computation techniques are required that are currently available only for fixed-order calculations.

Fixed-order theoretical calculations for fully differential cross sections for inclusive \ttbar production
are publicly available at NLO $O(\as^3)$ in the fixed-flavour number scheme~\cite{Mangano:1991jk},
and for $\ttbar$ production with one (NLO $O(\as^4)$)~\cite{Dittmaier:2007wz} and two (NLO $O(\as^5)$)~\cite{Bevilacqua:2010ve,Bevilacqua:2011aa} additional jets.
These calculations are used in the present analysis. Furthermore, NLO predictions for $\ttbar$ production with three additional jets exist~\cite{Hoche:2016elu}, but are not used in this paper because the sample of events with three additional jets is not large enough to allow us to measure multi-differential cross sections.
The exact fully differential NNLO $O(\as^4)$ calculations for inclusive \ttbar production have recently appeared in the literature~\cite{Czakon:2015owf,Czakon:2016dgf}, but these predictions have not been published yet {for multi-differential cross sections.} The NNLO calculations for \ttbar production with additional jets have not been performed yet.

{\tolerance=9600
In the case of the \njmttytttwo measurement, cross sections for inclusive $\ttbar$ and $\ttbar+1$ jet production in each bin of \mtt and \ytt are obtained in the following way.
The cross sections for inclusive $\ttbar+1$ jet production are taken from the $\nj \ge 1$ bins of the \njmttytttwo measurements. 
The cross sections for inclusive $\ttbar$ jet production are calculated from
the sum of the cross sections in the $\nj=0$ and $\nj \ge 1$ bins.
Statistical and systematic uncertainties and all correlations are obtained using error propagation.
Finally, the cross sections obtained for inclusive $\ttbar$ and $\ttbar+1$ jet production are 
compared to the NLO $O(\as^3)$ and NLO $O(\as^4)$ calculations, respectively.
For these processes, the ratios of NLO over LO predictions are about $1.5$ on average, 
and the requirement $\pt > 30\GeV$ for the jets ensures that logarithms of the ratio $\pt/\mt$ are not large, thereby demonstrating good convergence of the perturbation series.
Similarly, cross sections for inclusive $\ttbar$, $\ttbar+1$, and $\ttbar+2$ jets production are obtained using the \njmttyttthree measurement and compared to the NLO $O(\as^3)$, NLO $O(\as^4)$, and NLO $O(\as^5)$ calculations, respectively.
Thus, all cross sections are compared to calculations of the order in \as required for NLO accuracy.
For presentation purposes, the cross sections are shown in
Figs.~\ref{fig:xsec-nlo-nj2mttytt-pdfs}--\ref{fig:xsec-nlo-nj2mttytt-mt}
in the $\nj=0$ and  $\nj \ge 1$ bins used before for the \njmttytttwo measurements, one with $\nj=0$ and another with $\nj \ge 1$.
The measured cross sections
for  $\nj \ge 1$ are compared to the NLO calculation for inclusive $\ttbar+1$ jet production, while
those for $\nj = 0$ are compared to the difference of the
NLO calculations for inclusive $\ttbar$ and inclusive $\ttbar+1$ jet production.
The normalisation cross section is evaluated by integrating the differential cross sections over all bins, \ie it is given by the inclusive $\ttbar$ cross section. 
As discussed below, \chisq values are calculated for the comparisons of
data and NLO predictions and are also used for the extraction of parameter values. The total \chisq values obtained are identical for the comparisons based on 
inclusive $\ttbar$ and $\ttbar+1$ jet production cross sections and the ones based
on the \njmttytttwo results, shown in Figs.~\ref{fig:xsec-nlo-nj2mttytt-pdfs}--\ref{fig:xsec-nlo-nj2mttytt-mt},
because the \chisq values are invariant
under invertible linear transformations of the set of cross section values.

The NLO predictions are obtained using the \mg framework running in the fixed-order mode.
A number of the latest proton NLO PDF sets are used, namely:
ABMP16~\cite{Alekhin:2018pai}, CJ15~\cite{Accardi:2016qay}, CT14~\cite{Dulat:2015mca}, HERAPDF2.0~\cite{Abramowicz:2015mha}, JR14~\cite{Jimenez-Delgado:2014twa},
MMHT2014~\cite{Harland-Lang:2014zoa}, and NNPDF3.1~\cite{Ball:2017nwa},
available via the \lhapdf interface (version 6.1.5)~\cite{Buckley:2014ana}.
No \ttbar data were used in the determination of  the \cj, \ct, \herapdf and \jr PDF sets; only total \ttbar production cross section measurements were used to determine the ABMP16 and \mmht PDFs, and both total and differential (from LHC Run 1) \ttbar cross sections were used in the NNPDF3.1 extraction.
The number of active flavours is set to $n_{\mathrm{f}} = 5$,
an $m_{\PQt}^{\text{pole}} = 172.5$\GeV is used, and
$\as$ is set to the value used for the corresponding PDF extraction.
The renormalisation and factorisation scales
are chosen to be $\mu_\mathrm{r} = \mu_\mathrm{f} = H' / 2, H' = \sum_i {m_{\PQt,i}}$. Here the sum is running over
all final-state partons ($\PQt$, $\overline{\PQt}$, and up to three light partons in the $\ttbar + 2$ jet calculations)
and $m_{\PQt}$ denotes a transverse mass, defined as $m_{\PQt} = \sqrt{\smash[b]{m^2 + \pt^2}}$.
The theoretical uncertainty is estimated by varying $\mu_\mathrm{r}$ and $\mu_\mathrm{f}$ independently up and down by a factor of 2,
with the additional restriction
that the ratio $\mu_\mathrm{r} / \mu_\mathrm{f}$ stays between 0.5 and 2~\cite{Cacciari:2008zb}.
Additionally, an alternative scale choice $\mu_\mathrm{r} = \mu_\mathrm{f} = H / 2, H = \sum_i {m_{\PQt,i}}$, with the sum running only over $\PQt$ and $\overline{\PQt}$~\cite{Czakon:2016dgf}, is considered.
The scales are varied coherently in the predictions with different \nj.
The final uncertainty is determined as an envelope of all scale variations on the normalised cross sections.
This uncertainty is referred to hereafter as a scale uncertainty and is supposed to estimate the impact
of missing higher-order terms.
The PDF uncertainties are taken into account in the theoretical predictions for each PDF set.
The PDF uncertainties of \cj~\cite{Accardi:2016qay} and \ct~\cite{Dulat:2015mca}, evaluated at 90\% confidence level (\CL), are rescaled to the 68\% \CL for consistency with other PDF sets.
The uncertainties in the normalised \ttbar cross sections originating from $\as$ and \mt are estimated by varying them within $\asmz = 0.118 \pm 0.001$ and $\mt = 172.5 \pm 1.0\GeV$, respectively (for presentation purposes, in some figures larger variations of \asmz and \mt by $\pm 0.005$ and $\pm 5.0\GeV$, respectively, are shown).
\par}

To compare the measured cross sections to the NLO QCD calculations, the latter are further corrected from parton to particle level. The NLO QCD calculations are provided for parton-level jets and stable top quarks, therefore the corrections (further referred to as NP) are determined using additional \PowPyt MC simulations for \ttbar production with and without MPI, hadronisation and top quark decays, and defined as:
\begin{equation}
\mathcal{C}_{\text{NP}} = \frac{\sigma^{\text{particle}}_{\text{isolated from $\PQt \to \ell, \PQb$}}}{\sigma^{\text{parton}}_{\text{no MPI, no had., no \ttbar decays}}}.
\end{equation}
Here $\sigma^{\text{particle}}_{\text{isolated from $\PQt \to \ell,\PQb$}}$ is the cross section with MPI and hadronisation for jets built of particles excluding neutrinos and isolated from charged leptons and {\cPqb} quarks from the top quark decays, as defined in Section~\ref{sec:sel}, and $\sigma^{\text{parton}}_{\text{no MPI, no had., no \ttbar decays}}$ is the cross section without MPI and hadronisation for jets built of partons excluding $\PQt$ and $\PAQt$.
Both cross sections are calculated at NLO matched with parton showers.
The $\mathcal{C}_{\text{NP}}$ factors are used to correct the NLO predictions to particle level.
The NP corrections are determined in bins of the triple-differential cross sections as a function of \nj, \mtt, and \ytt, even though they depend primarily on \nj and have only weak dependence on the \ttbar kinematic properties.
For the cross sections with up to two extra jets measured in this analysis, the estimated NP corrections are close to 1, within 5\%.
The dependence of the NP corrections on MC modelling was studied using MC samples with varied hadronisation model, underlying event tune,
and ME and parton-shower scales, as detailed in Section~\ref{sec:systematics}. All resulting variations of  $\mathcal{C}_{\text{NP}}$ were found to be $\lesssim$1\%, therefore no uncertainties on the determined NP corrections are assigned.
To compare to the measured cross sections, the normalised multi-differential cross sections of the theoretical predictions are obtained by dividing the cross sections in specific bins by
the total cross section summed over all bins.

\begin{figure*}
    \centering
    \includegraphics[width=1.00\textwidth]{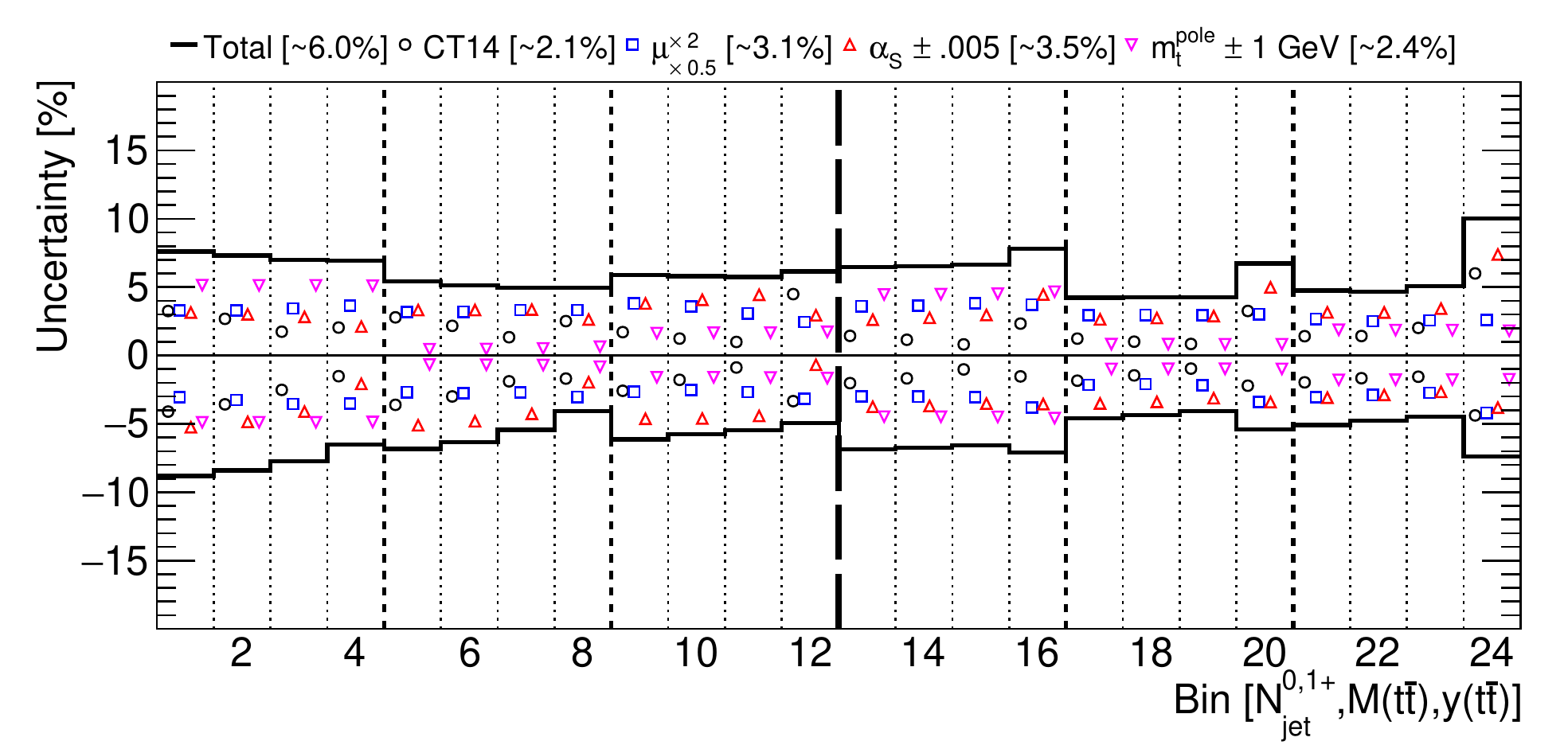}
    \includegraphics[width=1.00\textwidth]{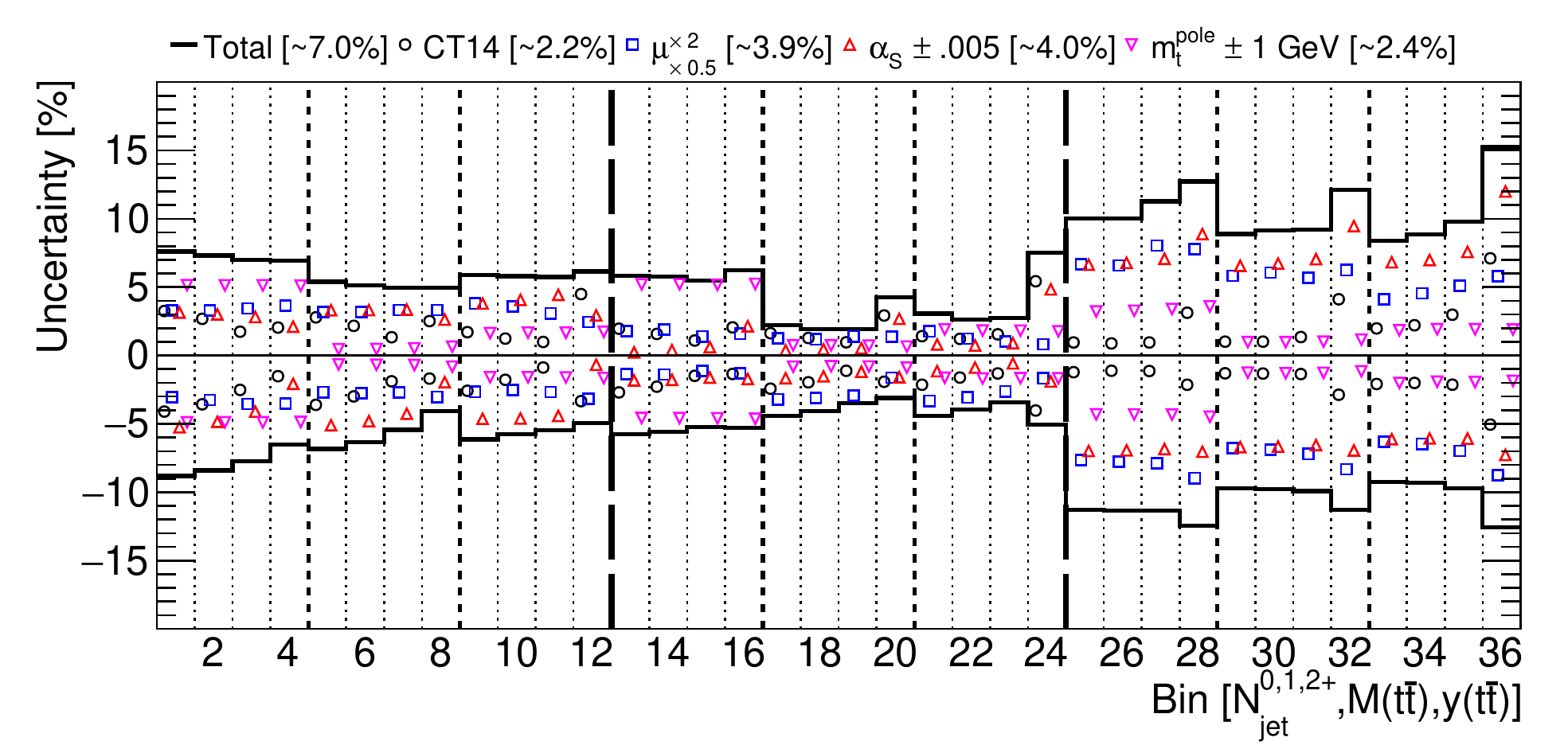}
    \caption{The theoretical uncertainties for \njmttytttwo (upper) and \njmttyttthree (lower) cross sections, arising from PDF, \asmz, and \mt variations,
    as well as the total theoretical uncertainties, with their bin-averaged values shown in brackets. The bins are the same as in Figs.~\ref{fig:xsec-mc-nj2mttytt} and \ref{fig:xsec-mc-nj3mttytt}.}
    \label{fig:thunc-nj23mttytt}
\end{figure*}

\begin{figure*}
    \centering
    {\includegraphics[width=1.00\textwidth]{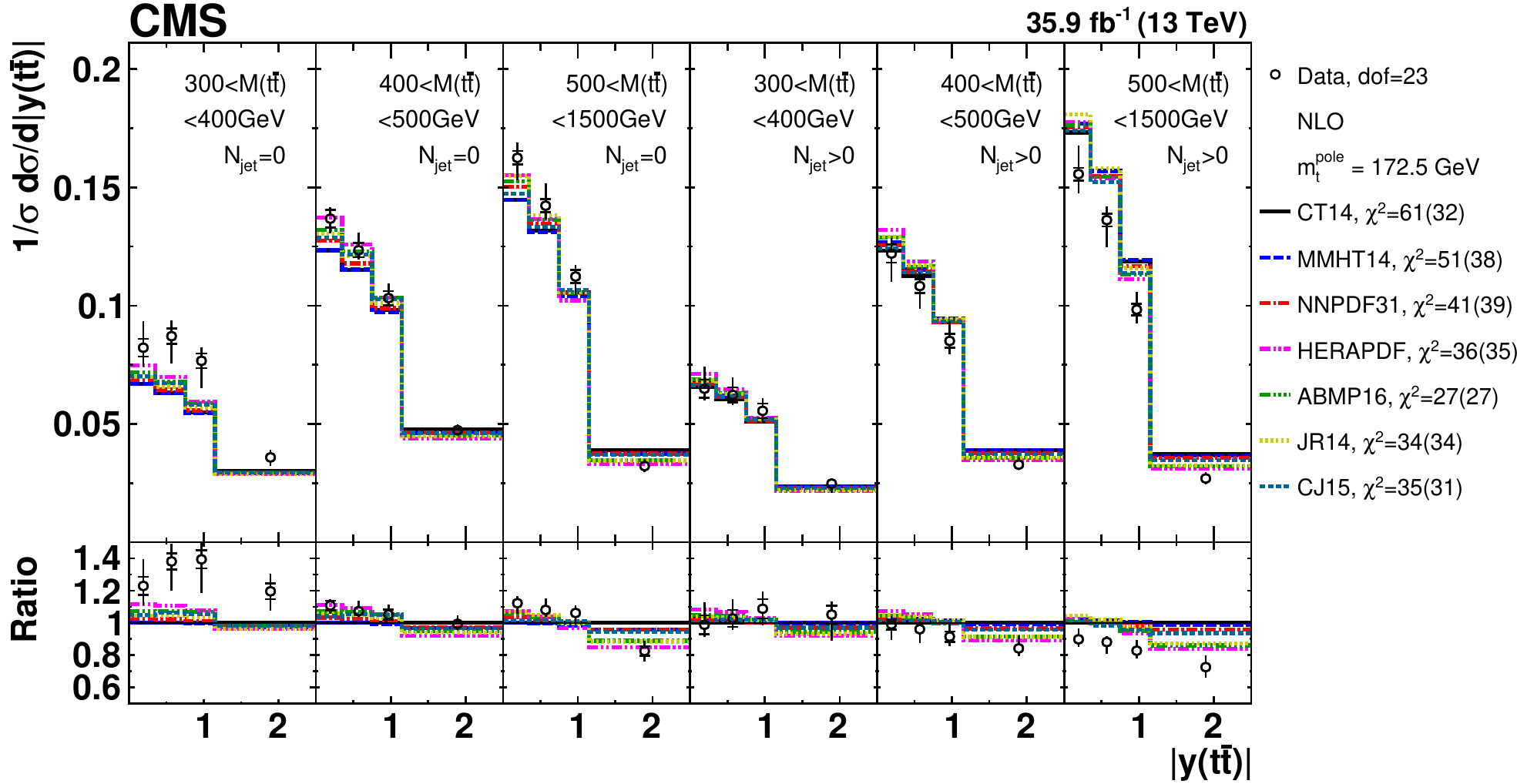}}
    \caption{Comparison of the measured \njmttytttwo cross sections to NLO predictions obtained using different PDF sets (further details can be found in Fig.~\ref{fig:xsec-mc-ytptt}).
    For each theoretical prediction, values of \chisq and \ndf for the comparison to the data are reported, while additional \chisq values that include PDF uncertainties are shown in parentheses.
    }
    \label{fig:xsec-nlo-nj2mttytt-pdfs}
\end{figure*}

\begin{figure*}
    \centering
    {\includegraphics[width=1.00\textwidth]{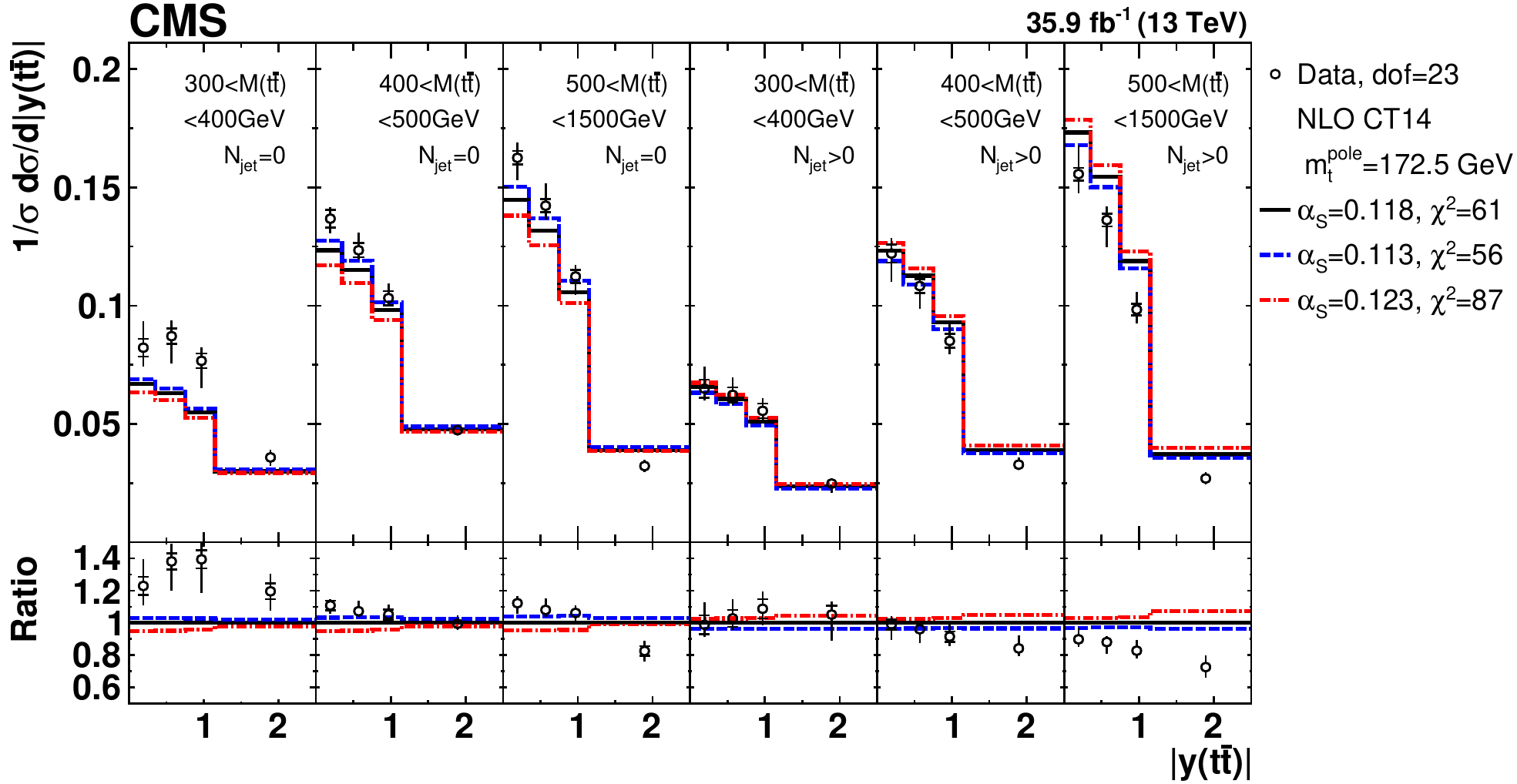}}
    \caption{Comparison of the measured \njmttytttwo cross sections to NLO predictions obtained using different $\asmz$ values (further details can be found in
    Fig.~\ref{fig:xsec-mc-ytptt}).
    For each theoretical prediction, values of \chisq and \ndf for the comparison to the data are reported. }
    \label{fig:xsec-nlo-nj2mttytt-as}
\end{figure*}

\begin{figure*}
    \centering
    {\includegraphics[width=1.00\textwidth]{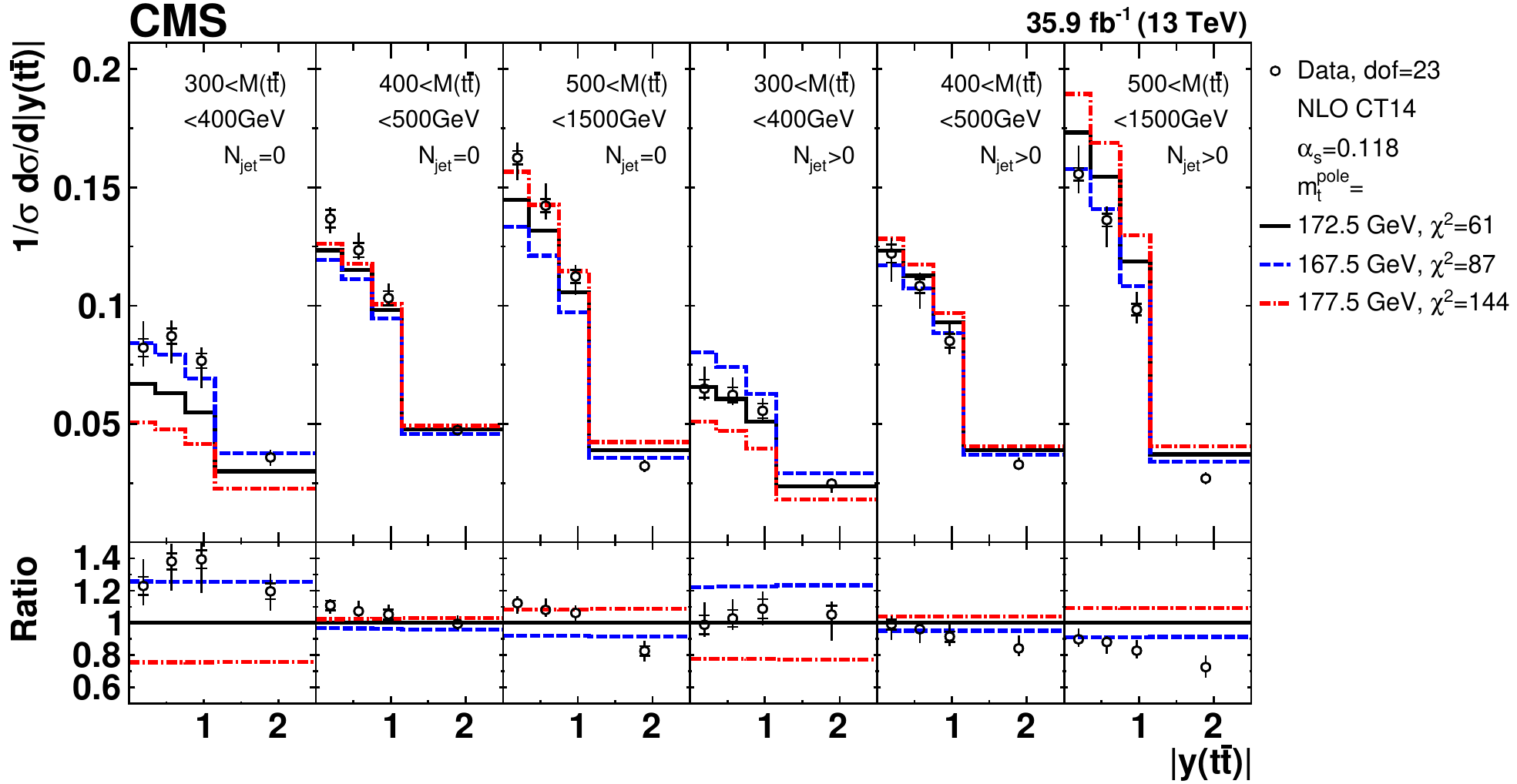}}
    \caption{Comparison of the measured \njmttytttwo cross sections to NLO predictions obtained using different \mt values (further details can be found in Fig.~\ref{fig:xsec-mc-ytptt}).
    For each theoretical prediction, values of \chisq and \ndf for the comparison to the data are reported. }
    \label{fig:xsec-nlo-nj2mttytt-mt}
\end{figure*}

{\tolerance=16000
The theoretical uncertainties for the \njmttytttwo and \njmttyttthree cross sections are illustrated in Fig.~\ref{fig:thunc-nj23mttytt}.
The CT14 PDF set with $\asmz = 0.118$, $\mt = 172.5\GeV$ is used as the nominal calculation. The contributions arising from the PDF, $\asmz$ ($\pm$0.005), and $\mt$ ($\pm 1$\GeV) uncertainties are shown separately. The total theoretical uncertainties are obtained by adding the
effects from PDF, \asmz, \mt, and scale variations in quadrature.
On average, the total theoretical uncertainties are 5--10\%. They receive similar contributions from PDF, \asmz, \mt, and scale variations.
This shows that the measured \njmttytttwo cross sections can be used for reliable and precise extraction of the PDFs and QCD parameters.
In this analysis the PDFs, $\asmz$, and \mt are extracted from the \njmttytttwo cross sections. These results are considered
to be the nominal ones and are checked by repeating the analysis using the \njmttyttthree cross sections.
\par}

In Figs.~\ref{fig:xsec-nlo-nj2mttytt-pdfs}--\ref{fig:xsec-nlo-nj2mttytt-mt}
the \njmttytttwo cross sections are compared to the predictions obtained
using different PDFs, $\asmz$, and \mt values.
For each comparison, a \chisq is calculated, taking into account the uncertainties of the data but ignoring uncertainties
of the predictions. For the comparison in Fig.~\ref{fig:xsec-nlo-nj2mttytt-pdfs},
additional \chisq values are determined, taking also PDF uncertainties in the predictions into account, i.e.\ Eq.~(\ref{eq:covmat}) becomes $\mathbf{Cov} = \mathbf{Cov}^\text{unf} + \mathbf{Cov}^\text{syst} + \mathbf{Cov}^\mathrm{PDF}$, where $\mathbf{Cov}^\mathrm{PDF}$ is a covariance matrix that accounts for the PDF uncertainties. Theoretical uncertainties from scale, \asmz, and $\mt$ variations are not included in this \chisq calculation.
Sizeable differences of the \chisq values
are observed for the predictions obtained using different PDFs.
These differences can be attributed to the different input data and methodologies that were used to extract these sets of PDFs as discussed elsewhere~\cite{Butterworth:2015oua,Accardi:2016ndt}.
Among the PDF sets considered, the best description of the data is provided by the ABMP16 PDFs.
This comparison also shows that the data prefer lower $\asmz$ and \mt value than in the nominal calculation using CT14.
The largest sensitivity to \mt is observed in the lowest \mtt region close to the threshold, while the sensitivity in the other \mtt bins occurs mainly because of the cross section normalisation.

The values of $\asmz$ and \mt are extracted by calculating a \chisq
between data and NLO predictions as a function of the input $\asmz$ or \mt value, and approximating the dependence with a parabola. The minimum of the parabola is taken as the extracted $\asmz$ or \mt value, while its uncertainty is estimated from the $\Delta\chi^2 = 1$ variation. This extraction is performed separately using different PDF sets, as well as different scale values. As for the additional \chisq values in Fig.~\ref{fig:xsec-nlo-nj2mttytt-pdfs}, the PDF uncertainties in the predictions are taken into account in these \chisq calculations.
Among the available PDF sets, only CT14, HERAPDF2.0, and ABMP16 provide PDF sets for enough different $\asmz$ values and are suitable for $\asmz$ extraction.
{Because the dependence of the measured \njmttytttwo cross sections on the \mtmc value (as well as on PDFs and \as) is much smaller than the sensitivity of the theoretical predictions to \mt (more details are given in Appendix~\ref{sec:appmtdep}), it is not taken into account in the extraction procedure.}

\begin{figure*}
    \centering
    \includegraphics[width=0.49\textwidth]{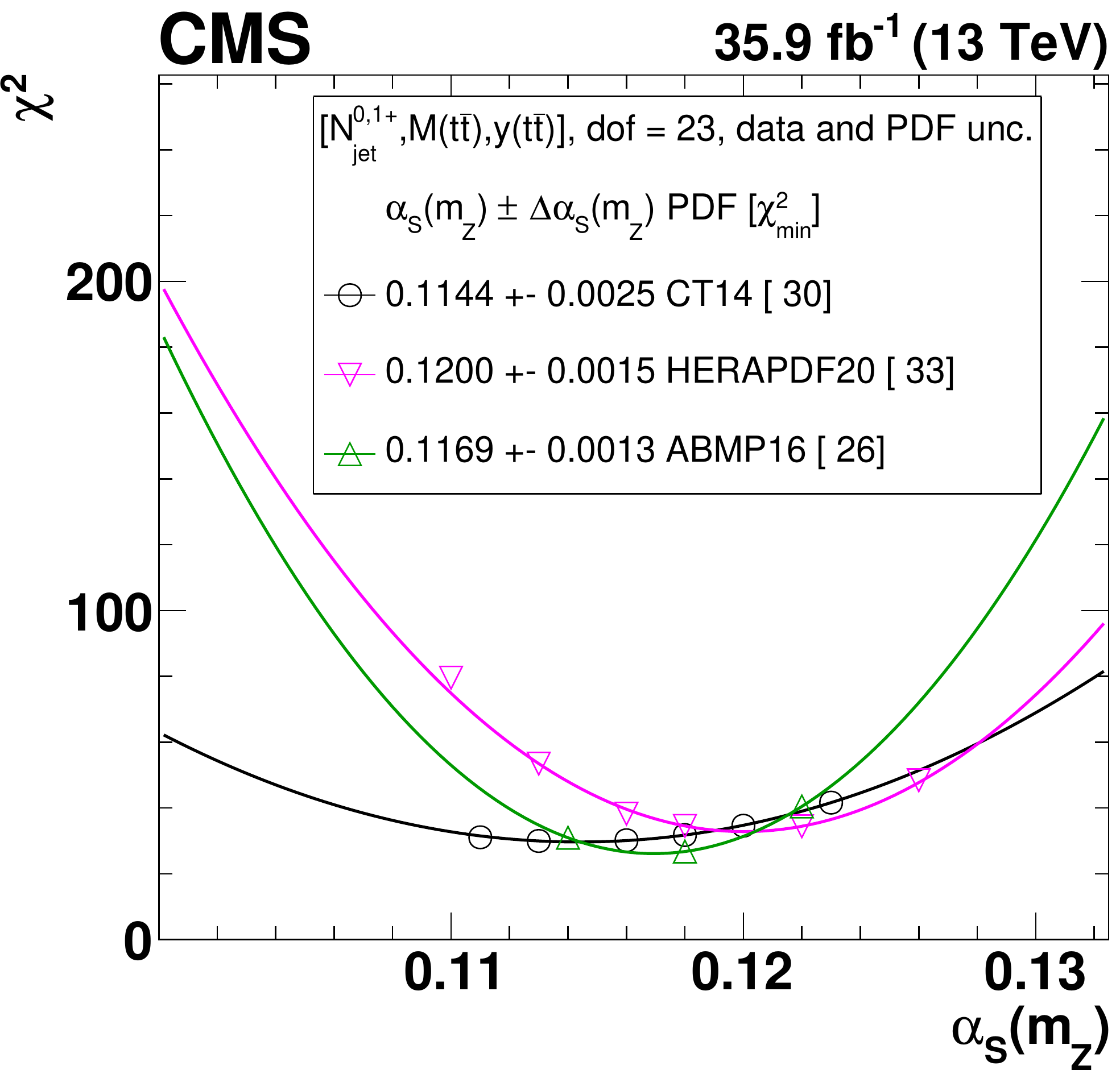}
    \includegraphics[width=0.49\textwidth]{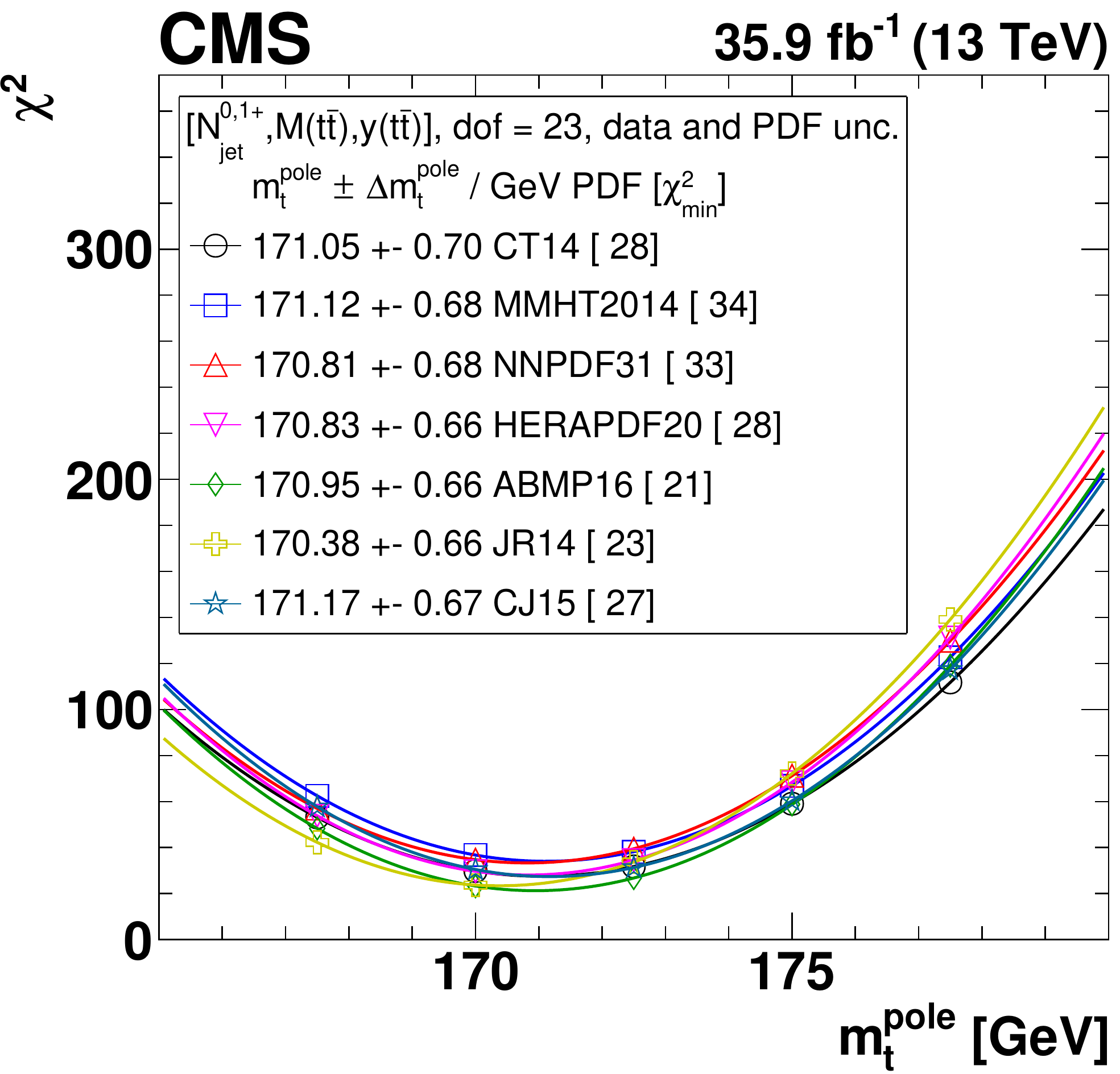}
    \caption{The $\asmz$ (left) and \mt (right) extraction at NLO from the measured \njmttytttwo cross sections using different PDF sets. The extracted $\asmz$ and \mt values are reported for each PDF set, and the estimated minimum \chisq value is shown in brackets. Further details are given in the text.}
    \label{fig:scan-asmt-nj2mttytt-pdfs}
\end{figure*}

The $\asmz$ and \mt scans for different PDF sets are shown in Fig.~\ref{fig:scan-asmt-nj2mttytt-pdfs}. The extracted $\asmz$ and \mt values are reported in the plots. Furthermore, the $\asmz$ (\mt) scans were performed using altered scale and $\mt$ (\asmz) settings and different PDF sets. For all input PDF sets, the impact of the scale variations is moderate and a weak positive correlation ($\sim$30\%) between $\asmz$ and $\mt$ is observed (the distributions are shown in Figs. \ref{fig:scan-as-nj2mttytt-varmumt} and \ref{fig:scan-mt-nj2mttytt-varmuas} in Appendix~\ref{sec:app}).

\begin{figure*}
    \centering
    \includegraphics[width=0.49\textwidth]{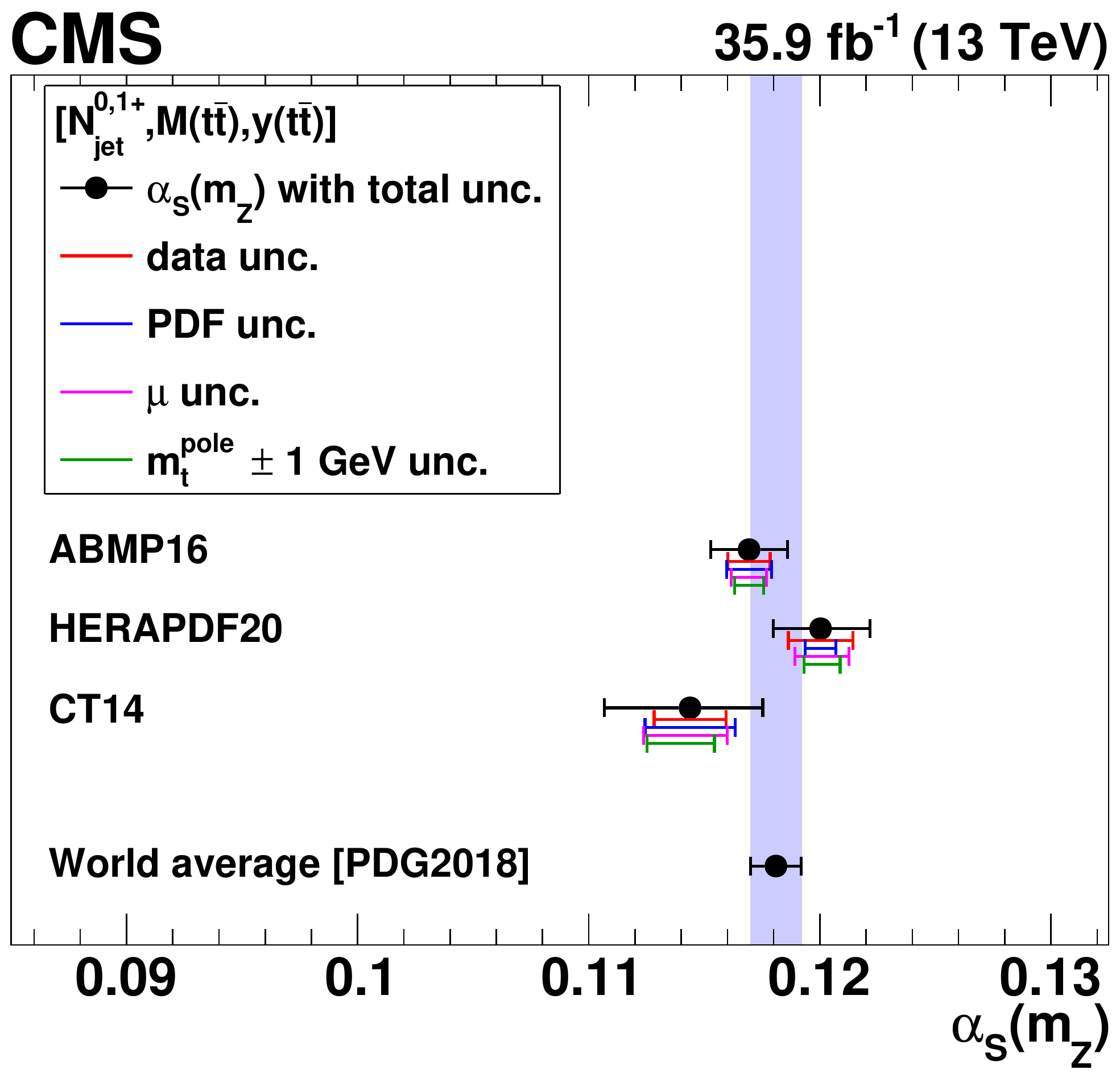}
    \includegraphics[width=0.49\textwidth]{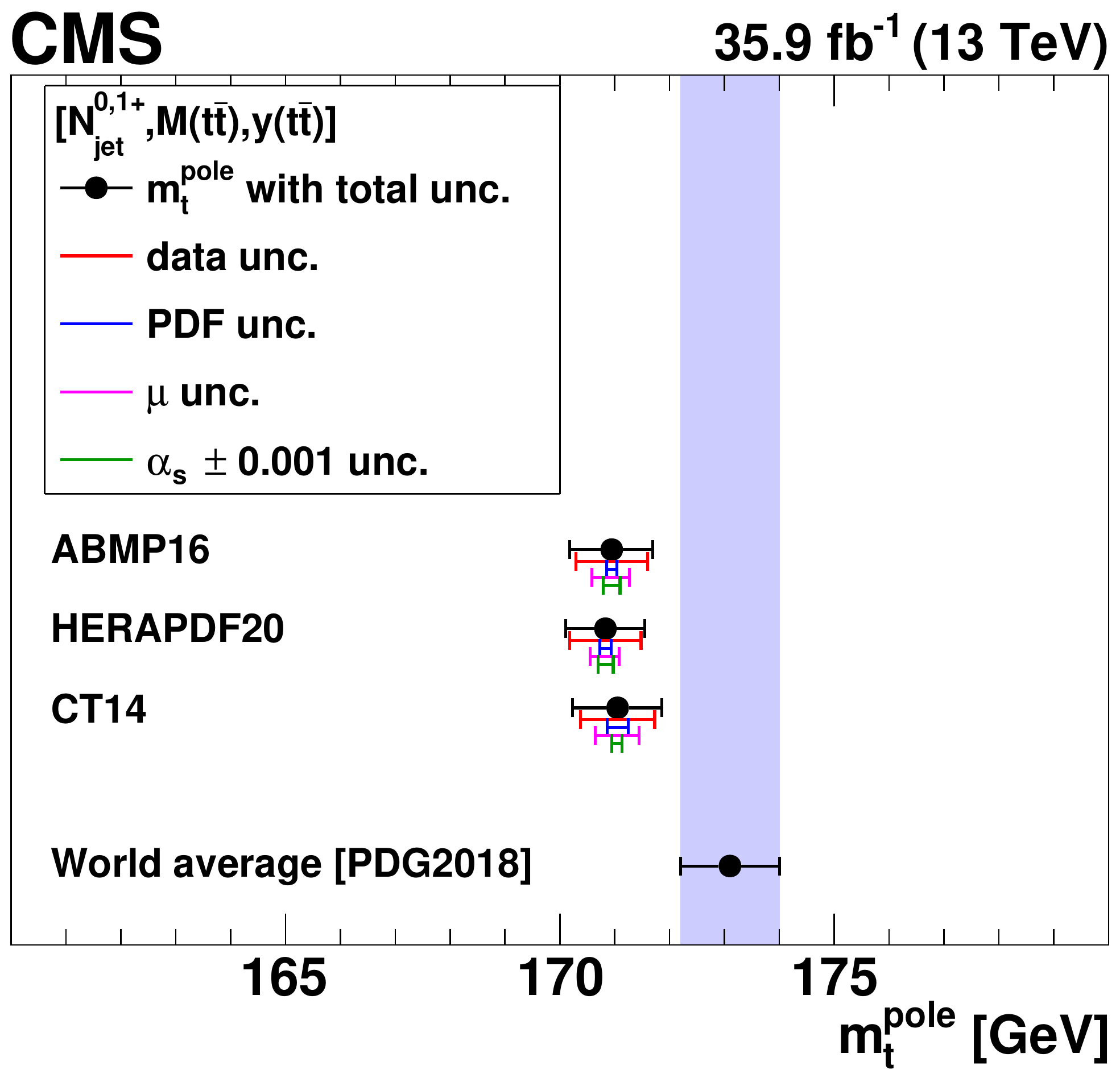}
    \caption{The \asmz (left) and $\mt$ (right) values extracted at NLO using different PDFs. 
The contributions to the total uncertainty arising from the data and from the theory prediction 
due to PDF, scale, and $\mt$ or $\asmz$ uncertainties are shown separately.
An additional theoretical uncertainty in the extracted $\mt$ (right) value 
of the order of $+1 \GeV$, due to missing
Coulomb and soft-gluon resummation near the \ttbar production threshold, is not shown.
The world average values $\asmz = 0.1181 \pm 0.0011$ and $\mt = 173.1 \pm 0.9\GeV$ from Ref.~\cite{pdg2018} are shown for reference.}
    \label{fig:fit-asmt-nj2mttytt}
\end{figure*}

The values of $\asmz$ and \mt, extracted at NLO, are compared in Fig.~\ref{fig:fit-asmt-nj2mttytt} to the world average~\cite{pdg2018} and reported in Tables~\ref{tab:aspdf} and \ref{tab:mtpdf}.
The contributions to the total uncertainty arising from the data and from the theoretical prediction due to PDF, scale, and $\mt$ or \asmz uncertainties are shown separately. For the extraction of \asmz, the experimental, PDF, scale, and $\mt$ uncertainties are comparable in magnitude. The size of the PDF uncertainties varies significantly for different PDF sets, and the extracted \asmz values depend on the input PDFs because of a strong correlation between $\as$ and the gluon distribution. This illustrates that precise and reliable $\asmz$ extractions from the observed data can be obtained only in a simultaneous PDF and $\asmz$ fit. For the \mt extraction, the total uncertainty is dominated by the data uncertainties. The world average~\cite{pdg2018} is computed based on extractions of \mt from inclusive \ttbar cross sections at NNLO+NNLL and differential distributions at NLO, and dominated by the inclusive cross section measurement and a measurement from leptonic distributions. For the combination, correlations were not taken into account. The world average \as value~\cite{pdg2018} is based on (at least) full NNLO QCD predictions.

\begin{table*}
    \topcaption{The \asmz values extracted at NLO using different PDFs, together with their fit, PDF, scale ($\mu$), and \mt uncertainties.}
    \label{tab:aspdf}
    \centering
  \renewcommand{\arraystretch}{1.4}
\centering
\begin{tabular}{ll}
PDF set & \asmz \\
\hline
       CT14&   $0.1144 \pm 0.0016(\textrm{fit}) \pm 0.0019(\textrm{PDF}) {}^{+0.0020}_{-0.0016}(\mu) {}^{+0.0019}_{-0.0011}(\mt)$ \\
  HERAPDF2.0&   $0.1200 \pm 0.0014(\textrm{fit}) \pm 0.0007(\textrm{PDF}) {}^{+0.0011}_{-0.0012}(\mu) {}^{+0.0007}_{-0.0009}(\mt)$ \\
     ABMP16&   $0.1169 \pm 0.0009(\textrm{fit}) \pm 0.0010(\textrm{PDF}) {}^{+0.0008}_{-0.0007}(\mu) {}^{+0.0006}_{-0.0006}(\mt)$ \\
\end{tabular}
\end{table*}

\begin{table*}
    \topcaption{The \mt values extracted at NLO using different PDFs, together with their fit, PDF, scale ($\mu$), and \as uncertainties.
An additional theoretical uncertainty of the order of $+1 \GeV$ due to missing
Coulomb and soft-gluon resummation near the \ttbar production threshold is not shown.
}
    \label{tab:mtpdf}
    \centering
  \renewcommand{\arraystretch}{1.4}
\centering
\begin{tabular}{ll}
PDF set & \mt [GeV] \\
\hline
       CT14&    $171.1 \pm  0.7(\textrm{fit}) \pm  0.2(\textrm{PDF}) {}^{+0.4}_{-0.4}(\mu) {}^{+0.1}_{-  0.1}(\as)$ \\
  HERAPDF2.0&   $170.8 \pm  0.6(\textrm{fit}) \pm  0.1(\textrm{PDF}) {}^{+0.3}_{-0.3}(\mu) {}^{+0.1}_{-  0.1}(\as)$ \\
     ABMP16&    $170.9 \pm  0.7(\textrm{fit}) \pm  0.1(\textrm{PDF}) {}^{+0.4}_{-0.3}(\mu) {}^{+0.2}_{-  0.2}(\as)$ \\
\end{tabular}
\end{table*}

Near the mass threshold, relevant for the \mt extraction, the {fixed-order} perturbation series should be improved with  
Coloumb and soft-gluon resummation that, however, is not available in the tools used to obtain theoretical predictions in this work. In Ref.~\cite{Kiyo:2008bv} these effects are found to be relevant only very close to the threshold (within a few GeV) and give a correction of about $+1\%$ to the total \ttbar cross section.
A more recent study for the total cross section shows that these corrections are presently known only with a large relative uncertainty~\cite{Piclum:2018ndt}.
Attributing a $+1\%$ total cross section correction to our first \mtt interval and assuming it to be independent of $\abs{\ytt}$ and \nj leads to an increase of the predicted cross section in each of the lowest \mtt bins by 5\%.
The effect of such a correction on the extraction of parameters has been tested 
for the simultaneous  PDF, $\as$, and \mt fit 
described in Section~\ref{sec:qcd}.  A shift of $+0.7 \GeV$ is observed for \mt with respect
to the nominal result listed in Eq.~\ref{eq:as-nom}, while shifts of $\as$ and PDFs are small.
No attempt was made to further quantify the effects on the
separate $\as$ and \mt extractions discussed in 
the present section.
In the future, theoretical calculations should include gluon resummation effects to accurately extract \mt from differential \ttbar cross sections.
For the time being one can assume an additional theoretical uncertainty in the performed \mt extraction of the order
of $+1 \GeV$ due to neglecting these effects. 

Furthermore, the impact of the parton shower was discussed in Ref.~\cite{Alioli:2013mxa}, where the predictions for $\ttbar+1$ jet production obtained at NLO and using $\POWHEG$ NLO calculations matched with the \PYTHIA parton shower have been compared (as shown in Fig.~1 of Ref.~\cite{Alioli:2013mxa})
and agreement between different approaches was found to be within 0.5\GeV for the extracted \mt.

Moreover, electroweak corrections can be significant in some regions of phase space~\cite{Kuhn:2013zoa,Czakon:2017wor},
but for the analysis in this paper no electroweak corrections were applied.
For inclusive \ttbar production in the kinematic region of this analysis,
electroweak corrections are calculated in ~\cite{Czakon:2017wor}
and found to be smaller than 2\% for any region of the \mtt distribution and smaller than 1\% for the \ytt distribution.

The \as and \mt extraction
is validated by repeating the procedure:
\begin{enumerate}
    \item \textit{Using single-differential \nj, \mtt, $\abs{\ytt}$ cross sections.} The plots are available in Appendix~\ref{sec:app}. The largest sensitivity to \as is observed when using the $\abs{\ytt}$ cross sections; the value for \as is, however, strongly dependent on the PDF set used. The \nj distribution provides a smaller \as sensitivity, but with little dependence on the PDFs. For \mt, the largest sensitivity is observed when using the \mtt cross sections. In fact, almost no sensitivity to \mt is present in $\abs{\ytt}$ or \nj single-differential cross sections. All determinations using the single-differential cross sections yielded \asmz and \mt values that are consistent with the nominal determination,

    \item \textit{Using triple-differential \njmttyttthree cross sections.} The distributions are available in Appendix~\ref{sec:app}. The extracted \asmz and \mt values are consistent with the nominal ones obtained using two $\nj$ bins and have similar precision with a slightly different uncertainty composition: smaller data uncertainties but larger scale uncertainties are present when using three \nj bins.
    The different uncertainties are expected since more \nj bins provide more sensitivity to \as, while the NLO theoretical prediction for the last \nj bins (two or more extra jets) have larger scale uncertainties compared to the other bins (as shown in Fig.~\ref{fig:thunc-nj23mttytt}).
This shows that NLO QCD predictions are able to describe \ttbar data with up to two hard extra jets, however higher-order calculations are desirable to match the experimental precision in order to achieve a most accurate \as and \mt determination.

    \item \textit{Using triple-differential \ptttmttytt cross sections with two \pttt bins.} The NLO calculations for inclusive \ttbar and $\ttbar+1$ jet production with an appropriate jet \pt threshold are used to describe the distribution in the two \pttt bins (see Appendix~\ref{sec:app:ptttmttytt} for further details). The extracted \asmz and \mt values (the plots are available in Appendix~\ref{sec:app:ptttmttytt}) are consistent with the nominal ones but have slightly larger experimental, PDF, and scale uncertainties compared to the nominal results based on the \njmttytttwo cross sections. Nevertheless, these results are an important cross-check, because the \ptttmttytt cross sections are provided at parton level and do not require non-perturbative corrections, which have to be applied for distributions involving \nj.

    \item \textit{Using unnormalised cross sections.} Consistent \asmz and \mt values are obtained, but with substantially larger experimental and scale uncertainties due to the increased scale dependence of the NLO predictions for the unnormalised cross sections and uncancelled normalisation uncertainties in the measured cross sections.

    \item {\tolerance=16000 \textit{Using NNLO/NLO factors as a function of  {\mtt}.} The ratios of NNLO over NLO calculations from Ref. \cite{Czakon:2016dgf} are used to multiply the NLO calculations. The NNLO/NLO corrections are obtained with the CT14, MMHT2014, and NNPDF3.0 PDF sets and $\mt=173.1$ GeV, and applied independently of \ytt or \nj (note that NNLO/NLO corrections for the \ytt distribution are generally smaller than for \mtt, and no NNLO corrections for the \nj distribution are available). The NNLO/NLO corrections for the \mtt bins used in this analysis do not exceed $2\%$, and the impact on the extracted \mt and \as values is $-0.2\GeV$, $-0.4\GeV$, $+0.01\GeV$, and $-0.0005$, $-0.0006$, $-0.0004$ using the CT14, MMHT2014, NNPDF3.0 PDFs, respectively, which is compatible with the uncertainties assigned to the NLO results. Because of the several assumptions explained above, this study should not be interpreted as an extraction of \mt and \as at NNLO, but only as a test of the scale uncertainties assigned to the NLO results.\par}

    \item \textit{Using a NLO calculation matched to parton showers.} The \PowPyt simulation with the NNPDF3.0 PDF set is used to determine the \mt value. The extracted value is lower by $0.4$ GeV than the nominal one determined using the NLO calculations, which is compatible with the uncertainties assigned to the NLO results.
\end{enumerate}

\section{Simultaneous PDF, \texorpdfstring{$\as$}{alphas}, and \texorpdfstring{$\mt$}{mt} fit}
\label{sec:qcd}

{\tolerance=1200
The triple-differential normalised \njmttytttwo cross sections are used in a simultaneous PDF, $\as$, and \mt fit at NLO (also referred to as a QCD analysis, or PDF fit), together with the combined HERA inclusive deep inelastic scattering (DIS) data~\cite{Abramowicz:2015mha}.
The \xfitter program (version 2.0.0)~\cite{Alekhin:2014irh}, an open-source QCD fit framework for PDF determination, is used.
The precise HERA DIS data, obtained from the combination of individual H1 and ZEUS results,
are directly sensitive to the valence and sea quark distributions and probe the gluon
distribution through scaling violations. Therefore, these data form the core of all PDF fits.
The measured \ttbar cross sections are included in the fit to constrain \as, \mt, and the gluon distribution at high values of $x$,
where $x$ is the fraction of the proton momentum carried by a parton.
The typical probed $x$ values can be estimated using the LO kinematic relation
\begin{equation}
x = \frac{\mtt}{\sqrt{s}}\re^{\pm y(\ttbar)}.
\end{equation}
The present measurement is expected to be mostly sensitive to $x$ values in the region $0.01 \lesssim x \lesssim 0.1$,
as estimated using the highest or lowest $\abs{\ytt}$ or \mtt bins and taking the low or high bin edge where the cross section is largest.
\par}

\subsection{Details of the QCD analysis}
\label{sec:qcdanalysis:details}
{\tolerance=1200
The scale evolution of partons is calculated through DGLAP
    equations~\cite{Dokshitzer:1977sg,Gribov:1972ri,Altarelli:1977zs,Curci:1980uw,Furmanski:1980cm,Moch:2004pa,Vogt:2004mw} at NLO,
    as implemented in the \qcdnum program~\cite{Botje:2010ay} (version 17.01.14).
    The Thorne--Roberts~\cite{Thorne:1997ga,Thorne:2006qt,Thorne:2012az} variable-flavour number scheme at NLO is used for the treatment of the heavy-quark contributions.
    The number of flavours is set to 5, with {\PQc} and {\cPqb} quark mass parameters $M_{\PQc}= 1.47$\GeV and $M_{\PQb} = 4.5$\GeV~\cite{Abramowicz:2015mha}.
    For the DIS data $\mu_\mathrm{r}$ and $\mu_\mathrm{f}$ are set to $Q$,
    which denotes the four-momentum transfer.
    The $Q^2$ range of the HERA data is restricted to $Q^2 > \qmin = 3.5\GeV^2$~\cite{Abramowicz:2015mha}.
    The theoretical predictions for the \ttbar cross sections are calculated as described in Section~\ref{sec:asmt}
    and are included in the fit using the \mg (version 2.6.0)~\cite{Alwall:2014hca} framework, interfaced with the \amcfast (version 1.3.0)~\cite{Bertone:2014zva} and \applgrid (version 1.4.70)~\cite{Carli:2010rw} programs.
    The \as and \mt are left free in the fit.
    Technically, ApplGrid tables have been produced for fixed values of \mt, while the theoretical predictions as a function of \mt were obtained by linear interpolation between two predictions using different \mt values. The results do not depend significantly on which particular \mt values are used for linear interpolation.
Consistent results were also obtained using a cubic spline interpolation.
\par}

{\tolerance=9600
The procedure for the determination of the PDFs follows the approach of \herapdf~\cite{Abramowicz:2015mha}.
    The parametrised PDFs are the gluon distribution $x\Pg(x)$, the valence quark distributions $x\cPqu_{\mathrm{v}}(x)$ and $x\cPqd_{\mathrm{v}}(x)$, and
    the $\cPqu$- and $\cPqd$-type antiquark distributions $x\overline{\mathrm{U}}(x)$ and $x\overline{\mathrm{D}}(x)$. At the initial QCD evolution scale $\mu_\mathrm{f0}^2 = 1.9\GeV^2$, the PDFs are parametrised as:\par}
        \begin{equation}\begin{aligned}
        x\Pg(x) &= A_{\Pg} x^{B_{\Pg}}\,(1-x)^{C_{\Pg}}\, (1 + E_{\Pg} x^2) - A'_{\Pg} x^{B'_{\Pg}}\,(1-x)^{C'_{\Pg}},\\
        x\cPqu_\mathrm{v}(x) &= A_{\cPqu_\mathrm{v}}x^{B_{\cPqu_\mathrm{v}}}\,(1-x)^{C_{\cPqu_\mathrm{v}}}\,(1+D_{\cPqu_\mathrm{v}}x) ,\\
        x\cPqd_\mathrm{v}(x) &= A_{\cPqd_\mathrm{v}}x^{B_{\cPqd_\mathrm{v}}}\,(1-x)^{C_{\cPqd_\mathrm{v}}},\\
        x\overline{\mathrm{U}}(x)&= A_{\overline{\mathrm{U}}}x^{B_{\overline{\mathrm{U}}}}\, (1-x)^{C_{\overline{\mathrm{U}}}}\, (1+D_{\overline{\mathrm{U}}}x), \\
        x\overline{\mathrm{D}}(x)&= A_{\overline{\mathrm{D}}}x^{B_{\overline{\mathrm{D}}}}\, (1-x)^{C_{\overline{\mathrm{D}}}},
        \end{aligned}
        \label{eq:dv}
        \end{equation}
    \tolerance=1200
    assuming the relations $x\overline{\mathrm{U}}(x) = x\cPaqu(x)$ and $x\overline{\mathrm{D}}(x) = x\cPaqd(x) + x\cPaqs(x)$.
    Here, $x\cPaqu(x)$, $x\cPaqd(x)$, and $x\cPaqs(x)$ are the up, down, and strange antiquark distributions, respectively.
    The sea quark distribution is defined as $x\Sigma(x)=x\cPaqu(x)+x\cPaqd(x)+x\cPaqs(x)$.
    The normalisation parameters $A_{\cPqu_{\mathrm{v}}}$, $A_{\cPqd_\mathrm{v}}$, and $A_{\cPg}$ are determined by the QCD sum rules.
    The $B$ and $B'$ parameters determine the PDFs at small $x$,
    and the $C$ parameters describe the shape of the distributions as $x\,{\to}\,1$.
    The parameter $C'_{\Pg}$ is fixed to 25 such that the term does not contribute at large $x$~\cite{Martin:2009iq,Abramowicz:2015mha}.
    Additional constraints $B_{\overline{\mathrm{U}}} = B_{\overline{\mathrm{D}}}$ and
    $A_{\overline{\mathrm{U}}} = A_{\overline{\mathrm{D}}}(1 - f_{\cPqs})$ are imposed to ensure the same normalisation
    for the $x\cPaqu$ and $x\cPaqd$ distributions as $x \to 0$.
    The strangeness fraction $f_{\cPqs} = x\cPaqs/( x\cPaqd + x\cPaqs)$ is fixed to
    $f_{\cPqs}=0.4$ as in the \herapdf analysis~\cite{Abramowicz:2015mha}.
    This value is consistent with the determination of the
    strangeness fraction when using the CMS measurements of $\PW$+${\PQc}$ production~\cite{Chatrchyan:2013mza}.

The $D$ and $E$ parameters are added for some distributions in order to provide a more flexible functional form.
The parameters in Eq.~(\ref{eq:dv}) are selected by first fitting with all $D$ and $E$ parameters set to zero,
and then including them independently one at a time in the fit.
The improvement in the $\chi^2$ of the fit is monitored
and the procedure is stopped when no further improvement is observed. This leads to a 15-parameter fit.
The $\chi^2$ definition used for the HERA DIS data follows that of Eq.~(32) in Ref.~\cite{Abramowicz:2015mha}.
It includes an additional logarithmic term that is relevant when the estimated statistical and uncorrelated systematic uncertainties in the data
are rescaled during the fit~\cite{Aaron:2012qi}.
For the \ttbar data presented here, a $\chi^2$ definition without such a logarithmic term is employed.
The treatment of the experimental uncertainties in the \ttbar double-differential cross section measurements follows the prescription given in Section~\ref{sec:results}.
The correlated systematic uncertainties are treated through nuisance parameters.
For each nuisance parameter
a Gaussian probability density function is assumed
and a corresponding penalty term is added to the \chisq.
The treatment of the experimental uncertainties for the HERA DIS data follows the prescription given
in Ref.~\cite{Abramowicz:2015mha}.

The uncertainties are estimated according to the general approach of \herapdf~\cite{Abramowicz:2015mha} in which
the fit, model, and parametrisation uncertainties are taken into account.
Fit uncertainties are determined using the criterion of $\Delta\chi^2 = 1$.
Model uncertainties arise from the variations in the values assumed for the {\PQc} quark mass parameter
of $1.41\leq M_{\PQc}\leq 1.53\GeV$,
the strangeness fraction $0.3 \leq f_{\cPqs} \leq 0.5$, and the value of $Q^2_{\text{min}}$ imposed on the HERA data.
The latter is varied within $2.5 \leq \qmin\leq 5.0\GeV^2$, following Ref.~\cite{Abramowicz:2015mha}.
The parametrisation uncertainty is estimated by varying the functional form in Eq.~(\ref{eq:dv})
of all parton distributions, with $D$ and $E$ parameters added or removed one at a time.
Additional parametrisation uncertainties are considered by using two other functional forms in Eq.~(\ref{eq:dv}): with $A'_{\Pg} = 0$ and $E_{\Pg} = 0$, since the $\chi^2$ in these variants of the fit are only a few units worse than that with the nominal parametrisation.
Furthermore, $\mu_\mathrm{f0}^2$ is changed from 1.9 to 1.6 and $2.2\GeV^2$.
The parametrisation uncertainty is constructed as an envelope, built from the maximal differences between the PDFs or QCD
parameters resulting from the central fit and all parametrisation variations.
For the PDFs, this uncertainty is valid in the $x$ range covered by the PDF fit to the data.
The total uncertainty is obtained by adding the fit, model, and parametrisation uncertainties in quadrature.
For \as and \mt extraction, the scale uncertainties in the theoretical predictions for \ttbar production are also considered.

A cross-check is performed using the MC method~\cite{Giele:1998gw,Giele:2001mr}.
It is based on analysing
a large number of pseudo-data sets called replicas. For this cross-check, 1000 replicas are
created by taking the data and fluctuating the values of the cross sections
randomly within their statistical and systematic uncertainties taking correlations into account.
All uncertainties are assumed to follow Gaussian distributions. The central values for the fitted parameters
and their uncertainties are estimated using the mean and RMS values over the replicas.
The obtained values of the PDF parameters, \asmz, and \mt and their fit uncertainties are in agreement with the nominal results.

\subsection{Fit results}
The resulting values of $\asmz$ and \mt extracted using NLO calculations are:
\ifthenelse{\boolean{cms@external}}
{
\begin{equation}
\begin{aligned}
 \asmz =& 0.1135 \pm 0.0016\,(\text{fit}) {}^{+0.0002}_{-0.0004}\,(\text{model}) \\
  &{}^{+0.0008}_{-0.0001}\,(\text{param}) {}^{+0.0011}_{-0.0005}\,(\text{scale})\\
  =&  0.1135{}^{+0.0021}_{-0.0017}\,(\text{total}),\\
 \mt =& 170.5 \pm 0.7\,(\text{fit}) \pm 0.1\,(\text{model}) {}^{+0.0}_{-0.1}\,(\text{param})\\
 & \pm 0.3\,(\text{scale})\GeV \\
 =& 170.5 \pm 0.8 (\text{total})\GeV.
 \label{eq:as-nom}
\end{aligned}
\end{equation}
}
{
\begin{equation}
\begin{aligned}
 \asmz &= 0.1135 \pm 0.0016 (\text{fit}) {}^{+0.0002}_{-0.0004} (\text{model}) {}^{+0.0008}_{-0.0001} (\text{param}) {}^{+0.0011}_{-0.0005} (\text{scale}) = 0.1135{}^{+0.0021}_{-0.0017} (\text{total}),\\
 \mt &= 170.5 \pm 0.7 (\text{fit}) \pm 0.1 (\text{model}) {}^{+0.0}_{-0.1} (\text{param}) \pm 0.3 (\text{scale})\GeV = 170.5 \pm 0.8 (\text{total})\GeV.
 \label{eq:as-nom}
\end{aligned}
\end{equation}
}
Here `fit', `model' and `param' denote the fit, model and parameter uncertainties discussed above.
The uncertainties arising from the scale variations are estimated by repeating the fit with altered values of the scales as described in Section~\ref{sec:asmt} and taking the differences with respect to the nominal result. The individual contributions to the uncertainties are listed in Table~\ref{tab:asmtunc}. The extracted \asmz and \mt values have only weak positive correlation $\rho(\asmz,\mt) = 0.3$, where the correlation was obtained from the data uncertainties propagated to the fit.
This shows that the two SM parameters can be simultaneously determined from these data to high precision with only weak correlation between them.
As discussed in Section~\ref{sec:asmt}, one expects an additional theoretical uncertainty 
in the extracted \mt value of the order of $+1 \GeV$ due to gluon resummation corrections
that are missing in the NLO calculation.

\begin{table*}
    \topcaption{The individual contributions to the uncertainties for the \asmz and \mt determination.}
    \label{tab:asmtunc}
    \centering
  \renewcommand{\arraystretch}{1.4}
\centering
\begin{tabular}{lccc}
    Parameter               &         Variation          &                     $\asmz$ & $\mt$ [\GeVns{}] \\
    \hline
    \multicolumn{4}{c}{Fit uncertainty}                      \\
    Total                   &      $\Delta\chi^2=1$      &              $\pm 0.0016$ & $\pm 0.7$ \\ [\cmsTabSkip]
    \multicolumn{4}{c}{Model uncertainty}                \\
    $f_\PQs$                   & $f_\PQs = 0.5$     &              ${+0.0001}$ & ${~~~0.0}$ \\
    $f_\PQs$                   & $f_\PQs = 0.3$     &              ${~~~0.0000}$ & ${~~~0.0}$ \\
    $Q^2_{\text{min}}$      & $Q^2_{\text{min}} = 5.0\GeV^2$ &        ${+0.0002}$ & ${+0.1}$ \\
    $Q^2_{\text{min}}$      & $Q^2_{\text{min}} = 2.5\GeV^2$ &        ${-0.0004}$ & ${-0.1}$ \\
    $M_{\PQc}$   &        $M_c = 1.49\GeV$ & ${+0.0001}$ & ${~~~0.0}$ \\
    $M_{\PQc}$   &        $M_c = 1.37\GeV$ & ${~~~0.0000}$ & ${~~~0.0}$ \\
    Total                   &                            &              $^{+0.0002}_{-0.0004}$ & $^{+0.1}_{-0.1}$ \\ [\cmsTabSkip]
    \multicolumn{4}{c}{PDF parametrisation uncertainty}  \\
    $\mu_\mathrm{f,0}^2$                 & $\mu_\mathrm{f,0}^2 = 2.2\GeV^2$ &        ${-0.0001}$ & ${~~~0.0}$ \\
    $\mu_\mathrm{f,0}^2$                 & $\mu_\mathrm{f,0}^2 = 1.6\GeV^2$ &        ${+0.0002}$ & ${~~~0.0}$ \\
    $A'_{\Pg}$               &        set to 0         &           { $+0.0002$} & { $-0.1$} \\
    $E_{\Pg}$               &        set to 0         &           { $+0.0008$} & {$~~~0.0$} \\
    Total                   &                            &              $^{+0.0008}_{-0.0001}$ & $-0.1$ \\ [\cmsTabSkip]
    \multicolumn{4}{c}{Scale uncertainty}  \\
    $\mu_\mathrm{r}$ variation                 & $\mu_\mathrm{r} = H'$ &        $+0.0004$ & $-0.2$ \\
    $\mu_\mathrm{r}$ variation                 & $\mu_\mathrm{r} = H'/4$ &        $+0.0007$ & $+0.1$ \\
    $\mu_\mathrm{f}$ variation                 & $\mu_\mathrm{f} = H'$ &        $-0.0002$ & $+0.3$ \\
    $\mu_\mathrm{f}$ variation                 & $\mu_\mathrm{f} = H'/4$ &        $+0.0001$ & $-0.3$ \\
    $\mu_\mathrm{r,f}$ variation               & $\mu_\mathrm{r,f} = H'$ &        $+0.0004$ & $+0.1$ \\
    $\mu_\mathrm{r,f}$ variation               & $\mu_\mathrm{r,f} = H'/4$ &        $+0.0011$ & $-0.2$ \\
    alternative $\mu_\mathrm{r,f}$                 & $\mu_\mathrm{r,f} = H/2$ &         $-0.0005$ & $+0.1$ \\
    Total                   &                            &              $^{+0.0011}_{-0.0005}$ & $^{+0.3}_{-0.3}$ \\
\end{tabular}
\end{table*}

\begin{figure*}
    \centering
    \includegraphics[width=1.00\textwidth]{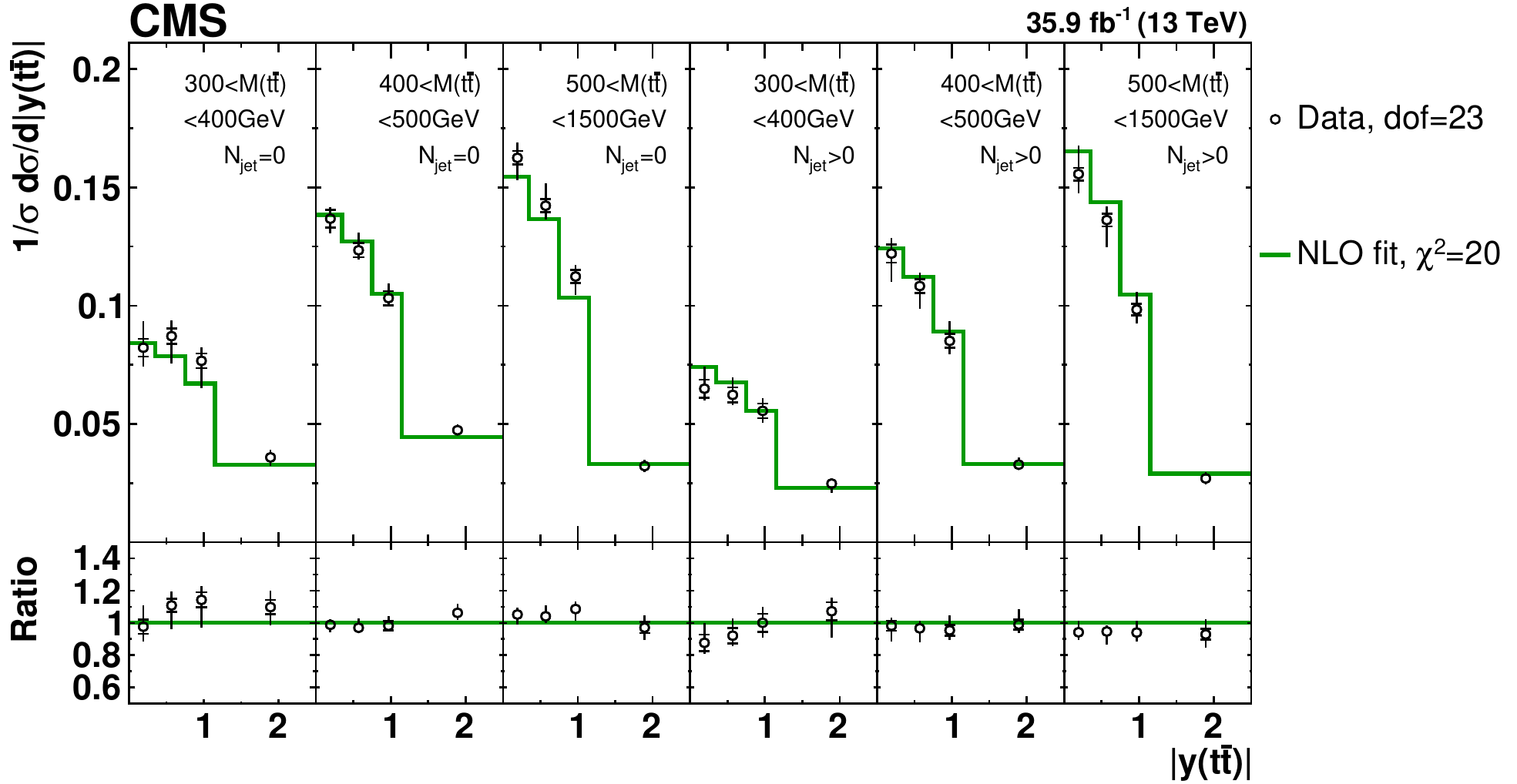}
    \caption{Comparison of the measured \njmttytttwo cross sections to the NLO predictions using the parameter values from the simultaneous PDF, \as, and \mt fit (further details can be found in Fig.~\ref{fig:xsec-mc-ytptt}).
        Values of \chisq and \ndf are reported.}
    \label{fig:bestfit}
\end{figure*}

\begin{figure*}
    \centering
    \includegraphics[width=1.00\textwidth]{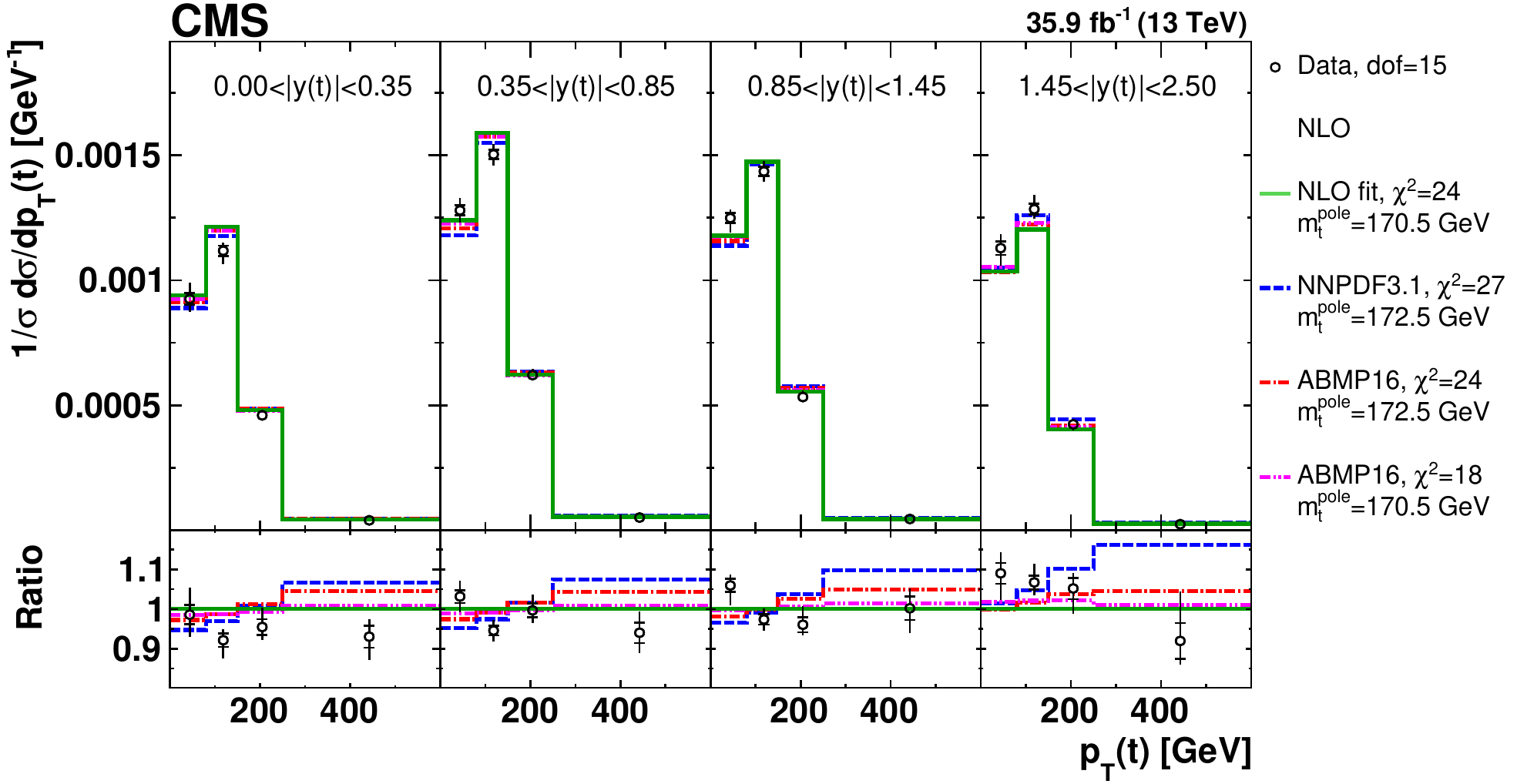}
    \caption{Comparison of the measured \ytptt cross sections to the NLO predictions using the parameter values from the simultaneous PDF, \as and \mt fit of the \njmttytttwo cross sections, as well as the predictions obtained using the NNPDF3.1 and ABMP16 PDF sets with different values of \mt (see Fig.~\ref{fig:xsec-mc-ytptt} for further details).
        In the lower panel, the ratios of the data and theoretical predictions to the predictions from the fit are shown.
        For each theoretical prediction, values of \chisq and \ndf for the comparison to the data are reported.}
    \label{fig:ytptt-topptproblem}
\end{figure*}

The global and partial \chisq values of the fit 
are listed in Table~\ref{tab:chi2fit}, illustrating the consistency of the input data with the fit model.
In particular, the \ttbar data are well described in the fit.
The DIS data show \chisqndf values slightly larger than unity,
similar to what is observed and investigated in Ref.~\cite{Abramowicz:2015mha}.
For the \ttbar data, the full \chisq (including uncorrelated and correlated data uncertainties) is 20 for 23 degrees of freedom.
The \ttbar cross sections are compared to the NLO predictions obtained after the fit in Fig.~\ref{fig:bestfit}.
{Furthermore, in Fig.~\ref{fig:ytptt-topptproblem} the \ytptt cross sections (which were not used in the fit) are compared to NLO predictions obtained using the fitted PDFs, \as and \mt, as well as other global PDF sets. The data are in satisfactory agreement with the predictions obtained in this analysis. In particular, these predictions or predictions obtained using the ABMP16 PDF set describe the slope of \ptt considerably better than the predictions obtained using the NNPDF3.1 PDF set, while the difference in the \chisq values is less significant.
Additionally, the predicted \ptt slope is sensitive to the \mt values used in the calculations.}

\begin{table*}
    \topcaption{The global and partial \chisqndf values for all variants of the QCD analysis.
        The variant of the fit that uses the HERA DIS only is denoted as `Nominal fit'.
        For the HERA measurements, the energy of the proton beam, $E_{\Pp}$, is listed for each data set, with the electron energy being $E_{\Pe}=27.5\GeV$,
        CC and NC standing for charged and neutral current, respectively.
        The correlated \chisq and the log-penalty $\chi^2$ entries refer to the \chisq contributions
        from the nuisance parameters and from the logarithmic term, respectively, as described in the text.
    }
    \label{tab:chi2fit}
    \renewcommand{\arraystretch}{1.4}
    \centering
    \begin{tabular}{lcc}
        \multirow{2}{*}{Data sets} & \multicolumn{2}{c}{\chisqndf} \\
        \cline{2-3}
        & Nominal fit & +\njmttytttwo  \\
        \hline
        CMS \ttbar     &  & $10 / 23$  \\
        CMS \ttbar\/ Correlated $\chi^2$    &  & $10$  \\
        HERA CC $\Pem\Pp$, $E_{\Pp}=920$\GeV & $55 / 42$& $55 / 42$   \\
        HERA CC $\Pep\Pp$, $E_{\Pp}=920$\GeV & $38 / 39$& $39 / 39$   \\
        HERA NC $\Pem\Pp$, $E_{\Pp}=920$\GeV & $218 / 159$& $217 / 159$  \\
        HERA NC $\Pep\Pp$, $E_{\Pp}=920$\GeV & $438 / 377$& $448 / 377$  \\
        HERA NC $\Pep\Pp$, $E_{\Pp}=820$\GeV & $70 / 70$& $71 / 70$  \\
        HERA NC $\Pep\Pp$, $E_{\Pp}=575$\GeV & $220 / 254$& $222 / 254$  \\
        HERA NC $\Pep\Pp$, $E_{\Pp}=460$\GeV & $219 / 204$& $220 / 204$ \\ [\cmsTabSkip]
        HERA Correlated $\chi^2$  & 82 & 80 \\
        HERA Log-penalty $\chi^2$  & $+2$& $-7$ \\ [\cmsTabSkip]
        Total \chisqndf & $1341 / 1130$& $1364 / 1151$ \\
    \end{tabular}	
\end{table*}

\begin{figure*}
     \centering
     \includegraphics[width=0.75\textwidth]{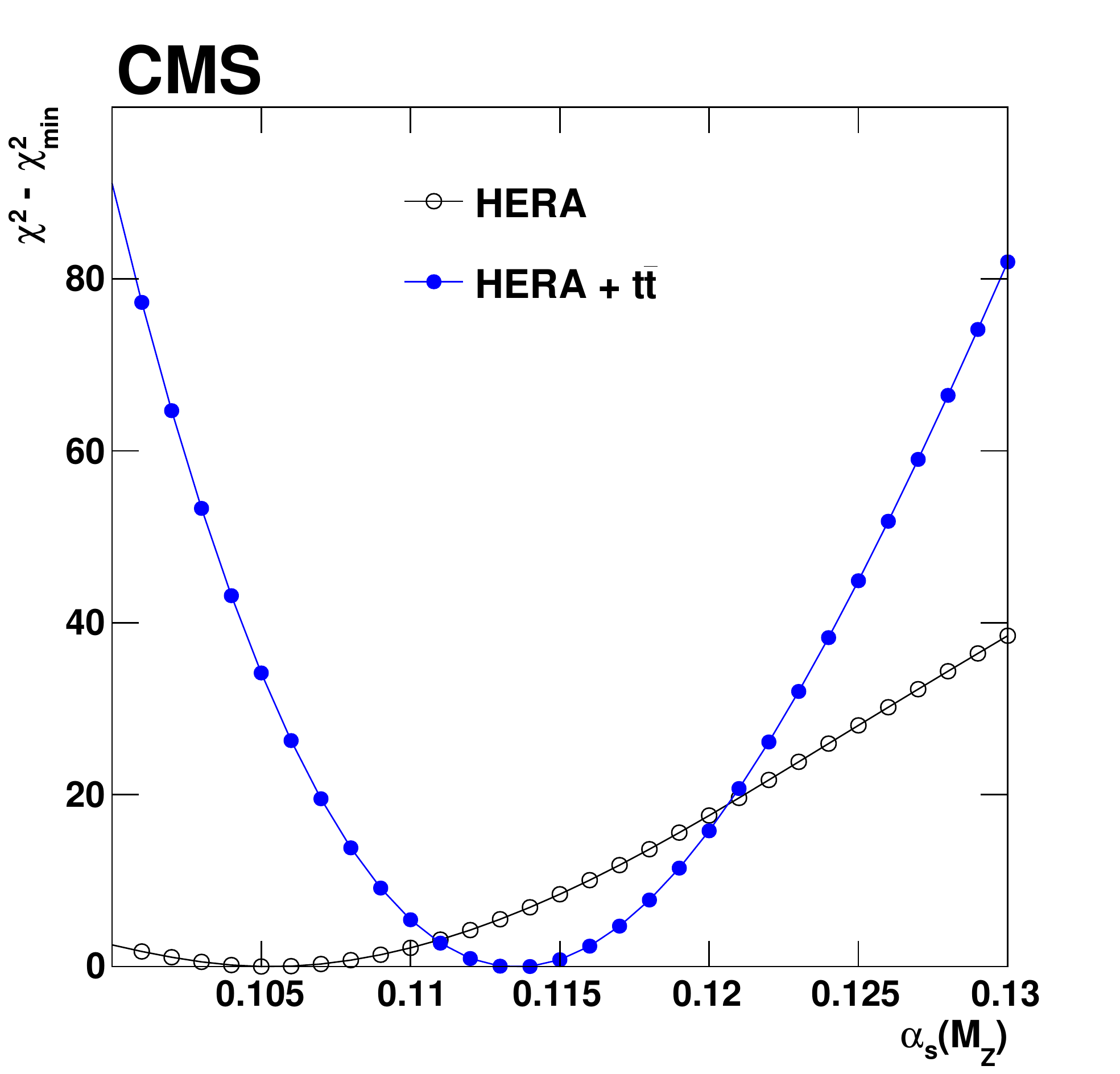}
     \caption{$\Delta \chi^2 = \chi^2 - \chi^2_{\text{min}}$ as a function of \asmz in the QCD analysis using the HERA DIS data only, or HERA and \ttbar data.}
     \label{fig:scanas}
 \end{figure*}

Fits were performed for a series of \asmz values ranging from $\asmz = 0.100$ to $\asmz = 0.130$ using only HERA DIS data, or HERA and $\ttbar$ data. The results are shown in Fig.~\ref{fig:scanas}.
A shallow \chisq dependence on \asmz is present when using only the HERA DIS data, similar to the findings of the HERAPDF2.0 analysis~\cite{Abramowicz:2015mha}.
Once the \ttbar data are included in the fit, a distinctly sharper minimum in \chisq is observed which coincides with the one found in the simultaneous PDF and \asmz fit {given in Eq.~(\ref{eq:as-nom}).}

Both the \ttbar and the HERA DIS data are sensitive to the \asmz value in the fit: while the constraints from the \ttbar data seem to be dominant, the residual dependence of \asmz on the HERA DIS data may remain nonnegligible. There is no way to assess the latter quantitatively because the HERA DIS data cannot be removed from the PDF fit. However, as was investigated in the HERAPDF2.0 analysis~\cite{Abramowicz:2015mha}, when using only HERA DIS the minima are strongly dependent on the \qmin threshold. As a cross-check, the extraction of \asmz was repeated for a larger threshold variation $2.5 \leq \qmin\leq 30.0\GeV^2$. In contrast to the results of Ref.~\cite{Abramowicz:2015mha} obtained using only HERA DIS data, when adding the \ttbar data the extracted values of \asmz show no systematic dependence on \qmin and are consistent with the nominal result of Eq.~(\ref{eq:as-nom}) within the total uncertainty.

    To demonstrate the added value of the \ttbar cross sections, the QCD analysis is first performed using only the HERA DIS data.
    In this fit, $\asmz$ is fixed to the value extracted from the fit using the \ttbar data, $\asmz = 0.1135$, and the
$\asmz$ uncertainty of $\pm 0.0016$ is added to the fit uncertainties.
    Then the \njmttytttwo measurement is added to the fit.
    The global and partial \chisq values for the two variants of the fit are listed in Table~\ref{tab:chi2fit}.

\begin{figure*}
    \centering
    \includegraphics[width=0.49\textwidth]{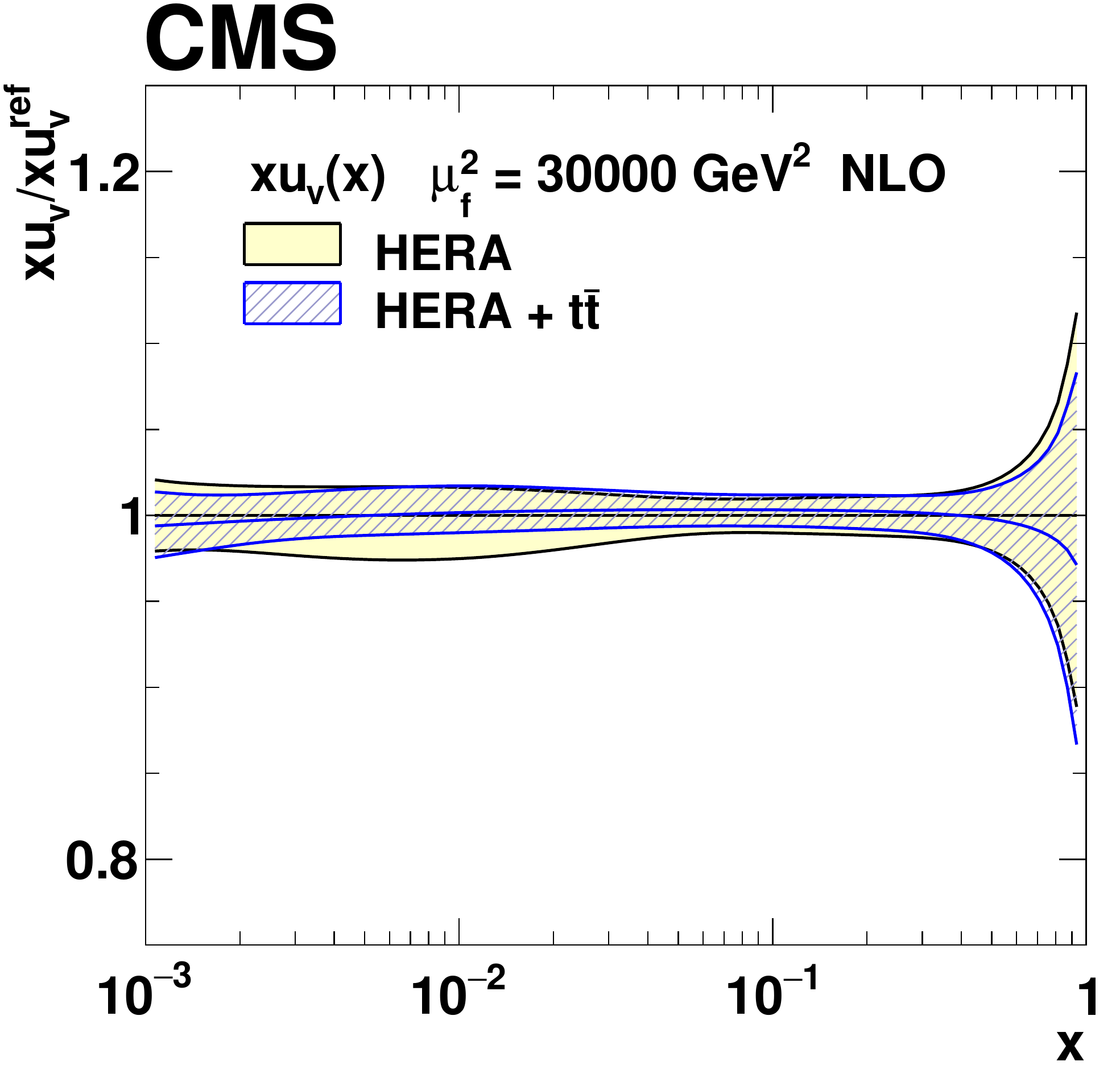}
    \includegraphics[width=0.49\textwidth]{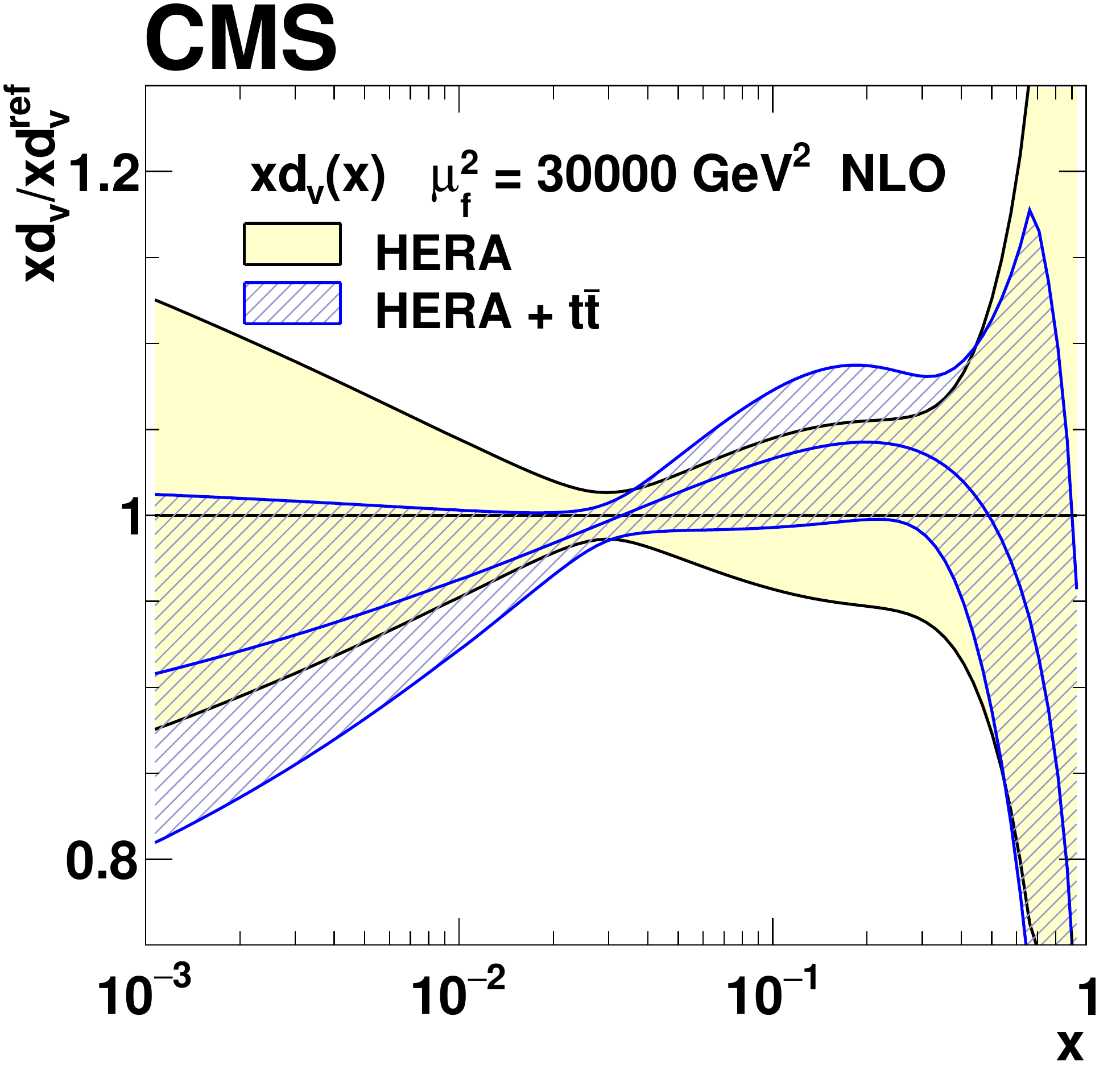}\\
    \includegraphics[width=0.49\textwidth]{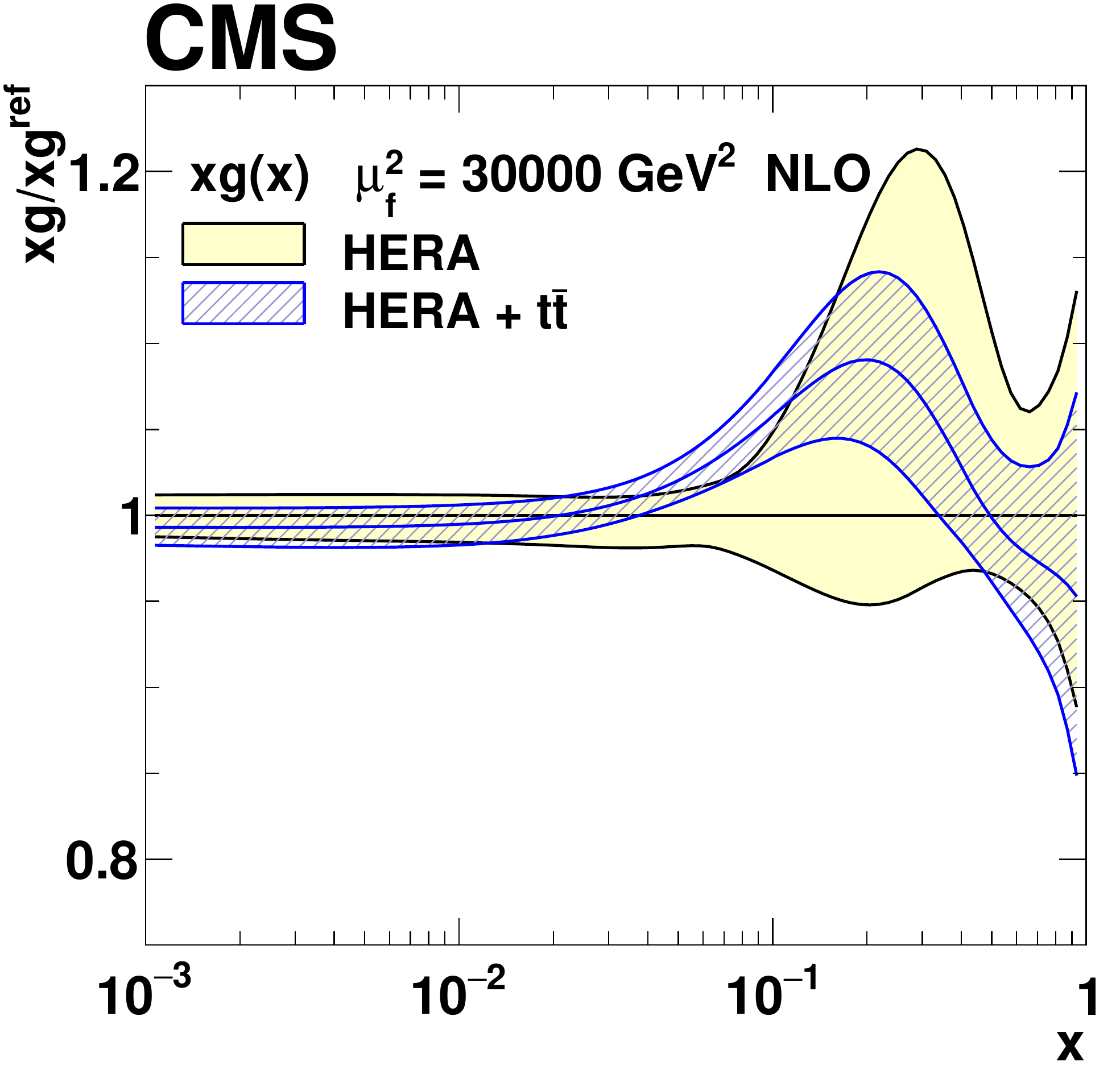}
    \includegraphics[width=0.49\textwidth]{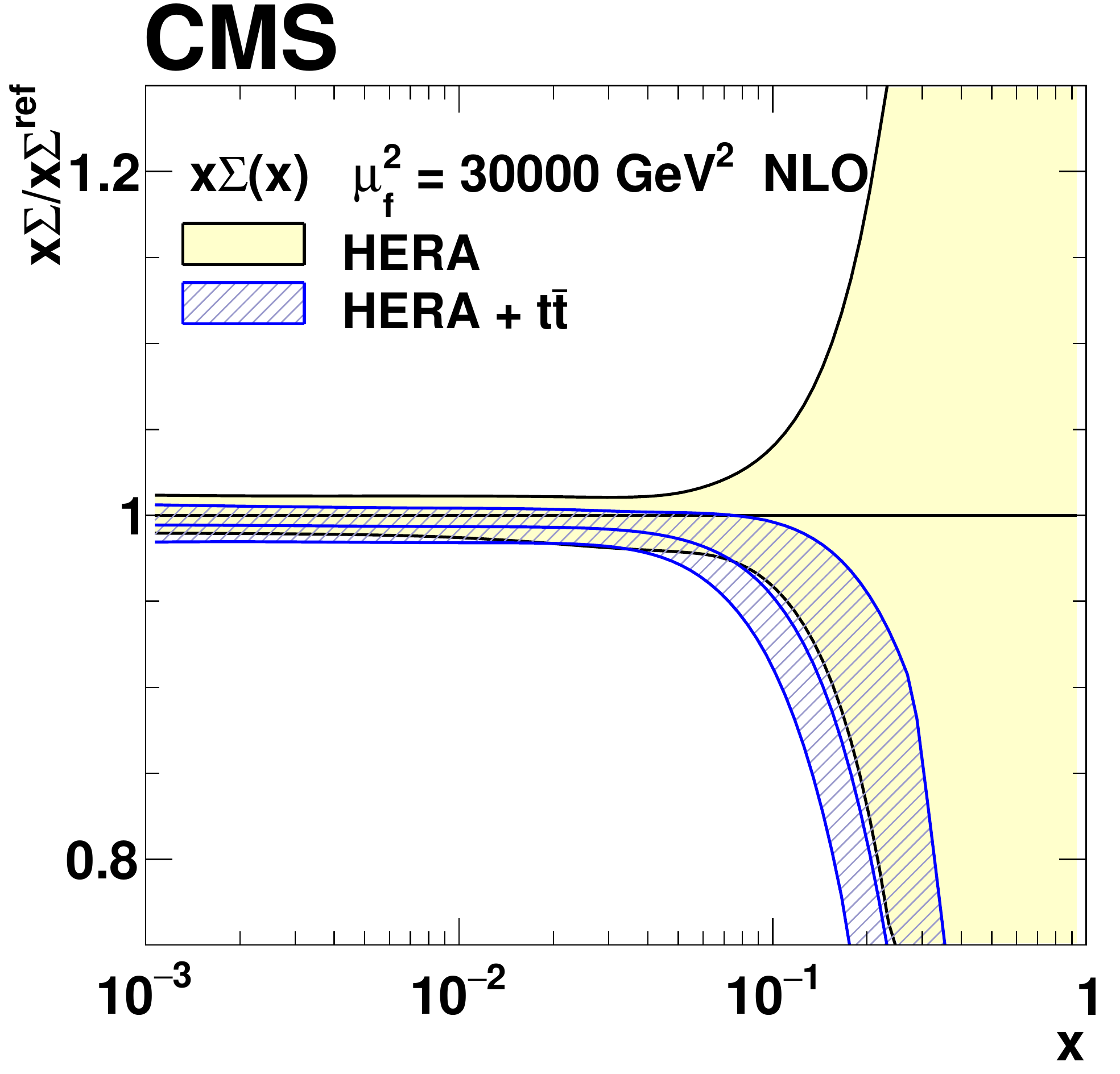}
    \caption{The PDFs with their total uncertainties in the fit using the HERA DIS data only, and the HERA DIS and \ttbar data. The results are normalised to the PDFs obtained using the HERA DIS data only.}
    \label{fig:pdfs-bands}
    \end{figure*}

    The corresponding PDFs are compared in Fig.~\ref{fig:pdfs-bands}. The largest impact of the \ttbar data is observed at $x \gtrsim 0.1$. In this region the gluon distribution lacks direct constraints in the fit using the HERA DIS data only. {The impact on the valence and sea quark PDFs is expected because of the correlations between the different distributions in the fit arising in the PDF evolution and from the momentum sum rule.}

\begin{figure*}
    \centering
    \includegraphics[width=0.49\textwidth]{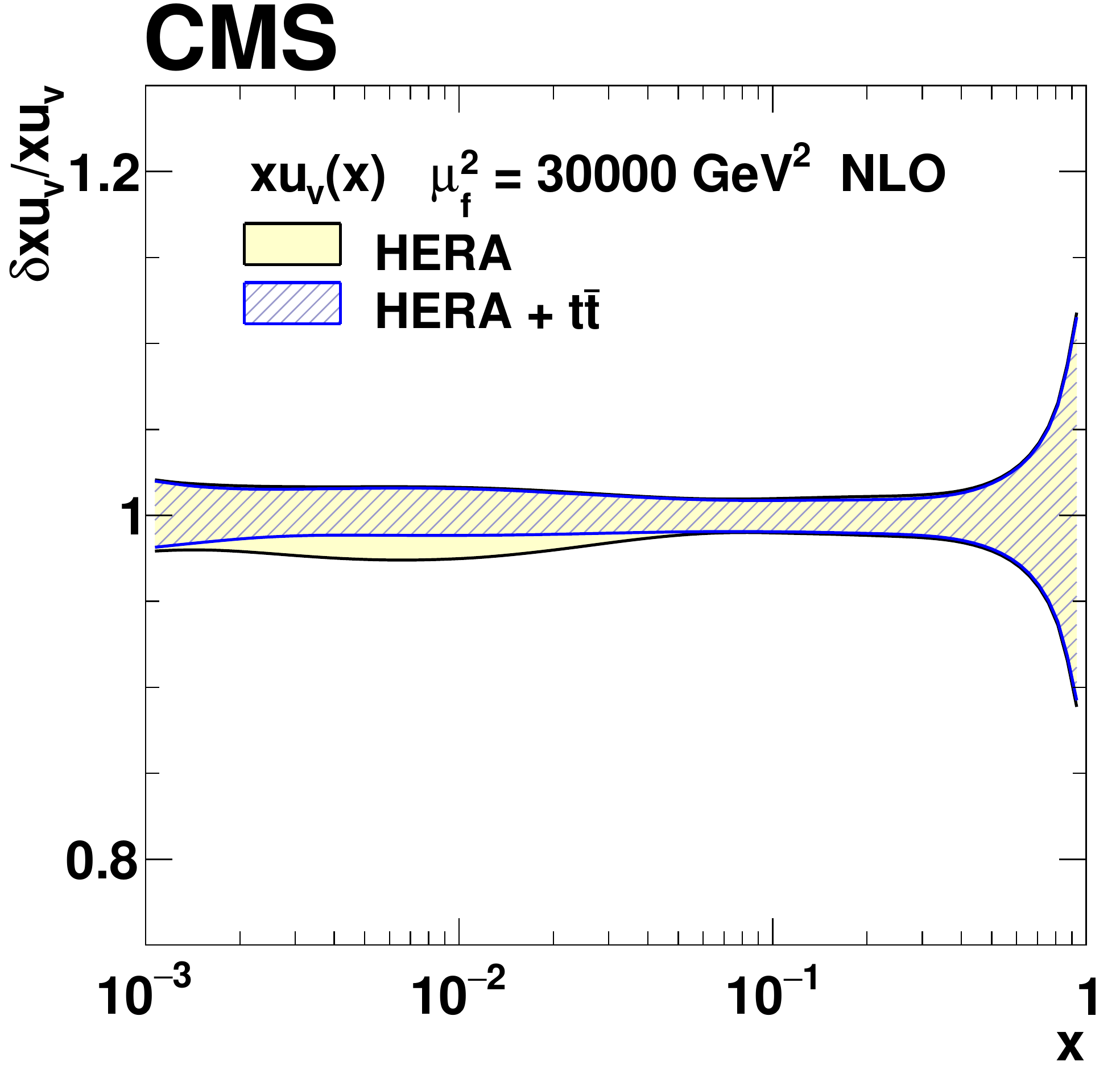}
    \includegraphics[width=0.49\textwidth]{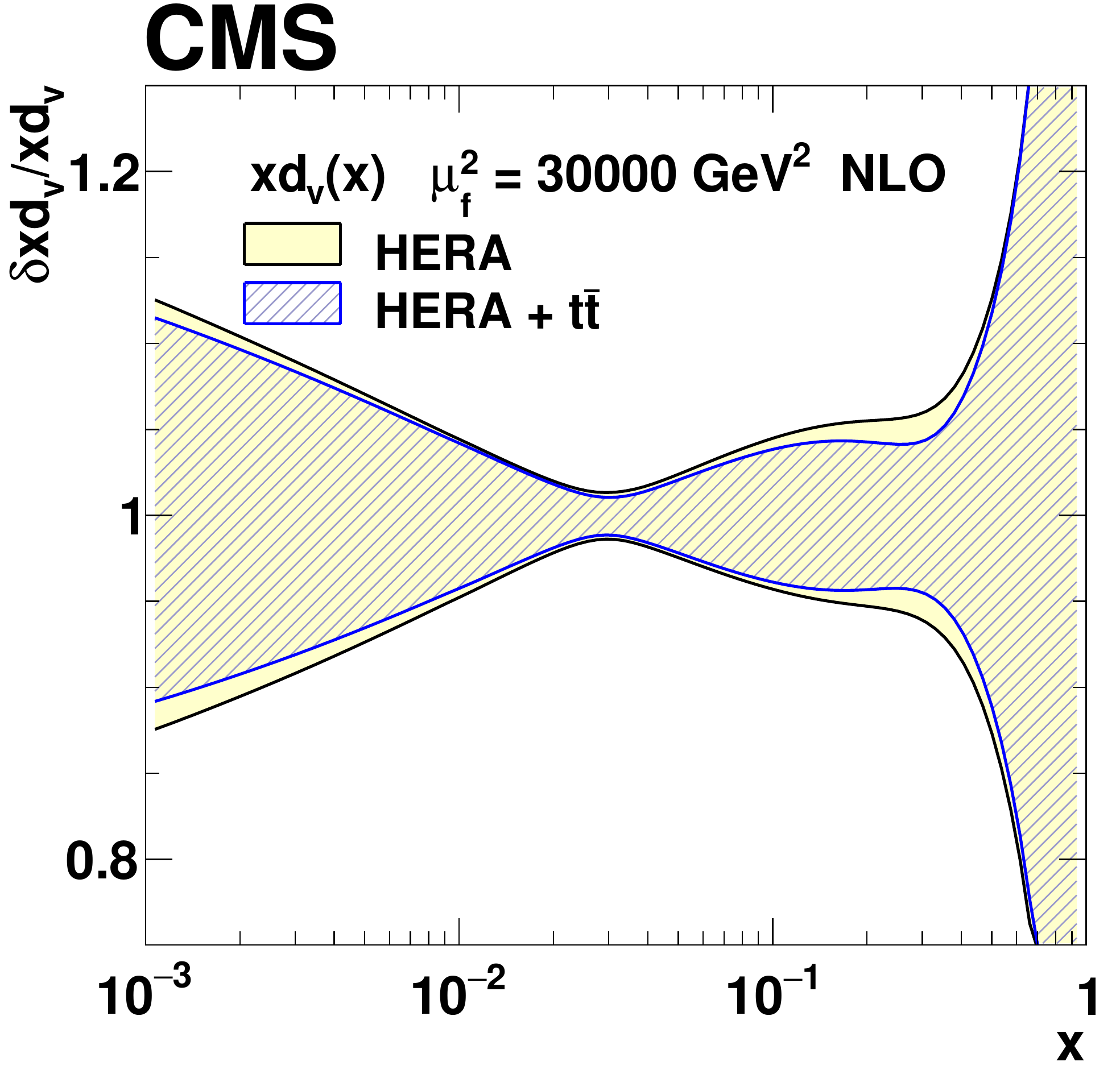}\\
    \includegraphics[width=0.49\textwidth]{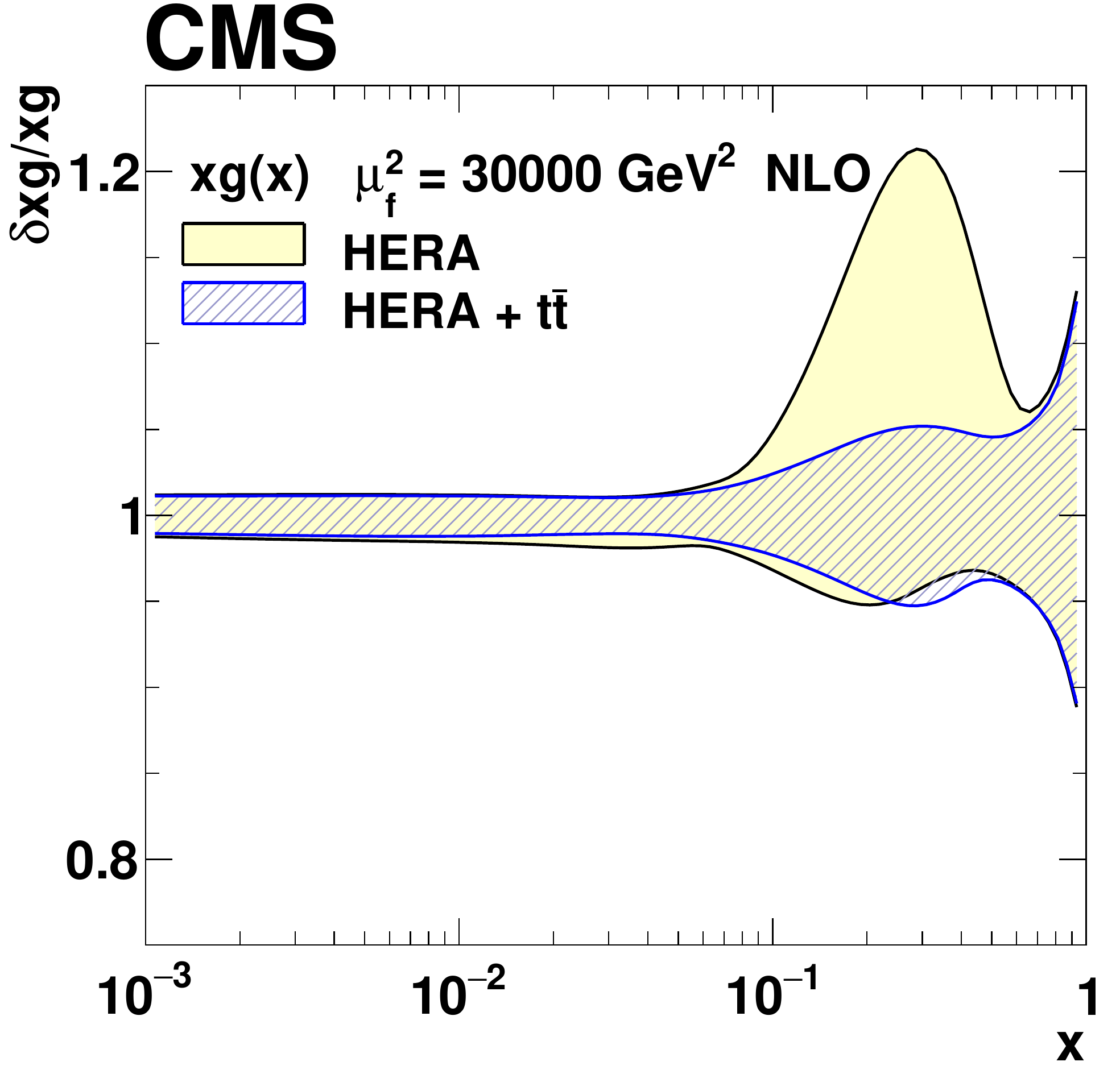}
    \includegraphics[width=0.49\textwidth]{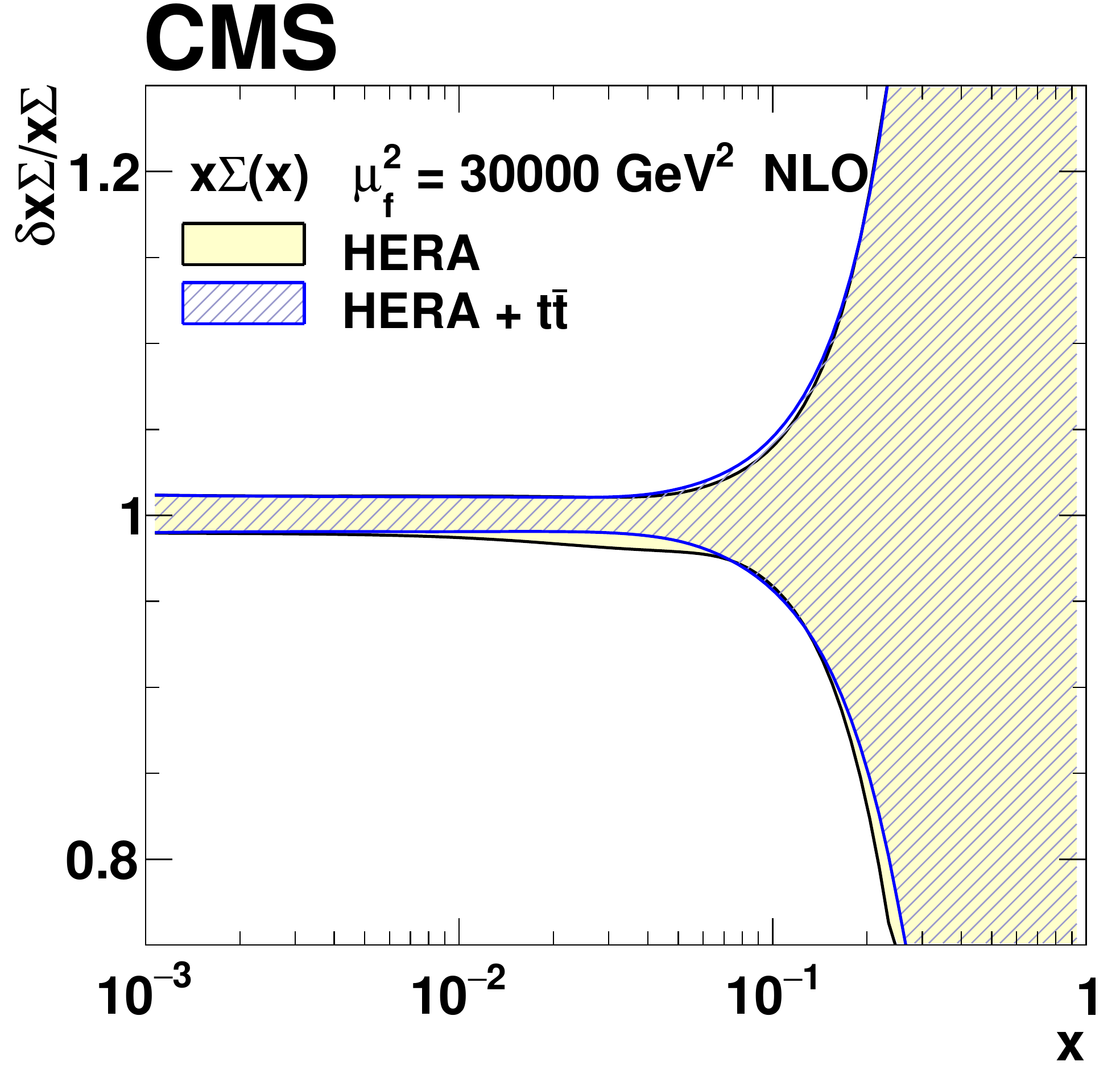}
    \caption{The relative total PDF uncertainties in the fit using the HERA DIS data only, and the HERA DIS and \ttbar data.}
    \label{fig:pdfs-totunc}
    \end{figure*}

In Fig.~\ref{fig:pdfs-totunc}
the total PDF uncertainties are shown for the two variants of the fits.
A reduction of uncertainties is observed for the gluon distribution, especially at $x \sim 0.1$ where the included \ttbar data are expected to provide constraints, while the improvement at $x \lesssim 0.1$ originates mainly from the reduced correlation between \asmz and the gluon PDF. A smaller uncertainty reduction is observed for other PDFs as well (valence and sea quark distributions), because of the correlations between the PDF distributions in the fit, as explained above.
  In addition to the fit uncertainty reduction, the \ttbar data constrain the large asymmetric model uncertainty of the gluon PDF at high $x$.
   This uncertainty originates from the variation of \qmin in the fit, using the HERA DIS data only, because of a lack of direct constraints from these data.

\begin{figure*}
    \centering
    \includegraphics[width=1.00\textwidth]{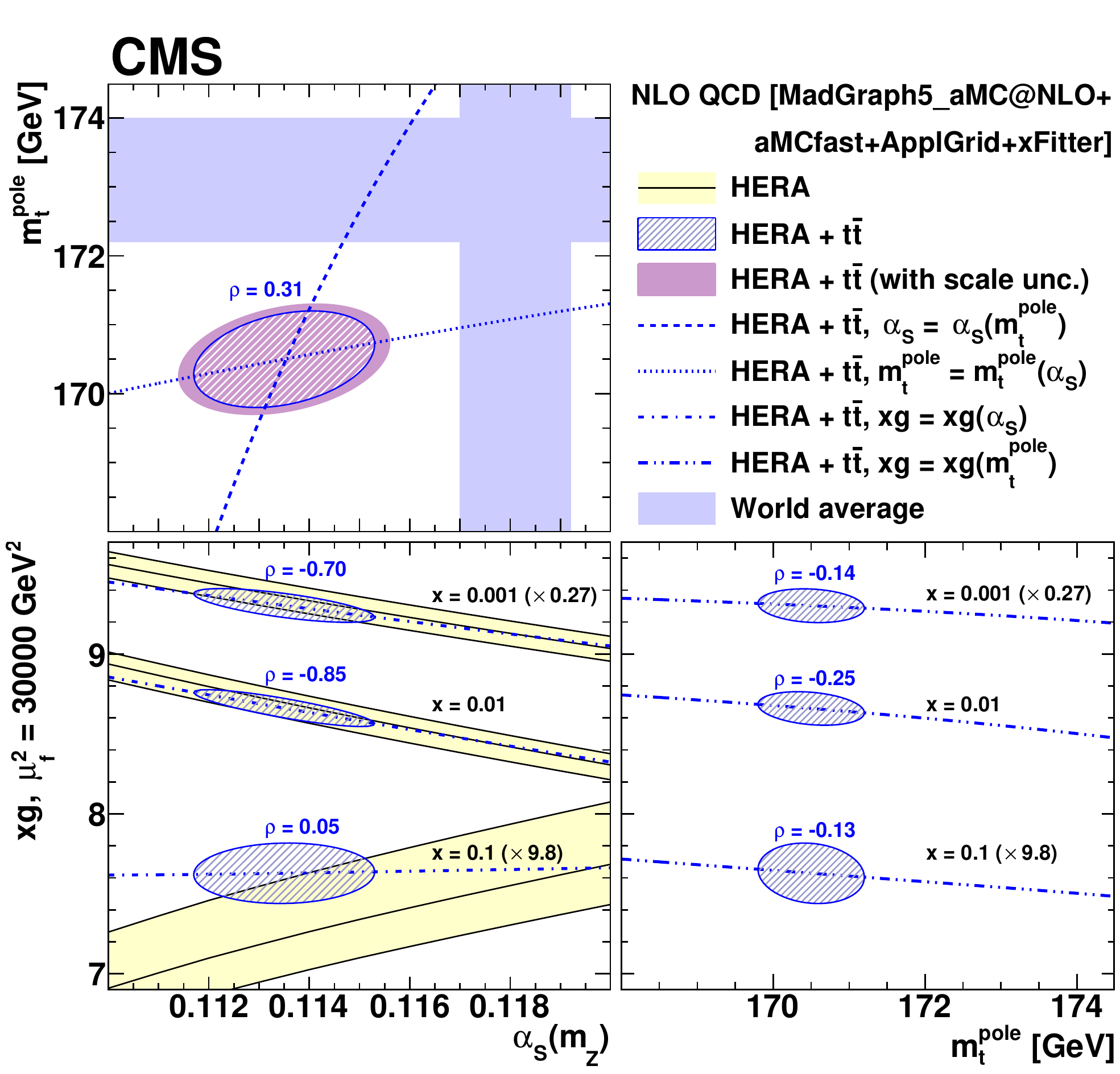}
    \caption{{The extracted values and their correlations for \as and \mt (upper left), \as and gluon PDF (lower left), and \mt and gluon PDF (lower, right). The gluon PDF is shown at the scale $\mu_\mathrm{f}^2 = 30\,000\GeV^2$ for several values of $x$.
    For the extracted values of \as and \mt, the additional uncertainties arising from the dependence on scale are shown (see Eq.~(\ref{eq:as-nom}) and Table~\ref{tab:asmtunc}).
    The correlation coefficients $\rho$ are also displayed. Furthermore, values of \as (\mt, gluon PDF) extracted using fixed values of \mt (\as) are displayed as dashed, dotted, or dash-dotted lines. The world average values $\asmz = 0.1181 \pm 0.0011$ and $\mt = 173.1 \pm 0.9\GeV$ from Ref.~\cite{pdg2018} are shown for reference.
    }}
    \label{fig:allcor}
    \end{figure*}

In Fig.~\ref{fig:allcor} the extracted \as, \mt, and gluon PDF at the scale $\mu_\mathrm{f}^2 = 30\,000\GeV^2$ for several values of $x$ are shown, together with their correlations.
  For this plot, the asymmetric \as and \mt uncertainties are symmetrised by taking the largest deviation, and the correlation of the fit uncertainties is assumed for the total uncertainties as well.
  The evolution of PDFs involves \asmz, therefore PDFs always depend on the \asmz assumed during their extraction.
  When using only the HERA DIS data, the largest dependence on \asmz is observed for the gluon distribution.
  The \ttbar data reduce this dependence, because they provide constraints on both the gluon distribution and \as, reducing their correlation.
  In addition, the multi-differential \njmttytttwo cross sections provide constraints on \mt.
  As a result, the gluon PDF, \asmz, and \mt can be determined simultaneously and their fitted values depend only
weakly on each other.
  This makes future PDF fits at NNLO, once corresponding theoretical predictions for inclusive \ttbar production with additional jets become available, very interesting.

 \section{Summary}
\label{sec:concl}

{\tolerance=9600
A measurement was presented of normalised multi-differential \ttbar production cross sections in $\pp$~collisions at $\sqrt{s}=13\TeV$,
performed using events containing two oppositely charged leptons (electron or muon).
The analysed data were recorded in 2016 with the CMS detector at the LHC, and correspond to an integrated luminosity of 35.9\fbinv.
The normalised \ttbar cross section is measured in the full phase space as a function of different pairs of kinematic variables
that describe either the top quark or the \ttbar system.
None of the central predictions of the tested Monte Carlo models is able to correctly describe all the distributions.
The data exhibit softer transverse momentum \ptt distributions than given by the theoretical predictions,
as was reported in previous single-differential and double-differential \ttbar cross section measurements.
The effect of the softer \ptt spectra in the data relative to the predictions is enhanced at larger values of
the invariant mass of the \ttbar system.
{The predicted \ptt slopes are strongly sensitive to the parton distribution functions (PDFs)
and the top quark pole mass \mt value used in the calculations, and the description of the data
can be improved by changing these parameters.}
\par}

The measured \ttbar cross sections as a function of the invariant mass and rapidity
of the \ttbar system, and the multiplicity of additional jets, have been incorporated into
two specific fits of QCD parameters at next-to-leading order,
after applying corrections for nonperturbative effects, together with the inclusive deep inelastic scattering data from HERA.
When fitting only $\asmz$ and \mt to the data, using external PDFs, the two
parameters are determined with high accuracy and rather weak correlation between them, however, the extracted $\asmz$
values depend on the PDF set. In a simultaneous fit of
$\as$, \mt, and PDFs, the inclusion of the new multi-differential \ttbar measurements has a significant impact
on the extracted gluon PDF at large values of $x$, where $x$ is the fraction of the proton
momentum carried by a parton, and at the same time allows an accurate determination of \as and \mt.
The values $\asmz = 0.1135{}^{+0.0021}_{-0.0017}$ and $\mt = 170.5 \pm 0.8 \GeV$ are obtained, which account for experimental and theoretical uncertainties.

The extraction of \mt performed in this paper exploits the sensitivity of the 
\ttbar invariant mass distribution to the value of \mt. The highest sensitivity comes from the region of low \ttbar masses.
Threshold corrections from Coulomb and soft-gluon resummation are expected to
affect this region. In Ref.~\cite{Kiyo:2008bv} an estimate of these effects
is provided, showing an expected increase of the total inclusive \ttbar production cross section by about 1\%.
A more recent study for the total cross section shows
that these corrections are presently known only with a large relative uncertainty~\cite{Piclum:2018ndt}.
Threshold corrections are neglected in the \mt extraction performed in the present analysis. A rough estimate shows that 
the inclusion of these corrections according to the size estimated in Ref.~\cite{Kiyo:2008bv} could lead to an increase of 
the extracted \mt value by up to $+0.7$ GeV. In the future, once precise calculations including
threshold corrections are available for differential \ttbar cross sections, 
such corrections should be included for an improved \mt extraction.
For the time being one can assume an additional 
theoretical uncertainty in the extracted \mt value of the order 
of $+1 \GeV$ due to neglected gluon resummation effects.

\begin{acknowledgments}
\hyphenation{Bundes-ministerium Forschungs-gemeinschaft Forschungs-zentren Rachada-pisek} We congratulate our colleagues in the CERN accelerator departments for the excellent performance of the LHC and thank the technical and administrative staffs at CERN and at other CMS institutes for their contributions to the success of the CMS effort. In addition, we gratefully acknowledge the computing centres and personnel of the Worldwide LHC Computing Grid for delivering so effectively the computing infrastructure essential to our analyses. Finally, we acknowledge the enduring support for the construction and operation of the LHC and the CMS detector provided by the following funding agencies: the Austrian Federal Ministry of Education, Science and Research and the Austrian Science Fund; the Belgian Fonds de la Recherche Scientifique, and Fonds voor Wetenschappelijk Onderzoek; the Brazilian Funding Agencies (CNPq, CAPES, FAPERJ, FAPERGS, and FAPESP); the Bulgarian Ministry of Education and Science; CERN; the Chinese Academy of Sciences, Ministry of Science and Technology, and National Natural Science Foundation of China; the Colombian Funding Agency (COLCIENCIAS); the Croatian Ministry of Science, Education and Sport, and the Croatian Science Foundation; the Research Promotion Foundation, Cyprus; the Secretariat for Higher Education, Science, Technology and Innovation, Ecuador; the Ministry of Education and Research, Estonian Research Council via IUT23-4, IUT23-6 and PRG445 and European Regional Development Fund, Estonia; the Academy of Finland, Finnish Ministry of Education and Culture, and Helsinki Institute of Physics; the Institut National de Physique Nucl\'eaire et de Physique des Particules~/~CNRS, and Commissariat \`a l'\'Energie Atomique et aux \'Energies Alternatives~/~CEA, France; the Bundesministerium f\"ur Bildung und Forschung, Deutsche Forschungsgemeinschaft, and Helmholtz-Gemeinschaft Deutscher Forschungszentren, Germany; the General Secretariat for Research and Technology, Greece; the National Research, Development and Innovation Fund, Hungary; the Department of Atomic Energy and the Department of Science and Technology, India; the Institute for Studies in Theoretical Physics and Mathematics, Iran; the Science Foundation, Ireland; the Istituto Nazionale di Fisica Nucleare, Italy; the Ministry of Science, ICT and Future Planning, and National Research Foundation (NRF), Republic of Korea; the Ministry of Education and Science of the Republic of Latvia; the Lithuanian Academy of Sciences; the Ministry of Education, and University of Malaya (Malaysia); the Ministry of Science of Montenegro; the Mexican Funding Agencies (BUAP, CINVESTAV, CONACYT, LNS, SEP, and UASLP-FAI); the Ministry of Business, Innovation and Employment, New Zealand; the Pakistan Atomic Energy Commission; the Ministry of Science and Higher Education and the National Science Centre, Poland; the Funda\c{c}\~ao para a Ci\^encia e a Tecnologia, Portugal; JINR, Dubna; the Ministry of Education and Science of the Russian Federation, the Federal Agency of Atomic Energy of the Russian Federation, Russian Academy of Sciences, the Russian Foundation for Basic Research, and the National Research Center ``Kurchatov Institute"; the Ministry of Education, Science and Technological Development of Serbia; the Secretar\'{\i}a de Estado de Investigaci\'on, Desarrollo e Innovaci\'on, Programa Consolider-Ingenio 2010, Plan Estatal de Investigaci\'on Cient\'{\i}fica y T\'ecnica y de Innovaci\'on 2013--2016, Plan de Ciencia, Tecnolog\'{i}a e Innovaci\'on 2013--2017 del Principado de Asturias, and Fondo Europeo de Desarrollo Regional, Spain; the Ministry of Science, Technology and Research, Sri Lanka; the Swiss Funding Agencies (ETH Board, ETH Zurich, PSI, SNF, UniZH, Canton Zurich, and SER); the Ministry of Science and Technology, Taipei; the Thailand Center of Excellence in Physics, the Institute for the Promotion of Teaching Science and Technology of Thailand, Special Task Force for Activating Research and the National Science and Technology Development Agency of Thailand; the Scientific and Technical Research Council of Turkey, and Turkish Atomic Energy Authority; the National Academy of Sciences of Ukraine, and State Fund for Fundamental Researches, Ukraine; the Science and Technology Facilities Council, UK; the US Department of Energy, and the US National Science Foundation.

Individuals have received support from the Marie-Curie programme and the European Research Council and Horizon 2020 Grant, contract Nos.\ 675440 and 765710 (European Union); the Leventis Foundation; the A.P.\ Sloan Foundation; the Alexander von Humboldt Foundation; the Belgian Federal Science Policy Office; the Fonds pour la Formation \`a la Recherche dans l'Industrie et dans l'Agriculture (FRIA-Belgium); the Agentschap voor Innovatie door Wetenschap en Technologie (IWT-Belgium); the F.R.S.-FNRS and FWO (Belgium) under the ``Excellence of Science -- EOS" -- be.h project n.\ 30820817; the Beijing Municipal Science \& Technology Commission, No. Z181100004218003; the Ministry of Education, Youth and Sports (MEYS) of the Czech Republic; the Lend\"ulet (``Momentum") Programme and the J\'anos Bolyai Research Scholarship of the Hungarian Academy of Sciences, the New National Excellence Program \'UNKP, the NKFIA research grants 123842, 123959, 124845, 124850, 125105, 128713, 128786, and 129058 (Hungary); the Council of Scientific and Industrial Research, India; the HOMING PLUS programme of the Foundation for Polish Science, cofinanced from European Union, Regional Development Fund, the Mobility Plus programme of the Ministry of Science and Higher Education, the National Science Center (Poland), contracts Harmonia 2014/14/M/ST2/00428, Opus 2014/13/B/ST2/02543, 2014/15/B/ST2/03998, and 2015/19/B/ST2/02861, Sonata-bis 2012/07/E/ST2/01406; the National Priorities Research Program by Qatar National Research Fund; the Programa de Excelencia Mar\'{i}a de Maeztu, and the Programa Severo Ochoa del Principado de Asturias; the Thalis and Aristeia programmes cofinanced by EU-ESF, and the Greek NSRF; the Rachadapisek Sompot Fund for Postdoctoral Fellowship, Chulalongkorn University, and the Chulalongkorn Academic into Its 2nd Century Project Advancement Project (Thailand); the Welch Foundation, contract C-1845; and the Weston Havens Foundation (USA). \end{acknowledgments}

\bibliography{auto_generated}
\clearpage
\numberwithin{table}{section}
\numberwithin{figure}{section}
\appendix
\section{Measured cross sections}
\label{sec:apptab}

Tables~\ref{tab:ytptt_xsec} to~\ref{tab:nj3mttyttthree_4_syst} provide the measured cross sections,
including their correlation matrices of statistical uncertainties and detailed breakdown of systematic uncertainties.
{The description of JES uncertainty sources can be found in Ref.~\cite{Khachatryan:2016kdb}.}

\begin{table*}[htb]
\centering
\topcaption{The measured \ytptt cross sections, along with their relative statistical and systematic uncertainties.}
\label{tab:ytptt_xsec}
\renewcommand*{\arraystretch}{1.55}
\tabcolsep2.0mm
\begin{tabular}{cccccc}
$\abs{y(\PQt)}$  & $p_\PQt(\PQt)$  [\GeVns{}]  & $\frac{1}{\sigma(\ttbar)} \frac{\rd\sigma}{\rd{}p_\PQt(\PQt) } [ \GeVns{}^{-1}]$ & stat. [\%] & syst. [\%] & bin\\
\hline
   0.00--0.35 &         0--80 & $9.251 \times 10^{-4}$ & 2.5 & ${}_{ -5.2}^{ +6.6}$ & 1 \\
   0.00--0.35 &       80--150 & $1.118 \times 10^{-3}$ & 1.8 & ${}_{ -4.6}^{ +2.4}$ & 2 \\
   0.00--0.35 &      150--250 & $4.596 \times 10^{-4}$ & 2.1 & ${}_{ -2.7}^{ +4.6}$ & 3 \\
   0.00--0.35 &      250--600 & $3.993 \times 10^{-5}$ & 2.9 & ${}_{ -5.6}^{ +3.7}$ & 4 \\
   0.35--0.85 &         0--80 & $1.279 \times 10^{-3}$ & 1.6 & ${}_{ -3.4}^{ +3.6}$ & 5 \\
   0.35--0.85 &       80--150 & $1.503 \times 10^{-3}$ & 1.3 & ${}_{ -2.8}^{ +2.5}$ & 6 \\
   0.35--0.85 &      150--250 & $6.206 \times 10^{-4}$ & 1.7 & ${}_{ -2.6}^{ +3.7}$ & 7 \\
   0.35--0.85 &      250--600 & $5.097 \times 10^{-5}$ & 2.8 & ${}_{ -4.7}^{ +4.9}$ & 8 \\
   0.85--1.45 &         0--80 & $1.248 \times 10^{-3}$ & 1.6 & ${}_{ -4.4}^{ +2.1}$ & 9 \\
   0.85--1.45 &       80--150 & $1.435 \times 10^{-3}$ & 1.3 & ${}_{ -2.7}^{ +2.6}$ & 10 \\
   0.85--1.45 &      150--250 & $5.328 \times 10^{-4}$ & 2.0 & ${}_{ -2.0}^{ +4.6}$ & 11 \\
   0.85--1.45 &      250--600 & $4.516 \times 10^{-5}$ & 2.9 & ${}_{ -5.4}^{ +4.0}$ & 12 \\
   1.45--2.50 &         0--80 & $1.128 \times 10^{-3}$ & 2.4 & ${}_{ -5.8}^{ +4.2}$ & 13 \\
   1.45--2.50 &       80--150 & $1.284 \times 10^{-3}$ & 1.6 & ${}_{ -2.5}^{ +4.0}$ & 14 \\
   1.45--2.50 &      150--250 & $4.238 \times 10^{-4}$ & 2.5 & ${}_{ -5.6}^{ +2.7}$ & 15 \\
   1.45--2.50 &      250--600 & $2.430 \times 10^{-5}$ & 4.9 & ${}_{ -4.2}^{+12.7}$ & 16 \\
\end{tabular}
\end{table*}
 \begin{table*}[htbp]
\centering
\topcaption{The correlation matrix of statistical uncertainties for the measured \ytptt cross sections. The values are expressed as percentages. For bin indices see Table~\ref{tab:ytptt_xsec}.}
\label{tab:ytptt_corr}
\tabcolsep2.0mm
\renewcommand*{\arraystretch}{1.45}
\cmsSwitchTable{%
\begin{tabular}{l|rrrrrrrrrrrrrrr}
bin & 2 & 3 & 4 & 5 & 6 & 7 & 8 & 9 & 10 & 11 & 12 & 13 & 14 & 15 & 16 \\
\hline
1 & $-22$ & $-11$ & $+3$ & $-29$ & $-31$ & $+5$ & $-2$ & $-43$ & $-4$ & $+5$ & $-3$ & $+5$ & $+4$ & $-5$ & 0 \\
2 & & $-20$ & $+1$ & $-28$ & $-3$ & $-17$ & $+5$ & $-4$ & $-12$ & $+3$ & 0 & $+1$ & $-3$ & 0 & 0 \\
3 & & & $-37$ & $+10$ & $-16$ & $-21$ & $+3$ & $+7$ & $+4$ & $-2$ & $+3$ & $-9$ & $-2$ & $+1$ & 0 \\
4 & & & & $-2$ & $+7$ & $+4$ & $-22$ & $-3$ & $+1$ & $+2$ & $+3$ & $-3$ & $-2$ & 0 & 0 \\
5 & & & & & $+4$ & $-12$ & $+4$ & $+36$ & $-17$ & $-3$ & $+2$ & $-43$ & $-10$ & $+9$ & $-2$ \\
6 & & & & & & $+2$ & $-5$ & $-19$ & $+23$ & $-18$ & $+5$ & $-15$ & $-15$ & $+3$ & $+1$ \\
7 & & & & & & & $-31$ & $-5$ & $-17$ & $-7$ & $-2$ & 0 & $-1$ & $-3$ & $+2$ \\
8 & & & & & & & & $+1$ & $+5$ & $-1$ & $-17$ & $-6$ & $-2$ & $+2$ & $+2$ \\
9 & & & & & & & & & $-9$ & $-13$ & $+3$ & $-18$ & $-32$ & $+5$ & $-1$ \\
10 & & & & & & & & & & $-3$ & $-4$ & $-34$ & $+2$ & $-12$ & $+5$ \\
11 & & & & & & & & & & & $-33$ & $-2$ & $-17$ & $-12$ & $+1$ \\
12 & & & & & & & & & & & & $-4$ & $+3$ & $+1$ & $-14$ \\
13 & & & & & & & & & & & & & $-14$ & $-28$ & $+6$ \\
14 & & & & & & & & & & & & & & $-12$ & $-7$ \\
15 & & & & & & & & & & & & & & & $-41$ \\
\end{tabular}}
\end{table*}
 \begin{table*}[p]
\centering
\topcaption{Sources and values of the relative systematic uncertainties in percent of the measured \ytptt cross sections. For bin indices see Table~\ref{tab:ytptt_xsec}.}
\label{tab:ytptt_syst}
\renewcommand*{\arraystretch}{1.75}
\tabcolsep2.0mm
\cmsTallTable{%
\begin{tabular}{l|
rrrrrrrrrrrrrrrr}
source / bin & 1 & 2 & 3 & 4 & 5 & 6 & 7 & 8 & 9 & 10 & 11 & 12 & 13 & 14 & 15 & 16 \\
\hline
\multirow{2}{*}{JER}& $+0.4$ & $-0.9$ & $-0.3$ & 0.0 & $+0.1$ & $+0.2$ & $+0.2$ & $-0.5$ & $-0.7$ & $-0.1$ & $+0.4$ & $-0.7$ & $+0.7$ & $+0.4$ & $-0.2$ & $-0.3$ \\
& $+0.7$ & $-0.3$ & $-0.2$ & $+0.8$ & $+0.2$ & $+0.1$ & $+0.3$ & $-0.2$ & $-0.6$ & $-0.1$ & $+0.7$ & $-0.1$ & $-0.9$ & $+0.3$ & $-0.3$ & $+1.0$ \\
\hline
\multirow{2}{*}{JESAbsoluteMPFBias}& $+0.6$ & $-0.3$ & $+0.2$ & $+0.2$ & $+0.3$ & 0.0 & 0.0 & $-0.3$ & 0.0 & $+0.1$ & $+0.1$ & 0.0 & $-1.2$ & $+0.3$ & $+0.1$ & $+0.8$ \\
& $+1.0$ & $-0.5$ & $+0.1$ & $-0.1$ & $+0.7$ & $-0.4$ & $+0.3$ & 0.0 & $-0.6$ & $+0.1$ & $+0.4$ & $-0.4$ & $-0.9$ & $+0.3$ & $-0.1$ & $+0.2$ \\
\hline
\multirow{2}{*}{JESAbsoluteScale}& $-0.5$ & $-0.5$ & $+0.4$ & 0.0 & $+0.7$ & $-0.1$ & $+0.1$ & $-0.1$ & $+0.4$ & $-0.5$ & $+0.1$ & $-0.5$ & $+0.2$ & $+0.1$ & $-0.9$ & $+2.1$ \\
& 0.0 & $-0.2$ & $+0.2$ & $-0.3$ & $+0.2$ & $-0.2$ & $+0.4$ & $-0.2$ & $-0.4$ & $-0.1$ & $+0.4$ & $+0.2$ & $+0.3$ & 0.0 & $-0.5$ & $+1.7$ \\
\hline
\multirow{2}{*}{JESAbsoluteStat}& $+1.2$ & $-0.1$ & $-0.8$ & $-0.2$ & $+0.1$ & 0.0 & $+0.1$ & $-0.8$ & $+0.3$ & 0.0 & $-0.3$ & $-0.1$ & $+0.1$ & $-0.3$ & $-0.8$ & $-0.5$ \\
\end{tabular}}
\end{table*}
 \begin{table*}[p]
\centering
\topcaption{Table~\ref{tab:ytptt_syst} continued.}
\label{tab:ytptt_1_syst}
\renewcommand*{\arraystretch}{1.75}
\tabcolsep2.0mm
\cmsTallTable{%
\begin{tabular}{l|
rrrrrrrrrrrrrrrr}
source / bin & 1 & 2 & 3 & 4 & 5 & 6 & 7 & 8 & 9 & 10 & 11 & 12 & 13 & 14 & 15 & 16 \\
\hline
\multirow{2}{*}{JESSinglePionHCAL}& $-0.1$ & $-0.3$ & $-0.2$ & 0.0 & $+0.2$ & $+0.3$ & $+0.2$ & $-0.6$ & $+0.2$ & $-0.1$ & 0.0 & 0.0 & $-0.1$ & $-0.1$ & 0.0 & $+0.1$ \\
& $+0.2$ & $+0.1$ & $-0.3$ & $+0.4$ & $-0.1$ & $-0.5$ & $+0.6$ & $-0.4$ & $+0.1$ & $-0.2$ & $+0.2$ & $+0.6$ & $-0.1$ & $+0.3$ & $-0.3$ & $+0.9$ \\
\hline
\multirow{2}{*}{JESTimePtEta}& $+0.8$ & $-0.7$ & $+0.6$ & $+0.1$ & $-0.6$ & $-0.3$ & $-0.4$ & $-0.2$ & $-0.7$ & $+0.2$ & $+0.3$ & 0.0 & $+0.5$ & $+0.8$ & $-0.5$ & $+1.5$ \\
& $-0.1$ & $+0.2$ & $+0.1$ & $+0.1$ & $+0.3$ & $-0.4$ & $+0.1$ & $+0.1$ & $-0.2$ & $-0.5$ & $+0.6$ & $-0.4$ & $+0.7$ & 0.0 & $-1.2$ & $+1.6$ \\
\hline
\multirow{2}{*}{$\ptvecmiss$}& $+0.7$ & $-0.6$ & $+0.6$ & $-0.1$ & $+0.2$ & $-0.1$ & $+0.1$ & $-0.6$ & $+0.3$ & $+0.1$ & $-0.3$ & $-0.1$ & $-0.6$ & $-0.1$ & $-0.3$ & $+0.5$ \\
& $-0.2$ & $+0.1$ & 0.0 & $+0.5$ & $-0.2$ & $+0.4$ & $-0.1$ & $+0.1$ & $-0.2$ & $-0.2$ & $+0.4$ & $+0.2$ & $+0.2$ & $-0.2$ & 0.0 & $+0.4$ \\
\hline
\multirow{2}{*}{lepton ID/ISO}& $-0.2$ & $+0.1$ & 0.0 & $-0.4$ & 0.0 & $+0.2$ & 0.0 & $-0.2$ & $-0.1$ & $+0.2$ & $-0.1$ & $-0.3$ & $-0.1$ & $+0.3$ & 0.0 & $-0.6$ \\
\end{tabular}}
\end{table*}
 \begin{table*}[p]
\centering
\topcaption{Table~\ref{tab:ytptt_syst} continued.}
\label{tab:ytptt_2_syst}
\renewcommand*{\arraystretch}{1.75}
\tabcolsep2.0mm
\cmsTallTable{%
\begin{tabular}{l|
rrrrrrrrrrrrrrrr}
source / bin & 1 & 2 & 3 & 4 & 5 & 6 & 7 & 8 & 9 & 10 & 11 & 12 & 13 & 14 & 15 & 16 \\
\hline
\multirow{2}{*}{PDF eigenvector 15}& $+0.1$ & 0.0 & 0.0 & $-0.1$ & 0.0 & 0.0 & 0.0 & $-0.1$ & 0.0 & 0.0 & 0.0 & $-0.1$ & 0.0 & 0.0 & 0.0 & $-0.1$ \\
& 0.0 & 0.0 & 0.0 & 0.0 & 0.0 & 0.0 & 0.0 & 0.0 & 0.0 & 0.0 & 0.0 & 0.0 & 0.0 & 0.0 & 0.0 & 0.0 \\
\hline
\multirow{2}{*}{PDF eigenvector 16}& 0.0 & 0.0 & 0.0 & 0.0 & 0.0 & 0.0 & 0.0 & 0.0 & 0.0 & 0.0 & 0.0 & 0.0 & 0.0 & 0.0 & 0.0 & 0.0 \\
& 0.0 & 0.0 & 0.0 & 0.0 & 0.0 & 0.0 & 0.0 & 0.0 & 0.0 & 0.0 & 0.0 & 0.0 & 0.0 & 0.0 & 0.0 & $-0.1$ \\
\hline
\multirow{2}{*}{PDF eigenvector 17}& 0.0 & 0.0 & 0.0 & 0.0 & 0.0 & 0.0 & 0.0 & 0.0 & 0.0 & 0.0 & 0.0 & 0.0 & 0.0 & 0.0 & 0.0 & $-0.1$ \\
& 0.0 & 0.0 & 0.0 & $+0.1$ & 0.0 & 0.0 & 0.0 & $+0.1$ & 0.0 & 0.0 & 0.0 & $+0.1$ & 0.0 & 0.0 & 0.0 & $+0.1$ \\
\hline
\multirow{2}{*}{PDF eigenvector 18}& $-0.1$ & 0.0 & 0.0 & $+0.1$ & $-0.1$ & 0.0 & 0.0 & $+0.1$ & 0.0 & 0.0 & 0.0 & $+0.1$ & $+0.1$ & 0.0 & $+0.2$ & $+0.1$ \\
\end{tabular}}
\end{table*}

\begin{table*}[htbp]
\centering
\topcaption{The measured \mttyt cross sections, along with their relative statistical and systematic uncertainties.}
\label{tab:mttyt_xsec}
\renewcommand*{\arraystretch}{1.55}
\tabcolsep2.0mm
\begin{tabular}{cccccc}
$M(\ttbar)$  [\GeVns{}]  & $\abs{y(\PQt)}$  & $\frac{1}{\sigma(\ttbar)} \frac{\rd\sigma}{\rd\abs{y(\PQt)} } $ & stat. [\%] & syst. [\%] & bin\\
\hline
     300--400 &    0.00--0.35 & $1.506 \times 10^{-1}$ & 3.1 & ${}_{ -5.4}^{ +6.6}$ & 1 \\
     300--400 &    0.35--0.85 & $1.511 \times 10^{-1}$ & 2.2 & ${}_{ -8.1}^{ +3.5}$ & 2 \\
     300--400 &    0.85--1.45 & $1.161 \times 10^{-1}$ & 2.2 & ${}_{ -6.2}^{ +4.2}$ & 3 \\
     300--400 &    1.45--2.50 & $4.794 \times 10^{-2}$ & 4.6 & ${}_{-10.1}^{+19.5}$ & 4 \\
     400--500 &    0.00--0.35 & $2.317 \times 10^{-1}$ & 1.9 & ${}_{ -3.1}^{ +3.9}$ & 5 \\
     400--500 &    0.35--0.85 & $2.181 \times 10^{-1}$ & 1.3 & ${}_{ -2.4}^{ +2.2}$ & 6 \\
     400--500 &    0.85--1.45 & $1.605 \times 10^{-1}$ & 1.5 & ${}_{ -3.0}^{ +4.2}$ & 7 \\
     400--500 &    1.45--2.50 & $7.230 \times 10^{-2}$ & 2.1 & ${}_{ -2.2}^{ +6.8}$ & 8 \\
     500--650 &    0.00--0.35 & $1.538 \times 10^{-1}$ & 2.2 & ${}_{ -2.2}^{ +5.9}$ & 9 \\
     500--650 &    0.35--0.85 & $1.438 \times 10^{-1}$ & 1.8 & ${}_{ -6.8}^{ +1.7}$ & 10 \\
     500--650 &    0.85--1.45 & $1.062 \times 10^{-1}$ & 1.9 & ${}_{ -2.5}^{ +3.7}$ & 11 \\
     500--650 &    1.45--2.50 & $5.345 \times 10^{-2}$ & 2.5 & ${}_{ -7.4}^{ +3.2}$ & 12 \\
    650--1500 &    0.00--0.35 & $6.963 \times 10^{-2}$ & 2.8 & ${}_{ -8.6}^{ +4.3}$ & 13 \\
    650--1500 &    0.35--0.85 & $7.045 \times 10^{-2}$ & 2.3 & ${}_{ -4.6}^{ +6.2}$ & 14 \\
    650--1500 &    0.85--1.45 & $6.765 \times 10^{-2}$ & 2.0 & ${}_{ -5.5}^{ +3.1}$ & 15 \\
    650--1500 &    1.45--2.50 & $4.155 \times 10^{-2}$ & 2.2 & ${}_{ -3.2}^{ +4.8}$ & 16 \\
\end{tabular}
\end{table*}
 \begin{table*}[htbp]
\centering
\topcaption{The correlation matrix of statistical uncertainties for the measured \mttyt cross sections. The values are expressed as percentages. For bin indices see Table~\ref{tab:mttyt_xsec}.}
\label{tab:mttyt_corr}
\tabcolsep2.0mm
\renewcommand*{\arraystretch}{1.45}
\cmsSwitchTable{%
\begin{tabular}{l|rrrrrrrrrrrrrrr}
bin & 2 & 3 & 4 & 5 & 6 & 7 & 8 & 9 & 10 & 11 & 12 & 13 & 14 & 15 & 16 \\
\hline
1 & $-34$ & $-23$ & $+8$ & $-28$ & $-29$ & 0 & $+3$ & $-8$ & $+6$ & $+4$ & $-6$ & $+2$ & $-1$ & $-3$ & $-2$ \\
2 & & $+18$ & $-32$ & $-19$ & $-1$ & $-23$ & $-7$ & $+10$ & $-12$ & $-3$ & $+9$ & $-1$ & $+3$ & $+5$ & $-2$ \\
3 & & & $-21$ & $+3$ & $-21$ & $-13$ & $-32$ & $+3$ & $-1$ & $-11$ & $+2$ & $-2$ & $+1$ & $+3$ & $+2$ \\
4 & & & & $-5$ & $-9$ & $-24$ & $-14$ & $-11$ & $-4$ & $-1$ & $-32$ & $-4$ & $-8$ & $-9$ & 0 \\
5 & & & & & $-5$ & $-11$ & $-2$ & $-19$ & $-22$ & $+1$ & $+3$ & $-2$ & $+7$ & $+2$ & $-2$ \\
6 & & & & & & $+19$ & $-15$ & $-17$ & $+4$ & $-22$ & $-1$ & $+8$ & $-9$ & $+4$ & $+2$ \\
7 & & & & & & & $+1$ & $+4$ & $-14$ & $-2$ & $-21$ & 0 & $+4$ & $-7$ & $+5$ \\
8 & & & & & & & & $-2$ & $-1$ & $-15$ & $-5$ & $-3$ & $-3$ & $+3$ & $-17$ \\
9 & & & & & & & & & $-17$ & $-3$ & $+2$ & $-38$ & $-4$ & $+4$ & $-1$ \\
10 & & & & & & & & & & $-2$ & $-5$ & $+2$ & $-28$ & $-12$ & $+2$ \\
11 & & & & & & & & & & & $-8$ & $+2$ & $-1$ & $-27$ & $-9$ \\
12 & & & & & & & & & & & & $-1$ & $+3$ & $+5$ & $-34$ \\
13 & & & & & & & & & & & & & $-24$ & $+2$ & $-1$ \\
14 & & & & & & & & & & & & & & $-19$ & $+2$ \\
15 & & & & & & & & & & & & & & & $-23$ \\
\end{tabular}}
\end{table*}
 \begin{table*}[p]
\centering
\topcaption{Sources and values of the relative systematic uncertainties in percent of the measured \mttyt cross sections. For bin indices see Table~\ref{tab:mttyt_xsec}.}
\label{tab:mttyt_syst}
\renewcommand*{\arraystretch}{1.75}
\tabcolsep2.0mm
\cmsTallTable{%
\begin{tabular}{l|
rrrrrrrrrrrrrrrr}
source / bin & 1 & 2 & 3 & 4 & 5 & 6 & 7 & 8 & 9 & 10 & 11 & 12 & 13 & 14 & 15 & 16 \\
\hline
\multirow{2}{*}{JER}& $-0.6$ & $+0.3$ & $-0.3$ & $+2.4$ & $+0.3$ & 0.0 & $-0.2$ & $-0.2$ & $+0.4$ & $-0.6$ & $+0.6$ & $-0.8$ & $-1.1$ & $-0.4$ & $-0.2$ & $-0.1$ \\
& $-0.2$ & $-1.1$ & $-0.7$ & $+3.9$ & $+0.7$ & $-0.2$ & $-0.3$ & $+0.8$ & 0.0 & $-0.7$ & $-0.1$ & $-1.6$ & $-0.2$ & $+0.3$ & $-0.2$ & $+0.7$ \\
\hline
\multirow{2}{*}{JESAbsoluteMPFBias}& $-0.9$ & $+0.1$ & $-0.7$ & $+0.1$ & $+0.9$ & 0.0 & $-0.4$ & $+1.4$ & $+0.3$ & $-1.1$ & $-0.3$ & 0.0 & $+0.1$ & $+0.2$ & $+0.2$ & $+0.3$ \\
& $-1.3$ & $-0.5$ & $-0.8$ & $+1.7$ & $+0.7$ & $+0.5$ & $+0.2$ & $+0.6$ & $+0.1$ & $-0.7$ & $+0.2$ & $-1.0$ & $-0.4$ & $-0.2$ & $-0.1$ & $+0.2$ \\
\hline
\multirow{2}{*}{JESAbsoluteScale}& $-0.2$ & $-0.8$ & $+0.7$ & $+2.1$ & $+0.7$ & $-0.2$ & $-0.3$ & $-0.4$ & $+0.3$ & $-0.4$ & 0.0 & $-1.2$ & 0.0 & $-0.6$ & 0.0 & $+0.9$ \\
& $-0.5$ & $-1.0$ & $-0.4$ & $+3.8$ & $+0.4$ & $-0.3$ & $+0.1$ & $+0.2$ & $+0.4$ & $-0.3$ & $-0.2$ & $-1.0$ & $-0.9$ & 0.0 & $-0.1$ & $+0.1$ \\
\hline
\multirow{2}{*}{JESAbsoluteStat}& $+0.3$ & $+0.1$ & 0.0 & $+1.8$ & $-0.4$ & $-0.4$ & $-0.3$ & $+0.5$ & $+0.5$ & $-0.6$ & $+0.4$ & $-0.7$ & $-0.4$ & $-0.1$ & $+0.1$ & $-0.2$ \\
\end{tabular}}
\end{table*}
 \begin{table*}[p]
\centering
\topcaption{Table~\ref{tab:mttyt_syst} continued.}
\label{tab:mttyt_1_syst}
\renewcommand*{\arraystretch}{1.75}
\tabcolsep2.0mm
\cmsTallTable{%
\begin{tabular}{l|
rrrrrrrrrrrrrrrr}
source / bin & 1 & 2 & 3 & 4 & 5 & 6 & 7 & 8 & 9 & 10 & 11 & 12 & 13 & 14 & 15 & 16 \\
\hline
\multirow{2}{*}{JESSinglePionHCAL}& 0.0 & $-0.7$ & $-0.1$ & $+2.4$ & 0.0 & $+0.2$ & $-0.4$ & $+0.5$ & $+0.5$ & $-0.6$ & 0.0 & $-0.5$ & $-0.8$ & $-0.4$ & 0.0 & $-0.1$ \\
& $+0.2$ & $-0.1$ & $-0.5$ & $+2.4$ & $-0.2$ & $-0.1$ & $-0.7$ & $+0.4$ & $+0.3$ & $-0.5$ & $+0.1$ & $-0.8$ & 0.0 & $-0.3$ & $+0.3$ & $+0.3$ \\
\hline
\multirow{2}{*}{JESTimePtEta}& $+1.4$ & $-1.2$ & $+0.7$ & $+2.3$ & $-0.5$ & $-0.2$ & $-0.5$ & $+0.7$ & $+0.8$ & $-0.8$ & $+0.4$ & $-1.2$ & $-0.4$ & $-0.7$ & 0.0 & $+0.4$ \\
& $-0.5$ & $-0.4$ & $+0.2$ & $+0.7$ & $+0.6$ & $-0.2$ & $-0.1$ & $+1.2$ & $+0.2$ & $-0.8$ & $-0.1$ & $-1.2$ & 0.0 & $+0.3$ & 0.0 & $+0.3$ \\
\hline
\multirow{2}{*}{$\ptvecmiss$}& $+0.7$ & $-0.2$ & $-0.5$ & $-2.2$ & $-0.1$ & $+0.3$ & $+0.5$ & $+1.0$ & $-0.2$ & $-0.1$ & $+0.2$ & $-0.1$ & 0.0 & $-0.2$ & $-0.2$ & 0.0 \\
& $-0.4$ & $-1.0$ & $-0.8$ & $+3.4$ & $+0.8$ & $+0.1$ & $+0.3$ & $+0.1$ & $-0.1$ & $-1.0$ & $-0.2$ & $-0.8$ & $-0.2$ & $+0.2$ & $+0.1$ & $-0.1$ \\
\hline
\multirow{2}{*}{lepton ID/ISO}& $-0.2$ & $-0.6$ & $-0.6$ & $-0.2$ & $+0.4$ & $+0.4$ & $+0.6$ & $+0.5$ & $-0.2$ & $-0.3$ & $-0.1$ & $-0.4$ & $-0.2$ & $+0.1$ & $-0.2$ & $+0.1$ \\
\end{tabular}}
\end{table*}
 \begin{table*}[p]
\centering
\topcaption{Table~\ref{tab:mttyt_syst} continued.}
\label{tab:mttyt_2_syst}
\renewcommand*{\arraystretch}{1.75}
\tabcolsep2.0mm
\cmsTallTable{%
\begin{tabular}{l|
rrrrrrrrrrrrrrrr}
source / bin & 1 & 2 & 3 & 4 & 5 & 6 & 7 & 8 & 9 & 10 & 11 & 12 & 13 & 14 & 15 & 16 \\
\hline
\multirow{2}{*}{PDF eigenvector 15}& $+0.1$ & 0.0 & 0.0 & $-0.1$ & 0.0 & 0.0 & 0.0 & 0.0 & 0.0 & $+0.1$ & $+0.1$ & 0.0 & 0.0 & 0.0 & 0.0 & 0.0 \\
& 0.0 & 0.0 & 0.0 & 0.0 & 0.0 & 0.0 & 0.0 & 0.0 & 0.0 & 0.0 & 0.0 & 0.0 & 0.0 & 0.0 & 0.0 & 0.0 \\
\hline
\multirow{2}{*}{PDF eigenvector 16}& 0.0 & 0.0 & 0.0 & 0.0 & 0.0 & 0.0 & 0.0 & 0.0 & 0.0 & 0.0 & 0.0 & 0.0 & 0.0 & 0.0 & 0.0 & 0.0 \\
& 0.0 & $-0.1$ & $-0.1$ & 0.0 & $+0.1$ & 0.0 & $+0.1$ & $+0.1$ & 0.0 & $-0.1$ & $+0.1$ & $-0.1$ & 0.0 & $+0.1$ & 0.0 & 0.0 \\
\hline
\multirow{2}{*}{PDF eigenvector 17}& $+0.1$ & $-0.1$ & $-0.1$ & 0.0 & $+0.1$ & 0.0 & $+0.1$ & $+0.1$ & 0.0 & $-0.1$ & $+0.1$ & $-0.1$ & 0.0 & $+0.1$ & 0.0 & 0.0 \\
& $-0.1$ & 0.0 & 0.0 & $+0.1$ & 0.0 & $+0.1$ & 0.0 & 0.0 & 0.0 & $-0.1$ & $-0.1$ & 0.0 & $-0.1$ & 0.0 & 0.0 & $+0.1$ \\
\hline
\multirow{2}{*}{PDF eigenvector 18}& $-0.1$ & $-0.2$ & $-0.1$ & $+0.3$ & $+0.1$ & 0.0 & $+0.1$ & $+0.1$ & $-0.1$ & $-0.2$ & 0.0 & $-0.1$ & 0.0 & $+0.1$ & $+0.1$ & $+0.1$ \\
\end{tabular}}
\end{table*}

\begin{table*}[htbp]
\centering
\topcaption{The measured \mttytt cross sections, along with their relative statistical and systematic uncertainties.}
\label{tab:mttytt_xsec}
\renewcommand*{\arraystretch}{1.55}
\tabcolsep2.0mm
\begin{tabular}{cccccc}
$M(\ttbar)$  [\GeVns{}]  & $\abs{y(\ttbar)}$  & $\frac{1}{\sigma(\ttbar)} \frac{\rd\sigma}{\rd\abs{y(\ttbar)} } $ & stat. [\%] & syst. [\%] & bin\\
\hline
     300--400 &    0.00--0.35 & $1.554 \times 10^{-1}$ & 2.5 & ${}_{ -4.5}^{ +6.7}$ & 1 \\
     300--400 &    0.35--0.75 & $1.485 \times 10^{-1}$ & 2.3 & ${}_{ -9.6}^{ +3.9}$ & 2 \\
     300--400 &    0.75--1.15 & $1.320 \times 10^{-1}$ & 2.4 & ${}_{ -5.6}^{ +6.3}$ & 3 \\
     300--400 &    1.15--2.50 & $5.856 \times 10^{-2}$ & 2.4 & ${}_{ -6.9}^{+10.9}$ & 4 \\
     400--500 &    0.00--0.35 & $2.506 \times 10^{-1}$ & 1.4 & ${}_{ -1.8}^{ +3.4}$ & 5 \\
     400--500 &    0.35--0.75 & $2.296 \times 10^{-1}$ & 1.3 & ${}_{ -3.9}^{ +1.4}$ & 6 \\
     400--500 &    0.75--1.15 & $1.889 \times 10^{-1}$ & 1.5 & ${}_{ -2.3}^{ +3.9}$ & 7 \\
     400--500 &    1.15--2.50 & $8.038 \times 10^{-2}$ & 1.2 & ${}_{ -3.9}^{ +3.7}$ & 8 \\
     500--650 &    0.00--0.35 & $1.833 \times 10^{-1}$ & 1.7 & ${}_{ -2.2}^{ +3.0}$ & 9 \\
     500--650 &    0.35--0.75 & $1.714 \times 10^{-1}$ & 1.6 & ${}_{ -5.2}^{ +2.0}$ & 10 \\
     500--650 &    0.75--1.15 & $1.340 \times 10^{-1}$ & 2.0 & ${}_{ -2.1}^{ +6.7}$ & 11 \\
     500--650 &    1.15--2.50 & $4.390 \times 10^{-2}$ & 2.3 & ${}_{ -7.7}^{ +2.7}$ & 12 \\
    650--1500 &    0.00--0.35 & $1.331 \times 10^{-1}$ & 1.6 & ${}_{ -4.5}^{ +3.0}$ & 13 \\
    650--1500 &    0.35--0.75 & $1.079 \times 10^{-1}$ & 2.0 & ${}_{ -4.2}^{ +5.2}$ & 14 \\
    650--1500 &    0.75--1.15 & $8.007 \times 10^{-2}$ & 2.6 & ${}_{ -7.9}^{ +5.1}$ & 15 \\
    650--1500 &    1.15--2.50 & $1.731 \times 10^{-2}$ & 4.7 & ${}_{ -6.3}^{+11.3}$ & 16 \\
\end{tabular}
\end{table*}
 \begin{table*}[htbp]
\centering
\topcaption{The correlation matrix of statistical uncertainties for the measured \mttytt cross sections. The values are expressed as percentages. For bin indices see Table~\ref{tab:mttytt_xsec}.}
\label{tab:mttytt_corr}
\tabcolsep2.0mm
\renewcommand*{\arraystretch}{1.45}
\cmsSwitchTable{%
\begin{tabular}{l|rrrrrrrrrrrrrrr}
bin & 2 & 3 & 4 & 5 & 6 & 7 & 8 & 9 & 10 & 11 & 12 & 13 & 14 & 15 & 16 \\
\hline
1 & $-23$ & $-20$ & $+2$ & $-33$ & $-24$ & $+3$ & 0 & $-15$ & $+4$ & $+4$ & $-5$ & $+5$ & $-1$ & $-4$ & $-2$ \\
2 & & $+13$ & $-27$ & $-20$ & $-6$ & $-25$ & $-5$ & $+5$ & $-15$ & $-3$ & $+7$ & 0 & $+4$ & $+3$ & $-4$ \\
3 & & & $-9$ & $+3$ & $-25$ & $-14$ & $-30$ & $+3$ & $-4$ & $-15$ & $-7$ & $-3$ & $+2$ & $+4$ & $+3$ \\
4 & & & & $-4$ & $-7$ & $-25$ & $-16$ & $-9$ & 0 & $-6$ & $-43$ & $-6$ & $-8$ & $-2$ & $+6$ \\
5 & & & & & $+2$ & $-7$ & $-6$ & $-11$ & $-18$ & $+1$ & $-1$ & $-6$ & $+4$ & $+1$ & $-3$ \\
6 & & & & & & $+20$ & $-15$ & $-18$ & $+11$ & $-16$ & $-6$ & $+6$ & $-10$ & $+1$ & $+1$ \\
7 & & & & & & & $+3$ & $+1$ & $-15$ & $+5$ & $-15$ & $+2$ & $+2$ & $-10$ & $-1$ \\
8 & & & & & & & & $-2$ & $-8$ & $-18$ & $+27$ & $-6$ & $-2$ & 0 & $-33$ \\
9 & & & & & & & & & $-11$ & $-4$ & 0 & $-31$ & $-5$ & $+4$ & $-2$ \\
10 & & & & & & & & & & $+4$ & $-11$ & $-4$ & $-20$ & $-12$ & $+3$ \\
11 & & & & & & & & & & & $-5$ & $+5$ & $-9$ & $-23$ & $-11$ \\
12 & & & & & & & & & & & & $-3$ & $+2$ & $-7$ & $-32$ \\
13 & & & & & & & & & & & & & $-26$ & $+1$ & $-1$ \\
14 & & & & & & & & & & & & & & $-16$ & $-3$ \\
15 & & & & & & & & & & & & & & & $-16$ \\
\end{tabular}}
\end{table*}
 \begin{table*}[p]
\centering
\topcaption{Sources and values of the relative systematic uncertainties in percent of the measured \mttytt cross sections. For bin indices see Table~\ref{tab:mttytt_xsec}.}
\label{tab:mttytt_syst}
\renewcommand*{\arraystretch}{1.75}
\tabcolsep2.0mm
\cmsTallTable{%
\begin{tabular}{l|
rrrrrrrrrrrrrrrr}
source / bin & 1 & 2 & 3 & 4 & 5 & 6 & 7 & 8 & 9 & 10 & 11 & 12 & 13 & 14 & 15 & 16 \\
\hline
\multirow{2}{*}{JER}& $+0.5$ & $-1.4$ & $+1.0$ & $+1.4$ & $+0.2$ & $-0.1$ & $+0.3$ & $-0.5$ & 0.0 & $-0.4$ & $+0.6$ & $-0.7$ & $-0.1$ & 0.0 & $-1.3$ & $+0.2$ \\
& $-0.3$ & $-1.1$ & $+0.1$ & $+1.1$ & $+0.2$ & $-0.2$ & $+0.1$ & $+0.1$ & 0.0 & $-0.8$ & $+0.1$ & $-0.5$ & $-0.1$ & $+0.6$ & $+0.2$ & $+1.0$ \\
\hline
\multirow{2}{*}{JESAbsoluteMPFBias}& $-0.2$ & $-0.8$ & 0.0 & $+0.5$ & $+0.6$ & $-0.1$ & $+0.4$ & $-0.2$ & $+0.1$ & $-1.0$ & $+0.2$ & $-0.8$ & $-0.1$ & $+0.9$ & $-0.1$ & $+1.2$ \\
& $-0.4$ & $-1.1$ & $-0.1$ & $+1.0$ & $+0.4$ & 0.0 & $+0.9$ & $-0.3$ & $+0.2$ & $-0.6$ & $+0.5$ & $-1.0$ & $-0.6$ & $+0.9$ & $-1.0$ & $+0.7$ \\
\hline
\multirow{2}{*}{JESAbsoluteScale}& $+0.6$ & $-1.0$ & $-0.3$ & $+1.4$ & $+0.2$ & 0.0 & $+0.3$ & $-0.2$ & 0.0 & $-0.4$ & $-0.6$ & $-1.0$ & $-0.4$ & $+0.1$ & $+0.3$ & $+1.6$ \\
& $-0.7$ & $-1.7$ & 0.0 & $+1.5$ & $+0.5$ & $+0.1$ & $+0.7$ & $-0.3$ & 0.0 & $-0.3$ & $+0.5$ & $-1.0$ & $-0.3$ & 0.0 & $-0.7$ & $+1.1$ \\
\hline
\multirow{2}{*}{JESAbsoluteStat}& $+0.6$ & $-1.0$ & $+0.4$ & $+1.4$ & $-0.1$ & $+0.2$ & $+0.2$ & $-0.5$ & $-0.1$ & 0.0 & $+0.3$ & $-1.4$ & $-0.2$ & $+0.5$ & $-0.5$ & $+0.2$ \\
\end{tabular}}
\end{table*}
 \begin{table*}[p]
\centering
\topcaption{Table~\ref{tab:mttytt_syst} continued.}
\label{tab:mttytt_1_syst}
\renewcommand*{\arraystretch}{1.75}
\tabcolsep2.0mm
\cmsTallTable{%
\begin{tabular}{l|
rrrrrrrrrrrrrrrr}
source / bin & 1 & 2 & 3 & 4 & 5 & 6 & 7 & 8 & 9 & 10 & 11 & 12 & 13 & 14 & 15 & 16 \\
\hline
\multirow{2}{*}{JESSinglePionHCAL}& $-0.3$ & $-1.3$ & $+0.1$ & $+1.4$ & $+0.5$ & $-0.1$ & $+0.6$ & $-0.1$ & $-0.2$ & $-0.3$ & $+0.5$ & $-1.1$ & $-0.2$ & $+0.2$ & $-1.2$ & $+0.8$ \\
& $-0.4$ & $-0.9$ & $-0.5$ & $+1.5$ & $+0.3$ & $-0.3$ & $+0.1$ & $+0.1$ & $+0.1$ & $-0.4$ & $+0.3$ & $-1.3$ & $+0.1$ & $+0.3$ & $-0.3$ & $+2.2$ \\
\hline
\multirow{2}{*}{JESTimePtEta}& $+1.4$ & $-1.0$ & $-0.1$ & $+1.4$ & $-0.2$ & $-0.6$ & $+0.1$ & 0.0 & $-0.1$ & $-0.5$ & $+0.7$ & $-1.2$ & $-0.2$ & $+0.5$ & $-0.8$ & $+1.2$ \\
& $-0.3$ & $+0.2$ & $+1.4$ & $+0.3$ & $-0.1$ & $-0.5$ & $+0.4$ & $-0.2$ & $+0.2$ & $-0.9$ & 0.0 & $-0.6$ & 0.0 & $+0.5$ & $+0.1$ & $+1.0$ \\
\hline
\multirow{2}{*}{$\ptvecmiss$}& $+1.7$ & $-0.3$ & $-0.4$ & 0.0 & $-0.3$ & $+0.4$ & $-0.1$ & 0.0 & $-0.1$ & $-0.3$ & $-0.1$ & $-0.2$ & $+0.1$ & $-0.1$ & $-0.1$ & $-0.4$ \\
& $-1.0$ & 0.0 & 0.0 & $+1.1$ & $+0.1$ & $-0.1$ & $+0.4$ & 0.0 & $-0.3$ & $-0.4$ & $-0.2$ & $-0.3$ & 0.0 & $+0.2$ & 0.0 & $+0.5$ \\
\hline
\multirow{2}{*}{lepton ID/ISO}& $-0.2$ & 0.0 & $-0.3$ & $-0.3$ & $+0.1$ & $+0.2$ & 0.0 & $+0.4$ & $-0.1$ & $+0.1$ & 0.0 & $-0.2$ & 0.0 & $-0.1$ & $-0.1$ & $-0.5$ \\
\end{tabular}}
\end{table*}
 \begin{table*}[p]
\centering
\topcaption{Table~\ref{tab:mttytt_syst} continued.}
\label{tab:mttytt_2_syst}
\renewcommand*{\arraystretch}{1.75}
\tabcolsep2.0mm
\cmsTallTable{%
\begin{tabular}{l|
rrrrrrrrrrrrrrrr}
source / bin & 1 & 2 & 3 & 4 & 5 & 6 & 7 & 8 & 9 & 10 & 11 & 12 & 13 & 14 & 15 & 16 \\
\hline
\multirow{2}{*}{PDF eigenvector 15}& 0.0 & 0.0 & 0.0 & 0.0 & 0.0 & 0.0 & 0.0 & $+0.1$ & 0.0 & 0.0 & 0.0 & 0.0 & $-0.1$ & 0.0 & 0.0 & 0.0 \\
& 0.0 & 0.0 & 0.0 & 0.0 & 0.0 & 0.0 & 0.0 & 0.0 & 0.0 & 0.0 & 0.0 & 0.0 & 0.0 & 0.0 & 0.0 & 0.0 \\
\hline
\multirow{2}{*}{PDF eigenvector 16}& 0.0 & 0.0 & $+0.1$ & 0.0 & 0.0 & $-0.1$ & $-0.1$ & 0.0 & 0.0 & 0.0 & $+0.1$ & 0.0 & 0.0 & 0.0 & 0.0 & 0.0 \\
& 0.0 & 0.0 & 0.0 & 0.0 & 0.0 & 0.0 & 0.0 & 0.0 & 0.0 & 0.0 & 0.0 & 0.0 & 0.0 & 0.0 & 0.0 & 0.0 \\
\hline
\multirow{2}{*}{PDF eigenvector 17}& 0.0 & 0.0 & 0.0 & 0.0 & 0.0 & 0.0 & 0.0 & 0.0 & 0.0 & 0.0 & 0.0 & 0.0 & 0.0 & 0.0 & 0.0 & 0.0 \\
& 0.0 & 0.0 & $-0.1$ & $-0.1$ & $+0.1$ & $+0.1$ & 0.0 & $-0.1$ & 0.0 & 0.0 & $-0.1$ & $-0.1$ & 0.0 & $+0.1$ & $+0.1$ & 0.0 \\
\hline
\multirow{2}{*}{PDF eigenvector 18}& 0.0 & 0.0 & 0.0 & $+0.1$ & 0.0 & 0.0 & 0.0 & $-0.1$ & $-0.1$ & 0.0 & $+0.1$ & $+0.1$ & $+0.1$ & $+0.1$ & 0.0 & 0.0 \\
\end{tabular}}
\end{table*}

\begin{table*}[htbp]
\centering
\topcaption{The measured \mttdetatt cross sections, along with their relative statistical and systematic uncertainties.}
\label{tab:mttdetatt_xsec}
\renewcommand*{\arraystretch}{1.55}
\tabcolsep2.0mm
\begin{tabular}{cccccc}
$M(\ttbar)$  [\GeVns{}]  & $\Delta\eta(\PQt,\overline{\PQt})$  & $\frac{1}{\sigma(\ttbar)} \frac{\rd\sigma}{\rd\Delta\eta(\PQt,\overline{\PQt}) } $ & stat. [\%] & syst. [\%] & bin\\
\hline
     300--400 &      0.0--0.4 & $1.874 \times 10^{-1}$ & 4.4 & ${}_{-14.4}^{ +5.0}$ & 1 \\
     300--400 &      0.4--1.2 & $1.260 \times 10^{-1}$ & 1.5 & ${}_{ -6.0}^{ +3.4}$ & 2 \\
     300--400 &      1.2--6.0 & $1.444 \times 10^{-2}$ & 4.4 & ${}_{ -7.2}^{+17.7}$ & 3 \\
     400--500 &      0.0--0.4 & $1.633 \times 10^{-1}$ & 2.7 & ${}_{ -6.1}^{ +7.3}$ & 4 \\
     400--500 &      0.4--1.2 & $1.587 \times 10^{-1}$ & 1.4 & ${}_{ -3.1}^{ +5.2}$ & 5 \\
     400--500 &      1.2--6.0 & $3.510 \times 10^{-2}$ & 1.1 & ${}_{ -3.4}^{ +3.1}$ & 6 \\
     500--650 &      0.0--0.4 & $5.389 \times 10^{-2}$ & 4.2 & ${}_{ -7.4}^{ +5.8}$ & 7 \\
     500--650 &      0.4--1.2 & $6.632 \times 10^{-2}$ & 2.6 & ${}_{ -4.3}^{ +6.4}$ & 8 \\
     500--650 &      1.2--6.0 & $3.601 \times 10^{-2}$ & 1.0 & ${}_{ -2.9}^{ +1.4}$ & 9 \\
    650--1500 &      0.0--0.4 & $1.803 \times 10^{-2}$ & 4.8 & ${}_{ -5.4}^{ +8.5}$ & 10 \\
    650--1500 &      0.4--1.2 & $1.991 \times 10^{-2}$ & 3.3 & ${}_{ -7.0}^{ +6.0}$ & 11 \\
    650--1500 &      1.2--6.0 & $2.575 \times 10^{-2}$ & 1.0 & ${}_{ -3.5}^{ +3.9}$ & 12 \\
\end{tabular}
\end{table*}
 \begin{table*}[htbp]
\centering
\topcaption{The correlation matrix of statistical uncertainties for the measured \mttdetatt cross sections. The values are expressed as percentages. For bin indices see Table~\ref{tab:mttdetatt_xsec}.}
\label{tab:mttdetatt_corr}
\tabcolsep2.0mm
\renewcommand*{\arraystretch}{1.45}
\begin{tabular}{l|rrrrrrrrrrr}
bin & 2 & 3 & 4 & 5 & 6 & 7 & 8 & 9 & 10 & 11 & 12 \\
\hline
1 & $+11$ & $-77$ & $-3$ & $-37$ & $-22$ & $-8$ & $-2$ & $+16$ & $+1$ & $-1$ & $-13$ \\
2 & & $-8$ & $-44$ & $+13$ & $-35$ & $+11$ & $-8$ & $-18$ & $-4$ & $-1$ & $-1$ \\
3 & & & $+2$ & $-2$ & $-7$ & $+5$ & $+6$ & $-29$ & $-2$ & $-1$ & $+11$ \\
4 & & & & $-19$ & $-17$ & $-21$ & $-25$ & $+2$ & $+4$ & $+10$ & $+1$ \\
5 & & & & & $+10$ & $-14$ & $+6$ & $-30$ & $+7$ & $-6$ & $+3$ \\
6 & & & & & & $+10$ & $-15$ & $+1$ & $-3$ & $+6$ & $-21$ \\
7 & & & & & & & $-34$ & $+1$ & $-44$ & $+6$ & $+3$ \\
8 & & & & & & & & $-12$ & $+10$ & $-36$ & $-5$ \\
9 & & & & & & & & & 0 & $+1$ & $-42$ \\
10 & & & & & & & & & & $-28$ & $+1$ \\
11 & & & & & & & & & & & $-10$ \\
\end{tabular}
\end{table*}
 \begin{table*}[p]
\centering
\topcaption{Sources and values of the relative systematic uncertainties in percent of the measured \mttdetatt cross sections. For bin indices see Table~\ref{tab:mttdetatt_xsec}.}
\label{tab:mttdetatt_syst}
\renewcommand*{\arraystretch}{1.75}
\tabcolsep2.0mm
\cmsTallTable{%
\begin{tabular}{l|
rrrrrrrrrrrr}
source / bin & 1 & 2 & 3 & 4 & 5 & 6 & 7 & 8 & 9 & 10 & 11 & 12 \\
\hline
\multirow{2}{*}{JER}& $-1.7$ & $-0.1$ & $+2.3$ & $+0.8$ & $+0.4$ & $-0.2$ & $-0.2$ & $+0.4$ & $-0.6$ & $-0.5$ & $-1.0$ & 0.0 \\
& $-0.2$ & $-1.3$ & $+0.4$ & $+1.6$ & $-0.2$ & $+0.1$ & $-0.6$ & $+0.3$ & $-0.4$ & $+1.1$ & $+0.1$ & $+0.6$ \\
\hline
\multirow{2}{*}{JESAbsoluteMPFBias}& $-1.1$ & $-0.5$ & $+1.2$ & $+0.8$ & $-0.1$ & $+0.2$ & $-0.2$ & $-0.3$ & $-0.4$ & $+0.4$ & $+0.2$ & $+0.5$ \\
& $-1.5$ & $-0.4$ & $+1.0$ & $+0.2$ & $+0.7$ & 0.0 & $+0.1$ & 0.0 & $-0.3$ & $+0.7$ & $+0.1$ & $+0.1$ \\
\hline
\multirow{2}{*}{JESAbsoluteScale}& $-0.3$ & $-0.5$ & $+2.2$ & $+0.4$ & $-0.2$ & $-0.3$ & $-0.5$ & $+0.5$ & $-0.4$ & $+0.6$ & $+0.2$ & $+0.1$ \\
& $-0.6$ & $-0.5$ & $+0.6$ & $+1.3$ & $-0.4$ & $-0.1$ & 0.0 & 0.0 & $+0.3$ & 0.0 & $+0.2$ & 0.0 \\
\hline
\multirow{2}{*}{JESAbsoluteStat}& $-2.1$ & $+0.1$ & $+2.8$ & $+0.2$ & $+0.3$ & $-0.4$ & $-0.7$ & $+0.7$ & $-0.5$ & $+1.0$ & $-0.2$ & $+0.2$ \\
\end{tabular}}
\end{table*}
 \begin{table*}[p]
\centering
\topcaption{Table~\ref{tab:mttdetatt_syst} continued.}
\label{tab:mttdetatt_1_syst}
\renewcommand*{\arraystretch}{1.75}
\tabcolsep2.0mm
\cmsTallTable{%
\begin{tabular}{l|
rrrrrrrrrrrr}
source / bin & 1 & 2 & 3 & 4 & 5 & 6 & 7 & 8 & 9 & 10 & 11 & 12 \\
\hline
\multirow{2}{*}{JESSinglePionHCAL}& $-1.2$ & $-0.6$ & $+2.2$ & $+0.4$ & $+0.1$ & $+0.1$ & $+0.9$ & $-0.1$ & $-0.4$ & $-0.6$ & $+0.1$ & $+0.1$ \\
& $-0.3$ & $-0.5$ & $+1.2$ & $+0.8$ & $-0.3$ & $-0.2$ & $-0.2$ & $-0.1$ & $-0.2$ & $+0.6$ & $+0.7$ & $+0.3$ \\
\hline
\multirow{2}{*}{JESTimePtEta}& $-1.2$ & $-0.2$ & $+2.4$ & $+0.8$ & 0.0 & $-0.3$ & $+0.1$ & $+0.1$ & $-0.6$ & 0.0 & $-0.1$ & $+0.3$ \\
& $-1.3$ & $-0.8$ & $+2.2$ & $+0.4$ & $+0.1$ & $+0.3$ & $+0.9$ & $-0.5$ & $-0.7$ & $+0.1$ & $+0.7$ & $+0.4$ \\
\hline
\multirow{2}{*}{$\ptvecmiss$}& $-0.4$ & 0.0 & $-0.3$ & $+0.7$ & $-0.1$ & $+0.2$ & $-0.1$ & $-0.2$ & 0.0 & $+0.4$ & $-0.3$ & 0.0 \\
& $-1.0$ & $+0.7$ & $+0.3$ & $-0.1$ & $+0.1$ & $-0.1$ & $+0.4$ & 0.0 & $-0.2$ & $+0.5$ & 0.0 & $+0.1$ \\
\hline
\multirow{2}{*}{lepton ID/ISO}& $-0.2$ & $+0.2$ & $-0.6$ & $-0.1$ & $+0.3$ & $+0.1$ & 0.0 & $-0.3$ & $+0.2$ & $-0.3$ & $-0.4$ & $-0.1$ \\
\end{tabular}}
\end{table*}
 \begin{table*}[p]
\centering
\topcaption{Table~\ref{tab:mttdetatt_syst} continued.}
\label{tab:mttdetatt_2_syst}
\renewcommand*{\arraystretch}{1.75}
\tabcolsep2.0mm
\cmsTallTable{%
\begin{tabular}{l|
rrrrrrrrrrrr}
source / bin & 1 & 2 & 3 & 4 & 5 & 6 & 7 & 8 & 9 & 10 & 11 & 12 \\
\hline
\multirow{2}{*}{PDF eigenvector 15}& $-0.1$ & 0.0 & $+0.2$ & 0.0 & 0.0 & 0.0 & 0.0 & 0.0 & $+0.1$ & 0.0 & $-0.1$ & 0.0 \\
& 0.0 & 0.0 & $-0.1$ & 0.0 & 0.0 & 0.0 & 0.0 & 0.0 & 0.0 & 0.0 & 0.0 & 0.0 \\
\hline
\multirow{2}{*}{PDF eigenvector 16}& 0.0 & 0.0 & $-0.1$ & 0.0 & 0.0 & 0.0 & 0.0 & 0.0 & 0.0 & 0.0 & 0.0 & 0.0 \\
& 0.0 & 0.0 & $+0.1$ & 0.0 & 0.0 & 0.0 & 0.0 & 0.0 & 0.0 & 0.0 & $-0.1$ & 0.0 \\
\hline
\multirow{2}{*}{PDF eigenvector 17}& $-0.1$ & 0.0 & $+0.1$ & 0.0 & 0.0 & 0.0 & 0.0 & 0.0 & 0.0 & 0.0 & $-0.1$ & 0.0 \\
& $+0.1$ & $+0.1$ & $-0.2$ & 0.0 & $+0.1$ & 0.0 & 0.0 & 0.0 & $-0.1$ & $+0.1$ & $+0.1$ & 0.0 \\
\hline
\multirow{2}{*}{PDF eigenvector 18}& $+0.3$ & 0.0 & $-0.2$ & $+0.1$ & $-0.1$ & 0.0 & 0.0 & $+0.1$ & $-0.1$ & 0.0 & $+0.1$ & $+0.1$ \\
\end{tabular}}
\end{table*}

\begin{table*}[htbp]
\centering
\topcaption{The measured \mttdphitt cross sections, along with their relative statistical and systematic uncertainties.}
\label{tab:mttdphitt_xsec}
\renewcommand*{\arraystretch}{1.55}
\tabcolsep2.0mm
\begin{tabular}{cccccc}
$M(\ttbar)$  [\GeVns{}]  & $\Delta\phi(\PQt,\overline{\PQt})$ [rad]  & $\frac{1}{\sigma(\ttbar)} \frac{\rd\sigma}{\rd\Delta\phi(\PQt,\overline{\PQt}) } [ \mathrm{\mathrm{rad}}^{-1}]$ & stat. [\%] & syst. [\%] & bin\\
\hline
     300--400 &    0.00--2.20 & $3.535 \times 10^{-2}$ & 3.0 & ${}_{-16.2}^{ +6.3}$ & 1 \\
     300--400 &    2.20--2.95 & $1.249 \times 10^{-1}$ & 1.9 & ${}_{ -5.6}^{ +9.7}$ & 2 \\
     300--400 &    2.95--3.14 & $4.024 \times 10^{-1}$ & 3.2 & ${}_{-15.5}^{+20.3}$ & 3 \\
     400--500 &    0.00--2.20 & $2.617 \times 10^{-2}$ & 3.0 & ${}_{-10.7}^{+13.6}$ & 4 \\
     400--500 &    2.20--2.95 & $1.957 \times 10^{-1}$ & 1.2 & ${}_{ -6.3}^{ +5.7}$ & 5 \\
     400--500 &    2.95--3.14 & $8.082 \times 10^{-1}$ & 1.3 & ${}_{ -5.6}^{ +6.7}$ & 6 \\
     500--650 &    0.00--2.20 & $1.439 \times 10^{-2}$ & 4.0 & ${}_{-14.4}^{ +4.9}$ & 7 \\
     500--650 &    2.20--2.95 & $1.250 \times 10^{-1}$ & 1.8 & ${}_{ -6.9}^{ +6.9}$ & 8 \\
     500--650 &    2.95--3.14 & $6.251 \times 10^{-1}$ & 1.4 & ${}_{ -6.2}^{ +5.9}$ & 9 \\
    650--1500 &    0.00--2.20 & $7.308 \times 10^{-3}$ & 4.8 & ${}_{ -7.6}^{+12.2}$ & 10 \\
    650--1500 &    2.20--2.95 & $7.475 \times 10^{-2}$ & 2.1 & ${}_{ -6.1}^{ +6.9}$ & 11 \\
    650--1500 &    2.95--3.14 & $3.923 \times 10^{-1}$ & 1.5 & ${}_{ -6.2}^{ +5.4}$ & 12 \\
\end{tabular}
\end{table*}
 \begin{table*}[htbp]
\centering
\topcaption{The correlation matrix of statistical uncertainties for the measured \mttdphitt cross sections. The values are expressed as percentages. For bin indices see Table~\ref{tab:mttdphitt_xsec}.}
\label{tab:mttdphitt_corr}
\tabcolsep2.0mm
\renewcommand*{\arraystretch}{1.45}
\begin{tabular}{l|rrrrrrrrrrr}
bin & 2 & 3 & 4 & 5 & 6 & 7 & 8 & 9 & 10 & 11 & 12 \\
\hline
1 & $+4$ & $-50$ & $-39$ & $-26$ & $+1$ & $-11$ & $-5$ & $+9$ & $+6$ & $+2$ & $-5$ \\
2 & & $-23$ & $-26$ & $+16$ & $-50$ & $+6$ & $-13$ & $+1$ & $-2$ & $-2$ & 0 \\
3 & & & $+18$ & $-37$ & $-12$ & $+4$ & $+4$ & $-19$ & $-6$ & $-3$ & $-1$ \\
4 & & & & $+1$ & $-10$ & $-30$ & $-18$ & 0 & $-1$ & $+2$ & $+1$ \\
5 & & & & & $-7$ & $-12$ & $+19$ & $-35$ & $+5$ & $-15$ & $+10$ \\
6 & & & & & & $+5$ & $-30$ & $-1$ & $-3$ & $+8$ & $-11$ \\
7 & & & & & & & $-9$ & $-4$ & $-43$ & $-8$ & $+4$ \\
8 & & & & & & & & $-24$ & $+1$ & $-25$ & $-6$ \\
9 & & & & & & & & & $+2$ & $+1$ & $-32$ \\
10 & & & & & & & & & & $-23$ & $+1$ \\
11 & & & & & & & & & & & $-34$ \\
\end{tabular}
\end{table*}
 \begin{table*}[p]
\centering
\topcaption{Sources and values of the relative systematic uncertainties in percent of the measured \mttdphitt cross sections. For bin indices see Table~\ref{tab:mttdphitt_xsec}.}
\label{tab:mttdphitt_syst}
\renewcommand*{\arraystretch}{1.75}
\tabcolsep2.0mm
\cmsTallTable{%
\begin{tabular}{l|
rrrrrrrrrrrr}
source / bin & 1 & 2 & 3 & 4 & 5 & 6 & 7 & 8 & 9 & 10 & 11 & 12 \\
\hline
\multirow{2}{*}{JER}& $-1.8$ & $-0.1$ & $+2.7$ & $+0.7$ & $-0.4$ & $+0.3$ & $-2.1$ & $+0.1$ & $+0.1$ & $-0.1$ & $+0.2$ & $-0.6$ \\
& $-0.8$ & $-0.1$ & $-1.1$ & $+0.5$ & $+0.8$ & $+0.1$ & $-2.4$ & $+0.7$ & $-0.6$ & $+1.9$ & $+0.4$ & $+0.3$ \\
\hline
\multirow{2}{*}{JESAbsoluteMPFBias}& $-3.0$ & $+1.1$ & $+1.1$ & $+0.9$ & $+0.7$ & $-0.1$ & $-2.9$ & 0.0 & $-0.5$ & $+1.7$ & 0.0 & $+0.3$ \\
& $-3.6$ & $+0.1$ & $+2.1$ & $+2.0$ & $+0.2$ & $+0.1$ & $+0.1$ & $+0.2$ & $-0.6$ & $-0.3$ & $+0.2$ & $-0.1$ \\
\hline
\multirow{2}{*}{JESAbsoluteScale}& $-1.0$ & $+0.3$ & $+0.9$ & 0.0 & $+0.3$ & 0.0 & $-2.1$ & $+0.7$ & $-0.8$ & 0.0 & $+0.6$ & 0.0 \\
& $-1.8$ & $+0.4$ & $+0.5$ & $-0.9$ & $+0.7$ & 0.0 & $-0.4$ & $+0.8$ & $-0.4$ & $+0.4$ & $+0.2$ & $-0.2$ \\
\hline
\multirow{2}{*}{JESAbsoluteStat}& $-4.8$ & $+0.7$ & $+6.8$ & $-4.1$ & $-1.8$ & $+2.6$ & $-1.9$ & $-2.2$ & $+1.8$ & $-1.7$ & $-2.8$ & $+1.7$ \\
\end{tabular}}
\end{table*}
 \begin{table*}[p]
\centering
\topcaption{Table~\ref{tab:mttdphitt_syst} continued.}
\label{tab:mttdphitt_1_syst}
\renewcommand*{\arraystretch}{1.75}
\tabcolsep2.0mm
\cmsTallTable{%
\begin{tabular}{l|
rrrrrrrrrrrr}
source / bin & 1 & 2 & 3 & 4 & 5 & 6 & 7 & 8 & 9 & 10 & 11 & 12 \\
\hline
\multirow{2}{*}{JESSinglePionHCAL}& $-1.6$ & $+1.0$ & $+0.8$ & $-0.2$ & $+0.3$ & $+0.1$ & 0.0 & $-0.5$ & $-0.2$ & $+0.5$ & $+0.2$ & $-0.2$ \\
& $-1.3$ & $+1.3$ & $+0.9$ & $-0.2$ & $-0.3$ & $-0.3$ & $-1.7$ & $-0.1$ & $+0.2$ & $-0.1$ & $+0.7$ & $+0.2$ \\
\hline
\multirow{2}{*}{JESTimePtEta}& $-1.1$ & $+0.4$ & $+0.7$ & $+0.3$ & $+0.2$ & $+0.1$ & $-1.7$ & $+0.1$ & $-0.2$ & $+1.3$ & $+0.3$ & $-0.2$ \\
& $-1.8$ & $+0.2$ & $+1.8$ & $+0.2$ & $-0.2$ & $+0.3$ & $-1.7$ & $-0.1$ & $-0.3$ & $+1.0$ & $+0.6$ & $+0.1$ \\
\hline
\multirow{2}{*}{$\ptvecmiss$}& $-5.6$ & $-2.7$ & $+10.1$ & $+0.3$ & $-3.4$ & $+3.3$ & $+0.1$ & $-3.8$ & $+2.4$ & $-0.3$ & $-1.5$ & $+0.5$ \\
& $+1.6$ & $+4.1$ & $-6.2$ & $-1.0$ & $+2.3$ & $-2.1$ & $-1.3$ & $+2.2$ & $-1.7$ & $-0.2$ & $+1.4$ & $-0.4$ \\
\hline
\multirow{2}{*}{lepton ID/ISO}& $-0.1$ & 0.0 & $-0.4$ & $-0.2$ & $+0.4$ & $+0.1$ & $+0.2$ & 0.0 & $-0.1$ & 0.0 & 0.0 & $-0.2$ \\
\end{tabular}}
\end{table*}
 \begin{table*}[p]
\centering
\topcaption{Table~\ref{tab:mttdphitt_syst} continued.}
\label{tab:mttdphitt_2_syst}
\renewcommand*{\arraystretch}{1.75}
\tabcolsep2.0mm
\cmsTallTable{%
\begin{tabular}{l|
rrrrrrrrrrrr}
source / bin & 1 & 2 & 3 & 4 & 5 & 6 & 7 & 8 & 9 & 10 & 11 & 12 \\
\hline
\multirow{2}{*}{PDF eigenvector 15}& 0.0 & 0.0 & 0.0 & 0.0 & 0.0 & 0.0 & 0.0 & 0.0 & 0.0 & $-0.1$ & 0.0 & 0.0 \\
& 0.0 & 0.0 & 0.0 & 0.0 & 0.0 & 0.0 & 0.0 & 0.0 & 0.0 & 0.0 & 0.0 & 0.0 \\
\hline
\multirow{2}{*}{PDF eigenvector 16}& 0.0 & 0.0 & 0.0 & 0.0 & 0.0 & 0.0 & 0.0 & 0.0 & 0.0 & 0.0 & 0.0 & 0.0 \\
& 0.0 & 0.0 & 0.0 & 0.0 & 0.0 & 0.0 & 0.0 & 0.0 & 0.0 & 0.0 & 0.0 & 0.0 \\
\hline
\multirow{2}{*}{PDF eigenvector 17}& 0.0 & 0.0 & 0.0 & 0.0 & 0.0 & 0.0 & 0.0 & 0.0 & 0.0 & 0.0 & 0.0 & 0.0 \\
& 0.0 & 0.0 & $-0.1$ & 0.0 & $+0.1$ & 0.0 & 0.0 & $-0.1$ & 0.0 & $+0.1$ & 0.0 & 0.0 \\
\hline
\multirow{2}{*}{PDF eigenvector 18}& 0.0 & $-0.1$ & 0.0 & $+0.1$ & $-0.1$ & 0.0 & 0.0 & 0.0 & 0.0 & $+0.1$ & $+0.1$ & $+0.1$ \\
\end{tabular}}
\end{table*}

\begin{table*}[htbp]
\centering
\topcaption{The measured \mttpttt cross sections, along with their relative statistical and systematic uncertainties.}
\label{tab:mttpttt_xsec}
\renewcommand*{\arraystretch}{1.55}
\tabcolsep2.0mm
\begin{tabular}{cccccc}
$M(\ttbar)$  [\GeVns{}]  & $p_\PQt(\ttbar)$  [\GeVns{}]  & $\frac{1}{\sigma(\ttbar)} \frac{\rd\sigma}{\rd{}p_\PQt(\ttbar) } [ \GeVns{}^{-1}]$ & stat. [\%] & syst. [\%] & bin\\
\hline
     300--400 &         0--30 & $3.671 \times 10^{-3}$ & 1.8 & ${}_{ -9.3}^{+12.4}$ & 1 \\
     300--400 &        30--75 & $1.660 \times 10^{-3}$ & 2.1 & ${}_{-14.1}^{+10.6}$ & 2 \\
     300--400 &       75--150 & $5.749 \times 10^{-4}$ & 3.3 & ${}_{ -8.4}^{ +9.7}$ & 3 \\
     300--400 &      150--500 & $5.519 \times 10^{-5}$ & 5.1 & ${}_{ -9.7}^{ +8.1}$ & 4 \\
     400--500 &         0--30 & $4.915 \times 10^{-3}$ & 1.1 & ${}_{ -8.0}^{ +9.8}$ & 5 \\
     400--500 &        30--75 & $2.609 \times 10^{-3}$ & 1.2 & ${}_{ -7.8}^{ +7.8}$ & 6 \\
     400--500 &       75--150 & $8.995 \times 10^{-4}$ & 1.9 & ${}_{ -8.0}^{ +7.1}$ & 7 \\
     400--500 &      150--500 & $8.845 \times 10^{-5}$ & 3.8 & ${}_{-10.8}^{ +6.4}$ & 8 \\
     500--650 &         0--30 & $3.107 \times 10^{-3}$ & 1.6 & ${}_{-11.3}^{+12.1}$ & 9 \\
     500--650 &        30--75 & $1.733 \times 10^{-3}$ & 1.7 & ${}_{-11.2}^{+10.1}$ & 10 \\
     500--650 &       75--150 & $6.394 \times 10^{-4}$ & 2.5 & ${}_{ -8.0}^{ +6.8}$ & 11 \\
     500--650 &      150--500 & $7.086 \times 10^{-5}$ & 3.6 & ${}_{ -5.2}^{ +5.8}$ & 12 \\
    650--1500 &         0--30 & $1.705 \times 10^{-3}$ & 2.6 & ${}_{-12.3}^{+13.7}$ & 13 \\
    650--1500 &        30--75 & $9.328 \times 10^{-4}$ & 3.3 & ${}_{-15.5}^{+13.2}$ & 14 \\
    650--1500 &       75--150 & $4.219 \times 10^{-4}$ & 2.9 & ${}_{ -6.2}^{ +7.0}$ & 15 \\
    650--1500 &      150--500 & $5.924 \times 10^{-5}$ & 2.7 & ${}_{ -5.1}^{ +6.1}$ & 16 \\
\end{tabular}
\end{table*}
 \begin{table*}[htbp]
\centering
\topcaption{The correlation matrix of statistical uncertainties for the measured \mttpttt cross sections. The values are expressed as percentages. For bin indices see Table~\ref{tab:mttpttt_xsec}.}
\label{tab:mttpttt_corr}
\tabcolsep2.0mm
\renewcommand*{\arraystretch}{1.45}
\cmsSwitchTable{%
\begin{tabular}{l|rrrrrrrrrrrrrrr}
bin & 2 & 3 & 4 & 5 & 6 & 7 & 8 & 9 & 10 & 11 & 12 & 13 & 14 & 15 & 16 \\
\hline
1 & $-28$ & $-27$ & $+2$ & $-19$ & $-35$ & $+5$ & $-1$ & $-27$ & $+1$ & $+9$ & $-5$ & 0 & 0 & $-8$ & $-2$ \\
2 & & $+12$ & $-9$ & $-37$ & $+2$ & $-29$ & $+4$ & 0 & $-13$ & $-5$ & $+1$ & $+2$ & $-2$ & $+3$ & $-3$ \\
3 & & & $-9$ & 0 & $-33$ & $-30$ & $-7$ & $+11$ & $-9$ & $-18$ & $+7$ & $-5$ & $+3$ & $+6$ & $-3$ \\
4 & & & & 0 & $+4$ & $-10$ & $-63$ & $-4$ & $+2$ & $+6$ & $+10$ & 0 & $-1$ & $-2$ & $-1$ \\
5 & & & & & $-3$ & $-14$ & $-1$ & $+13$ & $-34$ & $-4$ & $+1$ & $-17$ & $+6$ & $+3$ & $-4$ \\
6 & & & & & & $+21$ & $-7$ & $-34$ & $+25$ & $-20$ & $+3$ & $+7$ & $-11$ & $+2$ & $+1$ \\
7 & & & & & & & $+3$ & $-7$ & $-18$ & $-7$ & $-10$ & $+4$ & $+1$ & $-8$ & $+5$ \\
8 & & & & & & & & $+1$ & $+1$ & $-9$ & $-46$ & $-1$ & 0 & $+3$ & $+8$ \\
9 & & & & & & & & & $-21$ & $-16$ & $+2$ & $-16$ & $-13$ & $+12$ & $-3$ \\
10 & & & & & & & & & & $+5$ & $-5$ & $-16$ & $+2$ & $-17$ & $+6$ \\
11 & & & & & & & & & & & $-8$ & $+9$ & $-13$ & $-27$ & $+1$ \\
12 & & & & & & & & & & & & $-2$ & $+3$ & $-2$ & $-45$ \\
13 & & & & & & & & & & & & & $-57$ & $+3$ & $+1$ \\
14 & & & & & & & & & & & & & & $-30$ & 0 \\
15 & & & & & & & & & & & & & & & $-21$ \\
\end{tabular}}
\end{table*}
 \begin{table*}[p]
\centering
\topcaption{Sources and values of the relative systematic uncertainties in percent of the measured \mttpttt cross sections. For bin indices see Table~\ref{tab:mttpttt_xsec}.}
\label{tab:mttpttt_syst}
\renewcommand*{\arraystretch}{1.75}
\tabcolsep2.0mm
\cmsTallTable{%
\begin{tabular}{l|
rrrrrrrrrrrrrrrr}
source / bin & 1 & 2 & 3 & 4 & 5 & 6 & 7 & 8 & 9 & 10 & 11 & 12 & 13 & 14 & 15 & 16 \\
\hline
\multirow{2}{*}{JER}& $+0.4$ & $-0.8$ & $-0.1$ & $+0.5$ & $+0.7$ & 0.0 & $-0.4$ & $-1.8$ & $+0.9$ & $-0.2$ & $-0.8$ & $+0.4$ & $+0.2$ & $-1.2$ & $-0.1$ & $+0.1$ \\
& $-0.3$ & $-0.8$ & $+0.2$ & $+0.1$ & $-0.1$ & $+0.7$ & $+0.1$ & $-0.6$ & $-0.9$ & $+0.7$ & 0.0 & $+0.1$ & $-0.3$ & $+0.3$ & $+1.5$ & $+0.3$ \\
\hline
\multirow{2}{*}{JESAbsoluteMPFBias}& $-0.1$ & $-0.1$ & $-0.2$ & $-0.1$ & $+0.4$ & $+0.5$ & $-0.5$ & $-0.9$ & $-0.5$ & 0.0 & $-0.5$ & 0.0 & $+0.1$ & $+0.4$ & $+0.5$ & $-0.4$ \\
& $+0.3$ & $-0.5$ & $-0.5$ & $-3.0$ & $+0.2$ & $+0.2$ & $-0.2$ & $+1.6$ & 0.0 & $+0.2$ & $-0.4$ & $-0.1$ & $-0.2$ & $-0.1$ & $+0.7$ & 0.0 \\
\hline
\multirow{2}{*}{JESAbsoluteScale}& $+0.7$ & $-0.3$ & $-0.2$ & $-0.4$ & $+0.1$ & $+0.1$ & 0.0 & $-0.4$ & $-0.5$ & $+0.2$ & $-0.8$ & $-0.4$ & $-0.1$ & $+0.4$ & $+0.4$ & $-0.1$ \\
& $+0.1$ & $-0.1$ & $-0.8$ & $-1.9$ & $+0.2$ & $+0.3$ & $+0.1$ & $+0.5$ & 0.0 & $-0.1$ & $-0.1$ & $+0.3$ & $-0.6$ & $+0.2$ & $+0.2$ & 0.0 \\
\hline
\multirow{2}{*}{JESAbsoluteStat}& $+4.4$ & $-3.4$ & $-3.0$ & $-2.9$ & $+3.7$ & $-1.9$ & $-4.2$ & $-2.6$ & $+4.0$ & $-2.2$ & $-3.5$ & $-1.9$ & $+4.8$ & $-2.6$ & $-2.7$ & $-2.3$ \\
\end{tabular}}
\end{table*}
 \begin{table*}[p]
\centering
\topcaption{Table~\ref{tab:mttpttt_syst} continued.}
\label{tab:mttpttt_1_syst}
\renewcommand*{\arraystretch}{1.75}
\tabcolsep2.0mm
\cmsTallTable{%
\begin{tabular}{l|
rrrrrrrrrrrrrrrr}
source / bin & 1 & 2 & 3 & 4 & 5 & 6 & 7 & 8 & 9 & 10 & 11 & 12 & 13 & 14 & 15 & 16 \\
\hline
\multirow{2}{*}{JESSinglePionHCAL}& $+0.5$ & $-0.4$ & $-1.2$ & $-1.0$ & $+0.3$ & $+0.2$ & $-0.1$ & $+0.7$ & $-0.3$ & $+0.4$ & $-0.2$ & $-0.9$ & $-0.3$ & $-0.3$ & $+0.5$ & $+0.1$ \\
& $+0.5$ & $+0.4$ & $-1.0$ & $+0.2$ & 0.0 & $+0.1$ & $-0.7$ & $-0.4$ & $-0.1$ & $+0.1$ & $-0.5$ & $-0.6$ & $+0.6$ & $-0.4$ & $+0.7$ & $+0.5$ \\
\hline
\multirow{2}{*}{JESTimePtEta}& $+0.2$ & $+0.4$ & $-0.1$ & $-0.4$ & $+0.2$ & $+0.4$ & $-0.8$ & $-0.4$ & $-0.5$ & $+0.2$ & $-0.2$ & $-0.7$ & $+0.1$ & $-0.3$ & $+0.2$ & $+0.5$ \\
& $-0.1$ & $-0.1$ & $-0.6$ & $-0.2$ & $+0.4$ & $+0.6$ & $-0.5$ & $-0.7$ & $-0.3$ & $-0.1$ & $-0.9$ & $+0.8$ & 0.0 & $+0.2$ & $+0.9$ & $+0.2$ \\
\hline
\multirow{2}{*}{$\ptvecmiss$}& $+6.5$ & $-7.8$ & $-3.1$ & $-0.8$ & $+5.2$ & $-4.8$ & $-1.4$ & $-0.2$ & $+6.7$ & $-7.2$ & $-1.9$ & $-0.5$ & $+7.5$ & $-9.9$ & $+0.2$ & $-0.6$ \\
& $-4.0$ & $+5.6$ & $+1.0$ & $+0.4$ & $-3.6$ & $+4.2$ & $+0.3$ & $+0.1$ & $-5.0$ & $+5.7$ & $-0.4$ & $-0.3$ & $-4.8$ & $+6.7$ & $-0.2$ & $+0.5$ \\
\hline
\multirow{2}{*}{lepton ID/ISO}& $-0.2$ & $-0.3$ & $-0.1$ & 0.0 & $+0.2$ & $+0.3$ & $+0.1$ & 0.0 & $-0.1$ & 0.0 & 0.0 & 0.0 & $-0.1$ & $-0.1$ & $-0.1$ & $-0.2$ \\
\end{tabular}}
\end{table*}
 \begin{table*}[p]
\centering
\topcaption{Table~\ref{tab:mttpttt_syst} continued.}
\label{tab:mttpttt_2_syst}
\renewcommand*{\arraystretch}{1.75}
\tabcolsep2.0mm
\cmsTallTable{%
\begin{tabular}{l|
rrrrrrrrrrrrrrrr}
source / bin & 1 & 2 & 3 & 4 & 5 & 6 & 7 & 8 & 9 & 10 & 11 & 12 & 13 & 14 & 15 & 16 \\
\hline
\multirow{2}{*}{PDF eigenvector 15}& 0.0 & 0.0 & 0.0 & 0.0 & 0.0 & 0.0 & 0.0 & $-0.1$ & 0.0 & 0.0 & 0.0 & 0.0 & 0.0 & 0.0 & $-0.1$ & $-0.1$ \\
& 0.0 & 0.0 & 0.0 & 0.0 & 0.0 & 0.0 & 0.0 & 0.0 & 0.0 & 0.0 & 0.0 & 0.0 & 0.0 & 0.0 & 0.0 & 0.0 \\
\hline
\multirow{2}{*}{PDF eigenvector 16}& 0.0 & 0.0 & 0.0 & 0.0 & 0.0 & 0.0 & 0.0 & 0.0 & 0.0 & 0.0 & 0.0 & 0.0 & 0.0 & 0.0 & 0.0 & 0.0 \\
& 0.0 & 0.0 & 0.0 & 0.0 & 0.0 & 0.0 & 0.0 & 0.0 & 0.0 & 0.0 & 0.0 & 0.0 & 0.0 & 0.0 & 0.0 & 0.0 \\
\hline
\multirow{2}{*}{PDF eigenvector 17}& 0.0 & 0.0 & 0.0 & 0.0 & 0.0 & 0.0 & 0.0 & 0.0 & 0.0 & 0.0 & 0.0 & 0.0 & 0.0 & 0.0 & 0.0 & 0.0 \\
& $-0.1$ & 0.0 & 0.0 & $+0.1$ & 0.0 & $+0.1$ & 0.0 & $+0.1$ & $-0.1$ & 0.0 & $-0.1$ & 0.0 & 0.0 & 0.0 & 0.0 & $+0.1$ \\
\hline
\multirow{2}{*}{PDF eigenvector 18}& 0.0 & 0.0 & 0.0 & 0.0 & 0.0 & 0.0 & 0.0 & $+0.1$ & 0.0 & 0.0 & 0.0 & 0.0 & 0.0 & $+0.1$ & $+0.1$ & $+0.1$ \\
\end{tabular}}
\end{table*}

\begin{table*}[htbp]
\centering
\topcaption{The measured \mttptt cross sections, along with their relative statistical and systematic uncertainties.}
\label{tab:mttptt_xsec}
\renewcommand*{\arraystretch}{1.55}
\tabcolsep2.0mm
\begin{tabular}{cccccc}
$M(\ttbar)$  [\GeVns{}]  & $p_\PQt(\PQt)$  [\GeVns{}]  & $\frac{1}{\sigma(\ttbar)} \frac{\rd\sigma}{\rd{}p_\PQt(\PQt) } [ \GeVns{}^{-1}]$ & stat. [\%] & syst. [\%] & bin\\
\hline
     300--450 &        0--100 & $3.173 \times 10^{-3}$ & 0.6 & ${}_{ -2.5}^{ +2.7}$ & 1 \\
     300--450 &      100--180 & $1.440 \times 10^{-3}$ & 1.2 & ${}_{ -5.7}^{ +5.5}$ & 2 \\
     300--450 &      180--600 & $1.956 \times 10^{-5}$ & 6.5 & ${}_{-28.8}^{+27.8}$ & 3 \\
     450--600 &        0--100 & $1.199 \times 10^{-3}$ & 1.2 & ${}_{ -4.6}^{ +4.9}$ & 4 \\
     450--600 &      100--180 & $2.104 \times 10^{-3}$ & 0.7 & ${}_{ -4.6}^{ +4.4}$ & 5 \\
     450--600 &      180--600 & $1.424 \times 10^{-4}$ & 1.9 & ${}_{ -3.4}^{ +2.6}$ & 6 \\
    600--1500 &        0--100 & $5.409 \times 10^{-4}$ & 2.5 & ${}_{ -5.1}^{ +5.4}$ & 7 \\
    600--1500 &      100--180 & $8.166 \times 10^{-4}$ & 1.9 & ${}_{ -4.6}^{ +5.4}$ & 8 \\
    600--1500 &      180--600 & $2.187 \times 10^{-4}$ & 1.0 & ${}_{ -4.4}^{ +4.4}$ & 9 \\
\end{tabular}
\end{table*}
 \begin{table*}[htbp]
\centering
\topcaption{The correlation matrix of statistical uncertainties for the measured \mttptt cross sections. The values are expressed as percentages. For bin indices see Table~\ref{tab:mttptt_xsec}.}
\label{tab:mttptt_corr}
\tabcolsep2.0mm
\renewcommand*{\arraystretch}{1.45}
\begin{tabular}{l|rrrrrrrr}
bin & 2 & 3 & 4 & 5 & 6 & 7 & 8 & 9 \\
\hline
1 & $-19$ & $-12$ & $-8$ & $-51$ & $-9$ & $-36$ & $-7$ & $-2$ \\
2 & & $-7$ & $-41$ & 0 & $-26$ & $+3$ & $-16$ & $+1$ \\
3 & & & $+2$ & $-3$ & $-22$ & $-1$ & $+3$ & $+1$ \\
4 & & & & $+2$ & $-16$ & $-17$ & $-27$ & $-1$ \\
5 & & & & & $-5$ & $-15$ & $+5$ & $-16$ \\
6 & & & & & & $+7$ & $-2$ & $-30$ \\
7 & & & & & & & $-27$ & $-7$ \\
8 & & & & & & & & $-22$ \\
\end{tabular}
\end{table*}
 \begin{table*}[p]
\centering
\topcaption{Sources and values of the relative systematic uncertainties in percent of the measured \mttptt cross sections. For bin indices see Table~\ref{tab:mttptt_xsec}.}
\label{tab:mttptt_syst}
\renewcommand*{\arraystretch}{1.75}
\tabcolsep2.0mm
\cmsTallTable{%
\begin{tabular}{l|
rrrrrrrrr}
source / bin & 1 & 2 & 3 & 4 & 5 & 6 & 7 & 8 & 9 \\
\hline
\multirow{2}{*}{JER}& $+0.3$ & $+0.1$ & $-3.7$ & $-0.3$ & 0.0 & $+0.2$ & $+0.5$ & $-0.7$ & $-0.2$ \\
& $-0.2$ & 0.0 & $-0.9$ & $-0.1$ & $+0.1$ & $+0.1$ & $+0.8$ & 0.0 & $+0.3$ \\
\hline
\multirow{2}{*}{JESAbsoluteMPFBias}& 0.0 & $+0.3$ & $-1.7$ & $-0.1$ & $-0.2$ & 0.0 & $+0.3$ & $-0.2$ & $+0.3$ \\
& $-0.2$ & $+0.6$ & $-0.4$ & 0.0 & $-0.1$ & 0.0 & $-0.1$ & 0.0 & $+0.1$ \\
\hline
\multirow{2}{*}{JESAbsoluteScale}& $+0.2$ & $-0.2$ & $-0.1$ & 0.0 & $-0.2$ & $+0.1$ & $+0.1$ & 0.0 & $+0.1$ \\
& $+0.1$ & $-0.3$ & $-1.0$ & $+0.5$ & $-0.2$ & $+0.2$ & $-0.4$ & $-0.3$ & $+0.1$ \\
\hline
\multirow{2}{*}{JESAbsoluteStat}& $+0.6$ & $-0.8$ & $-8.2$ & $-0.3$ & $+0.4$ & $-0.5$ & 0.0 & $-0.4$ & $+0.1$ \\
\end{tabular}}
\end{table*}
 \begin{table*}[p]
\centering
\topcaption{Table~\ref{tab:mttptt_syst} continued.}
\label{tab:mttptt_1_syst}
\renewcommand*{\arraystretch}{1.75}
\tabcolsep2.0mm
\cmsTallTable{%
\begin{tabular}{l|
rrrrrrrrr}
source / bin & 1 & 2 & 3 & 4 & 5 & 6 & 7 & 8 & 9 \\
\hline
\multirow{2}{*}{JESSinglePionHCAL}& $-0.1$ & $+0.3$ & $-0.2$ & $+0.4$ & $-0.1$ & $-0.2$ & $+0.7$ & $-0.5$ & $-0.1$ \\
& $+0.1$ & $-0.3$ & $-0.1$ & $-0.2$ & $-0.2$ & $-0.3$ & $+0.3$ & $+0.1$ & $+0.4$ \\
\hline
\multirow{2}{*}{JESTimePtEta}& 0.0 & $+0.1$ & $-1.6$ & $-0.1$ & 0.0 & 0.0 & $+0.2$ & $-0.2$ & $+0.2$ \\
& $+0.1$ & $-0.1$ & $-1.4$ & $+0.1$ & $-0.3$ & $+0.4$ & 0.0 & 0.0 & $+0.2$ \\
\hline
\multirow{2}{*}{$\ptvecmiss$}& $+0.2$ & $-0.6$ & $-1.3$ & $+0.2$ & $+0.1$ & $-0.6$ & $-0.1$ & $-0.2$ & 0.0 \\
& 0.0 & $+0.4$ & $+0.7$ & $-0.3$ & $-0.2$ & $+0.1$ & $-0.2$ & $+0.4$ & $+0.1$ \\
\hline
\multirow{2}{*}{lepton ID/ISO}& $+0.2$ & $-0.7$ & $+0.5$ & $-0.7$ & $+0.6$ & $-0.3$ & $+0.7$ & 0.0 & $-0.3$ \\
\end{tabular}}
\end{table*}
 \begin{table*}[p]
\centering
\topcaption{Table~\ref{tab:mttptt_syst} continued.}
\label{tab:mttptt_2_syst}
\renewcommand*{\arraystretch}{1.75}
\tabcolsep2.0mm
\cmsTallTable{%
\begin{tabular}{l|
rrrrrrrrr}
source / bin & 1 & 2 & 3 & 4 & 5 & 6 & 7 & 8 & 9 \\
\hline
\multirow{2}{*}{PDF eigenvector 15}& 0.0 & 0.0 & $-0.2$ & 0.0 & 0.0 & 0.0 & 0.0 & 0.0 & 0.0 \\
& 0.0 & 0.0 & $+0.1$ & 0.0 & 0.0 & 0.0 & 0.0 & 0.0 & 0.0 \\
\hline
\multirow{2}{*}{PDF eigenvector 16}& 0.0 & 0.0 & $+0.1$ & $-0.1$ & 0.0 & 0.0 & 0.0 & 0.0 & 0.0 \\
& 0.0 & 0.0 & $-0.1$ & $+0.1$ & 0.0 & 0.0 & 0.0 & 0.0 & 0.0 \\
\hline
\multirow{2}{*}{PDF eigenvector 17}& 0.0 & 0.0 & $-0.1$ & $+0.1$ & $-0.1$ & 0.0 & 0.0 & $+0.1$ & 0.0 \\
& $+0.1$ & $-0.1$ & $+0.3$ & $-0.2$ & $+0.1$ & 0.0 & $-0.1$ & $-0.2$ & 0.0 \\
\hline
\multirow{2}{*}{PDF eigenvector 18}& 0.0 & $+0.1$ & $+0.3$ & 0.0 & $-0.1$ & 0.0 & 0.0 & $+0.2$ & $+0.1$ \\
\end{tabular}}
\end{table*}

\begin{table*}[htbp]
\centering
\topcaption{The measured \njmttytttwo cross sections, along with their relative statistical and systematic uncertainties, and NP corrections (see Section~\ref{sec:asmt}).}
\label{tab:nj2mttytttwo_xsec}
\renewcommand*{\arraystretch}{1.55}
\tabcolsep2.0mm
\cmsSwitchTable{%
\begin{tabular}{cccccccc}
$N_\mathrm{jet}$  & $M(\ttbar)$  [\GeVns{}]  & $\abs{y(\ttbar)}$  & $\frac{1}{\sigma(\ttbar)} \frac{\rd\sigma}{\rd\abs{y(\ttbar)} } $ & stat. [\%] & syst. [\%] & {$\mathcal{C}_{\text{NP}}$} & bin\\
\hline
              0 &      300--400 &    0.00--0.35 & $8.224 \times 10^{-2}$ & 4.6 & ${}_{ -8.5}^{+12.9}$ & 0.973 & 1 \\
              0 &      300--400 &    0.35--0.75 & $8.712 \times 10^{-2}$ & 3.7 & ${}_{-12.6}^{ +6.9}$ & 0.974 & 2 \\
              0 &      300--400 &    0.75--1.15 & $7.671 \times 10^{-2}$ & 4.1 & ${}_{-14.5}^{ +6.3}$ & 0.973 & 3 \\
              0 &      300--400 &    1.15--2.50 & $3.590 \times 10^{-2}$ & 4.0 & ${}_{ -9.4}^{ +8.4}$ & 0.975 & 4 \\
              0 &      400--500 &    0.00--0.35 & $1.368 \times 10^{-1}$ & 2.7 & ${}_{ -3.6}^{ +2.4}$ & 0.971 & 5 \\
              0 &      400--500 &    0.35--0.75 & $1.234 \times 10^{-1}$ & 2.4 & ${}_{ -2.2}^{ +5.5}$ & 0.971 & 6 \\
              0 &      400--500 &    0.75--1.15 & $1.032 \times 10^{-1}$ & 2.9 & ${}_{ -1.4}^{ +5.3}$ & 0.972 & 7 \\
\end{tabular}}
\end{table*}
 \begin{table*}[htbp]
\centering
\topcaption{The correlation matrix of statistical uncertainties for the measured \njmttytttwo cross sections. The values are expressed as percentages. For bin indices see Table~\ref{tab:nj2mttytttwo_xsec}.}
\label{tab:nj2mttytttwo_corr}
\scriptsize
\tabcolsep0.75mm
\renewcommand*{\arraystretch}{1.45}
\begin{tabular}{l|rrrrrrrrrrrrrrrrrrrrrrr}
bin & 2 & 3 & 4 & 5 & 6 & 7 & 8 & 9 & 10 & 11 & 12 & 13 & 14 & 15 & 16 & 17 & 18 & 19 & 20 & 21 & 22 & 23 & 24 \\
\hline
1 & $-17$ & $-17$ & $+2$ & $-40$ & $-23$ & $+9$ & $+3$ & $-6$ & $+7$ & 0 & $-5$ & $-20$ & $-4$ & $+4$ & $-2$ & $+4$ & $+7$ & $+1$ & $-4$ & $+3$ & $-3$ & $-4$ & 0 \\
2 & & $+16$ & $-23$ & $-20$ & $-10$ & $-31$ & $-4$ & $+9$ & $-12$ & $+3$ & $+5$ & $-3$ & $-12$ & $-10$ & $+5$ & $+7$ & $-2$ & $+3$ & $+4$ & $-2$ & $+3$ & $+3$ & $-5$ \\
3 & & & $-2$ & $+8$ & $-31$ & $-20$ & $-29$ & $+1$ & $+2$ & $-9$ & $+1$ & $+4$ & $-10$ & $-12$ & $-6$ & $+1$ & $+2$ & 0 & $+3$ & $-3$ & $+2$ & $+3$ & $+1$ \\
4 & & & & $-1$ & $-2$ & $-24$ & $+1$ & $-9$ & $-1$ & 0 & $-31$ & $-3$ & $+3$ & $-5$ & $-25$ & $-6$ & 0 & $+2$ & $-13$ & $-5$ & $-9$ & $-4$ & $+10$ \\
5 & & & & & $+3$ & $-9$ & $-5$ & $-21$ & $-10$ & $+3$ & $-1$ & $-2$ & $+4$ & $+2$ & $-3$ & $-12$ & $-4$ & $+1$ & $+1$ & $+2$ & $+4$ & 0 & $-1$ \\
6 & & & & & & $+22$ & $-13$ & $-10$ & $-6$ & $-15$ & $-1$ & $+5$ & $-5$ & 0 & $+3$ & $-4$ & $-7$ & $-6$ & 0 & $+5$ & $-1$ & $+3$ & 0 \\
7 & & & & & & & $+11$ & $+4$ & $-13$ & $-10$ & $-12$ & $+3$ & $+1$ & $-4$ & 0 & $+1$ & $-5$ & $-7$ & $-4$ & $+1$ & $+4$ & 0 & $+1$ \\
\end{tabular}
\end{table*}
 \begin{table*}[p]
\centering
\topcaption{Sources and values of the relative systematic uncertainties in percent of the measured \njmttytttwo cross sections. For bin indices see Table~\ref{tab:nj2mttytttwo_xsec}.}
\label{tab:nj2mttytttwo_syst}
\renewcommand*{\arraystretch}{1.75}
\tabcolsep0.70mm
\cmsTallTable{%
\begin{tabular}{l|
rrrrrrrrrrrrrrrrrrrrrrrr}
source / bin & 1 & 2 & 3 & 4 & 5 & 6 & 7 & 8 & 9 & 10 & 11 & 12 & 13 & 14 & 15 & 16 & 17 & 18 & 19 & 20 & 21 & 22 & 23 & 24 \\
\hline
\multirow{2}{*}{JER}& $+0.2$ & $+0.6$ & $+0.4$ & $+0.8$ & $+0.1$ & $+0.7$ & $+0.5$ & $+0.8$ & $+0.1$ & $+0.5$ & $+0.2$ & $+0.9$ & $-0.6$ & $-0.1$ & $+0.1$ & $-0.8$ & $-0.5$ & $-0.6$ & $+0.3$ & $-1.0$ & $-0.6$ & $-0.6$ & $-0.5$ & $-1.8$ \\
& $-0.4$ & $-1.0$ & $-2.1$ & $-0.8$ & $-0.7$ & $+0.2$ & $-0.2$ & $+0.1$ & $-0.3$ & $-0.4$ & $+0.2$ & $-0.1$ & $+2.1$ & 0.0 & $+0.7$ & $-0.4$ & $+0.4$ & $-0.2$ & $+0.5$ & $+1.2$ & $+0.6$ & $+0.6$ & $+0.1$ & $+0.4$ \\
\hline
\multirow{2}{*}{JESAbsoluteMPFBias}& $+0.8$ & $+0.4$ & $-0.1$ & $+0.7$ & $-0.1$ & $+0.4$ & $+0.4$ & $+0.4$ & $+0.3$ & 0.0 & $+0.2$ & $-0.5$ & $+1.2$ & $+0.7$ & $+0.3$ & $-0.2$ & $-0.4$ & $-0.6$ & $-0.6$ & $-0.2$ & $-0.5$ & $-0.4$ & $-0.3$ & $-1.0$ \\
& $-0.8$ & $-0.9$ & $-1.5$ & $-0.3$ & $-0.1$ & $+0.1$ & $+0.1$ & $+0.1$ & $-0.1$ & $+0.2$ & 0.0 & 0.0 & $-0.3$ & $-0.8$ & $+0.2$ & $-0.8$ & $+0.5$ & $+0.3$ & $+0.5$ & $+0.4$ & $+0.5$ & $+0.2$ & $+0.4$ & $+0.7$ \\
\hline
\multirow{2}{*}{JESAbsoluteScale}& $+0.6$ & 0.0 & 0.0 & $+0.6$ & $-0.1$ & $+0.4$ & $+0.3$ & $+0.3$ & $+0.1$ & $-0.1$ & $+0.2$ & 0.0 & $+0.5$ & $+0.7$ & 0.0 & $+0.1$ & $-0.2$ & $-0.4$ & $-0.5$ & $+0.2$ & $-0.3$ & $-0.4$ & $-0.3$ & $-1.3$ \\
& $-0.7$ & $-0.7$ & $-1.1$ & $-0.4$ & $-0.1$ & 0.0 & $+0.3$ & $+0.2$ & $-0.1$ & $+0.1$ & $+0.2$ & $+0.1$ & $-0.3$ & $-0.4$ & $+0.2$ & 0.0 & $+0.4$ & $+0.3$ & $-0.1$ & $+0.2$ & $+0.3$ & $+0.3$ & 0.0 & $+0.3$ \\
\hline
\multirow{2}{*}{JESAbsoluteStat}& $+0.4$ & $+1.0$ & $-1.1$ & $-0.3$ & $-0.4$ & 0.0 & $-0.2$ & $+0.5$ & $+0.2$ & $+0.2$ & $+0.7$ & $+0.5$ & $+0.6$ & $+1.0$ & $+0.1$ & $+0.6$ & $-0.7$ & $-0.6$ & $-0.7$ & $-0.6$ & $+0.1$ & $+0.3$ & 0.0 & $-1.4$ \\
\end{tabular}}
\end{table*}
 \begin{table*}[p]
\centering
\topcaption{Table~\ref{tab:nj2mttytttwo_syst} continued.}
\label{tab:nj2mttytttwo_1_syst}
\renewcommand*{\arraystretch}{1.75}
\tabcolsep0.70mm
\cmsTallTable{%
\begin{tabular}{l|
rrrrrrrrrrrrrrrrrrrrrrrr}
source / bin & 1 & 2 & 3 & 4 & 5 & 6 & 7 & 8 & 9 & 10 & 11 & 12 & 13 & 14 & 15 & 16 & 17 & 18 & 19 & 20 & 21 & 22 & 23 & 24 \\
\hline
\multirow{2}{*}{JESSinglePionHCAL}& $-0.3$ & $-0.9$ & $-0.6$ & $-0.1$ & $-0.4$ & $+0.1$ & $+0.2$ & $-0.1$ & $-0.3$ & $-0.2$ & $-0.6$ & $+0.2$ & $+0.2$ & $-0.3$ & $+0.1$ & $+0.2$ & $+0.6$ & $+0.5$ & $+0.5$ & $+0.2$ & $+0.2$ & $+0.2$ & $+0.2$ & 0.0 \\
& $+0.3$ & $-0.1$ & $-0.5$ & $+0.5$ & $+0.2$ & $+0.6$ & $+0.5$ & $+0.6$ & $+0.5$ & $+0.1$ & $-0.1$ & $-0.3$ & $+0.3$ & $+0.7$ & $+0.2$ & $-0.7$ & $-0.6$ & $-0.6$ & $-0.6$ & $-0.1$ & $-0.3$ & $-0.1$ & $+0.1$ & $-0.8$ \\
\hline
\multirow{2}{*}{JESTimePtEta}& $-0.1$ & $-0.1$ & $-0.4$ & $-0.1$ & $-0.2$ & $-0.1$ & $+0.3$ & 0.0 & $+0.3$ & $-0.1$ & $+0.1$ & 0.0 & 0.0 & $-0.1$ & 0.0 & $+0.3$ & $+0.2$ & $+0.1$ & $-0.4$ & 0.0 & $-0.3$ & 0.0 & $-0.1$ & $+0.3$ \\
& $-0.2$ & $+0.2$ & 0.0 & $+0.1$ & 0.0 & $+0.3$ & 0.0 & $-0.1$ & $+0.1$ & $-0.2$ & $+0.1$ & $+0.4$ & $+0.4$ & $+0.1$ & $+0.1$ & $+0.2$ & 0.0 & $-0.1$ & 0.0 & 0.0 & $-0.1$ & $+0.1$ & $-0.1$ & $-1.6$ \\
\hline
\multirow{2}{*}{$\ptvecmiss$}& $-2.4$ & $-0.3$ & $-1.1$ & $-0.2$ & $+0.3$ & $+0.3$ & $-0.3$ & $-0.2$ & $+0.1$ & $+0.1$ & $+0.6$ & $+0.3$ & $+1.2$ & $+0.3$ & $+0.2$ & $+0.5$ & $-0.4$ & 0.0 & $+0.1$ & 0.0 & $+0.3$ & $+0.1$ & $+0.2$ & $-0.1$ \\
& $-0.1$ & $-0.2$ & $+0.8$ & $+0.4$ & $-0.1$ & $-0.3$ & $+0.2$ & $+0.8$ & $-0.1$ & $+0.1$ & $-0.5$ & $-0.6$ & $-0.4$ & $+0.5$ & $-0.4$ & $+0.8$ & $-0.1$ & $+0.1$ & 0.0 & $-0.2$ & $-0.2$ & $-0.3$ & $-0.2$ & $+0.1$ \\
\hline
\multirow{2}{*}{lepton ID/ISO}& $-0.3$ & 0.0 & $-0.5$ & $-0.3$ & $+0.1$ & $+0.2$ & $+0.1$ & $+0.6$ & $-0.1$ & $-0.1$ & $-0.1$ & $-0.7$ & 0.0 & $+0.1$ & 0.0 & $+0.1$ & $+0.3$ & $+0.4$ & $+0.3$ & $+0.6$ & $-0.1$ & $-0.1$ & $-0.2$ & $-0.7$ \\
\end{tabular}%
}
\end{table*}

\begin{table*}[htbp]
\centering
\topcaption{The measured \njmttyttthree cross sections, along with their relative statistical and systematic uncertainties, and NP corrections (see Section~\ref{sec:asmt}).}
\label{tab:nj3mttyttthree_xsec}
\renewcommand*{\arraystretch}{1.55}
\tabcolsep2.0mm
\cmsTallTable{%
%
}
\end{table*}
 \begin{table*}[htbp]
\centering
\topcaption{The correlation matrix of statistical uncertainties for the measured \njmttyttthree cross sections. The values are expressed as percentages. For bin indices see Table~\ref{tab:nj3mttyttthree_xsec}.}
\label{tab:nj3mttyttthree_corr}
\tabcolsep0.3mm
\renewcommand*{\arraystretch}{1.45}
\cmsTable{%
%
}
\end{table*}

\begin{landscape}

\begin{table*}[p]
\caption{Sources and values of the relative systematic uncertainties in percent of the measured \njmttyttthree cross sections. For bin indices see Table~\ref{tab:nj3mttyttthree_xsec}.}
\label{tab:nj3mttyttthree_syst}
\renewcommand*{\arraystretch}{1.75}
\cmsLongTable{%
%
}
\end{table*}

 \begin{table*}[p]
\centering
\caption{Table~\ref{tab:nj3mttyttthree_syst} continued.}
\label{tab:nj3mttyttthree_1_syst}
\renewcommand*{\arraystretch}{1.75}
\tabcolsep0.75mm
\cmsLongTable{%
%
}
\end{table*}

 \begin{table*}[p]
\centering
\caption{Table~\ref{tab:nj3mttyttthree_syst} continued.}
\label{tab:nj3mttyttthree_2_syst}
\renewcommand*{\arraystretch}{1.75}
\tabcolsep0.75mm
\cmsLongTable{%
%
}
\end{table*}

 \begin{table*}[p]
\centering
\caption{Table~\ref{tab:nj3mttyttthree_syst} continued.}
\label{tab:nj3mttyttthree_3_syst}
\renewcommand*{\arraystretch}{1.75}
\tabcolsep0.75mm
\cmsLongTable{%
%
}
\end{table*}
\end{landscape}

  \section{Additional details and plots for \texorpdfstring{\as}{as} and \texorpdfstring{\mt}{mt} extraction using external PDFs}
\label{sec:app}

\begin{figure*}
    \centering
    \includegraphics[width=0.47\textwidth]{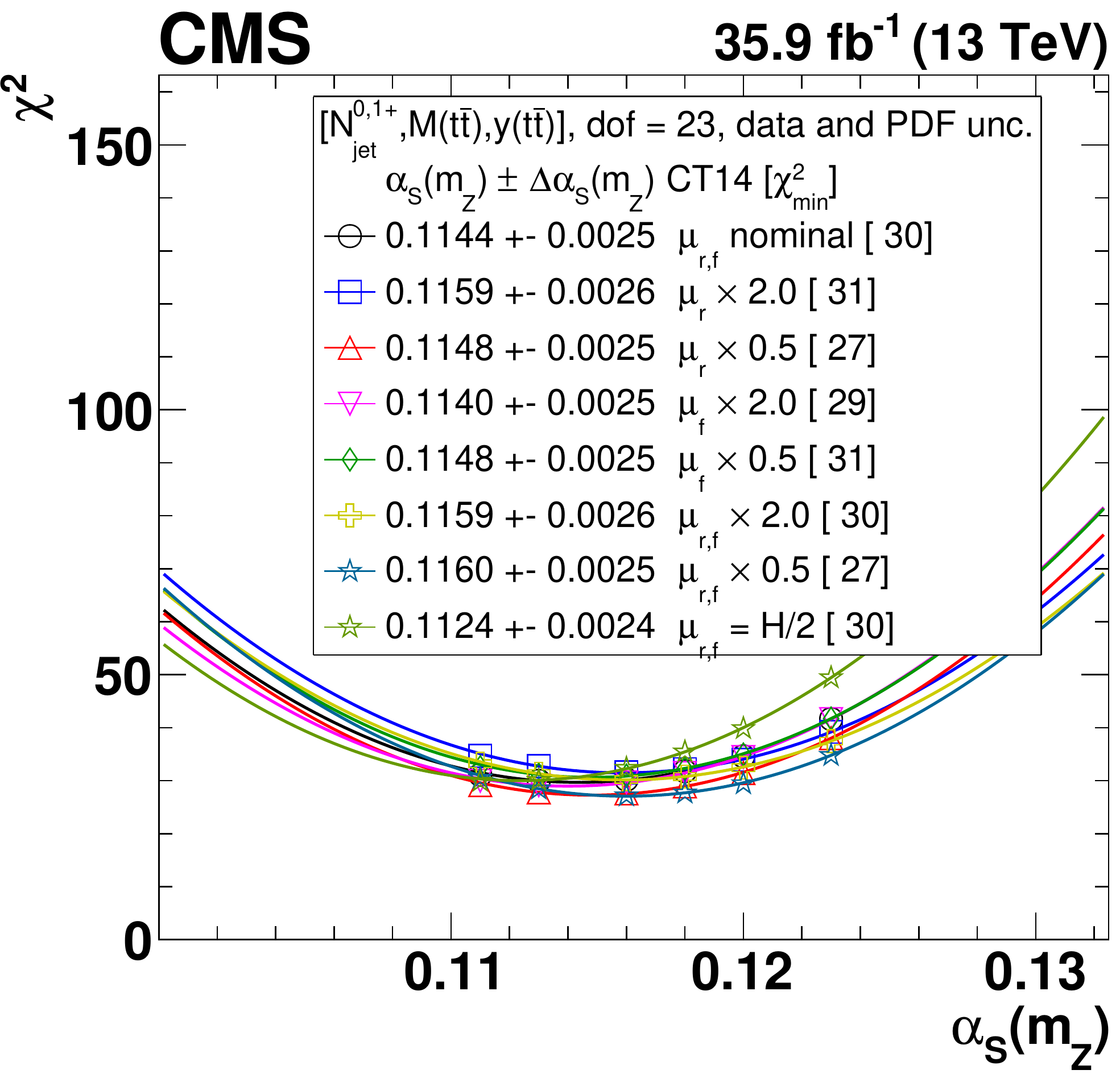}
    \includegraphics[width=0.47\textwidth]{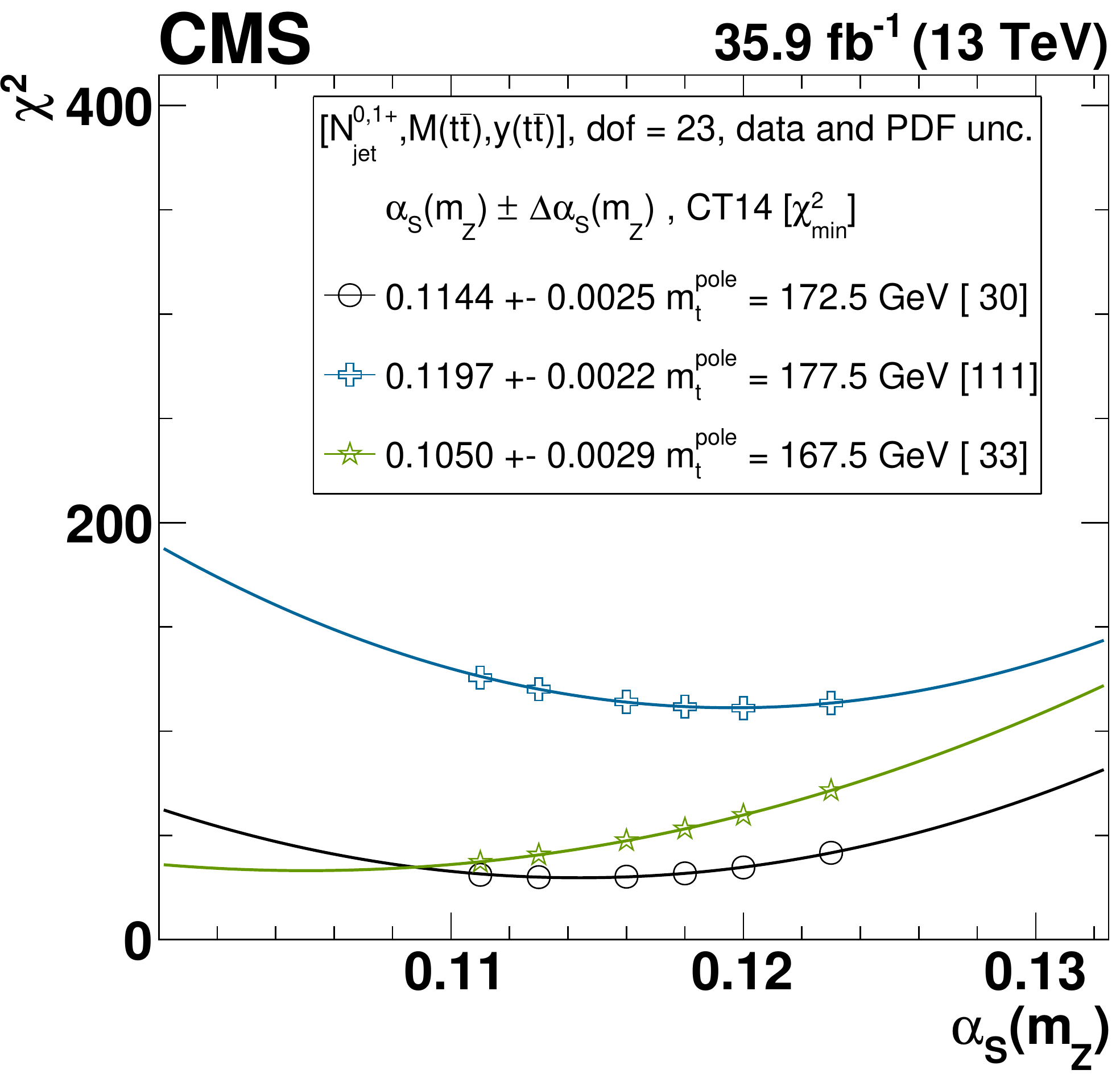}\\
    \includegraphics[width=0.47\textwidth]{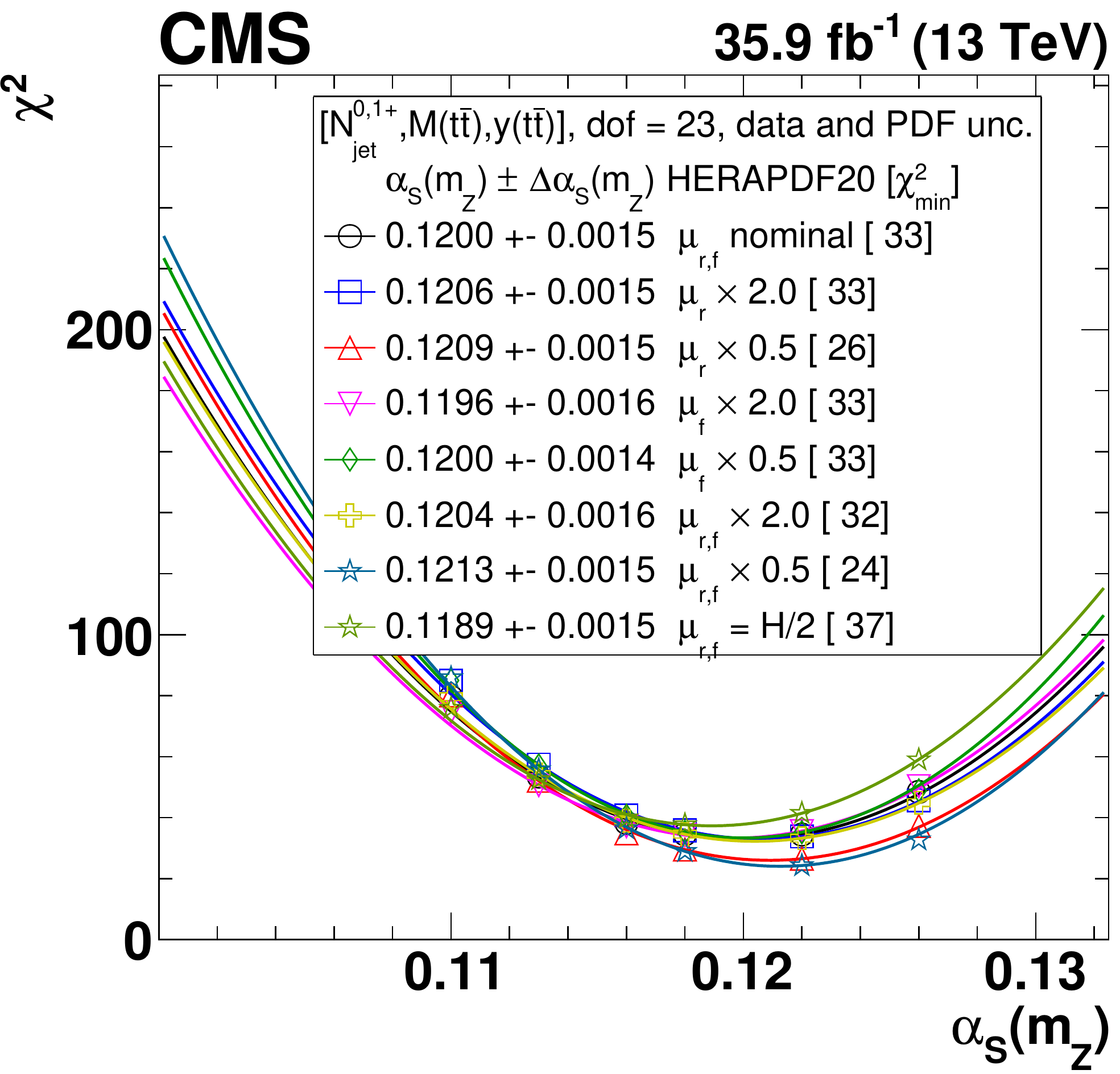}
    \includegraphics[width=0.47\textwidth]{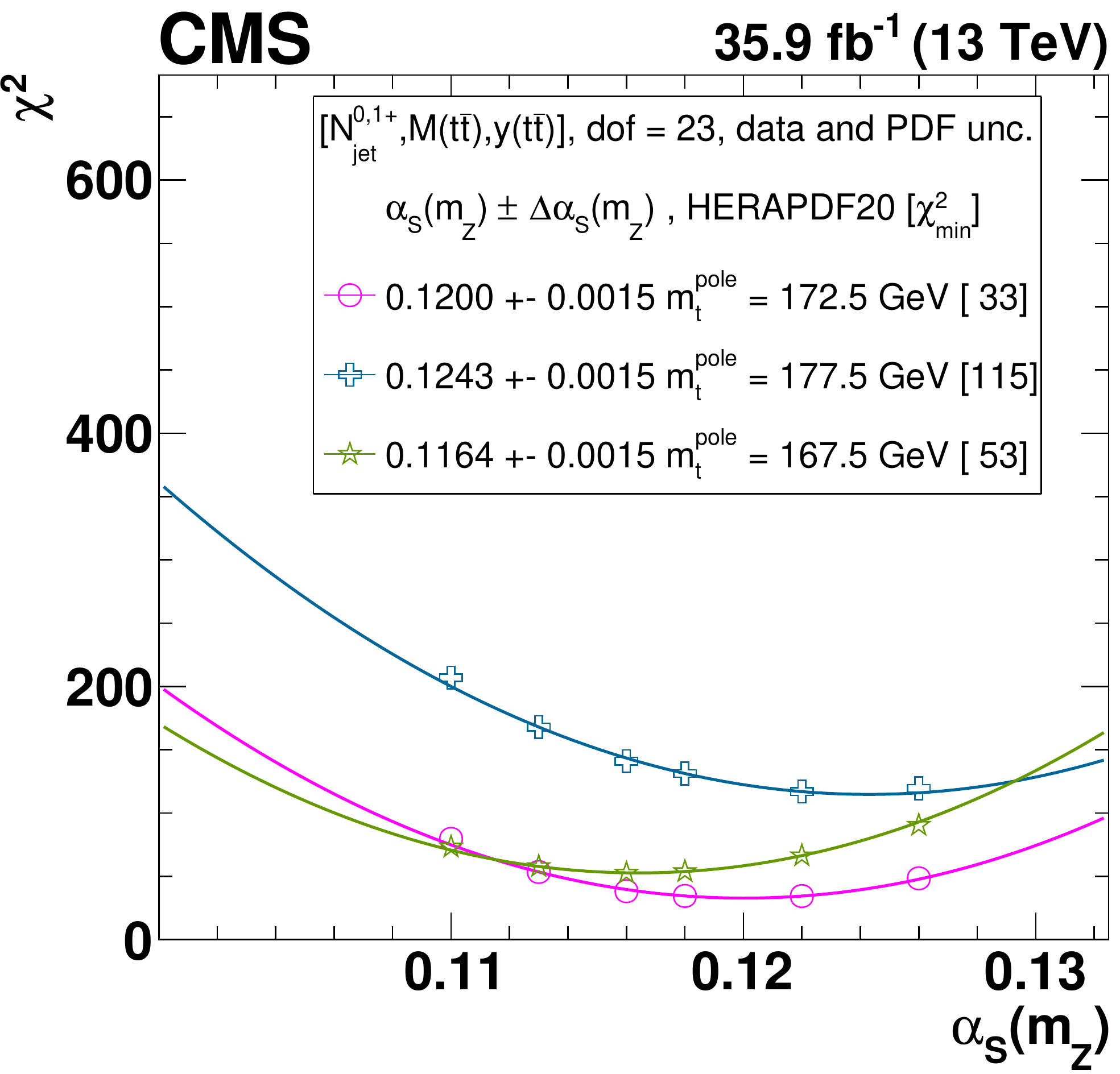}\\
    \includegraphics[width=0.47\textwidth]{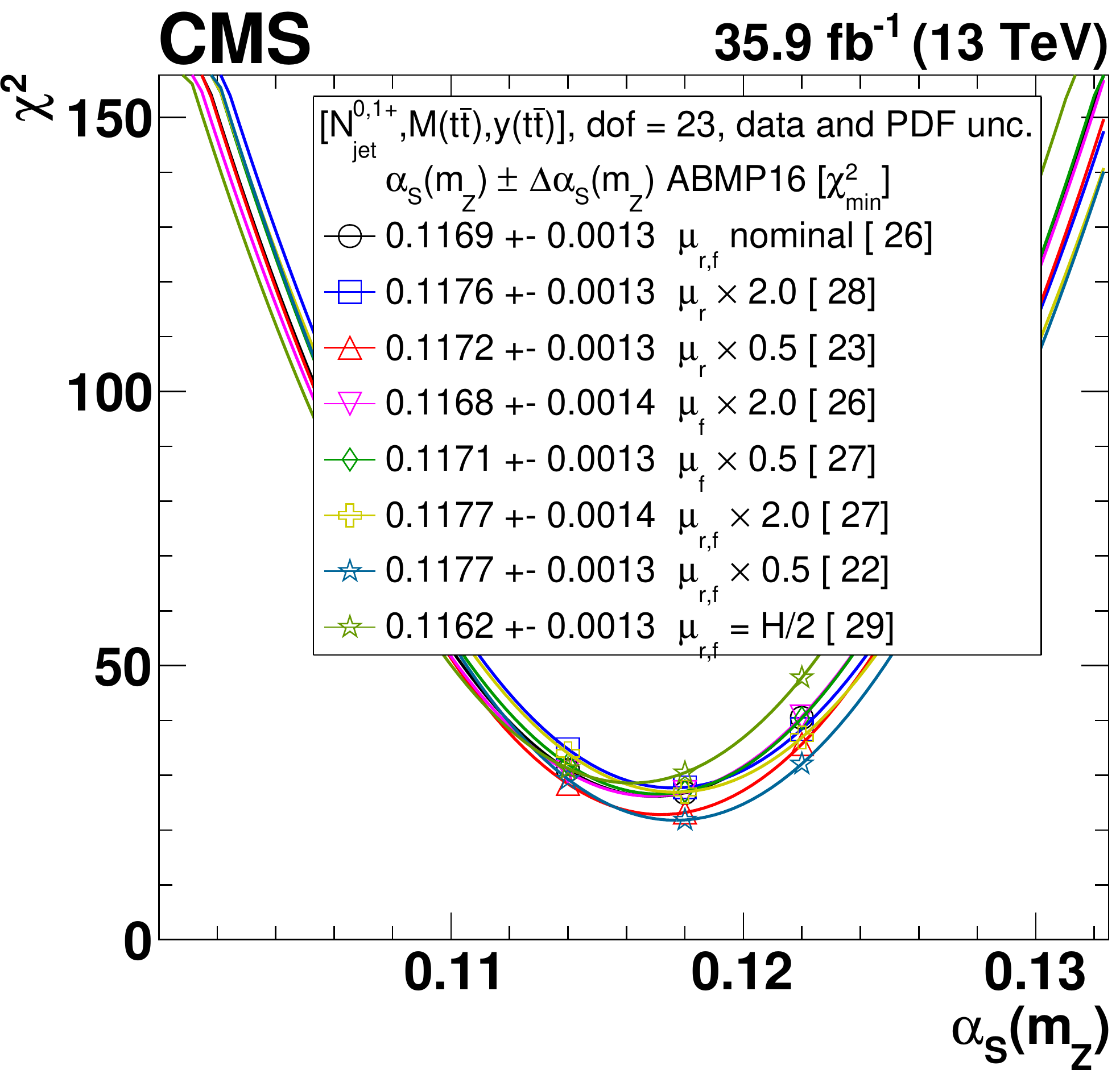}
    \includegraphics[width=0.47\textwidth]{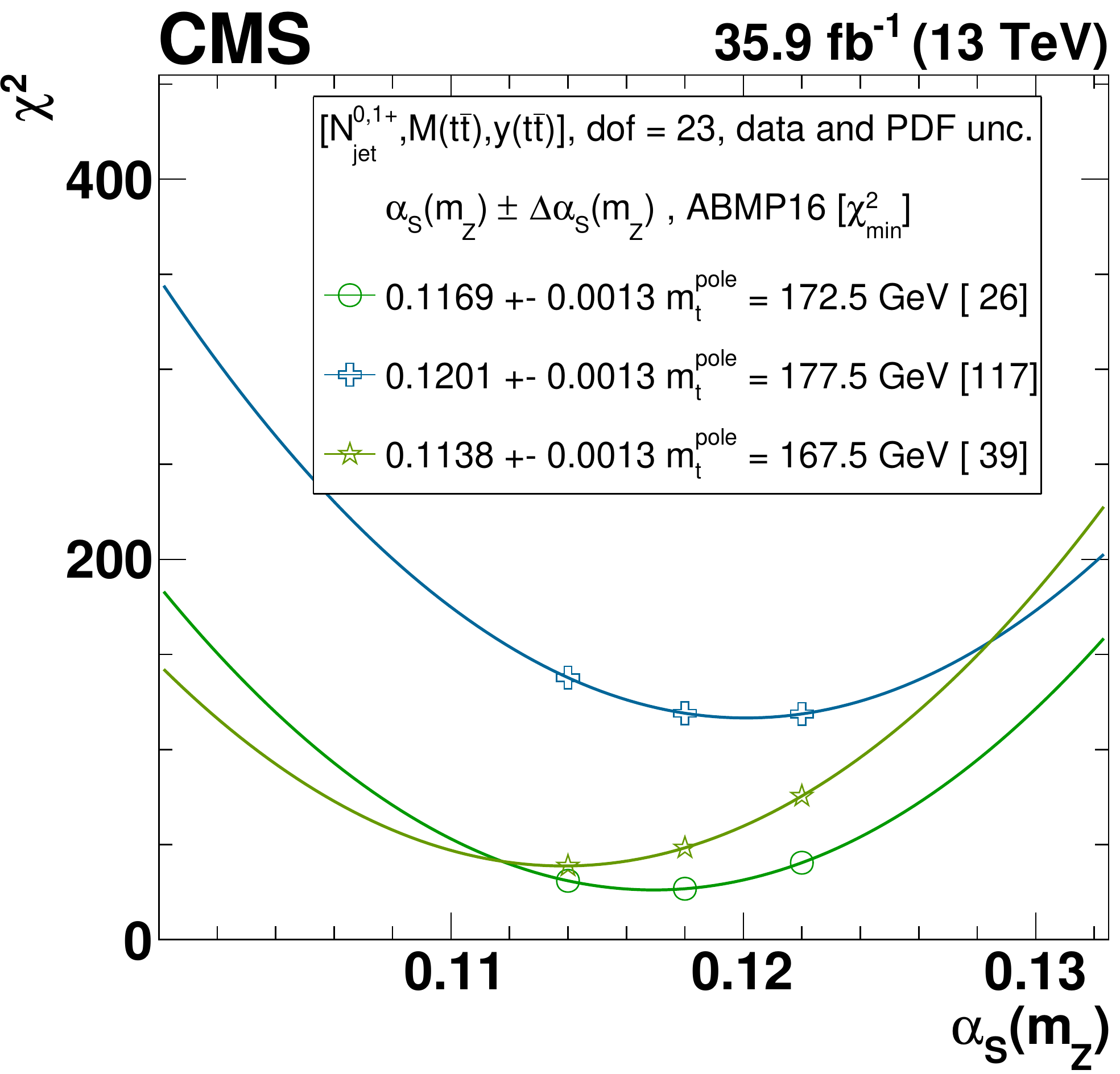}\\
    \caption{The $\asmz$ extraction from the measured \njmttytttwo cross sections using varied scale and \mt settings, and CT14 (upper), HERAPDF2.0 (middle), and ABMP16 (lower) PDF sets. Details can be found in the caption of Fig.~\ref{fig:scan-asmt-nj2mttytt-pdfs}.}
    \label{fig:scan-as-nj2mttytt-varmumt}
\end{figure*}

The $\asmz$ scans performed using altered scale and $\mt$ settings, and CT14, HERAPDF2.0, and ABMP16 PDF sets are shown in Fig.~\ref{fig:scan-as-nj2mttytt-varmumt}.

\begin{figure*}
    \centering
    \includegraphics[width=0.47\textwidth]{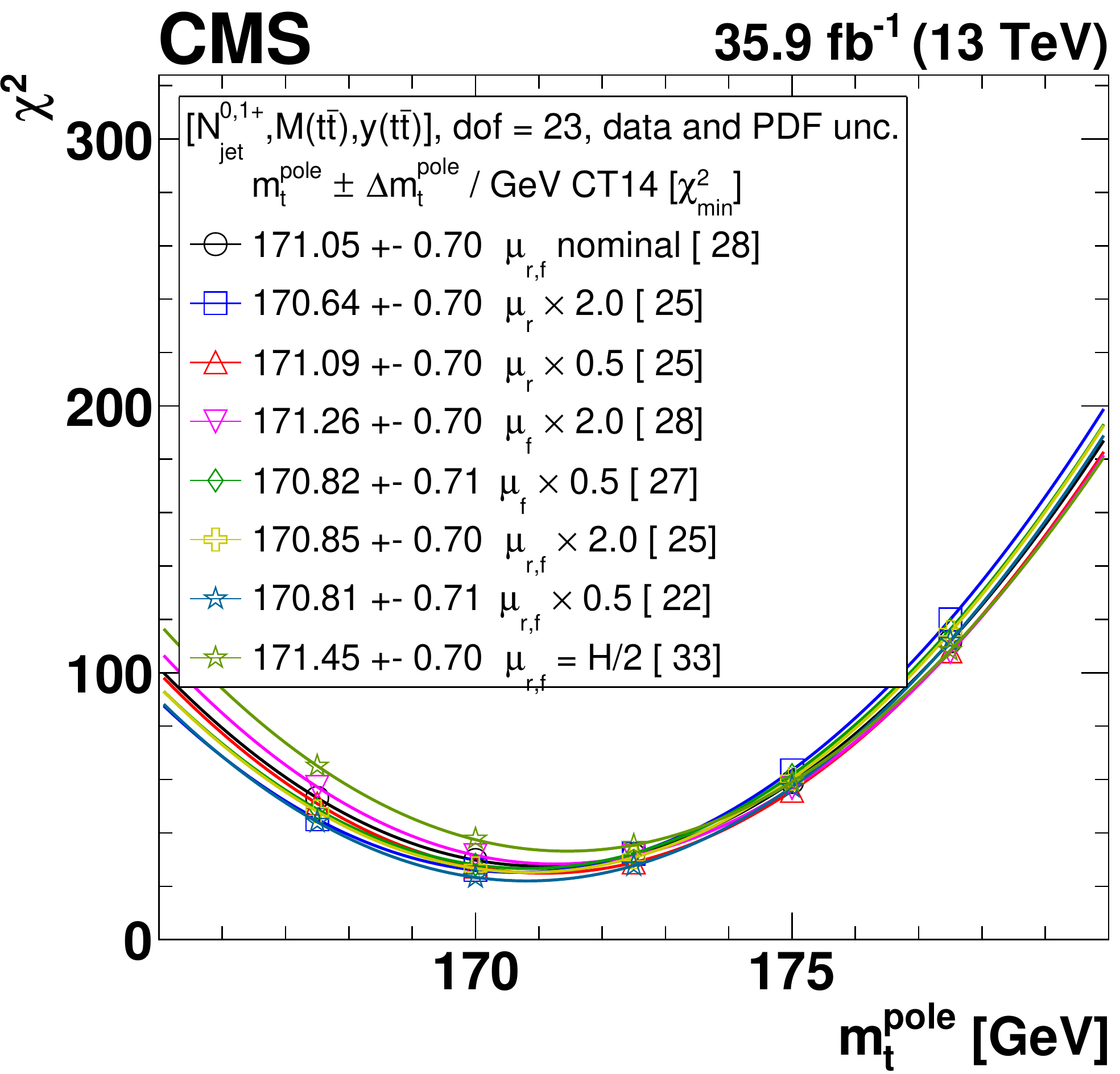}
    \includegraphics[width=0.47\textwidth]{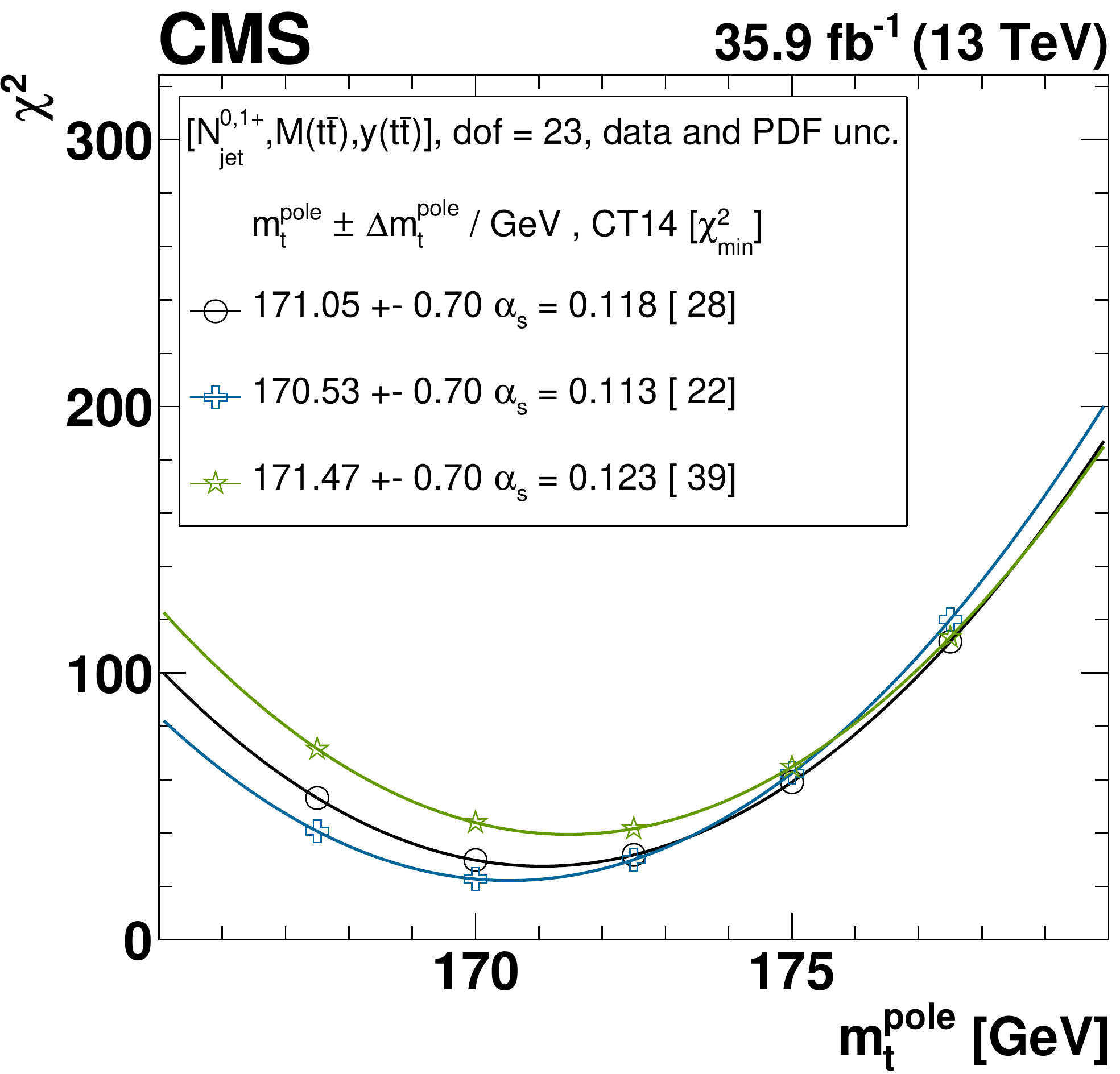}\\
    \includegraphics[width=0.47\textwidth]{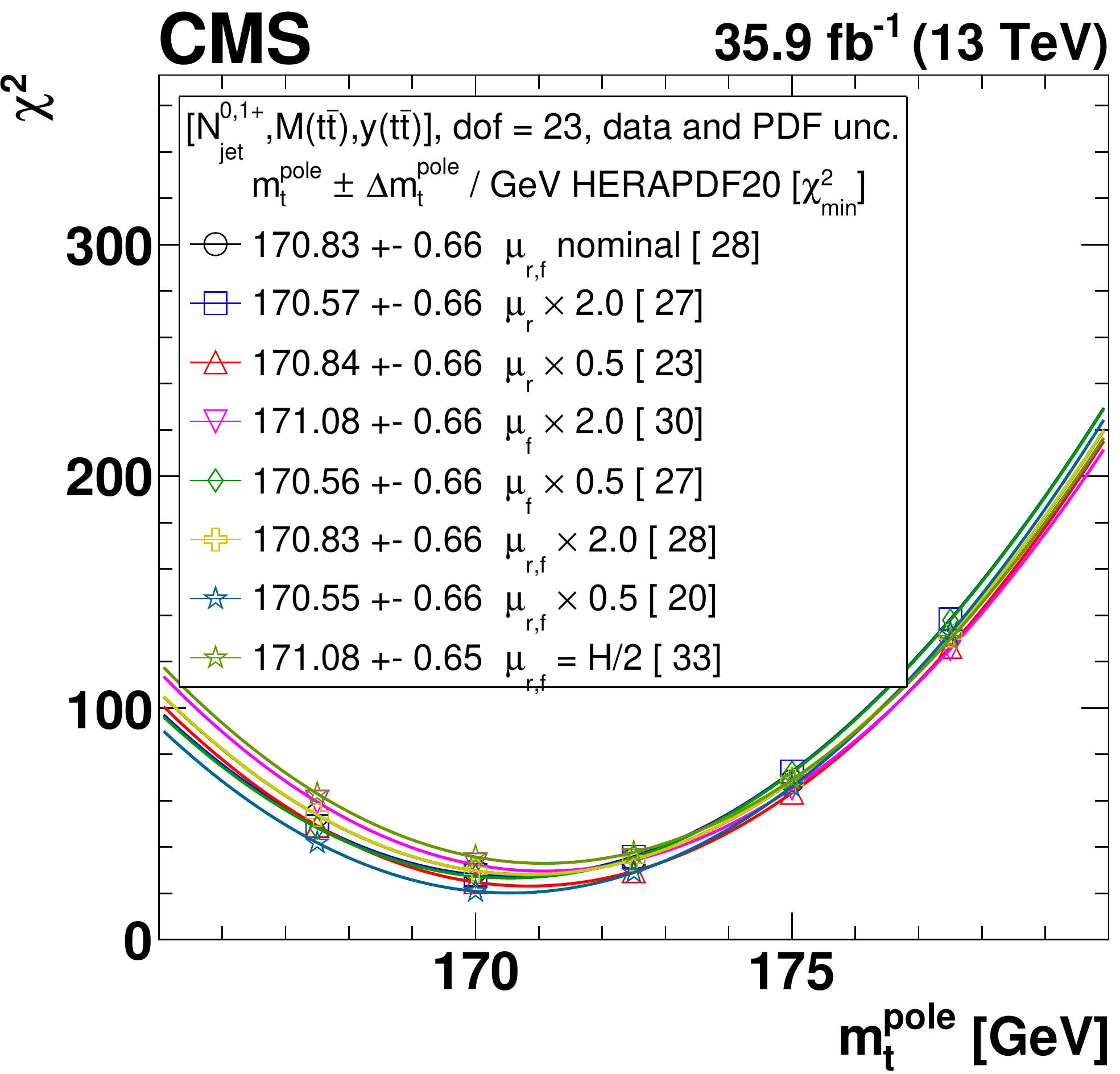}
    \includegraphics[width=0.47\textwidth]{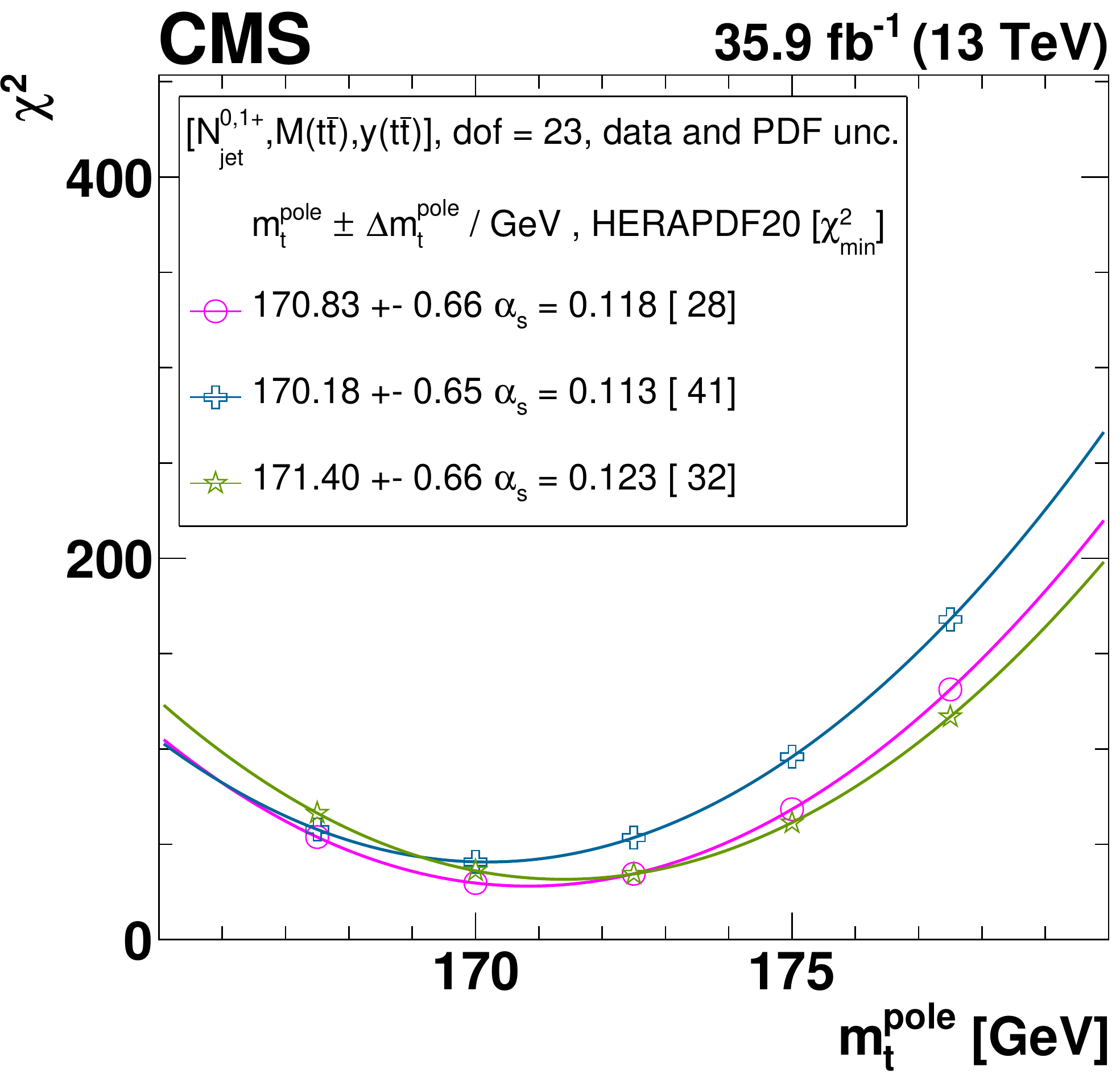}\\
    \includegraphics[width=0.47\textwidth]{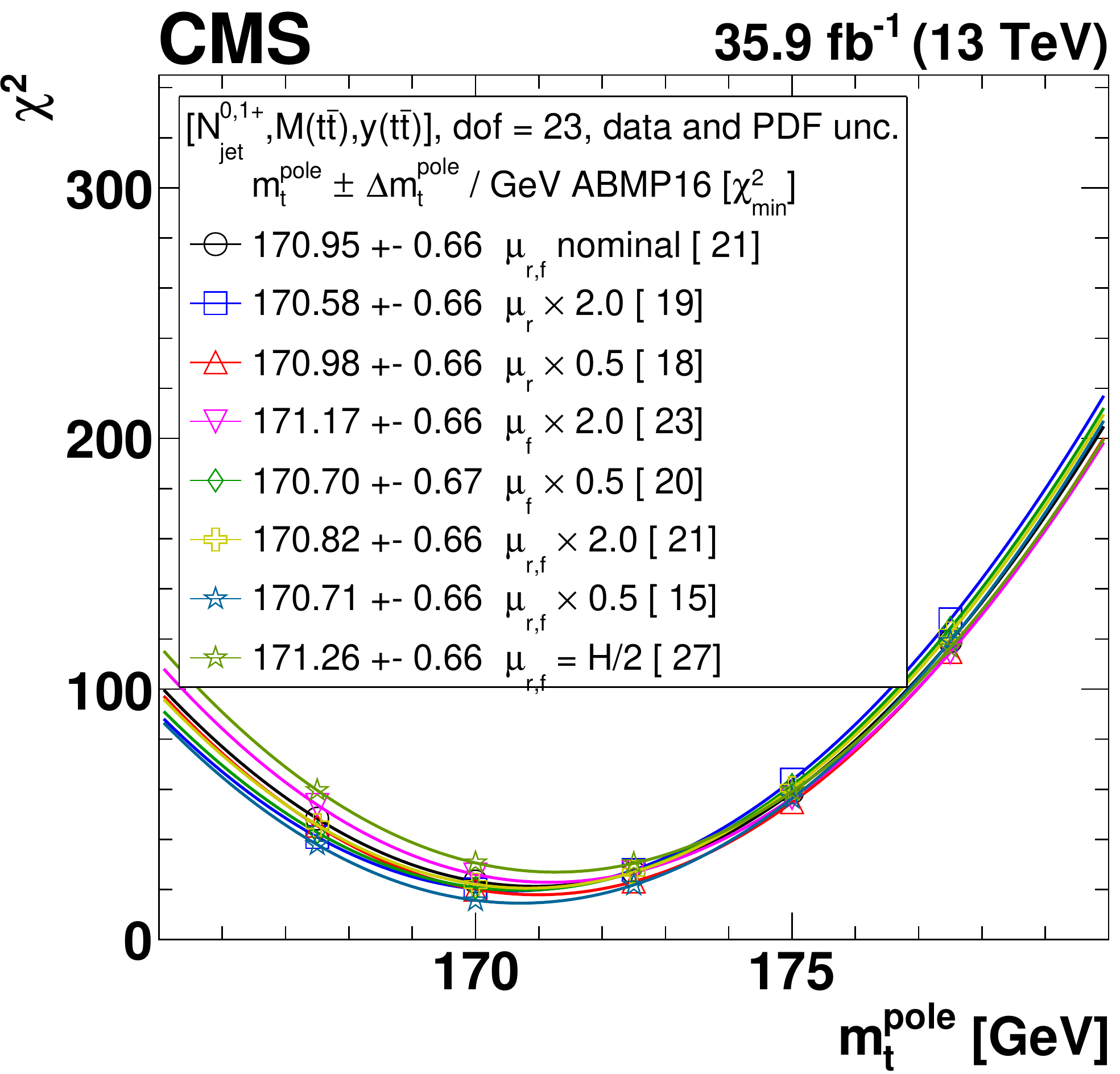}
    \includegraphics[width=0.47\textwidth]{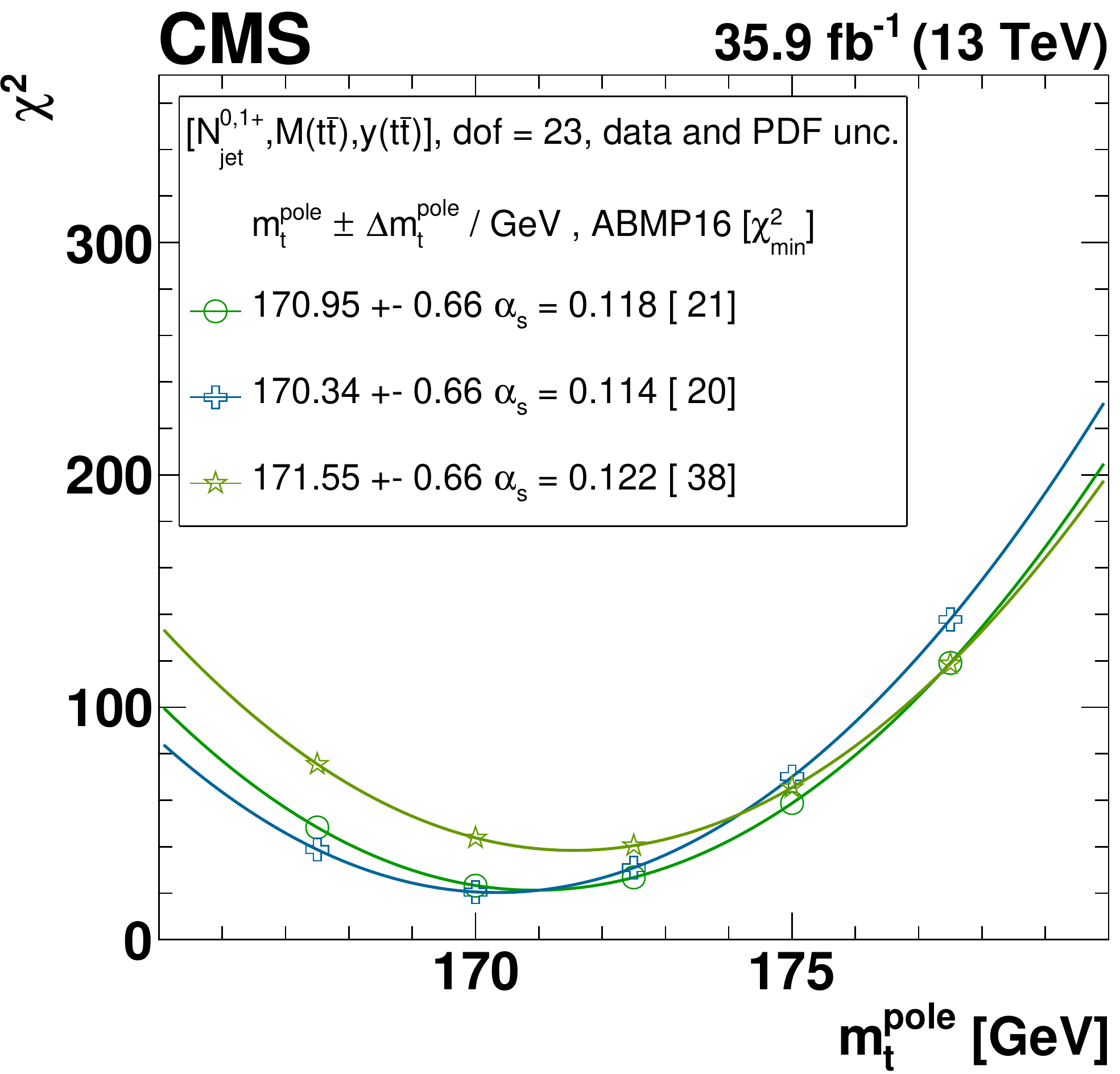}\\
    \caption{The $\mt$ extraction from the measured \njmttytttwo cross sections using varied scale and $\asmz$ settings, and CT14 (upper), HERAPDF2.0 (middle), and ABMP16 (lower) PDF sets. Details can be found in the caption of Fig.~\ref{fig:scan-asmt-nj2mttytt-pdfs}.}
    \label{fig:scan-mt-nj2mttytt-varmuas}
\end{figure*}

The $\mt$ scans for the same variations are shown in Fig.~\ref{fig:scan-mt-nj2mttytt-varmuas}.

\begin{figure*}
    \centering
    \includegraphics[width=0.47\textwidth]{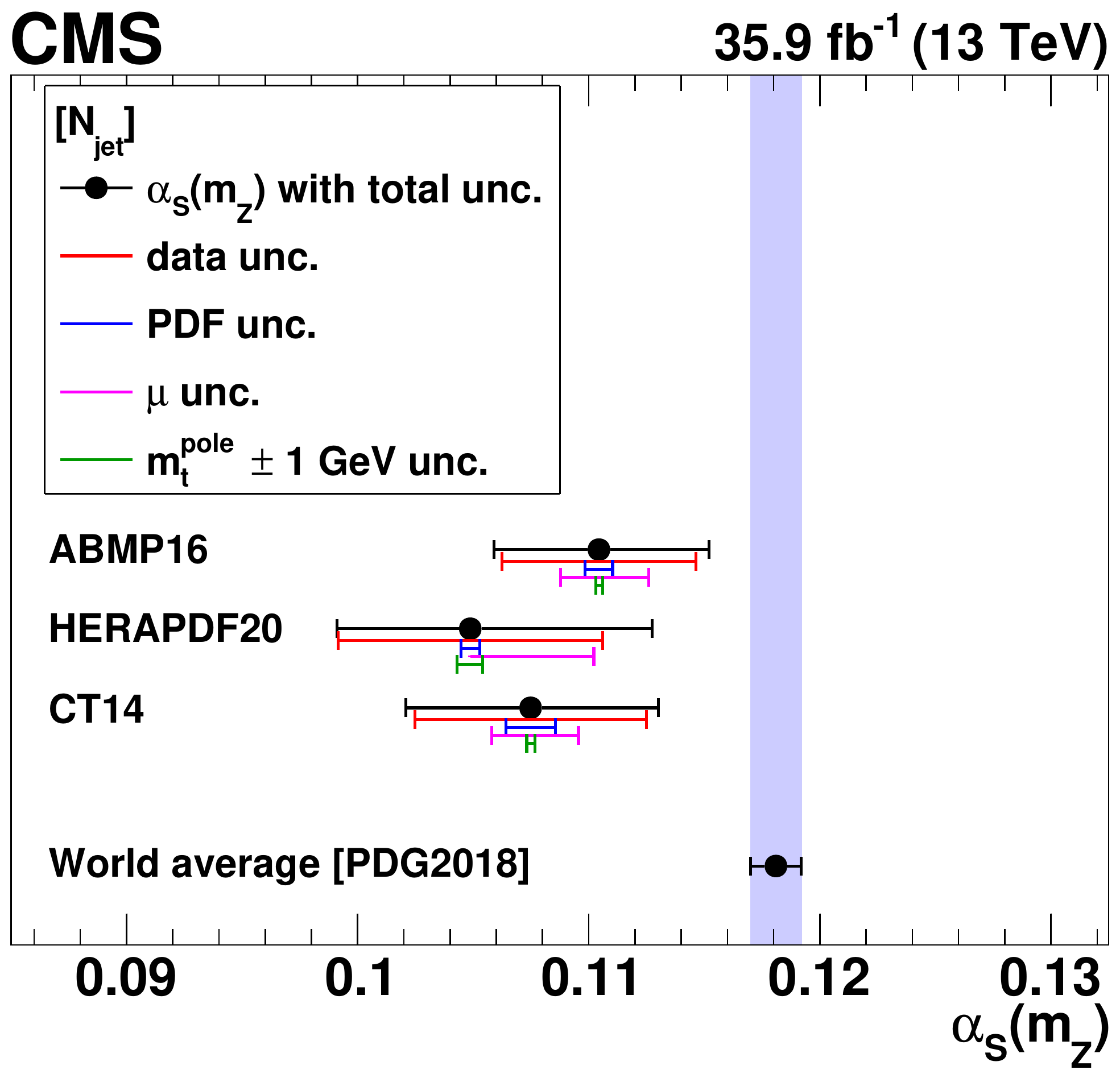}
    \includegraphics[width=0.47\textwidth]{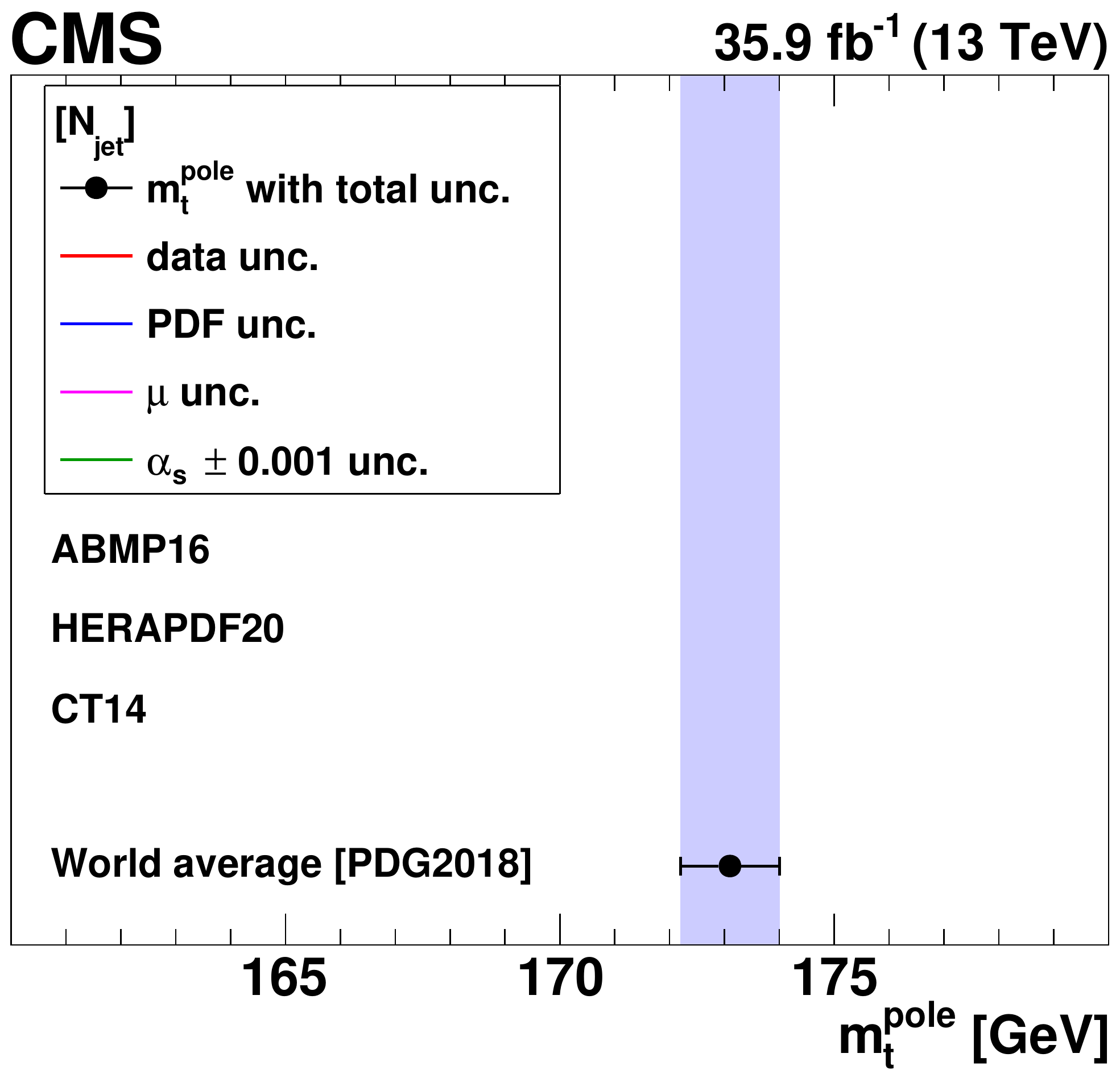}\\
    \includegraphics[width=0.47\textwidth]{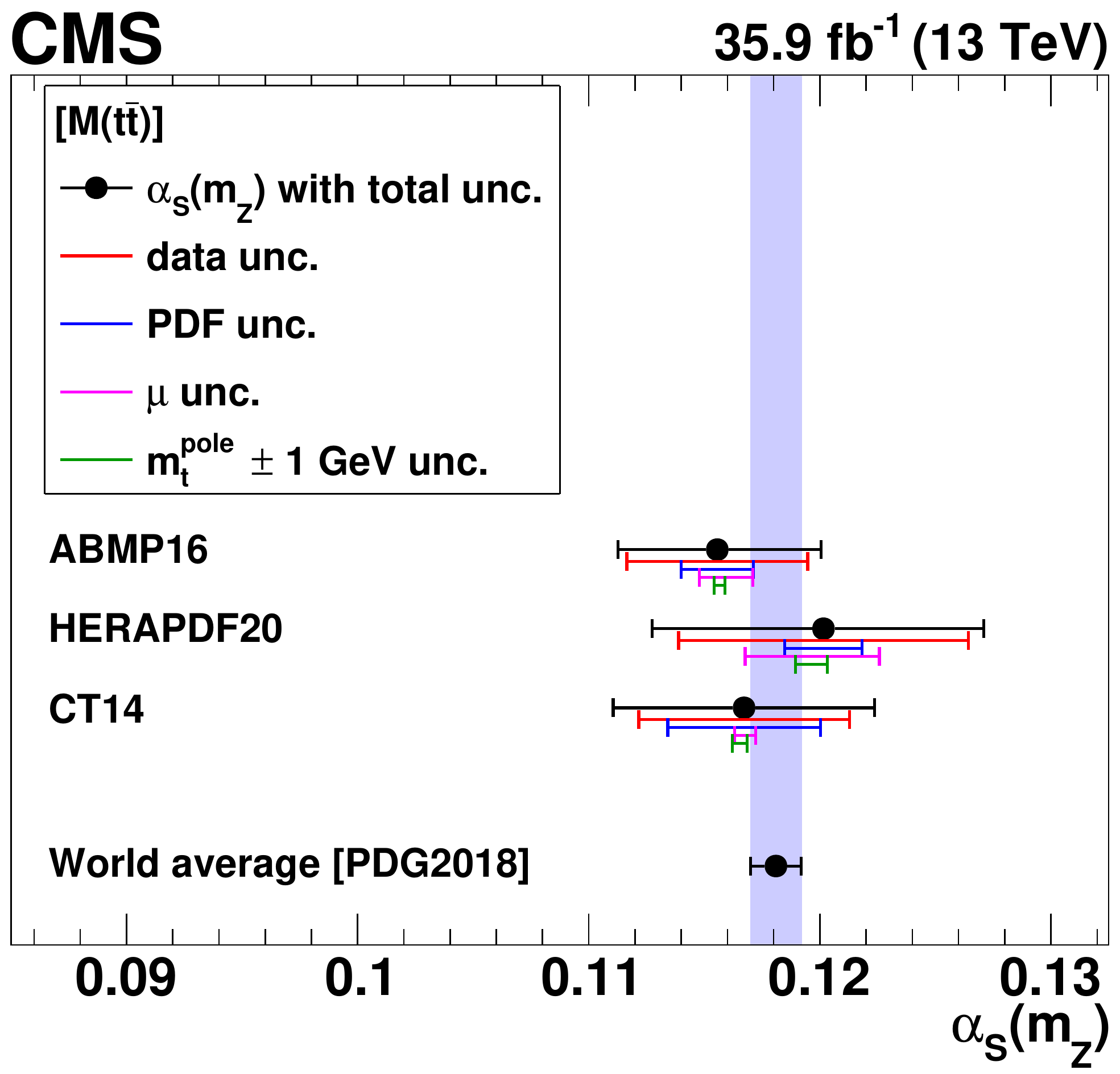}
    \includegraphics[width=0.47\textwidth]{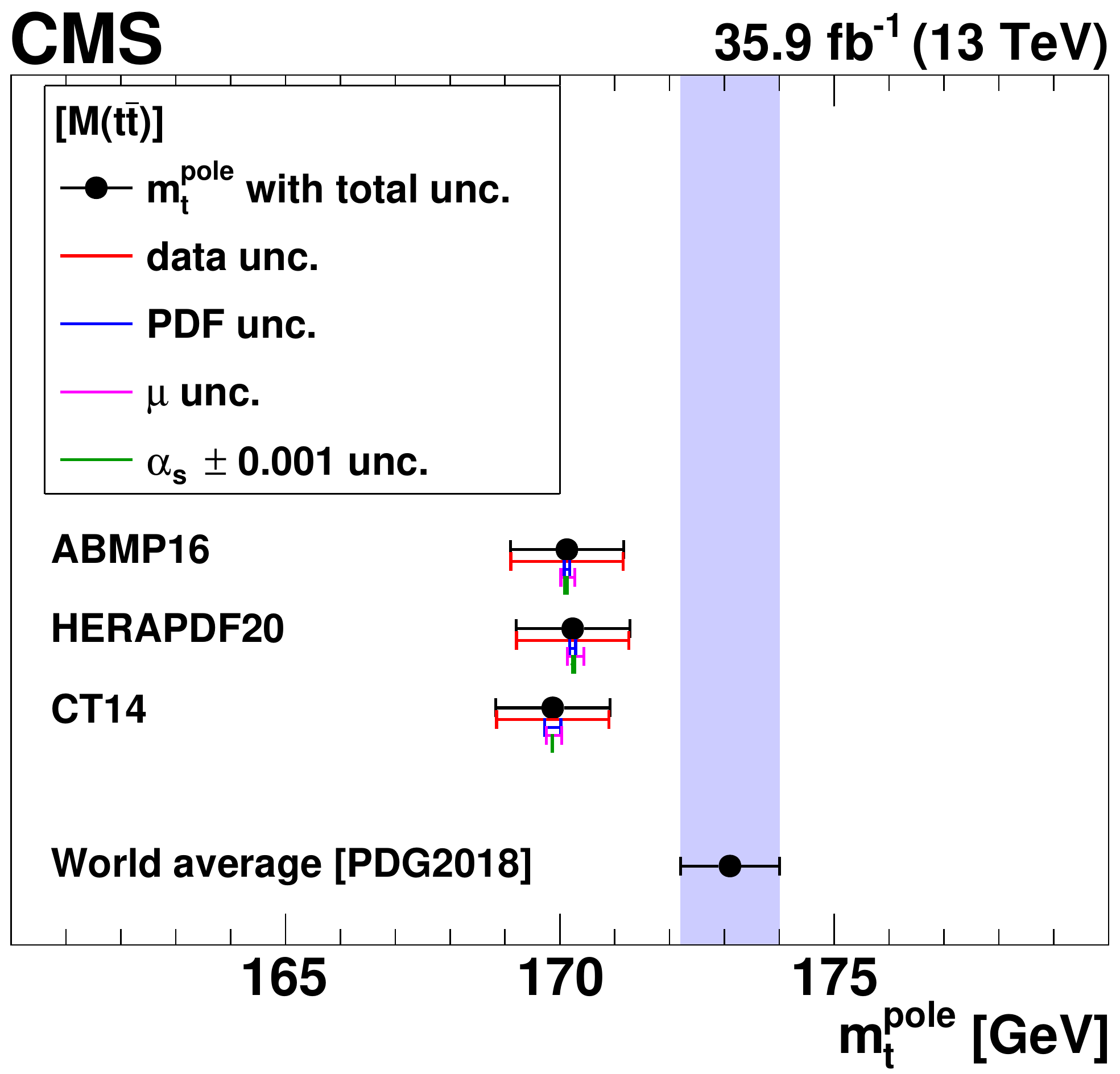}\\
    \includegraphics[width=0.47\textwidth]{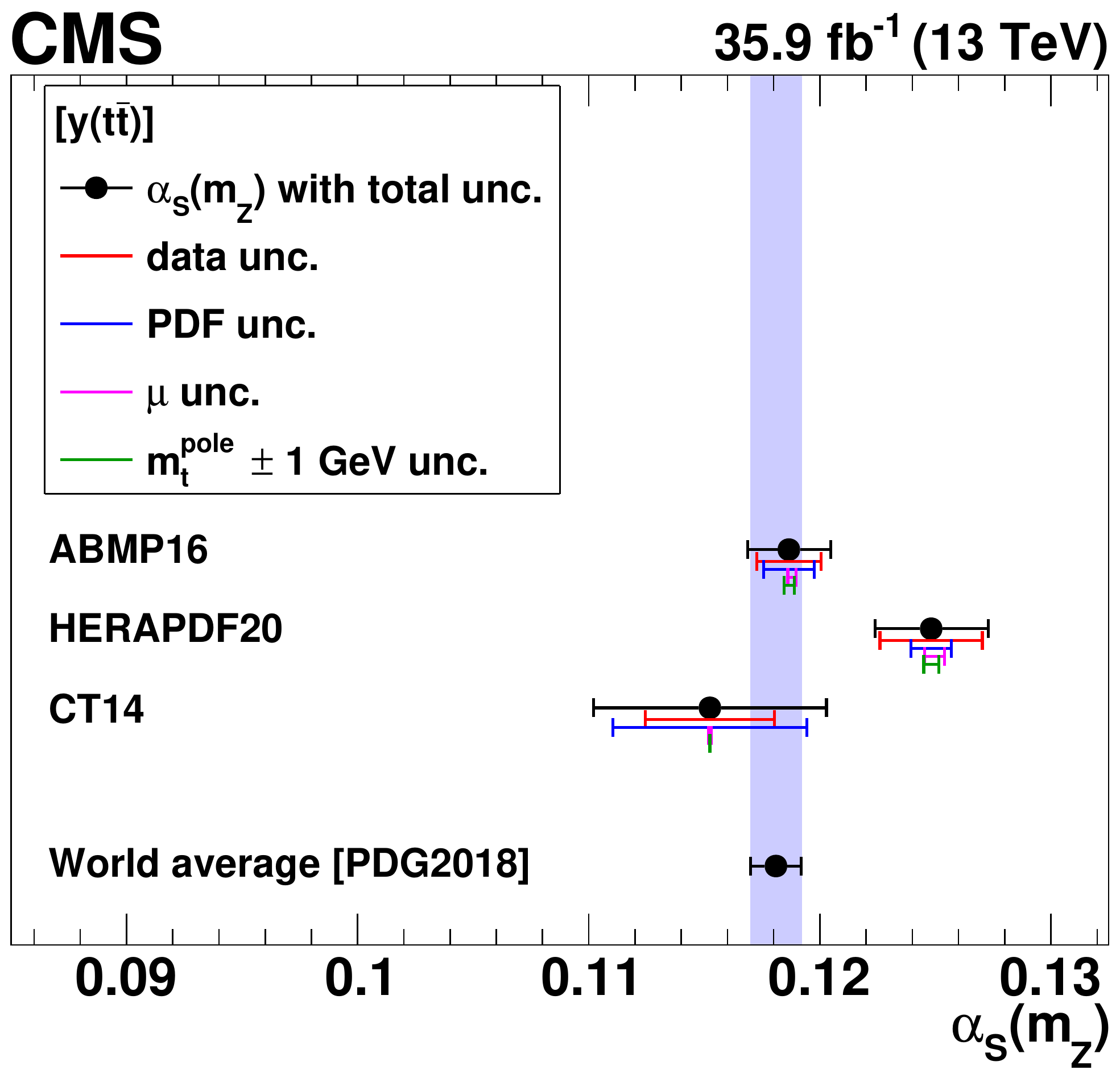}
    \includegraphics[width=0.47\textwidth]{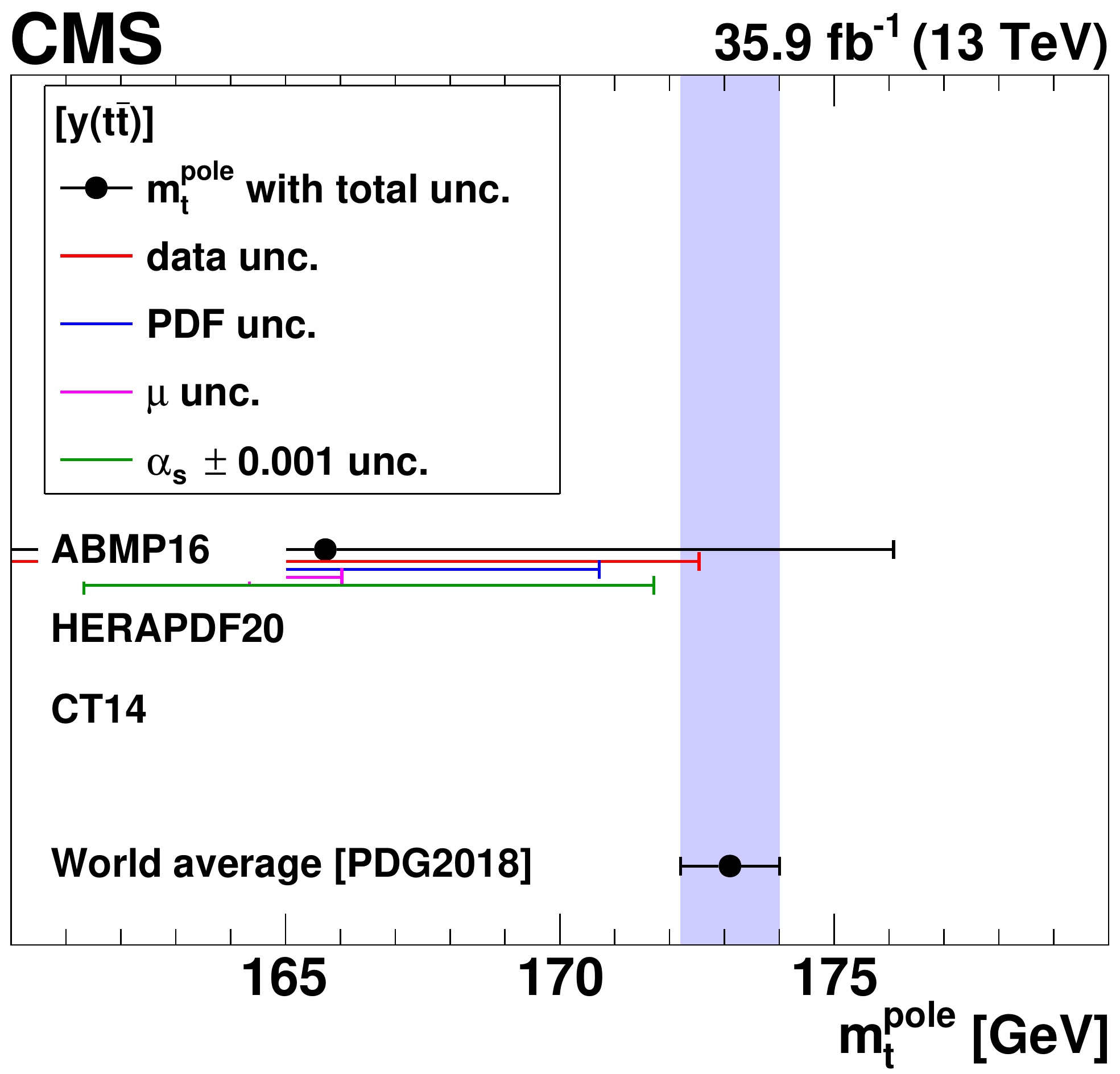}
    \caption{The \asmz (left) and \mt (right) values extracted using different single-differential cross sections,
for \nj (upper), \mtt (middle), and $\abs{\ytt}$ (lower) measurements.
For central values outside the displayed \mt range, no result is shown. Details can be found in the caption of Fig.~\ref{fig:fit-asmt-nj2mttytt}.}
    \label{fig:fit-asmt-1d}
\end{figure*}

The extracted values of \asmz and \mt obtained using single-differential cross sections as a function of \nj, \mtt, and $\abs{\ytt}$ are shown in Fig.~\ref{fig:fit-asmt-1d}.

\begin{figure*}
    \centering
    \includegraphics[width=0.47\textwidth]{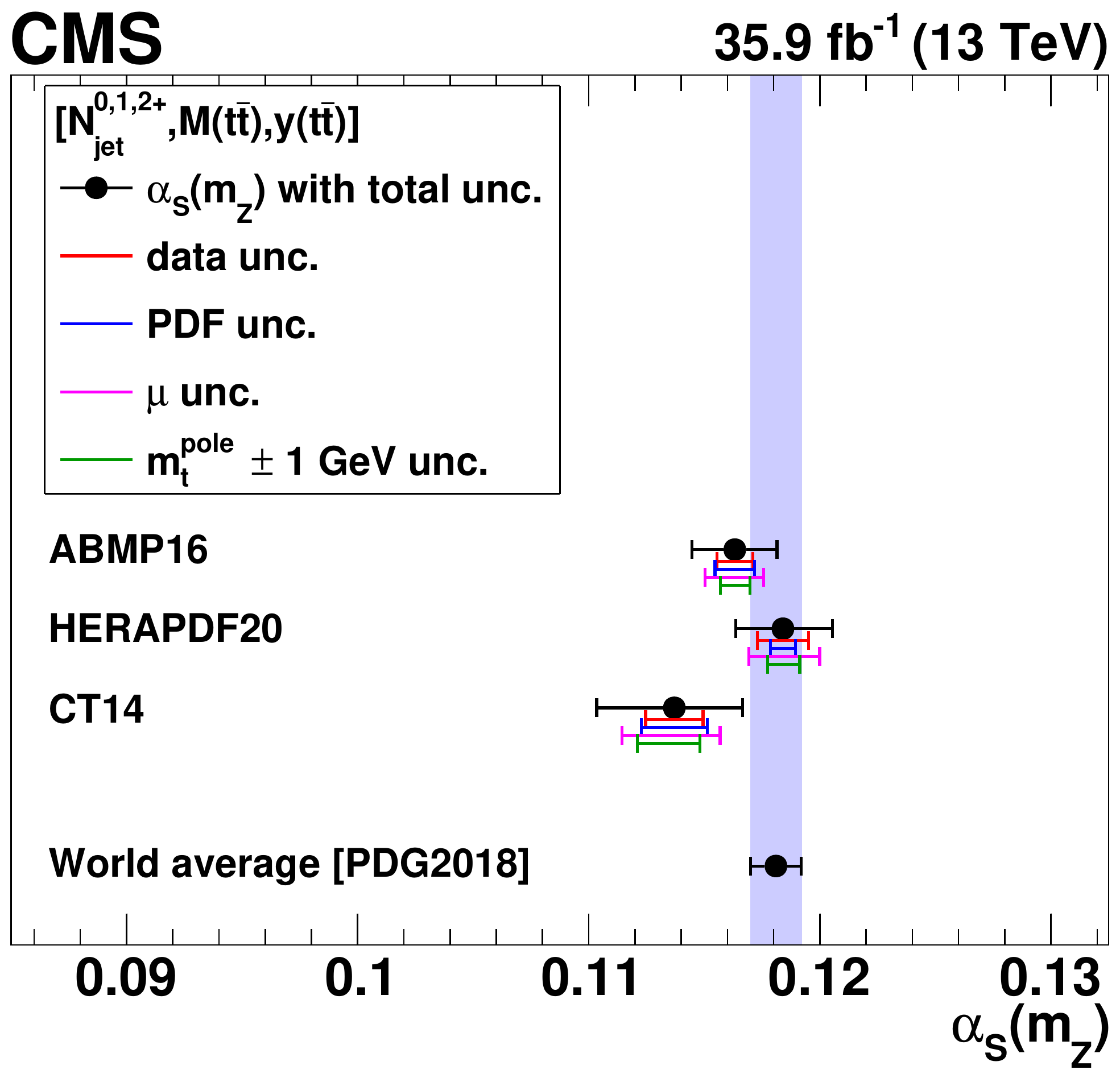}
    \includegraphics[width=0.47\textwidth]{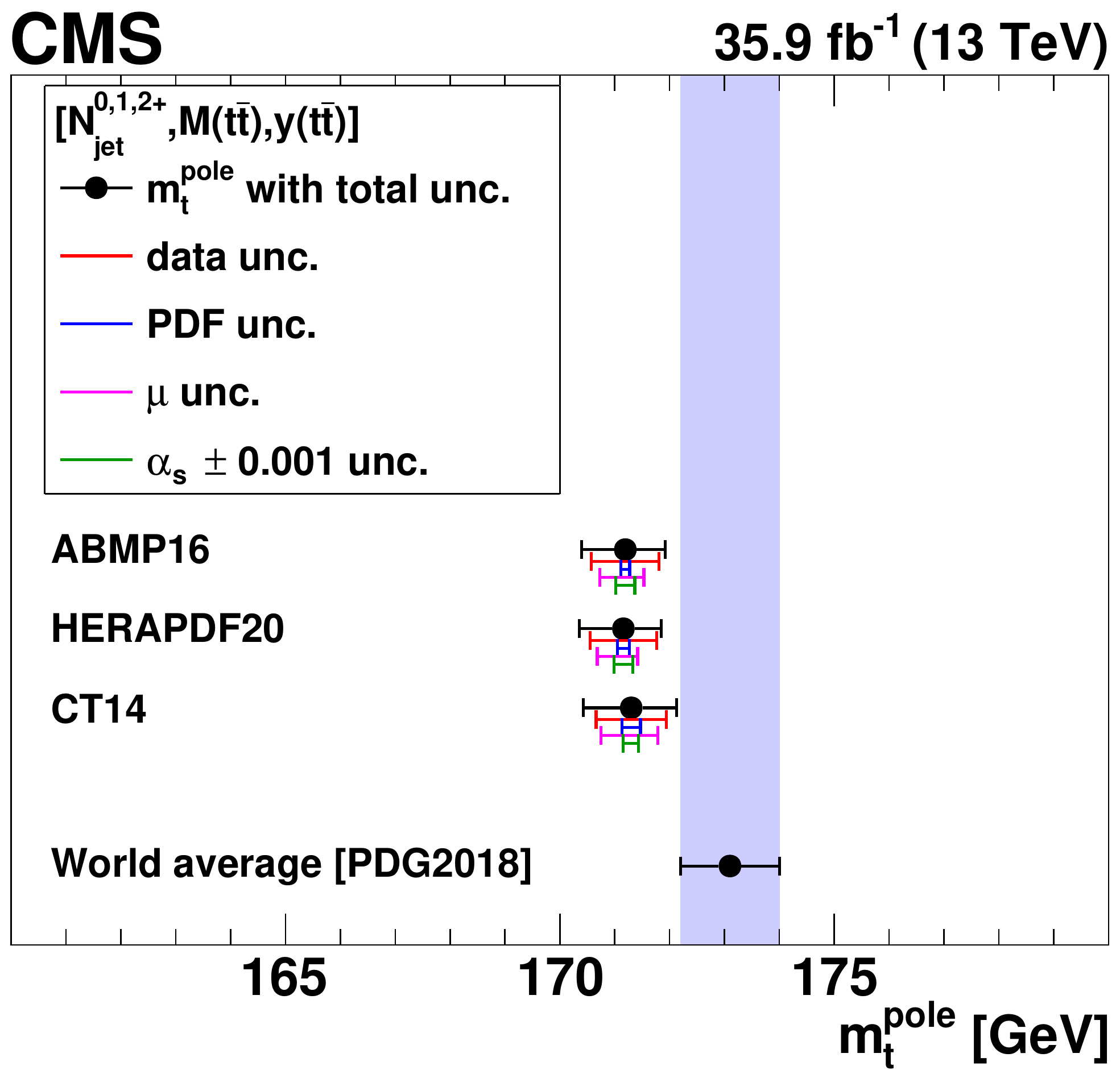}
    \caption{The \asmz (left) and $\mt$ (right) values extracted from the triple-differential \njmttyttthree cross sections. Details can be found in the caption of Fig.~\ref{fig:fit-asmt-nj2mttytt}.}
    \label{fig:fit-asmt-nj3mttytt}
\end{figure*}

{The extracted values of \asmz and \mt obtained using \njmttyttthree triple-differential cross sections are shown in Fig.~\ref{fig:fit-asmt-nj3mttytt}.}

\subsection{The \texorpdfstring{\as}{as} and \texorpdfstring{\mt}{mt} extraction using \texorpdfstring{\ptttmttytt}{[pttt,mtt,ytt]} cross sections with two \texorpdfstring{\pttt}{pttt} bins.}
\label{sec:app:ptttmttytt}

\begin{figure*}
    \centering
    \includegraphics[width=0.47\textwidth]{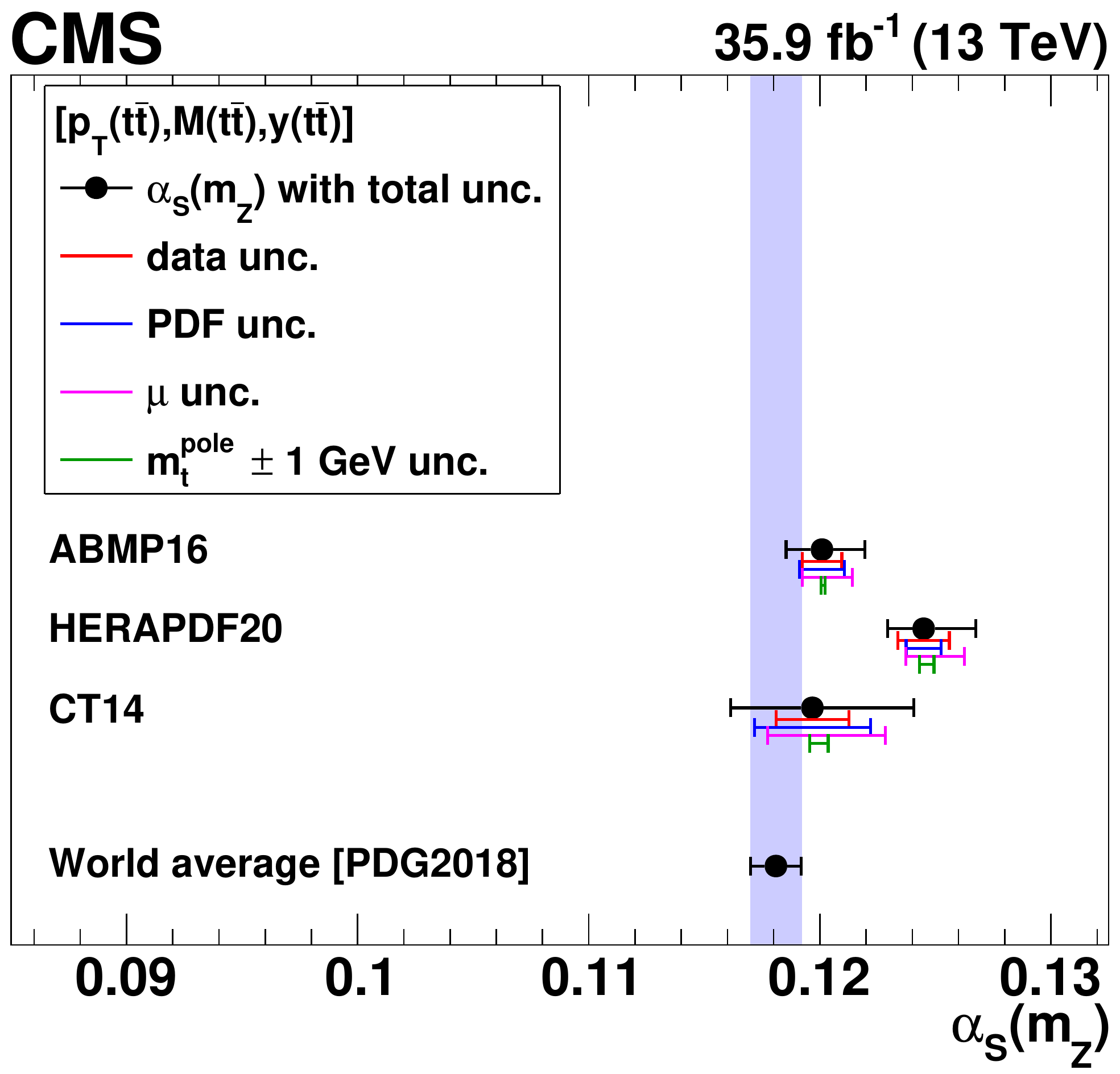}
    \includegraphics[width=0.47\textwidth]{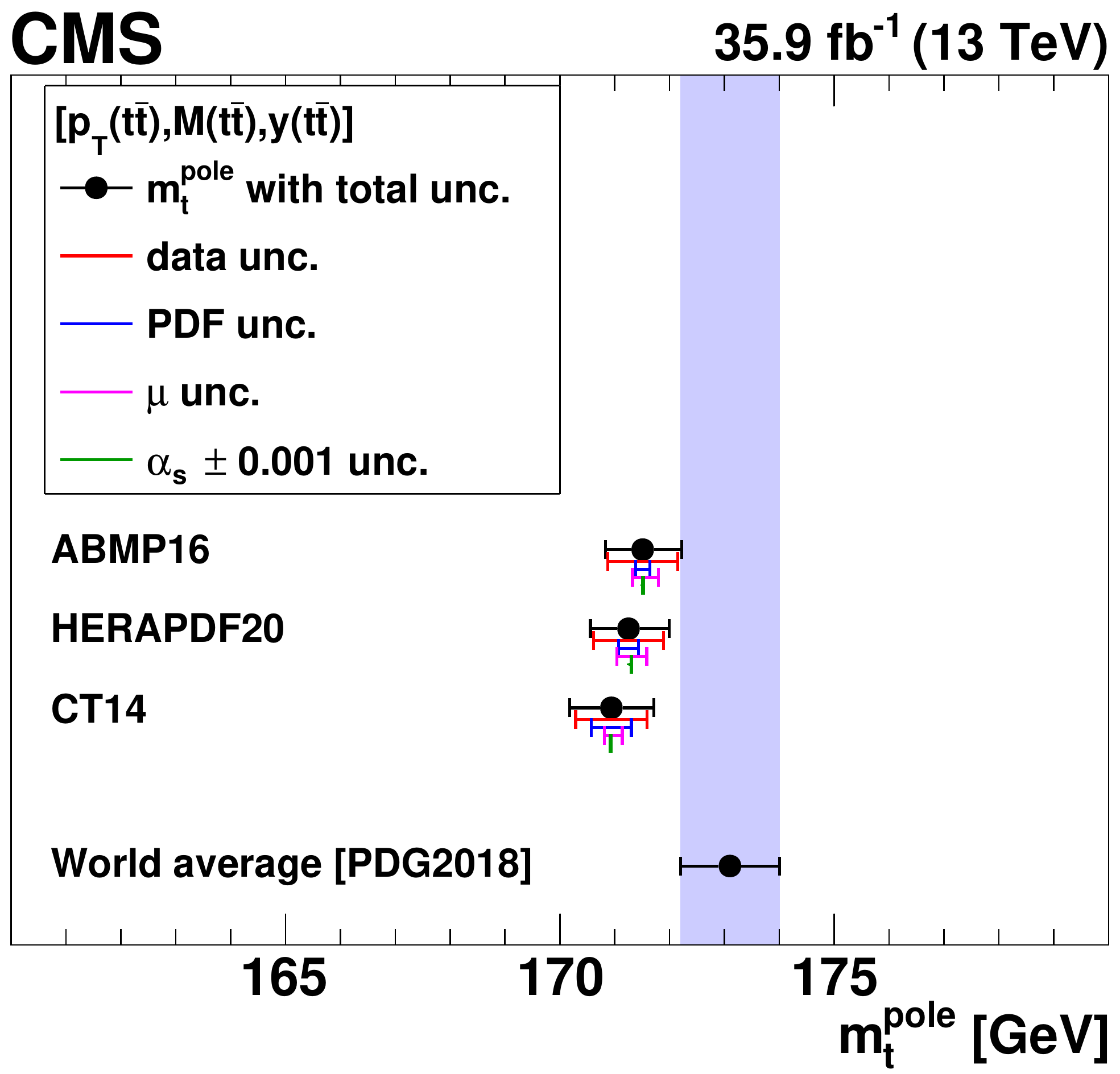}
    \caption{The \asmz (left) and $\mt$ (right) values extracted from the triple-differential \ptttmttytt cross sections. Details can be found in the caption of Fig.~\ref{fig:fit-asmt-nj2mttytt}.}
    \label{fig:fit-asmt-ptttmttytt}
\end{figure*}

The NLO calculations for inclusive \ttbar and $\ttbar+1$ jet production with an appropriate jet \pt threshold are used to describe the distribution in the two \pttt bins.
 Because the final state of the NLO calculation for \ttbar and at least one jet consists of at most two light partons, there can be up to two jets built from these partons, which balance the \ttbar transverse momentum. Therefore e.g. for $\pttt > 100\GeV$ there is at least one jet with $\pt > 50\GeV$ in the NLO calculation, and one can use
the $\ttbar+1$ jet calculation requiring $\pttt > 100\GeV$ without any requirement on the extra jet (if the jet \pt threshold is not larger than 50\GeV). In this analysis, the jet $\pt > 30\GeV$ was found to correspond approximately to $\pttt \gtrsim 50\GeV$, therefore the boundary of $50\GeV$ was chosen to split the data into two \pttt bins. The minimum jet \pt of 25\GeV was used in the NLO calculation for $\ttbar+1$ jet production (no selection on jet $\abs{\eta}$), and the predicted events were required to have $\pttt > 50\GeV$. This calculation was used for the bin with $\pttt > 50\GeV$, while for the bin with $\pttt < 50$\GeV the difference of the predictions for inclusive \ttbar production and $\ttbar+1$ jet production was used. The extracted values of \asmz and \mt are shown in Fig.~\ref{fig:fit-asmt-ptttmttytt}. They are consistent with the nominal ones (shown in Fig.~\ref{fig:fit-asmt-nj2mttytt}) but have slightly larger uncertainties.

\section{Dependence of measured cross sections on \texorpdfstring{\mtmc}{mtMC}}
\label{sec:appmtdep}

\begin{figure*}
    \centering
    {\includegraphics[width=1.00\textwidth]{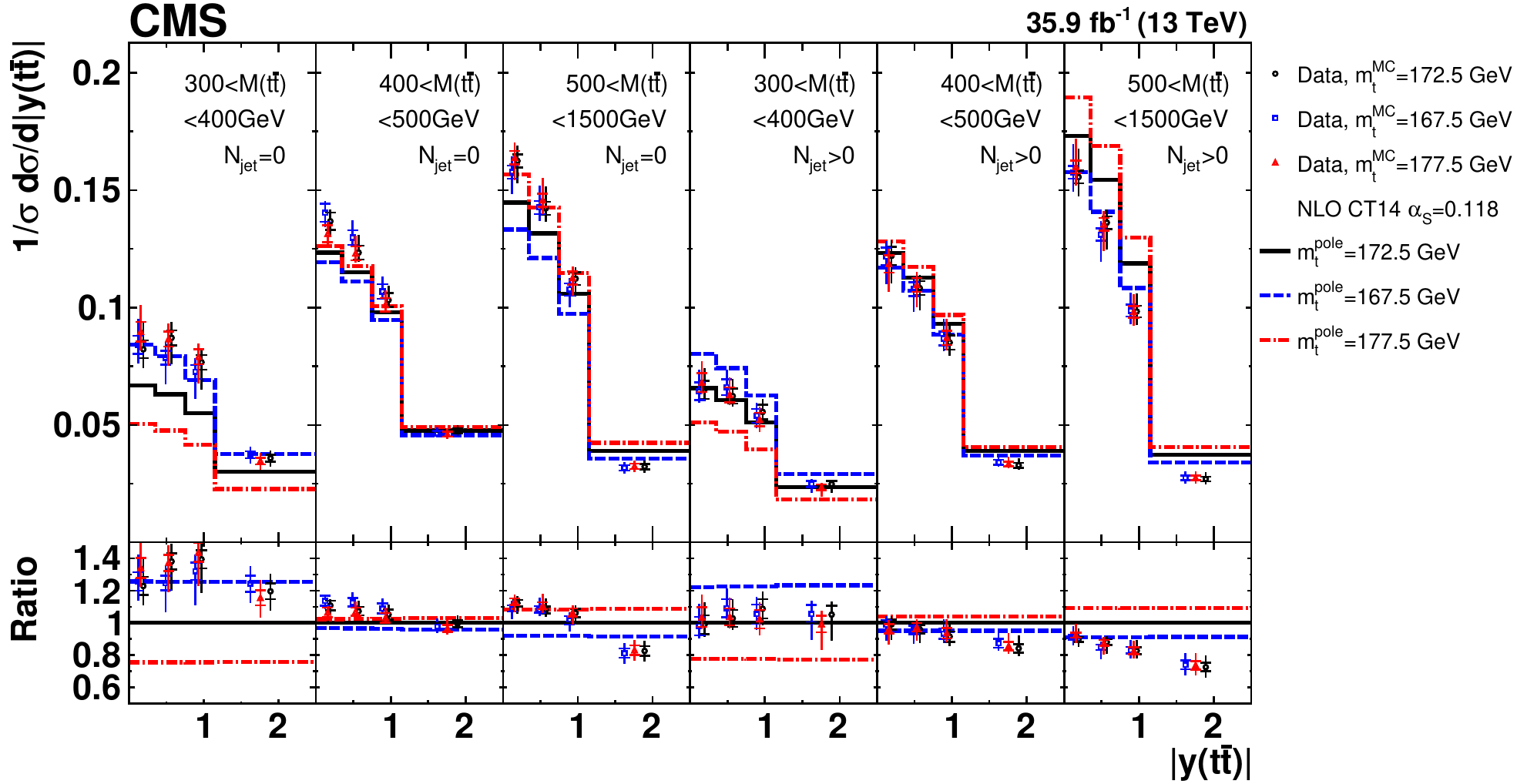}}
    \caption{Comparison of the measured \njmttytttwo cross sections obtained using different values of \mtmc to NLO predictions obtained using different \mt values
    (further details can be found in Fig.~\ref{fig:xsec-mc-ytptt}).}
    \label{fig:xsec-nlo-nj2mttytt-mt-mtdep}
\end{figure*}

{The dependence of the measured \njmttytttwo cross sections on the \mtmc value is shown in Fig.~\ref{fig:xsec-nlo-nj2mttytt-mt-mtdep}.
The cross sections are compared to the same theoretical predictions as in Fig.~\ref{fig:xsec-nlo-nj2mttytt-mt} with different values of \mt.
Compared to the sensitivity of the theoretical predictions to \mt, the dependence of the measured cross sections on \mtmc is negligible.}

\cleardoublepage \section{The CMS Collaboration \label{app:collab}}\begin{sloppypar}\hyphenpenalty=5000\widowpenalty=500\clubpenalty=5000\vskip\cmsinstskip
\textbf{Yerevan Physics Institute, Yerevan, Armenia}\\*[0pt]
A.M.~Sirunyan, A.~Tumasyan
\vskip\cmsinstskip
\textbf{Institut f\"{u}r Hochenergiephysik, Wien, Austria}\\*[0pt]
W.~Adam, F.~Ambrogi, E.~Asilar, T.~Bergauer, J.~Brandstetter, M.~Dragicevic, J.~Er\"{o}, A.~Escalante~Del~Valle, M.~Flechl, R.~Fr\"{u}hwirth\cmsAuthorMark{1}, V.M.~Ghete, J.~Hrubec, M.~Jeitler\cmsAuthorMark{1}, N.~Krammer, I.~Kr\"{a}tschmer, D.~Liko, T.~Madlener, I.~Mikulec, N.~Rad, H.~Rohringer, J.~Schieck\cmsAuthorMark{1}, R.~Sch\"{o}fbeck, M.~Spanring, D.~Spitzbart, W.~Waltenberger, J.~Wittmann, C.-E.~Wulz\cmsAuthorMark{1}, M.~Zarucki
% ...
\vskip\cmsinstskip
% ...
1:  Also at Vienna University of Technology, Vienna, Austria\\
% ...
\end{sloppypar}
\end{document}